\documentclass[12pt,a4wide]{book}

\usepackage{graphicx}
\usepackage{hyperref}
\usepackage{amsmath}
\usepackage{makecell}
\usepackage{parskip}
\usepackage{mathtools}
\usepackage{colortbl}
\usepackage{float}
\usepackage{algorithm}
\usepackage{algpseudocode}
\usepackage{enumerate}
\usepackage{adjustbox}
\usepackage{longtable}
\usepackage{enumitem}
\usepackage{booktabs}
\usepackage{amssymb}
\usepackage{pdfpages}
\usepackage{adjustbox}
\usepackage{rotating}
\usepackage{amssymb}
\usepackage{mathrsfs}
\usepackage{multirow}
\usepackage[font=small,labelfont=bf]{caption}
\usepackage[font=footnotesize,caption=false]{subfig}
\usepackage{ lscape}
\usepackage{etoolbox}
\usepackage{array, booktabs, caption}
\usepackage[font=small,labelfont=bf]{caption}
\usepackage[font=footnotesize,caption=false]{subfig}
\usepackage{makecell}

\usepackage[T1]{fontenc}
\usepackage{algpseudocode}
\usepackage[symbol]{footmisc}

\usepackage{makecell}
\usepackage{mathtools}
\usepackage[english]{babel}
\usepackage[T1]{fontenc}
\usepackage{enumerate}
\usepackage{longtable}
\usepackage{enumitem}
\usepackage{booktabs}
\usepackage{amssymb}
\usepackage{makecell}

\usepackage[T1]{fontenc}
\usepackage{color}
\usepackage{xcolor}

\usepackage{graphicx}
\usepackage{mathtools}
\usepackage{amsfonts}
\usepackage{amssymb}
\usepackage{graphics}
\usepackage{color}
\usepackage{pgfplots}
\usepackage{hyperref}
\usepackage{listings}

\usepackage{booktabs}
\usepackage{pgfplots}
\usepackage{amssymb}
\usepackage{mathrsfs}
\usepackage{amsmath}
\usepackage{hyperref}
\usepackage{longtable}
\usepackage{multirow}
\usetikzlibrary{arrows,automata}

\usepackage{float}
\usepackage[caption = false]{subfig}

\usepackage{listings}

\usepackage{url}
\usepackage{multirow}
\usepackage{multicol}
\usepackage{amssymb}
\usepackage{mathrsfs}
\usepackage{enumerate}
\usepackage{enumitem}
\newsavebox{\measurebox}

\newcommand{\floor}[1]{\left \lfloor #1 \right \rfloor}

\usepackage{cite}
\usepackage{epsf} 
\usepackage{latexsym} 
\usepackage{graphicx}
\usepackage{amssymb}
\usepackage{amsmath,amssymb}
\usepackage{amsfonts}
\usepackage{fancyhdr}
\usepackage{enumitem}
\usepackage{chngcntr}
\usepackage{floatpag}
\usepackage{latexsym}
\usepackage{tabularx, booktabs} 
\usepackage[title,titletoc,toc]{appendix}
\usepackage{subfig}
\usepackage{amsmath}
\usepackage{mathrsfs}
\usepackage{afterpage}
\newcommand\blankpage{%
    \null
    \thispagestyle{empty}%
    \addtocounter{page}{-1}%
    \newpage}
    
\makeatletter
\newcommand\fs@norules{\def\@fs@cfont{\bfseries}\let\@fs@capt\floatc@ruled
  \def\@fs@pre{}%
  \def\@fs@post{}%
  \def\@fs@mid{\kern3pt}%
  \let\@fs@iftopcapt\iftrue}      
 
\usepackage{longtable}  
  
\usepackage{pgfplots}  
\usepackage{multirow}
  
\usepackage{colortbl}

\definecolor{webgreen}{rgb}{0, 0.5, 0} 
\definecolor{webblue}{rgb}{0, 0, 0.5} 
\definecolor{webred}{rgb}{0.0, 0, 0} 
\definecolor{webred}{rgb}{0.5, 0, 0} 
\definecolor{webblack}{rgb}{0, 0, 0} 

\usepackage{booktabs}
\usepackage{makecell}

\usepackage{longtable}

\fancyhfoffset[L]{0.5cm}

\makeatother

\usepackage[toc,nonumberlist]{glossaries}
\makeglossaries
\glsdisablehyper

\newcolumntype{Y}{>{\centering\arraybackslash}X}
\newtheorem{theoremm}{Theorem}[chapter]
\newtheorem{eqed}{Example}[chapter]
\newtheorem {lemmaa}{Lemma}[chapter]

\newtheorem{defnn}{Definition}[chapter]
\newtheorem {corollaryy}{Corollary}[chapter]
\newtheorem {axiomm}{Axiom}[chapter]
\newtheorem {inferencee}{Inference}[chapter]

\newtheorem {hypothesiss}{Hypothesis}[chapter]
\newtheorem {conjecturee}{Conjecture}[chapter]

\newtheorem {prp}{Property}[chapter]
\newenvironment{example}{\begin{eqed} \rm}{\hfill$\Box$ \end{eqed}}

\newenvironment{infprf}{\noindent {\bf Informal Proof :\ } }{$\Box$ }

\newenvironment{Informal Proof}{\begin{prp} \sl}{\end{prp}}

\newenvironment{proof of correctness}{\noindent {\bf Proof of Correctness :\ } }{\hfill$\Box$ }

\newenvironment{sketch of proof}{\noindent {\bf Sketch of proof :\ } }{\hfill$\Box$ }

\newtheorem {definition}{Definition}

\newtheorem {procd}{Procedure}

\newcommand{\eat}[1]{}

\definecolor{mygreen}{rgb}{0,0.6,0}
\definecolor{mygray}{rgb}{0.5,0.5,0.5}
\definecolor{mymauve}{rgb}{0.58,0,0.82}
\begin{document}
\sloppy

\pagestyle{empty}
\pagestyle{empty} 
	\begin{center} 
	\huge{\textbf{\textsf{{Layered Cellular Automata}}}} \\ 
	\end{center}
	\begin{center} 
	\vspace{4.0cm}
	 \textbf{\normalsize{ABHISHEK DALAI}} \\
	\vspace{4.8cm}
        \begin{center}
 		 \includegraphics[height=4cm, width=4cm]{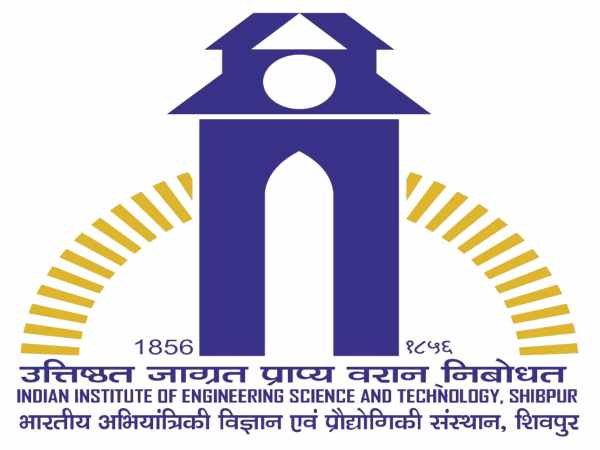}
        \end{center} 
        \vspace{1.2cm}     
      \scriptsize{\textbf{DEPARTMENT OF INFORMATION TECHNOLOGY}}\\ 
      \scriptsize\textbf{{INDIAN INSTITUTE OF ENGINEERING SCIENCE AND TECHNOLOGY, SHIBPUR}} \\
     \scriptsize{\textbf{HOWRAH, WEST BENGAL, INDIA-711103}}\\
               \vspace{0.5cm} 
   \large{\textbf{2023}} \\  
  	\end{center} 

\newpage


\pagestyle{empty} 
	\begin{center} 
	\Huge{\textbf{\textsf{{Layered Cellular Automata}}}} \\
        \vspace{1.5cm} 
        \Large{\textbf{Abhishek Dalai}}\\   
        	 \vspace{0.3cm} 
     \large{{Registration No. 2021ITM007}} \\
	     \vspace{1.2cm}
        \small{\em{A report submitted in partial fulfillment for the degree of}}\\     
        \vspace{0.12in} 
        \small{\textbf{Masters of Technology}}\\ 
        \small{\textbf{in}}\\ 
        \small{\textbf{Information Technology}}\\ 
        
     \vspace{1.0cm} 
	\large{\em{Under the supervision of}}\\ 
	\vspace{0.10cm} 
	\large{\textbf{Dr. Sukanta Das}}\\ 
	\small{Associate Professor}\\
	\small{Department of Information Technology}\\
	{Indian Institute of Engineering Science and Technology, Shibpur}\\       
        
        \vspace{0.45in} 
        \begin{center}
		\includegraphics[height=4cm, width=4cm]{logo}
        \end{center}
		\vspace{0.3cm}

	\normalsize{\textbf{Department of Information Technology}}\\ 
\normalsize{\textbf{Indian Institute of Engineering Science and Technology, Shibpur}}\\
	\normalsize{\textbf{Howrah, West Bengal, India -- 711103}}\\ 
    \normalsize{\textbf{2023}}\\ 
	\end{center} 

\newpage

\pagestyle{empty} 
\begin{center} 
\begin{figure}[h]
\vspace{-1.5cm} 
\centering
	\includegraphics[height=2.5cm, width=2.5cm]{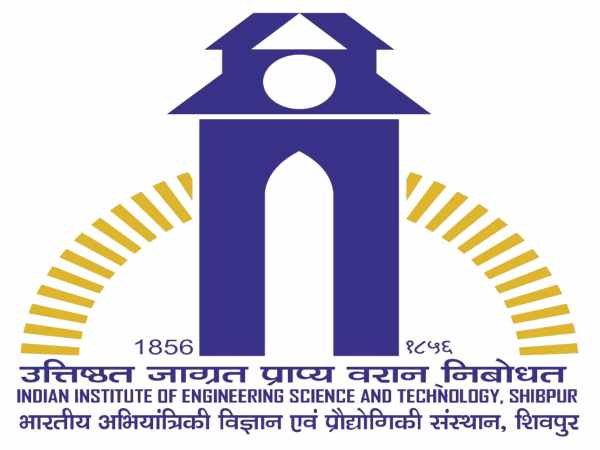}
\end{figure}
\small{\textbf{Department of Information Technology}}\\ 
	\small{\textbf{Indian Institute of Engineering Science and Technology, Shibpur}}\\
	\small{\textbf{Howrah, West Bengal, India -- 711103}}\\  
\vspace{0.40in} 

{\Large \bf CERTIFICATE OF APPROVAL}\\ 
\end{center} 
\vspace{0.22in} 
\normalsize{It is certified that, the thesis entitled \emph{\bf ``Layered Cellular Automata''}, is a record of bonafide work carried out under my guidance and supervision by 
\mbox{\textbf{Abhishek Dalai}} in the Department of Information Technology of Indian Institute of Engineering Science and Technology, Shibpur. 
 
In my opinion, the thesis has fulfilled the requirements for the degree of M.Tech in Information Technology of Indian Institute of Engineering Science and Technology, Shibpur. The work has reached the standard necessary for submission and, to the best of my knowledge, the results embodied in this thesis have not been submitted for the award of any other degree or diploma.

\vspace{1em} 
\begin{flushright}
	\includegraphics[height=4\baselineskip]{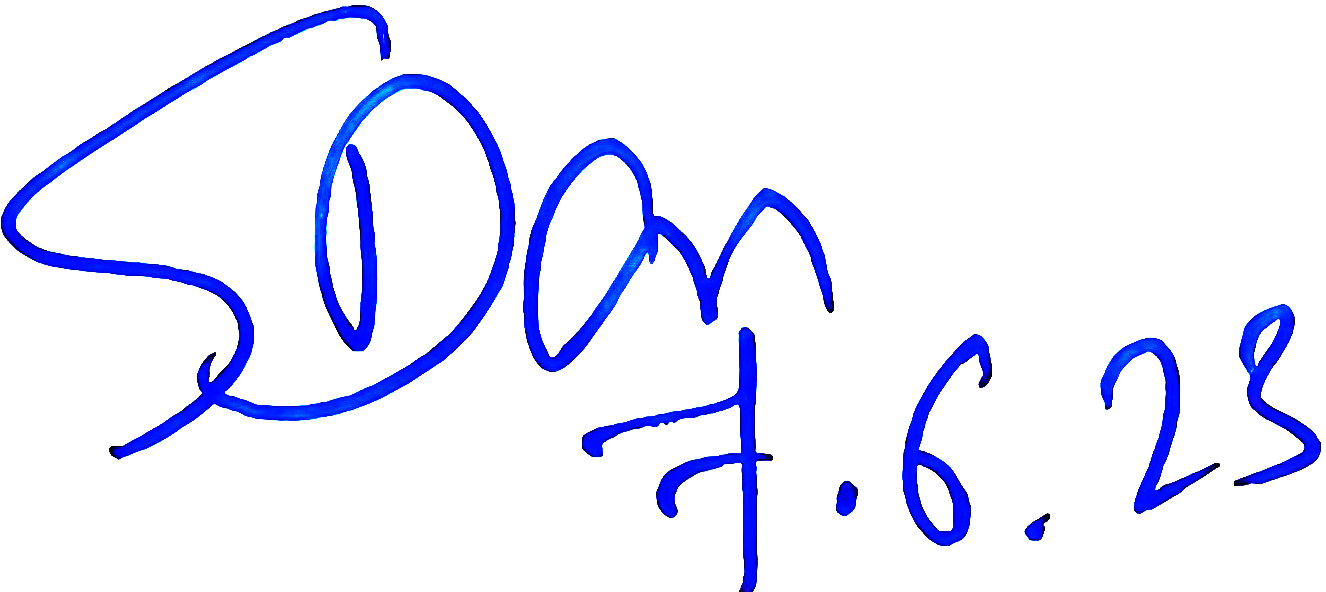} \par
			\textbf{(Dr. Sukanta Das)}  \\
		\small{Associate~ Professor} \\
	\small{Dept. of Information Technology} \\
		\small{Indian Institute of Engineering Science and Technology, Shibpur} \\
		\small{Howrah, West Bengal, India --711103} \\
\end{flushright}
\vspace{0.2in}
\hspace{-0.2in}{Counter signed by:}
\begin{flushright}
	\includegraphics[height=4\baselineskip]{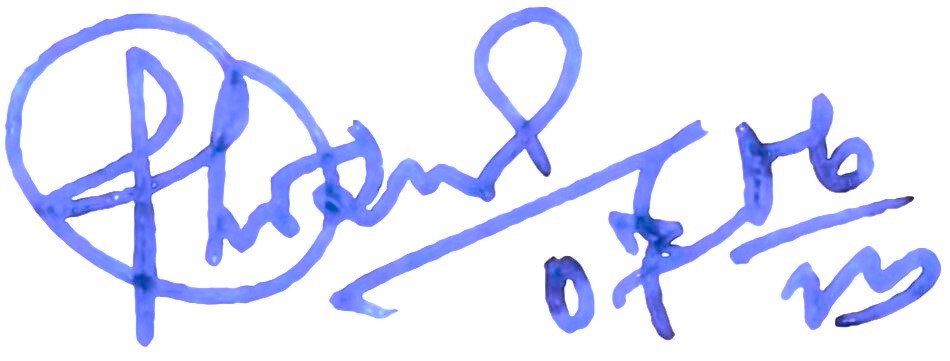} \par
	\textbf{(Dr. Prasun Ghosal)}  \\
	\small{Associate~ Professor \& Head} \\
	\small{Dept. of Information Technology} \\
	\small{Indian Institute of Engineering Science and Technology, Shibpur} \\
	\small{Howrah, West Bengal, India --711103} \\
\end{flushright}

	


\vspace{1mm}
} 
\newpage 

\newpage

\vspace*{\fill}  

\begin{quote}

\centering

\begin{large}


\textbf{\Large{\textit{Dedicated}}}\\

\vspace{0.25cm}

\textbf{\Large{\textit{to}}}\\

\vspace{0.25cm}
\textbf{\Large{\textit{My Parents}}}\\
\vspace{0.25cm}
\textbf{\Large{\textit{and}}}\\
\vspace{0.25cm}
\textbf{\Large{\textit{to all the struggling people over the world}}}\\
\vspace{0.25cm}

\vspace{0.25cm}


\vspace{0.25cm}


\end{large}

\end{quote}

\vspace*{\fill} 

\clearpage

\pagestyle{plain}
\pagenumbering{roman}
\afterpage{\blankpage}

\cleardoublepage
\phantomsection
\addcontentsline{toc}{chapter}{Acknowledgement}
\begin{center}
\vspace{-5.5cm}
 \textbf{\Large Acknowledgement}
\end{center}

I would like to express my sincere gratitude to my advisor Dr. Sukanta Das, Associate Professor in the Department of Information Technology at the Indian Institute of Engineering Science and Technology (IIEST), Shibpur. I am truly grateful for his unwavering support and assistance throughout the entire process of preparing this dissertation. I am also thankful for his patience, motivation, enthusiasm, and extensive knowledge in the field. His guidance has been invaluable in every stage of writing this thesis. I am truly fortunate to have had such an exceptional advisor and mentor. Throughout this journey, I have gained a wealth of knowledge from him, particularly in terms of research discipline and personal growth.

I would also like to acknowledge and extend my utmost respect and gratitude to Subrata Paul, Ph.D. scholar in the Department of Information Technology at the Indian Institute of Engineering Science and Technology (IIEST), Shibpur. I am truly thankful for his invaluable suggestions and advice, which have greatly contributed to my ability to approach research with a more analytical and rigorous mindset. His insights and guidance have been instrumental in applying scientific methodology effectively. 

All the reported works were accomplished through joint efforts. In the research ``Layered Cellular Automata and Pattern Classification'', where I worked with Dr. Sukanta Das and Subrata Paul to develop the theory of layered cellular automata. I've worked with them to investigate the dynamics and use of it in pattern classification. 

I am grateful for the financial support the Indian Institute of Engineering Science and Technology, Shibpur provided for my research during the tenure of my M.Tech.

I am grateful to the current head of department, Dr. Prasun Ghosal, of the Department of Information Technology at IIEST, Shibpur, as well as all the other respected professors for being so kind as to provide their support at various phases. In addition to my advisor, I would like to express my deepest appreciation to each and every member of the M.Tech committee for their insights and technical suggestions. I want to express my gratitude to the department's technical and non-technical employees (Malay-sir, Suman-da, and Dinu-da) for their support and dedication. 

I would like to extend my heartfelt gratitude to all of my friends for their unwavering support and encouragement throughout this journey. Among them, I would like to give special thanks to my dear friend Teena Tessa Mathew. Her constant support and encouragement have been invaluable, particularly during challenging times. Her presence by my side, motivating and inspiring me, has been instrumental in helping me overcome obstacles and stay focused. I am truly grateful for her unwavering friendship and support. I also thank my labmates Ganesh, Dipanjan, Spandan, Rishab, Ranit, Arghaya and Shankhadip for their support throughout the last two years. Finally, I would like to express my heartfelt gratitude to Anjali and Sandeep for being there for me when I needed it the most. During difficult times, their presence has provided comfort and reassurance, reminding me that I am not alone in facing life's challenges. Their friendship has been a constant source of joy and positivity, lifting my spirits and reminding me of the importance of having someone to rely on.

I would like to extend my utmost gratitude and deep respect to my parents, Mr. Akshaya Kumar Dalai and Mrs. Khulana Dalai and my sister Ms. Adarshi Dalai, for their unwavering support, sacrifices, and continuous inspiration throughout my academic journey. From the very beginning, they have been my pillars of strength, providing me with endless love, guidance, and encouragement. Thank you, Maa Bapa, for being my constant source of inspiration and for shaping me into the person I am today. I am eternally grateful for your love, guidance, and unwavering belief in me.

\vspace{35mm}

\noindent{\bf Dated:$~$}  
 \includegraphics[height=1.8\baselineskip]{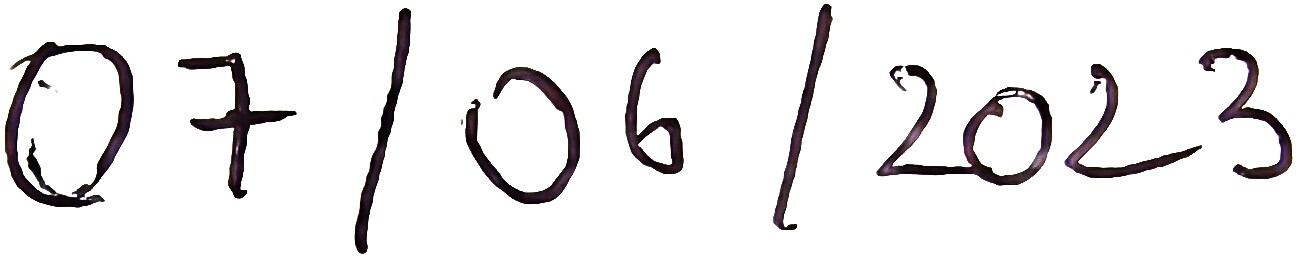}\hspace{1.3 in} \includegraphics[height=3.5\baselineskip]{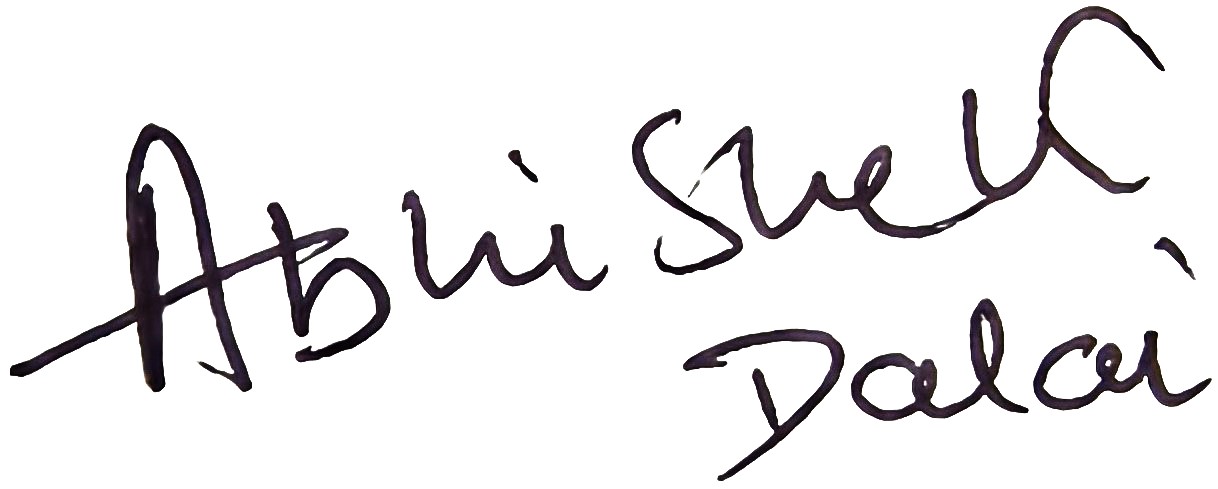} 
\\
{\bf Indian Institute of Engineering Science} \hspace*{0.4in} \dotfill \\
{\bf and Technology, Shibpur} \hspace*{2.0in} {\bf(Abhishek Dalai)} \\
{\bf Howrah, West Bengal, India} \hspace*{3.3cm} {\textbf{Reg. No.: 2021ITM007}}


\newpage
\cleardoublepage
\phantomsection
\addcontentsline{toc}{chapter}{Abstract}
\afterpage{\blankpage}
\chapter*{Abstract}
\label{abstr}

\textbf{C}ellular automata (CAs) has emerged as powerful computational models for studying dynamic systems across various scientific domains. Traditional CA models, however, have limitations in capturing the complexity and dynamics of real-world phenomena. To address this, a novel approach called Layered Cellular Automata (LCAs) has been introduced, incorporating an additional layer of computation to enhance the modeling capabilities.

\textbf{I}n this thesis, we delve into the concept of LCAs and explore its potential in capturing intricate behaviors and emergent properties. We begin by providing an overview of cellular automata, discussing their applications and limitations. We then introduce the notion of layering in CA and outline various LCA models, such as averaging, maximization, minimization, modified ECA neighborhood and LCA based on game of life.

\textbf{T}he dynamics of different LCA models are analyzed and their behaviors classified. We identify subsets of LCAs that are influenced by interlayer rules, showcasing variations in their dynamics compared to the parent CA. Additionally, we discover LCAs that are sensitive to changes in block size, leading to phenomena like phase transition and class transition.

\textbf{F}urthermore, we investigate the applicability of convergent LCAs for pattern recognition tasks. Through extensive experiments, we identify specific LCAs that exhibit convergence to fixed points from any initial configuration, which can be utilized in the design of two-class pattern classifiers. Our proposed LCA-based pattern classifier demonstrates competitive performance when compared to existing algorithms.

\textbf{L}ayered Cellular Automata offers a promising framework for modeling and understanding complex systems. This research opens up avenues for further exploration into the dynamics, emergent behavior, and practical applications of LCA. By overcoming the limitations of traditional CA models, LCA provides researchers with a versatile tool for studying and simulating intricate phenomena in diverse scientific domains.


\cleardoublepage
\tableofcontents

\cleardoublepage
\phantomsection
\addcontentsline{toc}{chapter}{\listfigurename}
\listoffigures

\cleardoublepage
\phantomsection
\addcontentsline{toc}{chapter}{\listtablename}
\listoftables


%
\clearpage

\pagestyle{fancy}
\pagenumbering{arabic} 
\chapter{Introduction}
\label{chap1}

Throughout history, the marvels of nature have always intrigued the advancement of science. Nature's operations are highly erratic, with every living organism contributing a unique role in influencing the collective behavior. Nonetheless, the fundamental mathematical model employed in the early days of modern computers, and even in von Neumann's computer design, was based on the Turing Machine~\cite{turing1937computable, Turing}. This machine's computation was governed by a central control tape head, while the CPU also acted as a centralized control mechanism.

Starting from the first computer to the current generation of smartphones, all computing systems have functioned in a centralized manner. Although we can detect patterns in nature, such as the shapes of snowflakes, the movement of ants, and the structure of seashells, which seem to suggest the emergence of centralized control, the reality is different. For instance, in a colony of ants, a leader may appear to be in control, but in reality, each ant independently makes its own decisions and carries out its assigned tasks. 

In the early 1900s, a novel field of research called Network Science emerged to explore individuality and parallelism in computing. Over the years, numerous models were introduced, many of which took inspiration from biological systems and enabled distributed and decentralized computing. One of the most significant breakthroughs in this field was the development of cellular automata. Decentralization has been a prevalent concept in computing since the advent of the first widely-used distributed systems like Ethernet~\cite{Clark21, Maarten7}. With the rise of the distributed system, the internet caused a paradigm shift, and the idea of decentralization has since gained widespread recognition across many domains of human activity.

An array of networked, yet independent, computer components make up a distributed system. These components only communicate with one another to coordinate their functions. From the standpoint of a process, a distributed system may be seen as a collection of geographically scattered processes that only communicate via message exchange. As a result, the processes in the system can only speak to one another while doing a computational task. In a distributed computing architecture, the supervision and control of the computation are not exercised by a single entity. The components and processes of a distributed computation may be recognized by their unique identifiers. A central organization is required for a system with detectable individual identities, or a ``non-anonymous system'', in order to give the processes their distinctive individuality. The fundamental tenets of distributed control are violated by this. As a result, a distributed system must be anonymous by definition~\cite{Maarten7, tel73}. A number of formal models have already been published for distributed systems~\cite{consys1,  Milner, Reisig}, providing useful insights. Conversely, because of their innate parallelism, cellular automata (CAs) can always be a natural choice for distributed computing frameworks.

A distributed system is composed of an array of independent computer components that are networked and communicate only for the purpose of coordinating their functions. To a process, a distributed system may seem like a group of geographically dispersed processes that can only communicate through message exchange when performing a computational task. In a distributed computing architecture, there is no single entity that supervises or controls the computation. The components and processes in a distributed computation are identified by their unique identifiers. A central organization is required for non-anonymous systems where individual identities can be detected, but this violates the basic principles of distributed control. Therefore, by definition, a distributed system must be anonymous~\cite{Maarten7, tel73}. Several formal models have been published for distributed systems~\cite{consys1,  Milner, Reisig} that provide valuable insights. However, due to their inherent parallelism, cellular automata (CAs) are always a natural choice for distributed computing frameworks.


Jon von Neumann introduced the concept of self-reproducing automata~\cite{Neuma66} in the early 1950s, which later came to be known as ``Cellular Automata''. He introduced constructive universality in cellular automata to study the feasibility of self-reproducing machines and the concept of computational universality. A computing machine is considered computationally universal if it can simulate any other computing machine. Von Neumann's universal constructor was capable of emulating other machines that could be embedded in its cellular automaton. Although computational universality and constructive universality are conceptually related to each other, a machine does not necessarily require a universal computer to be a universal constructor. Von Neumann demonstrated that a Turing machine could be implemented in his cellular automaton, although he highlighted that a Turing machine is not a necessary component of the universal constructor.

Artificial Life, introduced by Christopher Langton, has become a focal point for researchers in various fields, including science, philosophy, and technology, who are interested in studying biological phenomena. Artificial Life uses cellular automata (CA) as a basis for the artificial life model. Some CAs have inherited characteristics of biological systems, such as self-replication, self-organization, and self-healing. The Game of Life~\cite{Gardner70} by Conway is a significant example of such a CA depicting the behaviors of biological systems. 

The answer to the question ``How intelligent has a machine become?'' depends on how we define and measure intelligence in machines. Alan Turing developed the Turing Test~\cite{TT1950} to analyze the machine's intelligence based on the fact that how the system performs to a set of questions and passing them infer that the machine is intelligent. If we take a functional approach, where intelligence is evaluated based on a machine's ability to perform tasks, then we can say that machines have become increasingly intelligent as they are now capable of performing complex tasks that were previously thought to be exclusive to humans. For example, machines can now beat human champions in complex games like chess and Go, perform complex calculations and analysis, recognize and classify objects in images, and even generate creative works like music and art.
However, if we take a more holistic approach and define intelligence as a set of cognitive and behavioral traits that are characteristic of living systems, then the question becomes more complex. While machines have certainly become better at performing specific tasks, they still lack many of the complex cognitive and behavioral abilities that are associated with intelligence in living systems, such as consciousness, self-awareness, emotions, and creativity. Therefore, it can be argued that machines have not yet achieved true intelligence in the sense that they do not possess the same level of complexity, flexibility, and adaptability as living systems.
In summary, the answer to the question ``How intelligent has a machine become?'' depends on how we define and measure intelligence in machines. While machines have certainly made significant progress in performing specific tasks, they still lack many of the cognitive and behavioral traits that are associated with intelligence in living systems. Therefore, the question of whether machines can truly be considered intelligent remains a subject of ongoing debate and research.

\section{Motivation and Objective of the thesis}

The main objective of this thesis is to explore the computational capabilities of decentralized models of distributed computing, specifically focusing on distributed computing on cellular automata. In this context, each cell in a cellular automaton (CA) consists of a finite automaton that interacts with its neighboring cells to determine its next state~\cite{Neuma66}. The CA is distributed across a regular grid, and its appeal lies in the fact that complex global behavior can emerge from simple local interactions.
Cellular automata have been proposed as a potential mechanism for quantum information processing~\cite{PhysRevLett.88.237901}, with some researchers suggesting that nature itself operates as a quantum information processing system, utilizing cellular automata for its computational functions~\cite{Zuse1982,wolfram2002new}. Moreover, cellular automata have been employed as models for concurrency and distributed systems, and they have been used to computationally address various issues related to distributed systems~\cite{Cor,Smith71}. In this work, our aim is to study a new kind of cellular automata model named \textbf{Layered Cellular Automata} (LCA).
We explore the utilization of layered cellular automata (LCA) to address the following challenges:
\begin{itemize}
	\item Examine the behavior of a system under the influence of noise using layered cellular automata.
	\item Layered cellular automata can effectively classify patterns within a system.
\end{itemize}
Layered cellular automata (LCA) is a computational model that extends the concept of cellular automata (CA) by introducing additional layers of cells that influence cells in the lower layer. This advanced framework allows for more complex and dynamic simulations, enabling the study of intricate systems and phenomena.

Traditional cellular automata (CA) have been widely used for modeling and simulating complex systems across various scientific domains. However, they exhibit limitations in capturing the full complexity and dynamics of real-world systems. These limitations arise from two main factors: restricted local interactions and single rule.

Firstly, traditional CA models typically rely on local interactions, where the state of each cell is updated based only on the states of its immediate neighboring cells. This limited scope of interactions can fail to capture long-range dependencies and global patterns that are prevalent in many real-world systems. For example, in social networks or ecological systems, the behavior of an individual or a species may be influenced by individuals or species located far away. Traditional CA models struggle to incorporate such distant interactions, resulting in an incomplete representation of the system's dynamics.

Secondly, traditional CA models utilize a single rule that governs the evolution of cell states. While these rules can exhibit interesting and complex behaviors, they lack the flexibility to adapt to different problem domains or capture diverse patterns and dynamics. This restricts their applicability in various scientific domains and pattern recognition tasks.

To overcome these limitations, layered cellular automata (LCA) have been proposed as an advanced computational framework. LCA introduces an additional layer of computation, allowing for more complex simulations and capturing a broader range of system dynamics. By dividing the grid into blocks and introducing two separate rules for each layer, LCA models can incorporate interdependencies and interactions between different aspects of the system.

Moreover, LCA enables the modeling of systems with long-range interactions by allowing cells to receive influences from distant neighbors. This global influence allows for the capture of emergent behaviors and global patterns that are essential for understanding real-world systems. Additionally, the flexibility of LCA in defining different rules for each layer enhances their adaptability and versatility, enabling researchers to modify the behavior of each layer to specific problem domains or desired patterns.

By addressing the limitations of traditional CA, layered cellular automata provide a more powerful and flexible framework for simulating and studying complex systems. They extend the capabilities of CA models, enabling them to better capture the complexity and dynamics of real-world systems, and offering improved applicability in various scientific domains.

\section{Contribution of the thesis}
The study was conducted with the aim of achieving the aforementioned objective. The research activities yielded significant findings, which can be summarized as follows:
\begin{itemize}

\item In our research, we focused on exploring the behaviors of layered cellular automata, specifically examining the dynamic interactions between elementary cellular automata (ECAs) in the lower layer and the proposed rules in the upper layer. The key objective of our investigation was to understand how the dynamics of the lower layer can be influenced and modified by the presence of the upper layer.
By incorporating the upper layer with its own set of rules, we introduced an additional level of complexity and interaction within the layered cellular automata system. This allowed us to observe how the behavior of the lower layer, which is governed by ECAs, can be influenced and shaped by the dynamics of the upper layer.
We examined various configurations and combinations of ECAs and rules in the layered cellular automata. Our aim was to understand the effects of these interactions on the overall behavior and emergent properties of the system.
Next, we focused on exploring the behaviors of layered cellular automata in 2D model, specifically examining the dynamic interactions between Game of Life in the lower layer and the proposed rules in the upper layer.
 
\item Based on our research findings, it has been observed that certain layered cellular automata remain resilient to the influence of noise, exhibiting consistent behavior. On the other hand, some layered cellular automata undergo phase transition and class transition when exposed to noise, leading to significant changes in their dynamics. This highlights the varying sensitivity of different cellular automata models to the impact of noise. 

\item After a thorough examination of the dynamics, we have identified convergent layered cellular automata (LCAs) suitable for developing a two-class pattern classifier.
The proposed design of the LCA-based classifier has shown promising results in terms of its performance. When compared to existing approaches commonly used in pattern classification, the LCA-based classifier performs competitively. This suggests that the utilization of LCAs provides a viable alternative for pattern classification tasks.
By leveraging the inherent properties of LCAs, such as their ability to handle noise and exhibit dynamic behaviors, the proposed classifier can effectively distinguish between two different classes of patterns. This indicates the potential of LCAs as a powerful tool in pattern recognition and classification tasks.

\end{itemize}

\section{Organization of the thesis}
In this section, we present the structure of the thesis and provide a brief overview of each chapter. The thesis contributes to the understanding of the topic by offering a comprehensive exploration of layered cellular automata.

\begin{itemize}
	\item \textbf{Chapter~\ref{chap2}.} This serves as a survey of cellular automata (CAs). It provides a foundational understanding of CAs, their basic principles, and their applications in various fields. This chapter lays the groundwork for the subsequent chapters by familiarizing the reader with the fundamental concepts of CAs.	
	\item \textbf{Chapter~\ref{chap3}.} In this chapter, we explore a unique variant of cellular automata (CA) known as Layered Cellular Automata (LCA). LCA introduces an additional layer of computation that operates alongside the traditional CA layer. The upper layer has its own set of rules, represented by two distinct rules: $f$ and $g$. While $f$ is considered the default rule for the CA, $g$ is applied to the blocks, which are entities in the upper layer. The introduction of this layered structure forms the basis of LCA and allows for the interaction and influence between the two layers.
	\item \textbf{Chapter~\ref{chap4}.} This chapter focuses on examining the impact of the layered approach on the overall dynamics of traditional Cellular Automata (CAs). We investigate how the introduction of an additional layer influences the dynamical behavior of CAs. Through our experiments, we observe that certain Layered Cellular Automata (LCA) exhibit a change in their dynamical behavior, while others remain unaffected. This highlights the potential of the layered approach in modifying and shaping the dynamics of CAs.
	\item \textbf{Chapter~\ref{chap5}.} In this chapter, we explore an application of Layered Cellular Automata (LCA) by discussing its convergence property. We investigate how LCA can be utilized as a pattern classifier. Using different standard datasets, we deploy various convergent LCAs and evaluate their performance in classifying patterns. Our analysis reveals that the proposed LCA-based classifier demonstrates competitive performance when compared to existing classifier algorithms. This highlights the potential of LCA as an effective tool for pattern classification tasks.
       \item \textbf{Chapter~\ref{chapconclusion}.} In this concluding chapter, we summarize the key findings and contributions of the thesis. Additionally, we highlight a few unresolved problems that could serve as future research directions. By identifying these open issues, we provide opportunities for further exploration and development in the field of layered cellular automata.
	
\end{itemize}

\chapter{Survey on Cellular Automata}
\label{chap2}
A Cellular Automaton (CA) is a computational system that is abstract and discrete in nature. It comprises a network of finite state automata called cells, arranged in a regular pattern. Each cell has a local update rule that determines its state change based on the states of its neighboring cells. The update rule is applied simultaneously to all cells, resulting in a synchronized state change of the entire system.


Cellular automata have been used in a wide variety of domains, ranging from physics and chemistry to biology and social sciences. Some examples of cellular automata applications in different domains:
\begin{itemize}
	\item Physics: Cellular automata have been used to model a wide range of physical systems, including fluid dynamics, magnetohydrodynamics, and solid-state physics. For example, the Ising model is a well-known cellular automaton used to study magnetic materials~\cite{margolus1986cellular,montgomery1987magnetohydrodynamic,hoekstra2010introduction,domany1984equivalence}.

\item Chemistry: Cellular automata have been used to model chemical reactions and diffusion processes. For example, the reaction-diffusion model, first proposed by Alan Turing, is a well-known cellular automaton used to study pattern formation in chemical systems~\cite{menshutina2020cellular,kier2005cellular,scalise2016emulating}.

\item Biology: Cellular automata have been used to model a wide range of biological processes, including the spread of diseases, population dynamics, and the behavior of neural networks. For example, the Game of Life, a classic cellular automaton, has been used to model the evolution of populations of organisms~\cite{ermentrout1993cellular,ruxton1998need,chowdhury2022cellular}.

\item Social Sciences: Cellular automata have been used to model a wide range of social phenomena, including the spread of rumors, traffic patterns, and the emergence of social norms. For example, the Schelling model is a cellular automaton used to study segregation in cities~\cite{nowak1996modeling,maerivoet2005cellular,Das2022,bhattacharjee2021computation}.

\item Computer Science: Cellular automata have been used to study computation and algorithms, including the design of cellular automata-based cryptographic systems. For example, the Cellular Automaton Encryption Algorithm (CAEA) is a symmetric key encryption algorithm based on cellular automata~\cite{kari2005theory,nandi1994theory,Sethi2016}.

\item Image Processing: Cellular automata have been used in image processing applications, such as image compression and filtering. For example, the Cellular Automata-based Image Compression Algorithm (CAICA) uses cellular automata to compress digital images~\cite{popovici2002cellular,rosin2005training,paul1999cellular,Sadeghi12,ShawDS06}.

\item Robotics: Cellular automata have been used to model the behavior of robots and to control their movements. For example, the Cellular Neural Network (CNN) controller is a type of cellular automaton used to control the motion of a robot~\cite{popovici2002cellular,rosin2005training,nandi20141,paul1999cellular}. 

\item Environmental Modeling: Cellular automata have been used to model environmental systems, such as the spread of forest fires and the growth of vegetation. For example, the Forest Fire model is a cellular automaton used to study the spread of forest fires~\cite{ghosh2017application,vourkas2012fpga,karafyllidis1997model,colasanti2007simple}.

\item Materials Science: Cellular automata have been used to study the behavior of materials, such as the growth of crystals and the behavior of polymers. For example, the Crystal Growth model is a cellular automaton used to study the growth of crystals~\cite{krivovichev2004crystal,raabe2004mesoscale}.

\item Music: Cellular automata have been used in music composition, such as generating musical patterns and rhythms. For example, the Music Box model is a cellular automaton used to generate melodies~\cite{burraston2005cellular,bilotta2002synthetic}.
\end{itemize}

\section{von Neumann's Universal Constructor}

The history of von Neumann's universal constructor begins in the 1940s, when mathematician and physicist John von Neumann was working on a variety of mathematical and computational problems. One area of interest for von Neumann was the study of self-replication in biological systems, which he believed could be used as a model for the development of self-replicating machines.

In 1948, von Neumann published a paper titled ``The General and Logical Theory of Automat''~\cite{von2017general} in which he outlined a theoretical machine that he called the universal constructor. The machine was designed to be a self-replicating automaton, capable of building copies of itself using raw materials from its environment.

\begin{figure}[hbt!]\centering 
	\includegraphics[width=5.8in]{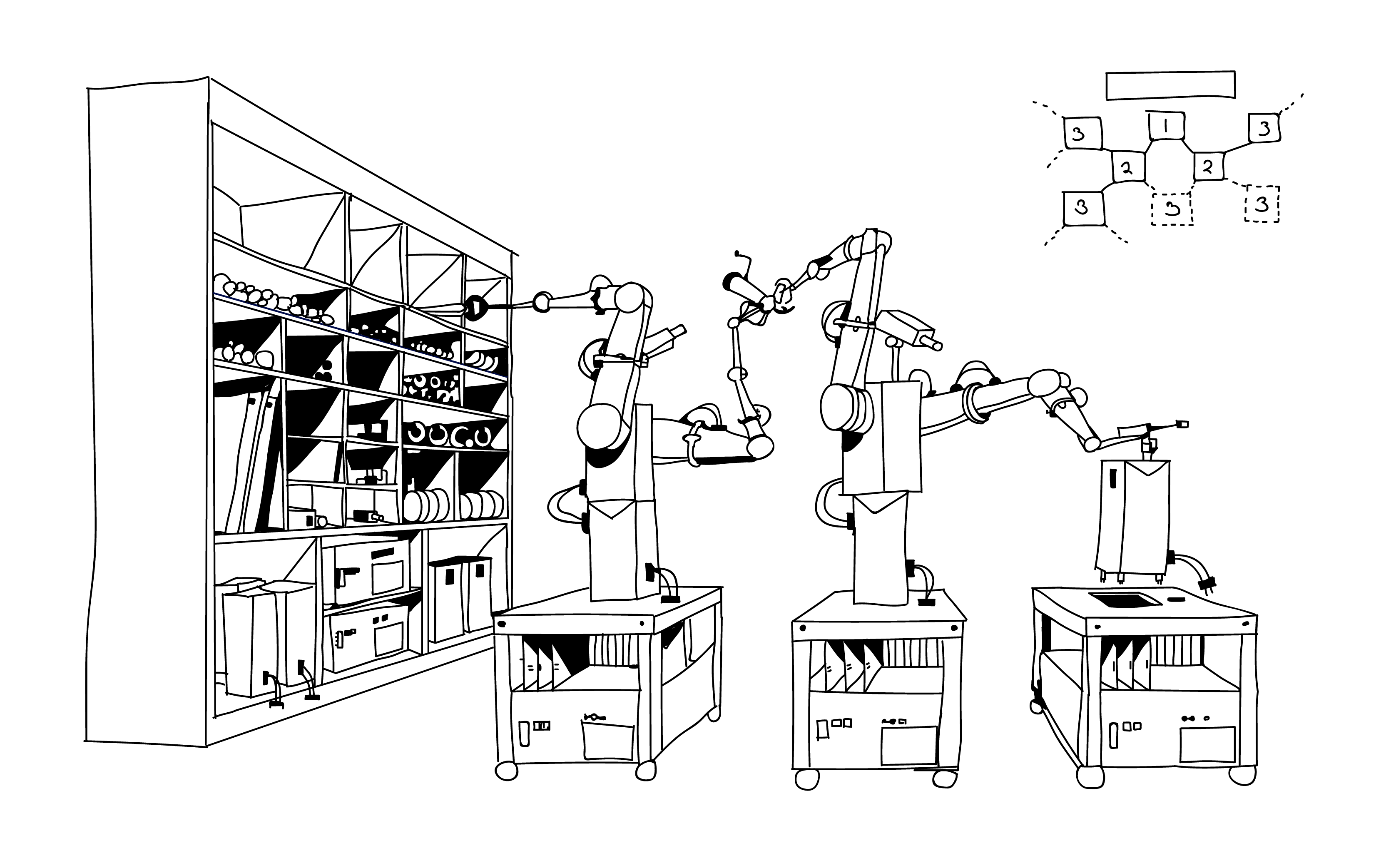} 
	\caption{ A robotic representation of self-replicating machines, inspired by the von Neumann self-reproducing automata~\cite{Neuma66}}
	\label{von_neuman}
\end{figure}

The universal constructor consisted of a grid of cells, similar to those used in cellular automata, that could store information and perform logical operations. The machine was designed to be programmable, allowing it to perform a wide range of tasks depending on the instructions it was given.

One of the key features of the universal constructor was its ability to build copies of itself. This was achieved through a process of self-replication, in which the machine would use its own components to create a duplicate of itself. Once the new machine was complete, it would be capable of building further copies of itself, leading to an exponential growth in the number of machines.

Von Neumann's ideas for the universal constructor were highly influential in the field of computer science and artificial intelligence. The concept of a self-replicating machine captured the imagination of researchers and inspired a generation of scientists to explore the possibilities of artificial life and self-replicating machines.

Despite the excitement surrounding von Neumann's ideas, however, the universal constructor was never actually built. This was due in part to the difficulty of constructing a machine that was capable of self-replication, as well as the practical challenges of designing a machine that could operate autonomously in the real world.

Despite these challenges, the concept of the universal constructor continues to be an important area of research in the field of artificial life and self-replicating machines. Researchers continue to explore ways to design machines that are capable of self-replication and that can be programmed to perform a wide range of useful tasks.

The von Neumann universal constructor remains a significant milestone in the history of computing and an inspiration for researchers in the field of artificial intelligence. Its impact can be seen in a wide range of applications, from the design of self-replicating robots for space exploration to the development of advanced manufacturing technologies that rely on self-replicating systems.

\section{Cellular Automata}
A cellular automaton (CA) is composed of a regular network of cells, with each cell being a finite automaton that utilizes a finite set of states, called S. These CAs undergo changes at specific times and locations, and the state of a cell evolves based on its neighboring cells. This means that a cell's current state is updated using a next-state function, also known as a local rule, with the cell's neighboring states serving as inputs to the function. The collection of all cell states at any given time is called the configuration of the CA. As the CA evolves, it transitions between configurations.
\begin{definition}
	A cellular automaton is a quadruple ($\mathcal{L}$, $\mathcal{S}$, $\mathcal{N}$, $\mathcal{R}$) where,
	\begin{itemize}
		\item $\mathcal{L} \subseteq \mathbb{Z}^\mathcal{D}$ is the lattice, where the cellular space is $\mathcal{D}-$dimensional. A lattice is a contiguous network of connected cells.
		\item  $\mathcal{S}$ is the finite set of states; e.g. $\mathcal{S} = \{0, 1,\cdots, d-1\}$.
		\item $\mathcal{N}=(\vec{v_1},\vec{v_2},\cdots,\vec{v_m})$ is the neighborhood vector of each cell $\vec{v}\in\mathcal{L}$ where $(\vec{v}+\vec{v}_i)\in\mathcal{L}$ and $m$ represents the number of neighbors of a cell.
		\item $\mathcal{R}:\mathcal{S}^m\rightarrow\mathcal{S}$ is the local transition rule for each cells in the lattice. Suppose $\mathcal{S}_{\vec{v}}$ is the current state of a cell $\vec{v}\in\mathcal{L}$, then the next state of the cell is $\mathcal{R}(s_{\vec{v}+\vec{v_1}},s_{\vec{v}+\vec{v_2}},\cdots,s_{\vec{v}+\vec{v_m}})$.
		\end{itemize}   
\end{definition}
Cellular automata can take various forms, with one of their fundamental properties being the type of grid they evolve. The most basic types of grids used for cellular automata are one-dimensional arrays. For two-dimensional cellular automata, square, triangular, or hexagonal shaped grids can be utilized. Furthermore, cellular automata can be constructed on Cartesian grids in any number of dimensions, with the integer lattice in multiple dimensions being a common choice. For instance, Wolfram's elementary cellular automata are implemented on a one-dimensional integer lattice. Similarly, Conway's game of life are implemented on a two-dimensional integer lattice.

The three fundamental characteristics of a classical cellular automaton (CA) are locality, synchronicity, and uniformity. 
Locality means that the computation in a CA using a rule is performed locally. This means that a cell's state changes based only on its local interactions with neighboring cells.
Synchronicity refers to the simultaneous updating of all cells in the CA. All cells change their states at the same time using the same update rule.
Uniformity is the use of the same local rule by all cells in the CA. This means that the same rule is applied throughout the entire lattice to perform the actual calculation.

The radius of a CA represents the number of neighboring cells that a cell depends for update in a particular direction. For example, if the radius of a CA is 3, then a cell depends on its three neighboring cells to the left and three neighboring cells to the right. In this case, the CA is a (3+3+1) = 7-neighborhood CA.
\begin{figure}[hbt!]\centering 
	\includegraphics[width=0.50\textwidth]{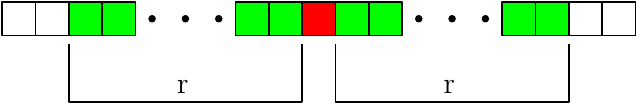} 
	\caption{Neighborhood dependence in a one-dimensional cellular automaton with radius $r$.}
	\label{r}
\end{figure}

 Figure.~\ref{r} illustrates the dependence on the state of the neighboring cells of a one-dimensional cellular automaton. To transition to the next state, a cell uses r neighboring cells to the left and r neighboring cells to the right of the cell that will be updated. Based on the state of the left and right cells, the next generation of the selected cell is generated.

When working with two-dimensional cellular automata, von Neumann and Moore neighborhood dependencies are commonly used to define a cell's neighborhood. The von Neumann neighborhood, coined by John von Neumann, is used in two-dimensional CA where the shape of grid is square. Each cell in this model CA model has one of the possible 29 states. In this neighborhood dependency, a cell's next state is determined by its current state and the states of its four neighbors as depicted in the Figure.~\ref{von}. 

The Moore neighborhood, on the other hand, includes a cell and its eight neighboring cells, including the four orthogonal neighbors and the four diagonal neighbors (see Figure.~\ref{moor}). This creates a nine-neighborhood dependency for the CA. The state of the cells reduced compared to von Neuman's model without compromising the ability to  reproduce itself and the computational capability that was showcased by von Neuman's model; see~\cite{Arbib66,Codd68,Banks71,thatcher1964,langton1984self,banks1970universality,Smith71,iirgen1987simple,Morita89,Culik90,Goucher2010a,martin1994universal,DUBACQ0202,ollinger2002quest,cook2004universality} for further details. John Conway's proposed Game of Life where each cells has 2 states, also uses Moore's neighbor dependency for evolution of states~\cite{Gardner70}.

\begin{figure}[hbt!]
	\subfloat[]{
		\begin{minipage}[c][1\width]{
				0.30\textwidth}
			\label{von}
			\centering
			\includegraphics[width=1\textwidth]{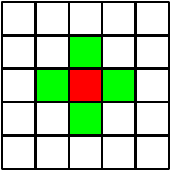}
	\end{minipage}}
	\hfill 	
	\subfloat[]{
		\begin{minipage}[c][1\width]{
				0.30\textwidth}
			\label{moor}
			\centering
			\includegraphics[width=1\textwidth]{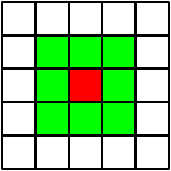}
	\end{minipage}}
	\caption{The neighborhood dependencies for two-dimensional cellular automata; (a) Von numann neighborhood; (b) Moore neighborhood.}
\label{fig:neighbors}
\end{figure}

\begin{figure*}[hbt!]
	\begin{center}
		\begin{tabular}{cc}
			\includegraphics[width=0.4\textwidth]{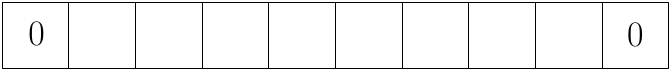}  &   \includegraphics[width=0.4\textwidth]{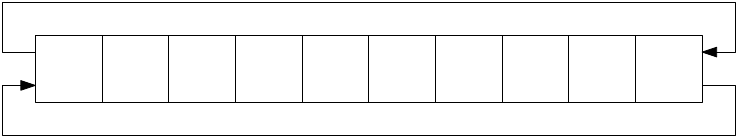}  \\ 
			
			(a) Null Boundary   & (b) Periodic Boundary \\
			\includegraphics[width=0.4\textwidth]{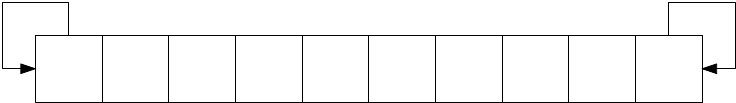}&
			
			\includegraphics[width=0.4\textwidth]{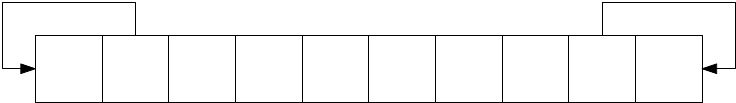} \\
			
			(c) Adiabatic Boundary & (d) Reflexive Boundary \\
			 \includegraphics[width=0.4\textwidth]{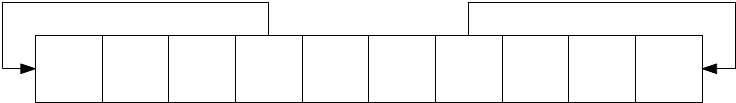} & \\   
			 
			  (e) Intermediate Boundary &\\
		\end{tabular}
		\caption{Boundary schemes of one-dimensional finite CAs. (a) Null Boundary (b) Periodic Boundary	(c) Adiabatic Boundary (d) Reflexive Boundary (e) Intermediate Boundary. }
		\label{neigh}
	\end{center}
\end{figure*}

The cellular space is infinite in nature. While the assumption of an infinite cellular space may be convenient for theoretical purposes, it is often not the case in practical applications of CAs, and thus the study of finite CAs with boundary conditions is necessary. We explore finite CA with mainly two types of boundary conditions, named as periodic and open boundary. Periodic boundary conditions are particularly useful for studying the behavior of CAs because they eliminate the boundary effects that can arise with open boundary conditions. With periodic boundary conditions, the CAs can be viewed as if they are on a torus or a sphere, where the boundary cells are considered to be adjacent to each other. This means that a CA with periodic boundary conditions is mathematically equivalent to an infinite CA with no boundary effects. This allows researchers to study the behavior of the CA without having to worry about the effects of the boundary. On the other hand, in case of open boundary condition, as the CA space is finite, the end cells in both directions will not have neighbors. Hence, these end cells are given a fixed state. Null boundary is a type of open boundary condition where the end cells are assigned with $0$ state. Following ~\cite{Wolfr83, Horte89a,Tsali91, ppc1,das2010scalable} are some of the usage of the working of null boundary condition. The choice of boundary condition can have a significant impact on the dynamics of a CA, particularly in finite CAs. For example, in the Game of Life, the choice of boundary condition can determine whether certain patterns, such as gliders, will move across the CA indefinitely or eventually collide with the boundary and be destroyed. Thus, the choice of boundary condition must be carefully considered when analyzing and simulating finite CAs. In this research work, we primarily worked with periodic boundary condition. Research related to periodic boundary condition in higher dimensional CA is mentioned in following~\cite{Palas1,Jin2012538,uguz2013reversibility} works. Figure.~\ref{neigh} shows different types of boundary conditions.

Wolfram~\cite{Wolfr83,wolfram2002new} popularized the concept of Elementary Cellular Automata (ECA) during 1980s. This model particularly works with one-dimensional cellular space, where cell has two states and simple neighbor scheme where each cell depends on its adjacent left and right cells. Due to its simplicity yet the ability to showcase the complex behavior upon local interactions, made it popularized among the researchers around the world towards this model. 

\subsection{Elementary Cellular Automata}

Elementary cellular automata is a 1-dimensional cellular space model where a cells interacts with its immediate neighbors in order to generate the next generation. Each cells can exist either of the two states. Each cell updates its state at discrete time steps based on the states of its nearest neighbors, according to a fixed rule table. Even though having a simple structure, it has the ability to display some of the complex behavior by interacting with neighbors.
ECAs has been applied in various fields, including:

Cryptography: ECA have been used to generate random sequences for use in encryption algorithms~\cite{BHATTACHARJEE2022100471,bhattacharjee2017pseudo,corona2022cryptographic,rajeshwaran2019cellular,xia2009data}.

Pattern recognition: ECA have been used to detect patterns in images and signals~\cite{wongthanavasu2013cellular,moran2019energy,braga1995pattern,MajiPhd}.

Pattern classification: ECA have been used for classifying a dataset into distinct classes~\cite{maji2003theory}.

Artificial life: ECA have been used to model the behavior of living systems, such as the evolution of populations and the emergence of self-organizing systems\cite{langton1986studying,li1990structure,chan2018lenia}.

Art and music: ECA have been used as a creative tool for generating patterns and structures in visual art and music~\cite{kojima2022application,adamatzky2016designing,holden2021survey}.

Overall, ECA have proven to be a versatile and powerful tool for modeling complex systems and exploring the dynamics of simple rules.

In elementary CAs, a cell's evolution is based on the states of its immediate left neighbor, itself, and immediate right neighbor (3-neighborhood). That is, the distance of neighbor cells $(r)$ is $1$ and has two states either 0 or 1. A cell is subjected to a function $f$ or local rule, depending on this rule, a cell's next state is decided. Table~\ref{Table:ECA} shows the state transition of a cell in the next state in two states 0 or 1. The rules are mentioned in binary format along with the corresponding decimal format has been mentioned.


\begin{table}[!htpb] 
	\centering	
	\scriptsize
	\begin{adjustbox}{width=0.8\columnwidth,center}
	\begin{tabular}{|cccccccccc|} \hline  
		   &  111 & 110 & 101 & 100 & 011 &  010 &  001 &  000 & Rule \\    
		RMT & 7 & 6 & 5 & 4 & 3 & 2 & 1 & 0 & \\ \hline  
		Next State:    & 0  &  0 &  0 &  1 & 1  & 0  &  0  &  1  & 25 \\  
		Next State:   & 0  &  0 &  1  &  1  &   0  &  0  &   1  &   0   & 50\\    
		Next State:   & 1  &  1 &  0  &  0  &   1  &  0  &   0  &   0   & 200\\       
		
		\hline
	\end{tabular}
\end{adjustbox}
	\caption{ECAs rules 25, 50 and 200.\label{Table:ECA}}  
\end{table}


On the basis of 3-neighborhood criteria, there are $2^{2^3}=256$ elementary cellular automata rules exist. Table~\ref{Table:ECA} represents some of the rules $(25,50,200)$ from 256. RMT or Rule Mean Term is the representation of 3-state binary configuration in decimal format i.e. 101 represented as RMT (5), where 1s are the two neighbors of the cell with state 0.  When a rule is applied on a 3-state binary configuration, the output determines whether RMT is \emph{active} or \emph{passive}. If middle cell changes its state in the next generation after rule is applied, then we can say RMT is active else it is RMT is passive. For example, In ECA rule 50, the RMTs $1(001), 4(100), 5(101)$ are passive as the middle cells changes it's state to 1 when rule 50 is applied. Rest RMTs are in passive state.

%

\subsection{Wolfram's Classification of Cellular Automata}

Elementary cellular automata (ECA) are simple yet powerful models that can exhibit a wide range of dynamic behaviors based on its interaction with neighbors and local rules. Due to these advantages, ECA is very popular among the researchers to use this model for studying some of the phenomena that can be observed in nature. On the basis of the dynamical behavior of ECA, Wolfram categorized ECAs into several class.  Different researchers further investigated the theoretical and practical  developments~\cite{ppc1,FSSPCA3,mitch98a,Sarkar00Survey,binder1993phase,ninagawa2014classifying,KARI20053,BhattacharjeeNR16}.  
Since ECA demonstrate the emergence of complex patterns and behaviors from simple rules, which makes them an interesting and valuable tool for studying the dynamics of complex systems. Some of the interesting dynamics that can be observed in ECA:

\textbf{Chaotic behavior}: Some ECA rules exhibit chaotic behavior, meaning that small changes in the initial conditions can lead to vastly different outcomes. This makes them useful for generating random numbers for use in cryptography. For example, rule 30 produces a complex and seemingly random pattern.

\textbf{Emergent patterns}: ECA rules can give rise to emergent patterns, such as oscillations, waves, and self-similar structures. For example, rule 18 produces a repeating pattern of three cells. These patterns can be seen as a form of computation, where the initial state of the automaton serves as the input and the resulting pattern is the output.

\textbf{Phase transitions}: ECA can undergo phase transitions, where small changes in the rule or initial conditions can lead to a qualitative change in the behavior of the system. For example, some rules transition from a uniform, static state to a complex, oscillating state as the initial density of ones is increased.

\textbf{Symmetry}: Some ECA rules exhibit symmetry, where the system has a repeating pattern that is symmetric about a central axis. For example, rule 90 produces a symmetric pattern that resembles a Sierpinski triangle.

\textbf{Self-organization}: ECA can exhibit self-organizing behavior, where local interactions between neighboring cells lead to the emergence of global structures and patterns. For example, rule 110 can create a variety of complex and interesting patterns, including fractal-like structures. Many behavior based on self-organization can be observed in many natural systems, such as the formation of crystals and the behavior of flocks of birds.

\textbf{Universality}: Some ECA rules are Turing-complete, meaning that they can simulate any computable function. This makes them a powerful tool for studying the properties of computation.


 In the paper~\cite{wolfram84b}, Stephen Wolfram proposed a classification system for cellular automaton rules based on the outcomes of their evolution from a random initial configuration. The basic building block of an elementary cellular automaton consists of a finite automaton defined over a one-dimensional array with two states either 0 or 1, and their states are updated synchronously based on their own state and the states of their two nearest neighbors. Wolfram's classification system categorizes cellular automaton rules into four types.
\begin{enumerate}
	\item[] \textbf{Class I. Homogeneous behavior}: In this class, the system quickly reaches a homogeneous state, where all cells are in the same state and remain so over time. This is true regardless of the initial configuration of the system.  
	\item[] \textbf{Class II. Periodic behavior}: In this class, the system evolves into a periodic pattern, where the pattern repeats after a fixed number of time steps. The period can be simple, such as a single repeating unit, or more complex, with multiple interacting components. Class II ECA exhibit regular, repetitive behavior that can be easily predicted and described.
	\item[] \textbf{Class III. Chaotic behavior}: In this class, the system evolves into a complex, aperiodic pattern, where the pattern never repeats and appears to be random or chaotic. Class III ECA exhibit behavior that is difficult to predict or describe, and small changes in the initial conditions can lead to vastly different outcomes~\cite{Supreeti_2018_chaos}.
	\item[] \textbf{Class IV. Complex behavior}: In this class, the system evolves into a complex, aperiodic pattern that exhibits a high degree of structure and organization. Class IV ECA exhibit behavior that is both complex and predictable, and they are often associated with emergent phenomena and self-organization.
\end{enumerate}

Stephen Wolfram's classification of cellular automata, includes not only ECAs but also two-dimensional CA like \emph{Game of Life}~\cite{Packa85b}, which is based on the patterns of behavior they exhibit over time, rather than on the specific rules or mechanics of the automaton.

According to Wolfram's classification, the Game of Life is a \emph{Class IV} automaton, which means that it exhibits complex and unpredictable behavior that is both structured and organized. In other words, the Game of Life is capable of producing a wide range of dynamic patterns or structures that are difficult to predict or understand. These structures can be self-replicating, self-organizing, and even self-healing in some cases. In the Game of Life, for example, certain initial configurations can give rise to stable, recurring patterns such as still lifes, oscillators, and spaceships. Other initial configurations, however, can give rise to highly complex and unpredictable behavior, including gliders, guns, and other intricate patterns that can interact with each other in surprising ways.


In addition to their rules and structure, the number of colors or unique states that a cellular automaton can exhibit must also be specified. This number is often an integer, with binary automata commonly having two colors labeled as ``white'' and ``black'' for the states 0 and 1 respectively. However, continuous range cellular automata can also be considered, allowing for a larger range of colors or states.

\begin{figure}[hbt!]\centering 
	\includegraphics[width=0.50\textwidth]{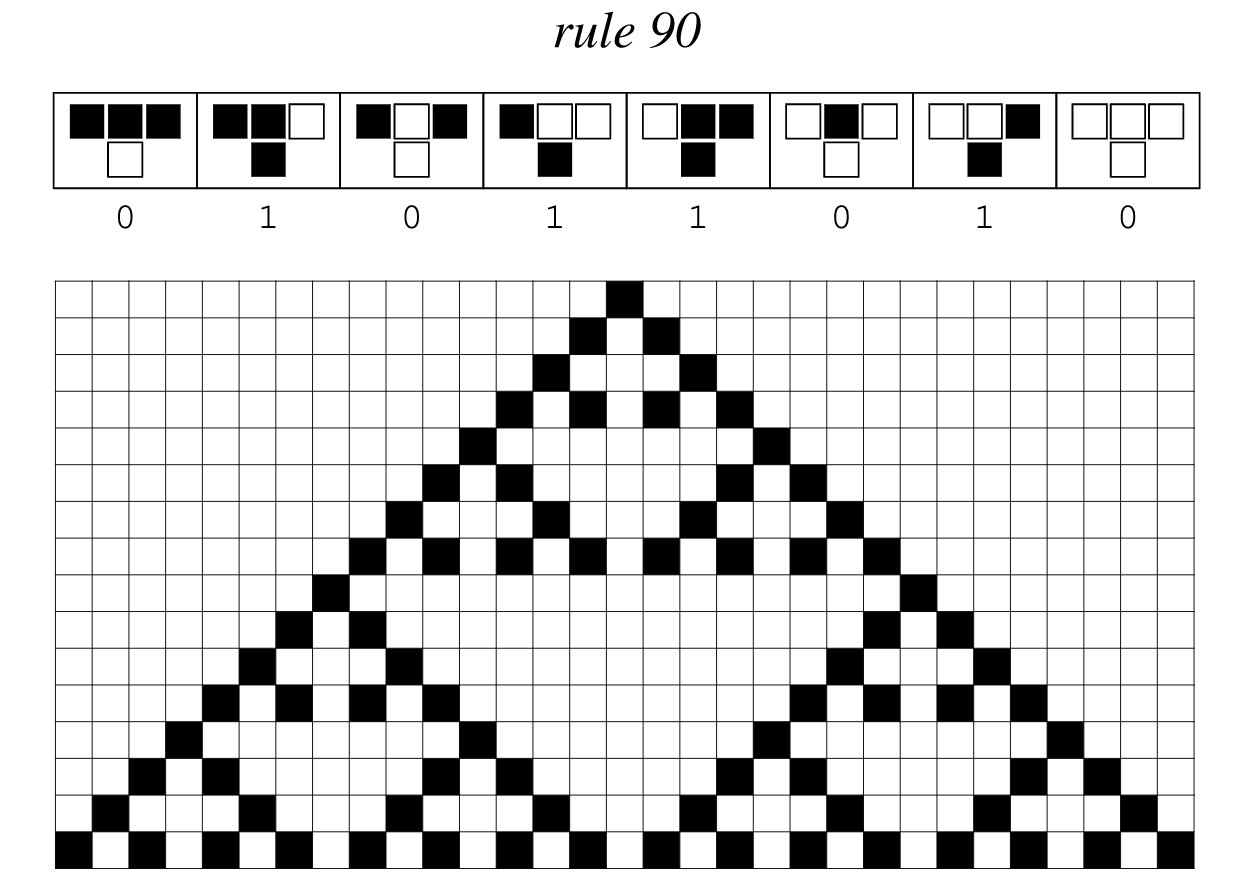} 
	\caption{Space-time diagram of Wolfram's Rule 90.}
	\label{rule90}
\end{figure}

Figure.~\ref{rule90} shows space-time diagram of  ECA $90$, where the initial configuration of the CA starts with a single $1$  from the middle of the cellular space. White cells represents 0 state, whereas 1 represents black cells.

\subsection{Li-Packard Classification}

Wolfram made the first attempt to classify the behavior of each CA rule based on observations, which provided a new perspective for researchers to better comprehend the dynamics of CAs. However, this classification does not fully distinguish the rules of one class from another, as some rules exhibit two types of behavior, such as chaotic behavior in some regions or two chaotic sections separated by a barrier. Local chaos is one example of such behavior. Li and Packard revised Wolfram's classification in 1990 and divided the ECA rules into five categories: null, fixed point, periodic, locally chaotic, and chaotic, based on rule space analysis that evaluates the probability of a rule being connected to another rule. The categories are presented in Table~\ref{Table:Li_Packard}.

%

\begin{table}[ht]
	\begin{center}
		\caption{Classification for 88 minimal representative ECAs by Li-Packard~\cite{Li90thestructure}.}
		\label{Table:Li_Packard}
		  \begin{adjustbox}{width=0.7\columnwidth,center}
			\begin{tabular}{|c|ccc|c|c|}
				\toprule 
				\textbf{Class I} & & \textbf{Class II}  &  & \textbf{Class III} & \textbf{Class IV} \\ Null
				& Fixed-Point & Periodic & Locally-Chaotic & Chaotic & Complex \\
				\midrule 
				0 & 2 & 1 & 26 & 18 & 41\\
				8 & 4 & 3 & 73 & 22 & 54\\
				32 & 10 & 5 & 154 & 30 & 106\\
				40 & 12 & 6 & & 45 & 110 \\
				128 & 13 & 7 && 60 &\\
				136 & 24 & 9 && 90 &\\
				160 & 34 & 11 && 105 &\\
				168 & 36 & 14 && 122 &\\
				& 42 & 15 && 126 &\\
				& 44 & 19 && 146 &\\
				& 46 & 23 && 150 &\\
				& 56 & 25 &&  &\\
				& 57 & 27 &&  &\\
				& 58 & 28 &&  &\\
				& 72 & 29 &&  &\\
				& 76 & 33 &&  &\\
				& 77 & 35 &&  &\\
				& 78 & 37 &&  &\\
				& 104 & 38 &&  &\\
				& 130 & 43 &&  &\\
				& 132 & 50 &&  &\\
				& 138 & 51 &&  &\\
				& 140 & 62 &&  &\\
				& 152 & 74 &&  &\\
				& 162 & 94 &&  &\\
				& 164 & 108 &&  &\\
				& 170 & 134 &&  &\\
				& 172 & 142 &&  &\\
				& 184 & 156 &&  &\\
				& 200 & 178 &&  &\\
				& 204 &  &&  &\\
				& 232 &  &&  &\\
				\bottomrule 
			\end{tabular}
			 \end{adjustbox}
	\end{center}
\end{table}

Among the five classes of ECA rules, a new class known as locally chaotic has been introduced. This class of CA behavior is intriguing because chaos is typically caused by infinitely long CAs. In contrast, locally chaotic ECAs display chaos within a small space. In this class, a cell's neighbors receive information that contributes to its state, but information cannot cross a wall, which acts as a blocking word. The behavior inside this defined region is always predictable, but looking at how information spreads among the cells there characterizes it as chaotic. This class of ECAs is different from the other classes, which are similar to Wolfram's original classes. An example of a locally chaotic ECA rule is Rule 26. Understanding and categorizing such behavior provides researchers with new insights into the dynamics of CAs, which can be useful in a variety of fields such as physics, mathematics, and computer science.

\subsection{Non-Uniformity in cellular automata}
Traditionally, all variations of Cellular Automata (CAs) exhibit three fundamental properties: uniformity, synchronicity, and locality. Uniformity refers to the fact that each cell in the CA is updated using an identical local rule. Synchronicity means that all cells are updated simultaneously, ensuring a coordinated update across the entire system. Locality implies that the rules governing cell updates act locally, with dependencies on neighboring cells being consistent.

It is worth emphasizing that cellular automata (CAs) carry out computations in a local manner, and the overall behavior of the system emerges from these local computations. Synchronicity, specifically, embodies a distinct form of uniformity in which all cells are updated simultaneously and uniformly. In essence, uniformity permeates the entire framework of CAs, encompassing the local rule, cell updates, and the lattice structure.
To summarize, uniformity in CAs can be understood in the following aspects:

\begin{itemize}
    \item The principle of uniformity extends to the updating process of cellular automata, where all cells undergo simultaneous updates during each discrete time step.
    \item Uniformity is also observed in the lattice structure and neighborhood dependency of cellular automata, as the lattice structure remains uniform and each cell exhibits a consistent neighborhood dependency.
    \item Uniformity is present in the local rule of cellular automata, where each cell follows the identical rule to update its state.
\end{itemize}

In the field of research, classical cellular automata (CA) have been extensively employed as a modeling tool. However, it has become evident that numerous natural phenomena, such as chemical reactions within living cells, exhibit non-uniform characteristics. This realization has necessitated the development of a new variant of CA that accommodates these non-uniform modeling requirements. Consequently, non-uniformity has been introduced into cellular automata, allowing for more accurate representations of diverse natural processes. The relaxation of the uniformity constraints in cellular automata (CA) has given rise to three main variants of non-uniformity. These variants are as follows:

\begin{itemize}
    \item Asynchronous cellular automata (ACAs): In ACAs, cells are not updated simultaneously at the same discrete time step. Instead, they can be independently updated, breaking the constraint of uniform updates.
    \item Automata Network: In this variant, the CA operates on a network, and the evolution of node states is influenced by the neighborhood defined by the network structure. This breaks the constraint of uniform neighborhood dependencies.
    \item Hybrid or non-uniform cellular automata: In this variant, cells are allowed to assume different local transition functions, resulting in varying rules for updating their states. This breaks the constraint of a uniform local rule across all cells.
\end{itemize}

These three variants of non-uniformity in CAs enable modeling approaches that can better capture the complexity of natural phenomena, accommodating scenarios where uniformity constraints may not hold~\cite{BhattacharjeeNR16}.

\subsubsection{Asynchronous cellular automata (ACAs)}
In contrast to the natural assumption of a global clock in synchronous systems, cellular automata (CA) also operate under the assumption of a global clock, which enforces simultaneous updates of all cells. However, this global clock assumption is not always realistic and has been relaxed in the case of asynchronous cellular automata (ACAs). The concept of ACAs and their computational capabilities were initially developed by Nakamura in 1974~\cite{naka}, and subsequent studies by Golze (1978)~\cite{GOLZE1978176}, Nakamura (1981)~\cite{Nakamura22}, Hemmerling (1982)~\cite{Hem82}, Ingerson (1984)~\cite{Ingerson84}, and Le Caër (1989)~\cite{LeC89} have further explored and refined the understanding of ACAs. R. Cori (1993)~\cite{Cor} extended the application of ACAs to a two-dimensional grid to investigate concurrent situations arising in distributed systems.

Asynchronous cellular automata (ACAs) introduce independence among cells, allowing them to evolve and update independently during the system's evolution. The application of asynchronism in ACAs encompasses various interpretations, but at its core, it involves breaking the perfect synchronous update scheme. In the literature, two main asynchronous updating schemes are commonly discussed: fully asynchronous updating and $\alpha$-asynchronous updating. These schemes represent different approaches to handling the timing and order of cell updates in an asynchronous manner. 

\begin{itemize}
    \item In the fully asynchronous updating scheme, the update process involves selecting a single cell uniformly and randomly at each time step. Consequently, only one cell is updated during each step, chosen independently of the other cells in the system.
    \item In the $\alpha$-asynchronous updating scheme, each cell is updated with a probability denoted as $\alpha$. This means that the cell applies the rule and transitions to a new state with a probability of $\alpha$. Conversely, with a probability of $1-\alpha$, the cell does not apply the rule and remains in its current state.
\end{itemize}

The parameter $\alpha$ in the $\alpha$-asynchronous updating scheme is commonly referred to as the synchrony rate. When the synchrony rate $\alpha$ is equal to 1, the cellular automaton (CA) operates synchronously. In this regard, classical CAs can be seen as a special case of $\alpha$-asynchronous CAs.
Building upon this concept, Bouré (2012)~\cite{probing12} introduced other asynchronous updating schemes, namely $\beta$-asynchronism and $\gamma$-asynchronism. Subsequently, Dennunzio (2013)~\cite{Dennunzio13} further expanded the range of asynchronous updating methods by introducing an m-asynchronous CA. This work aimed to generalize and encompass the various updating approaches used in prior research.

In the work of Blok and Bergersen (1999)~\cite{PhysRevE.55.6249}, they examined the effects of updating sites with a specific probability on the behavior of the Game of Life cellular automaton. Ruxton (1998)~\cite{RUXTON98} conducted an analysis of the sensitivity of ecological systems, modeled using simple stochastic cellular automata, to spatio-temporal ordering. In the studies by Tomassini and Venzi (2002) and Fatès (2013)~\cite{Nazim1414}, asynchronous rules were employed to address the density classification problem. Biswanath Sethi and Sukanta Das (2013)~\cite{Sethi6641432} explored the application of asynchronous cellular automata for pattern classification. They explored the use of asynchronous cellular automata in symmetric key cryptography~\cite{Biswanath2016}. They used asynchronism to study the reversibility of elementary CA where cells are updated asynchronously~\cite{Sethi14}. Raju Hazari and Sukanta Das (2018)~\cite{Raju18} studied ECA based number conservation.

\subsubsection{Automata Network}
In traditional cellular automata, a regular network structure with uniform local neighborhood dependency is commonly assumed. However, in the case of automata networks (also known as cellular automata networks), this strict requirement of uniform local neighborhood dependency is relaxed. In automata networks, the rules governing cellular automata allow for cells to have an arbitrary number of neighbors, enabling the application of various network topologies. This flexibility is demonstrated in studies of Marr (2009)~\cite{Marr20}.

It is important to note that the rules in automata networks are not always strictly local. Consequently, the behavior exhibited by automata networks with non-local rules can differ from that of conventional local rule-based cellular automata. Researchers such as Boccara (1994)~\cite{boccara1994some} have conducted studies exploring the implications and behaviors arising from non-local rules in automata networks.

\subsubsection{Hybrid or non-uniform cellular automata}
Among the models mentioned above, the Hybrid CA or Non-uniform CA stands out as the most popular and extensively studied model. In this type of cellular automaton, cells have the ability to employ different local rules. The exploration of non-uniform CAs began with the work of Pries (1986)~\cite{Pries86}, where they investigated the group properties of 1-dimensional finite CAs under null and periodic boundary conditions. 
Reversibility~\cite{Acri04,Sarkar12,Sethi14,Fats2018,das2009characterization} and convergence~\cite{DasMNS09,royCP15,Sethi14aaa,das2008exploring,adak2016synthesis} being his field of interest,  Sukanta Das (2007)~\cite{SukantaTH} has provided a generalized definition for non-uniform cellular automata (CAs). This definition allows cells to follow different rules with varying neighborhood dependencies. Supreeti Kamilya (2019)~\cite{KAMILYA2019116} studied the implication of chaos in non-uniform CA.

\subsection{Temporally Stochastic Cellular Automata}

Temporally stochastic cellular automata~\cite{subrata2022} involves the use of two elementary cellular automata rules, denoted as $f$ and $g$. In this context, rule $f$ serves as the default rule governing the system's behavior. However, rule $g$ is introduced as a temporal component that is applied to the entire system with a certain probability denoted as $\tau$. The role of $\tau$ in this context is to introduce noise into the system, impacting the evolution of the cellular automaton.

Essentially, the temporally stochastic cellular automata model incorporates randomness through the intermittent application of rule $g$, while rule $f$ remains the primary governing rule. The probability $\tau$ determines the frequency or likelihood of applying rule $g$ as opposed to rule $f$. This stochastic aspect introduces an element of unpredictability or variability into the behavior of the cellular automaton, reflecting real-world scenarios where external noise or disturbances can influence system dynamics.

The primary objective in the study of TSCAs is to explore the possibility of combining periodic and chaotic rules to observe chaotic or periodic dynamics. Researchers have undertaken extensive classifications of TSCAs to comprehensively understand the various dynamic behaviors exhibited by these automata. Special attention is given to phase transitions and the different types of class transition dynamics that arise in TSCAs. By conducting these investigations, researchers aim to elucidate the underlying mechanisms responsible for the emergence of distinct dynamical phenomena in TSCAs. 

Beyond characterizing the dynamics of TSCAs, researchers have also explored the computational capabilities of these automata and their potential applications. The affinity classification problem, which extends the classical density classification problem, has been introduced in the context of TSCAs. By leveraging the stochastic application of rules, TSCAs demonstrate promising performance as pattern classifiers when applied to standard datasets. Furthermore, TSCAs have been utilized in modeling self-healing systems, showcasing their applicability in various real-world scenarios.

In addition to studying the dynamics and computational abilities of TSCAs, researchers have proposed a novel model of computing units based on cellular automata. This model aims to alleviate the computational burden on Central Processing Units (CPUs) by distributing the workload across a network of cellular automata. Each cell in the computing unit represents a tiny processing element with attached memory. This cellular structure, implemented on the Cayley Tree, has shown potential in efficiently solving diverse computational problems. Notably, the model has been successfully applied to address Searching problems, demonstrating its effectiveness in solving complex computational tasks.

\subsection{Conway’s Game of Life}

Von Neumann's proposed cellular automata model~\cite{Neuma66} involved each cell having 29 possible states, resulting in high computational complexity for determining each cell's state. As a result, researchers aimed to reduce this complexity by decreasing the number of states without sacrificing the self-replication property of the machine. Conway's Game of Life is one of the most famous examples of cellular automata and was first introduced by mathematician John Horton Conway in 1970. The game is played on a two-dimensional grid of cells, with each cell either ``alive'' or ``dead''. The cells evolve based on a set of simple rules, which determine whether a cell will be alive or dead in the next generation.

The rules of the game are as follows:
\begin{itemize}
	\item A live cell will die if it has less than two live neighbors, resembling under-population.
	\item A live cell will survive to the next generation if it has either two or three live neighbors.
	\item A live cell will die if it has more than three live neighbors, resembling over-population.
	\item A dead cell will become a live cell if it has exactly three live neighbors, mimicking the process of reproduction.
\end{itemize} 

These simple rules give rise to complex and sometimes unexpected patterns of behavior. For example, certain patterns can repeat indefinitely, while others can grow and change over time.

The game was originally developed as a mathematical model to study the behavior of populations, but it quickly became popular among computer scientists and hobbyists as a programming challenge. The game's simple rules make it easy to implement in software, and its complex behavior has made it a popular subject for study and experimentation.

\begin{figure}[hbt!]\centering 
	\includegraphics[width=0.7\textwidth]{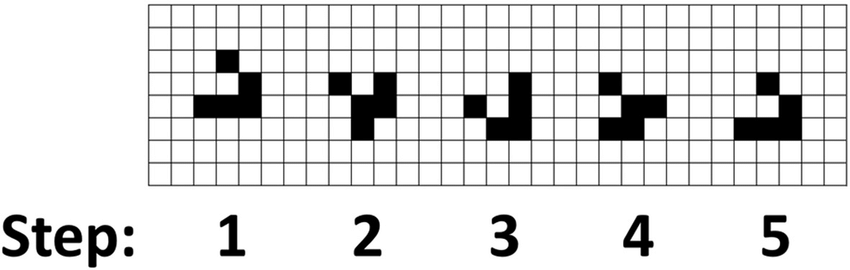} 
	\caption{Glider pattern on Game of Life in different time-steps}
	\label{gol}
\end{figure}

In the decades since its creation, the Game of Life has inspired a wide range of research and applications, from the study of self-replicating machines to the design of computer algorithms. It has also become a popular tool for exploring the relationship between simple rules and complex behavior, and has been used to study everything from the evolution of organisms to the behavior of physical systems.

\section{Artificial life}

Christopher Langton is widely regarded as one of the pioneers of the field of artificial life. He was a computer scientist and mathematician who played a key role in organizing the early workshops on artificial life that helped to establish the field as a distinct area of scientific inquiry.
Langton's work focused on developing computational models of living systems, with a particular emphasis on cellular automata and other simple, rule-based systems. One of his most famous contributions to the field was the development of the concept of ``self-organization'', which refers to the ability of complex systems to spontaneously organize themselves into coherent structures and patterns.
In his early work, Langton explored the behavior of simple cellular automata systems and discovered that they could exhibit a wide range of complex and unpredictable behaviors, including the emergence of complex patterns, the evolution of new structures, and the ability to process and transmit information. Researchers used these findings to develop new computational models of self-organizing systems, which was applied to a variety of fields, including biology, ecology, and social systems~\cite{copeland2004essential,Gotts09,Jeanson2008E,banda2,Bersini94,Sughimura14}.
One of Langton's most famous contributions to the field of artificial life was the development of the concept of ``artificial life as a platform for discovering the principles of living systems.'' This idea proposed that by creating and studying artificial life systems, researchers could gain a deeper understanding of the fundamental principles of living systems and their evolution.
Today, the legacy of Christopher Langton and his contributions to the field of artificial life can be seen in the many ongoing research projects and applications in fields such as robotics, biotechnology, and environmental monitoring. His work continues to inspire new generations of scientists and researchers to explore the frontiers of artificial life and to push the boundaries of our understanding of living systems.

\subsection{Langton's loop}
Langton's Loop is a self-replicating structure that was discovered by Christopher Langton in the 1984. It is an example of a cellular automaton, which is a simple mathematical model that consists of a grid of cells that can exist in one of 8 states~\cite{langton1984self}. The replicating pattern in Langton's loop is composed of a loop that contains genomic information. The genetic information is a series of cells that flow through the arm of the loop and eventually merge to form another loop. Some components of the genetic code are responsible for making the loop turn left three times before closing or ceasing to reproduce. This self-replication process has no size limitations and can be replicated infinitely in a two-dimensional space. This system is considered an excellent example of synthetic self-reproduction~\cite{langton86, LangtonII,langton90}.

\begin{figure}[hbt!]
	\subfloat[]{
		\begin{minipage}[c][1\width]{
				0.40\textwidth}
			\label{langtonloop}
			\centering
			\includegraphics[width=1\textwidth]{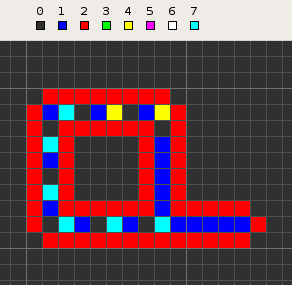}
	\end{minipage}}
	\hfill 	
	\subfloat[]{
		\begin{minipage}[c][1\width]{
				0.40\textwidth}
			\label{rule124255}
			\centering
			\includegraphics[width=1\textwidth]{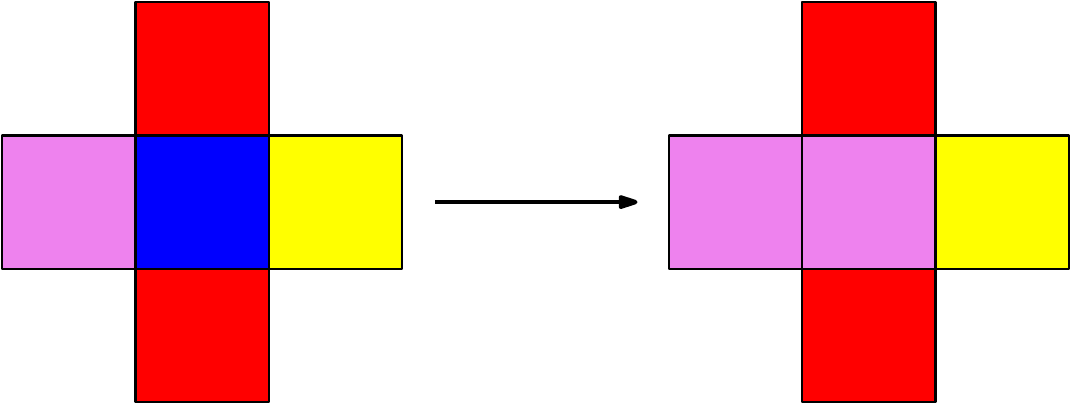}
	\end{minipage}}
	\caption{(a) Self reproduction in Langton's loop; (b) Rule 124255 transition.}
	\label{langtonLoop}
\end{figure}

In conway's game of life where each cells has two states either 0-white and 1-black and the neighborhood is considered using Moore's neighborhood schemes, whereas in Langton's loop, each cells has either of the 8 states which can be represented as color codes: $0-black, 1-blue, 2-red, 3-green, 4-yellow, 5-magenta, 6-white, 7-cyan$. It considers von Neumann's neighborhood schemes where a cell has 4 neighbors eliminating the diagonal cells. Researchers have been studying cellular automata that allow figures to make copies of themselves over the decades they have been trying to simplify how many states and how many rules needed to create self-replicating figures. In 1984, Christopher Langton said I only need eight states and 219 rules and then the figure~\ref{langtonloop} will self-replicate. These rules can be written as 6 digit number, for example take rule 124255~\ref{rule124255}, where each digit represents the state of the center, top, right, bottom and left cells. The last digit represents the state in which the center cell will exist if it satisfyies other state of other cells. According to rule 124255, if center-blue(1), top-red(2), right-yellow(4), bottom-red(2), left-magenta(5) then the center cell will change to magenta(5).
\subsection{Langton's Ant}

Langton's Ant is a simple two-dimensional cellular automaton that is named after its creator, Christopher Langton. It is an example of an agent-based system, where a simple set of rules applied to a single agent can generate complex behavior and patterns.

The Langton's Ant algorithm works as follows:
\begin{itemize}
	\item Start with a two-dimensional grid of cells, where each cell is either ``on'' or ``off''.
	\item Place an ``ant'' on the grid, facing in any direction.
	\item At each step, the ant follows two rules:
	\begin{itemize}
		\item If the ant is on an ``off'' cell, it turns right 90 degrees and flips the cell to ``on''.
		\item If the ant is on an ``on'' cell, it turns left 90 degrees and flips the cell to ``off''.		
   \end{itemize}
   \item The ant then moves forward one cell in the direction it is facing.
   \item Repeat this process for a specified number of steps, or until the ant reaches the edge of the grid.
\end{itemize}

The behavior of the ant is deterministic, meaning that if you start with the same initial conditions (grid, ant position, and direction), the ant will follow the same sequence of moves every time. However, the resulting pattern can be highly complex and unpredictable.

Langton's Ant is interesting because it produces emergent behavior that is difficult to predict from the simple rules that govern the ant's movement. After a certain number of steps (approx. 10000), the ant's path becomes periodic, creating a repeating pattern. However, the length of the period and the resulting pattern depend on the initial configuration and are not predictable.

Langton's Ant has been studied in both mathematics and computer science, and it has been used as an example of emergent behavior in complex systems. It is also a popular subject for computer simulations and games, and it has been implemented in various programming languages and platforms.

\begin{figure}\centering    
	\includegraphics[width=0.4\textwidth]{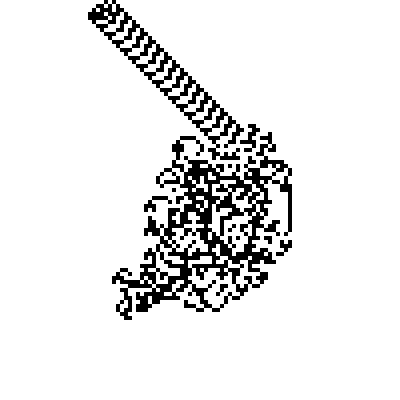}
	\captionof{figure}{Langton's Ant after $12000$ iterations.}
	\label{langtonAnt}
\end{figure}

\section{Summary}
This chapter provides an overview of cellular automata (CAs), which are computational models that have been used extensively to investigate dynamical systems in various fields. Researchers have developed various models using CAs to better understand the behavior of complex systems. To classify CAs based on their dynamical behaviors, several approaches have been proposed. For example, some researchers have focused on the topology of self-replicating cellular automata to gain insights into artificial life.

Parametrization is another useful tool for forecasting the behavior of CAs. By defining specific parameters, researchers can partially describe the behavior of CAs. However, it is worth noting that some CAs may not be identified accurately using parameters alone. This is because certain types of CAs, such as homogeneous, periodic, or chaotic CAs, may exhibit similar behaviors even if they have different parameters. Therefore, developing more precise parameters is an ongoing challenge for the research community.

While traditional CAs have been well-studied, relatively little is known about the behavior of layered cellular automata (LCAs). LCAs offer a new area for researchers to explore the dynamics of 1-Dimensional two-state cellular automata and develop new parameters for them. By gaining a better understanding of LCAs, researchers may be able to apply these models to various real-world scenarios and gain new insights into complex systems.

\chapter{Layered Cellular Automata : Definition}
\label{chap3}

\section{Intoduction}

Cellular automata (CAs) are mathematical models that have been used extensively in various fields of science to study complex systems. A traditional cellular automaton(CA)~\cite{framfram94} follows the same rule to update each cell of the lattice to generate its next state, where a cell's state is updated based on its current state and its nearest neighbors. 

In this work, we depart from traditional CA and add an additional influence, where the CA update is divided into two levels, each of which is updated according to separate rules. The next configuration of the CA is generated by applying both rules, with the lower layer follows rule $f$ and the upper layer follows rule $g$. The lower layer rule $f$ is typically a predefined rule, similar to traditional CA model, while the upper layer rule $g$ is the proposed rule which is applied on blocks of cells. We also take into account the CA's finite nature, which employs periodic boundary condition. This means that cells on the edge of the lattice are considered to be adjacent to cells on the opposite edge. We have named this type of CAs as Layered Cellular Automata (LCAs) where each cell interacts with its neighbors to generate the next state configuration but along with that the cell also admits some kind of outer world influence (influence from the distant neighbors).

Currently, CA has proven to be highly beneficial in many fields of science, including physics, biology, sociology, etc~\cite{margolus1986cellular,montgomery1987magnetohydrodynamic,kier2005cellular,scalise2016emulating,ruxton1998need,chowdhury2022cellular,nowak1996modeling}. The CAs that converge to fixed points from any seed have been widely employed for designing pattern classifiers~\cite{DasMNS09,subrata2022,ACAPattern}. The LCAs have many potential applications, including pattern classification. In this thesis, we study the dynamical behavior of this variant of CAs through extensive experiments and use the LCAs for identifying a set of LCAs that converge to fixed points from any initial configuration. We then develop a two-class pattern classifier using convergent LCAs. While pattern classification methods using CAs have been developed in the past, our approach with LCAs offers competitive results. The LCAs proposed in this work introduce an outer world influence in addition to the cell's immediate neighbors, making them a new and promising class of CA models for pattern classification and other applications in various fields of science.

Overall, the LCAs proposed in this work offer a new and promising way of modeling complex systems that incorporate an additional influence beyond the cell's immediate neighbors. The experiments conducted demonstrate the potential of LCAs in pattern classification and other applications.

\section{The model}

A Layered Cellular Automaton (LCA) is composed of a regular network of cells, with each cell being a finite automaton utilizes a finite set of states, called $\mathcal{S}$. The lattice is divided into equal sized blocks called $\mathcal{B}$. These LCAs undergo changes at specific times and locations, and the state of a cell evolves based on its neighboring cells along with distant cells. The collection of all cell states at any given time is called the configuration of the LCA. As the LCA evolves, it transitions between configurations.

\begin{defnn}
	\label{chap2:def1}
	A layered cellular automaton is a 8-tuple ($\mathcal{L},\mathcal{S},\mathcal{N}_0,\mathcal{R}_0,\mathcal{C},\mathcal{B},\mathcal{N}_1,\mathcal{R}_1$), where
	\begin{itemize}
		\item $\mathcal{L}\subseteq\mathbb{Z}^{\mathcal{D}}$ is the lattice, where $\mathcal{D}$ is the dimension. Each element of $\mathcal{L}$ is called a cell. 
		
		\item $\mathcal{S}$ is the finite set of states.
		
		
		\item $\mathcal{N}_0 = (\vec{v_1}, \vec{v_2}, \cdots, \vec{v_m})$ is the neighborhood vector of each cell $\vec{v}\in\mathcal{L}$ where ($\vec{v}$+$\vec{v_{i}}$)$\in$$\mathcal{L}$ and $m$ is the number of neighbors of a cell. 
		
		\item $\mathcal{R}_0:\mathcal{S}^{m}\rightarrow \mathcal{S}$ is the local transition rule for cells. If $s_{\vec{v}}$ is the present state of cell $\vec{v}\in\mathcal{L}$, then the next state of the cell is $\mathcal{R}_0(s_{\vec{v}+\vec{v_1}},s_{\vec{v}+\vec{v_2}},\cdots,s_{\vec{v}+\vec{v_m}})$. We call $\mathcal{R}_0$ as the rule of layer 0.
		
		\item $\mathcal{C}:\mathcal{B}^{p}\rightarrow\{TRUE,FALSE\}$ is a condition.
		
		\item $\mathcal{B}$ is the set of connected cells called block. 
		
		
		\item $\mathcal{N}_1 = (\vec{x_1}, \vec{x_2}, \cdots, \vec{x_p})$ is the neighborhood vector of each block $\vec{x}$. So a cell$\in\mathcal{L}$ and ($\vec{x}$+$\vec{x_{i}}$)$\in$$\mathcal{L}$, where $p$ is the number of neighbor blocks.
		

		
		\item $\mathcal{R}_1:\mathcal{B}^{p}\rightarrow \mathcal{B}$ is the local transition rule for blocks. If $\mathcal{C}$ is true, then $\mathcal{R}_1$ is applied on blocks.  We call $\mathcal{R}_1$ as the rule of layer 1.
		

		
	\end{itemize}
\end{defnn}

The difference between classical CA and LCA is, a classical CA uses a single rule for evolution, whereas an LCA uses two different rules in two layers. Let $f$ be a rule used as $\mathcal{R}_0$ in \emph{layer 0} at each time-step and rule $g$ used as $\mathcal{R}_1$ in \emph{layer 1} when $\mathcal{C}$ holds. Hence, if $\mathcal{C}$ is false for each step, then only $f$ is applied. This implies that classical CAs are a special case of LCAs. A collection of cells at a given time-step is known as configuration i.e. $c:\mathcal{L}\rightarrow\mathcal{S}$, which can be represented as, $c=(s_{\vec{v}})_{\vec{v}\in\mathcal{L}}$, where $s_{\vec{v}}$ is the state of cell $\vec{v}\in\mathcal{L}$. The set of all possible configurations is denoted as $c=\mathcal{S}^{\mathcal{L}}$. 

The idea behind this model is to represent the dynamics of the society. Layered cellular automata (LCA) can be used to model societies with hierarchical structures, capturing the dynamics of power, social roles, and interactions within such systems. In an LCA model of a society, each cell represents an individual. Each individual's socioeconomic status, preferences, opinions, ideologies or any other relevant characteristics influence social interactions and behaviors. These characteristics are represented by state of a cell. Rule $f$ governs the interactions and behaviors within layer 0, capturing how individuals or groups within the society interact with one another. These rules define how the state of a cell evolves based on the states of its neighboring cells within the same layer. They can encompass social dynamics such as friendship formation, influence, formation of organization etc. Rule $g$ capture the interactions and relationships between different layers within the society. These rules define how information, influence, or resources flow between layers, representing various social dynamics. For example, interlayer rules can govern the transmission of information, power, or decisions from higher layers (e.g., government, institutions) to lower layers (e.g., individuals, local communities). Through the interaction between two layers dynamics, LCA can simulate emergent behaviors that arise from the interactions of individuals or groups within the society. These emergent behaviors can include the formation of social networks, the spread of opinions or beliefs, cooperation or competition, or the formation of social hierarchies. For example, when a government announce some policies, certain section of society support or oppose it. Which may lead to several consequences such as protest, demonstration, social movement, revolution. The partition of India is one such example which was a series of protests and riots due to the decision of British government to divide India into two nations on the basis of faith. Some were in favor of partition such as All India Muslim League, Hindu Mahasabha and activists whereas idea of partition was opposed by Indian National Congress, Secular Nationalists and some Muslim leader and activists~\cite{talbot2009partition}. Eventually, India was divided into two nations.

LCA provides a framework for capturing the complex interactions and behaviors that occur within a society. By simulating the model and analyzing its outputs, researchers can gain a deeper understanding of social dynamics, inform policy decisions, and explore hypothetical scenarios to better comprehend the complexities of real-world societies.
In the following sections we discuss about different types of LCA model we used in our research work.

\section{LCA based on counting}
In this work, we consider one-dimensional cellular automata, where $\mathcal{L}=\mathbb{Z}/n\mathbb{Z}$ is the set of indices that represent the cells, where $n$ is the total number of cells. A state from $\mathcal{S}=\{0,1\}$ is allocated to a cell at each time step $t\in\mathbb{N}$. The proposed LCA consists of two layers -- {\em layer 0} and {\em layer 1}. {\em Layer 0} behaves like a traditional elementary cellular automata (ECA), whereas {\em layer 1}, formed by the cells of {\em layer 0}, influences the behavior of cells of {\em layer 0}. {\em Layer 0} and {\em layer 1} represent the lower layer and upper layer respectively. In the case of {\em layer $0$}, we consider three neighborhood structure and consider ECA rules for $f$, that is, a cell updates depending on self, left neighbor and right neighbor using an ECA rule. However, at block level update (\emph{layer} $1$), three neighborhood structure is followed. That is, a block is updated depending on self, left and right blocks and using a local transition rule $\mathcal{R}_1$ (say $g$). This $g$ is applied when the condition $\mathcal{C}$ for a block becomes $\emph{TRUE}$.


Let us first discuss the {\em layer $0$} rule, The changes of states of each cell are performed synchronously at each time step in accordance with a local rule $f:\mathcal{S}^3\rightarrow \mathcal{S}$. Given a set of cells $\mathcal{L}$ and local function $f$, one can define the global transition function for $f$, $G_f:\mathcal{S}^\mathcal{L}\rightarrow\mathcal{S}^\mathcal{L}$, that is, the image $y=(y_i)_{i\in\mathcal{L}}=G_f(c)$ of a configuration $c=(c_i)_{i\in\mathcal{L}}\in\mathcal{S}^\mathcal{L}$ is given by,
\begin{align*}\label{eq1}
	\forall i\in\mathcal{L},y_i=f(c_{i-1},c_i,c_{i+1})
\end{align*} 

Each rule $f$ is associated with a `decimal code' $w$, where $w$ = $f$($0,0,0$) $\cdot$ $2^0$ + $f$($0,0,1$) $\cdot$ $2^1$ + $\cdots$ + $f$($1,1,1$) $\cdot$ $2^7$, for the naming purpose. There are $2^8$ = $256$ ECA rules in two-state three-neighborhood dependency.

\begin{example}\label{ex2}
	Let us assume, $f$ is ECA rule $90$--
	\begin{align*}
		s^{t+1}_{i} & = 90(s^t_{i-1}, s^t_{i},  s^t_{i+1})\\
		& =   (s^t_{i-1}+  s^t_{i+1})\;mod\;2\\
		& = s^t_{i-1} \oplus s^t_{i+1}\\
		& = \begin{cases}
			0  \text{ if } s^t_{i-1}\text{ and } s^t_{i+1} \text{ are same} \\
			1 \text{ if } s^t_{i-1} \text{ and } s^t_{i+1} \text{ are different}
		\end{cases}
	\end{align*}
	where, $\oplus$ is the XOR operation between the left neighboring state the and right neighboring state. That means, if the cell has the same state in both the left and right neighborhood then the next state will be $0$ else it will be $1$. 
\end{example}

%

In LCAs, a cell not only gets influenced by its adjacent neighbors but also gets influenced by distant neighbors. Here the second rule $g$ is the external influencer for the model.  Let us discuss the second rule ({\em layer $1$}). Let us consider the lattice is divided into equal size blocks and each block consists of $b$ number of cells in each block. Each block is updated based on self, left block and right block using $g$. The changes are performed synchronously at each time-step in accordance with a local transition rule $g:\mathcal{B}^3\rightarrow\mathcal{B}$, given a set of blocks $\mathcal{L}$ and local function $g$, one can define the global transition function $G_g:(\mathcal{S}^{\mathcal{B}})^{\mathcal{L}/\mathcal{B}}\rightarrow(\mathcal{S}^{\mathcal{B}})^{\mathcal{L}/\mathcal{B}}$.

The main motivation behind rule g is to introduce some influence from distant cells. In order to update a block at \emph{layer 1}, modified ECA rule logic is applied on each block. That is, the image $z = (\mathcal{B}_i)_{i \in \mathcal{L}/\mathcal{B}} = G_g(c)$ of a configuration $c = (\mathcal{B}_i)_{i \in \mathcal{L}/\mathcal{B}} \in (\mathcal{S}^{\mathcal{B}})^{\mathcal{L}/\mathcal{B}}$ is given by,
\begin{equation}
	\nonumber
	\forall i \in \mathcal{L}/\mathcal{B}, \mathcal{B}_i = g(\mathcal{B}_{i-1},\mathcal{B}_i,\mathcal{B}_{i+1})
\end{equation} 

\begin{figure}[!htpb]
	
	\centering
		\includegraphics[width=0.7\textwidth]{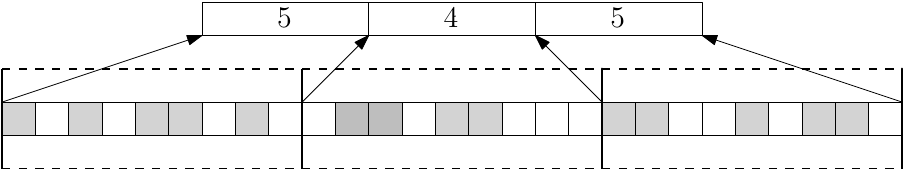}
	
	\caption{Rule \emph{g}, where $5,4$ and $5$ represent the number of $1$s in the blocks.}
	\label{fig:2}
	
\end{figure}

\begin{figure}[!htpb]
	
	\centering
		\includegraphics[width=0.7\textwidth]{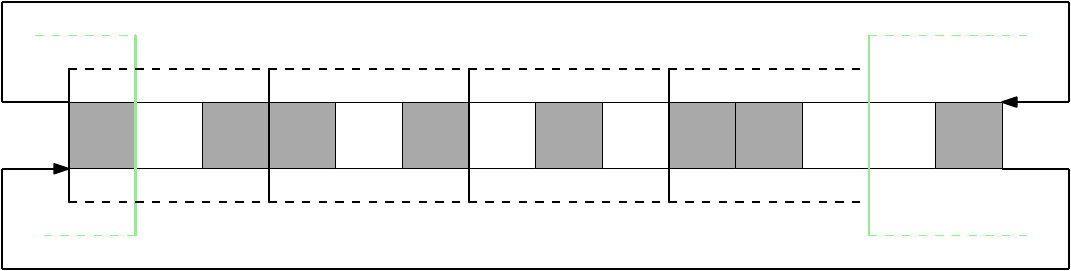}
	
	\caption{CA size $n=14$, block size $b=3$. The right-most block consists (marked as {\em green}) of two remaining cells and one cell from the left-most block.}
	
	\label{fig:3}	
\end{figure}

We have taken variable length $b$ ranges from $1$ to $n-1$, where $n$ is the CA size. We assume that the CA is a one-dimensional array of cells and the division of blocks is taken into account starting from the configuration's left-most cell (starts from $0^{th}$ index). In this direction, we consider two types of division, -- for $n\% b = 0$, there is no cell, such that the cell belongs to two different blocks, -- for $n\% b \neq 0$, the remaining $n\%b$ number of cells, along with $b-(n\%b)$ number of cells from the left most block, make up the right-most block, see the block marked in {\em green}, in Figure.~\ref{fig:3}. In the second type of division, certain cells are presented in both the left-most and right-most blocks. Figure.\ref{fig:2} shows that each block has its left neighbor block and right neighbor block. Based on the number of $1$'s present in the left and right blocks, modification in the current block is carried out. 

In layer 1, we update a block by modifying selected cells present in each block based on rule $g$. In next section we discuss about three schemes of rule $g$ which is based on counting of cells in a block and based on counting, if $\mathcal{C}$ is satisfied then only we apply the rule $g$.

\subsection{Averaging}
\label{averageof1s}
The main crux of LCA is a cell not only gets influenced by its adjacent neighbors but also gets influenced by some distant neighbors. In this first counting scheme, we implement a hierarchy in which many of the local decision at {\em layer 0} is influenced by the upper layer decision which is based on averaging. The configuration $c^{t+1}$ is made up of applying $f$ and $g$ on $c^{t}$ in sequence.

As per the definition of LCA, the rule $f$ is applied at the cell level, whereas the rule $g$ is applied at the block level. Rule $f$ is the classical implementation of ECA rules in layer 0. Rule $g$ is the proposed rule which performs the \emph{average} function. The main motivation behind rule $g$ is to balance the density of $1$s in the current block ($\mathcal{B}^{t+1}_{i}$) at time $t+1$ by averaging the number of $1$s in the left block ($\mathcal{B}^{t}_{i-1}$), current block ($\mathcal{B}^{t}_{i}$) and right block ($\mathcal{B}^{t}_{i+1}$) at time $t$.

Rule $g$ is defined as the number of $1$s that need to be dropped or added in the current block at time $t+1$, which is governed by $K$. Here $K$ corresponds to $\mathcal{C}$, which must be satisfied to apply rule $g$. Essentially, the value of $K$ determines whether the number of $1$s in the current block is greater than or less than the average number of $1$s in the left, current, and right blocks. If the number of $1$s in the current block is greater than the average, some $1$s need to be dropped to balance it out. On the other hand, if the number of $1$s in the current block is less than the average, some $1$s need to be added to balance it out. The value of $K$ determines how many $1$s need to be dropped or added to balance the density of $1$s across neighboring blocks.

\begin{equation}
	K  = \floor {|\mathcal{B}^{t}_{i}|_{\#1} -\frac{(|\mathcal{B}^{t}_{i-1}|_{\#1}+|\mathcal{B}^{t}_{i}|_{\#1}+|\mathcal{B}^{t}_{i+1}|_{\#1})}{3}}
\end{equation}

\begin{itemize}
	\item If $K$ $<$ $0$, then $\vert K \vert$ number of $1$s re-spawn in the places of cells with state $0$ in the block $\mathcal{B}^{t+1}_{i}$. Selection of state $0$ cells are chosen from left to right in the block.
	\item If $K$ $>$ $0$, then $K$ number of $1$s will be dropped from the cells in the block $\mathcal{B}^{t+1}_{i}$. Selection of state $1$ cells are chosen from left to right in the block.
	\item If $K$ = $0$, then there will be no change in the block $\mathcal{B}^{t+1}_{i}$.
\end{itemize}

\begin{figure*}[hbt!]
	\begin{center}
		\begin{adjustbox}{width=\columnwidth,center}
			\begin{tabular}{cccc}
				(a) ($46,g^{500}$) & (b) ($46,g^{400}$) & (c) ($46,g^{200}$) & (d) ($46,g^{50}$) \\ [6pt]
				\includegraphics[width=31mm]{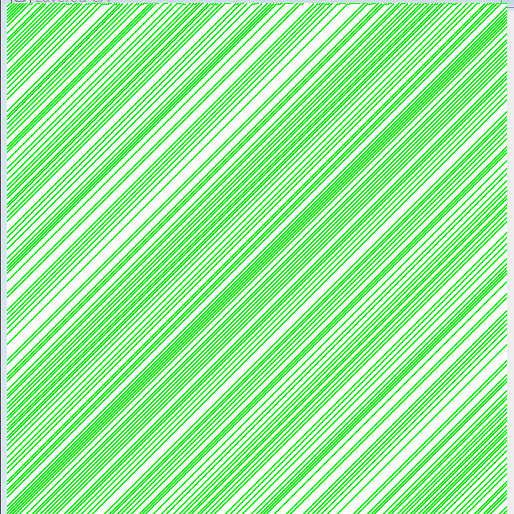} & \includegraphics[width=31mm]{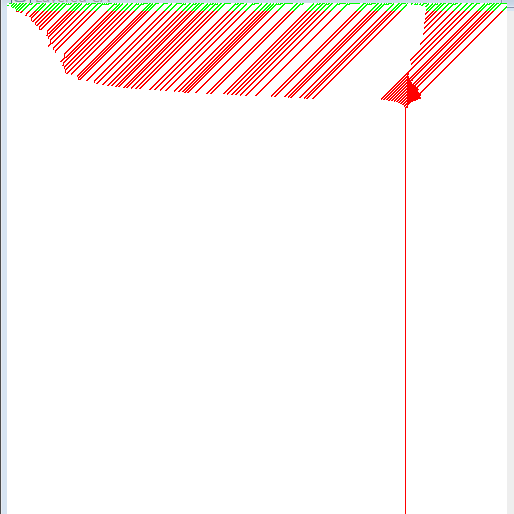} &   \includegraphics[width=31mm]{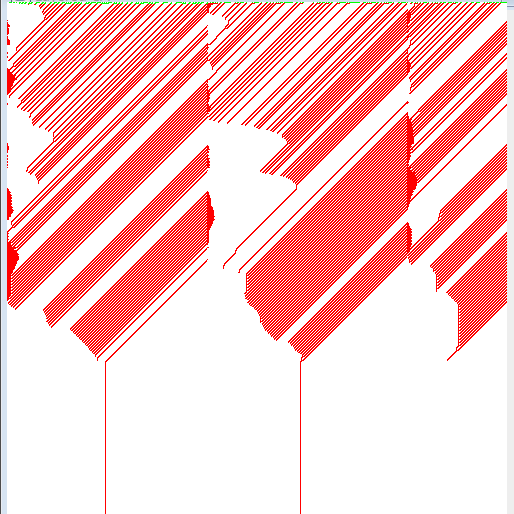} &   \includegraphics[width=31mm]{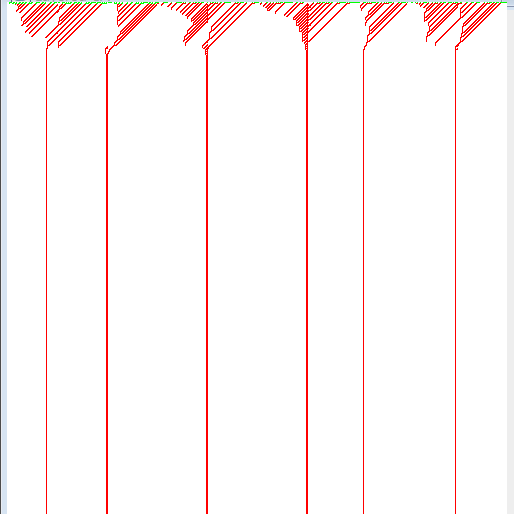}  \\		
					
			\end{tabular}
		\end{adjustbox}
		\caption{Space-time diagram of LCA($46,g^b$), for different values of $b$.}
		\label{rule46}
	\end{center}
\end{figure*}

Here $|\mathcal{B}|_{\#1}$ represents number of 1s present in the block and $K$ is the number of $1$s to be dropped or needed to be introduced in the block to balance the density of $1$s. When $n=b$, then the dynamic of the LCA is the same as traditional ECA of rule $f$, see in Figure.~\ref{rule46}(a) , where $f$ is ECA $46$, as the current, left and right blocks become same. Hence, the value of $K$ becomes zero. In Figure.~\ref{rule46}, we can observe the changes in the dynamical behavior in the space-time diagram, as we reduce the value of $b$ gradually. An LCA is denoted as ($f,g^b$), where $f$ works on lower layer (say \emph{layer 0}) and $g$ works on upper layer (say \emph{layer 1}), $b$ is the number of cells in each block (block size).

\begin{example}\label{ex1}

	Let us consider a LCA($46,g^{50}$), where $f=46$ and the considered block size $b=50$. Figure.~\ref{rule46}(d) shows the space-time diagram of ECA $46$ and block $50$ starting with random initial configuration with CA size $n=500$, where $green/red$ cell denotes $state-1$ and $white$ cell denotes $state-0$. When $g^b$ modifies the states then the cells of $state-1$ are marked as $red$ and the cells of  $state-0$ are marked as $white$, the cells of $state-1$ are marked as $green$ and the cells of  $state-0$ are marked as $white$ otherwise. In Figure.~\ref{rule46}(a) shows the space-time diagram of  LCA($46,g^{500}$), which is similar to the ECA $46$. Here $b=n$; that is, no modification occurs in the blocks by $g^b$, and the dynamics show {\em green}, one can say that when $b=n$  then the LCA behaves like ECA. ECA is a special case of LCA when considered block size $b=n$, i.e., LCA($f,g^n$) is similar to ECA $f$ where $n$ is the size of CA.	
\end{example}

In summary, this first counting scheme is designed to average the density of 1s in the current block at time $t+1$ by counting the number of 1s in the left, right, and current blocks at time $t$, and then adding or removing appropriate number of 1s from the current block at time $t+1$. The concept of averaging has various applications such as in image processing, LCA with averaging scheme can be used for smoothing out the image and reducing noise, which can enhance the accuracy of object recognition algorithms. In signal processing, averaging scheme can help in reducing noise in signals and enhancing their quality, which can improve the performance of algorithms that rely on these signals. In climate modeling, it can help in predicting the average behavior of these weather over time, which can aid in forecasting weather patterns and other climate-related phenomena.

 \subsection{Maximization}
 
This second counting scheme is designed to maximize the number of 1s in the current block at time $t+1$ by counting the number of 1s in the left, right, and current blocks at time $t$. Specifically, the scheme counts the number of 1s in the left block ($\mathcal{B}^{t}_{i-1}$), current block ($\mathcal{B}^{t}_{i}$), and right block ($\mathcal{B}^{t}_{i+1}$) at time $t$, and then calculates the difference between the block having maximum count of 1s and count of 1s present in the current block. The difference value is then used to determine the number of 1s that need to be added to the current block at time $t+1$ to achieve maximization. 

Rule $g$ is the proposed rule that operates at the block level, while rule $f$ (the classical implementation of ECA rules in layer 0) is applied at the cell level. This means that rule $g$ considers the properties of entire blocks when deciding how many 1s to add to the current block, while rule $f$ only looks at the individual cells within each block. Rule $g$ governs the number of 1s that need to be added to the current block to perform maximization. We take $Mx$ as maximization parameter, which calculate the number of 1s that should be added to the current block at time $t+1$.

\begin{equation}
	Mx  = MAX(|\mathcal{B}^{t}_{i-1}|_{\#1},|\mathcal{B}^{t}_{i}|_{\#1},|\mathcal{B}^{t}_{i+1}|_{\#1}) - |\mathcal{B}^{t}_{i}|_{\#1} +1
\end{equation}

$Mx$ number of $1$s re-spawn in the places of cells with state $0$ in the block $\mathcal{B}^{t+1}_{i}$. Selection of state $0$ cells are chosen from left to right in the block. $Mx$ ensures that the current block has always one cell as state 1 more than the previous maximum value of 1s. However in order to perform maximization of the current block, $\mathcal{C}$ must be satisfied.


Essentially, $\mathcal{C}$ is a set of conditions that must be satisfied for rule $g$ to be applied. The conditions ensure that the current block is not already maximized and that adding 1s to it will not cause it to become identical to its left or right neighbor blocks. If the following conditions $\mathcal{C}$ is satisfied, then rule $g$ is applied, and the specified number of 1s are added to the current block at time $t+1$. 

\begin{figure*}[hbt!]
	\begin{center}
		\begin{adjustbox}{width=\columnwidth,center}
			\begin{tabular}{ccccc}
				(a) Rule $128$ & (b) ($128,g^{50}$) & (c) ($128,g^{65}$) & (d) ($128,g^{100}$) & (e) ($128,g^{150}$) \\ [6pt]
				\includegraphics[width=31mm]{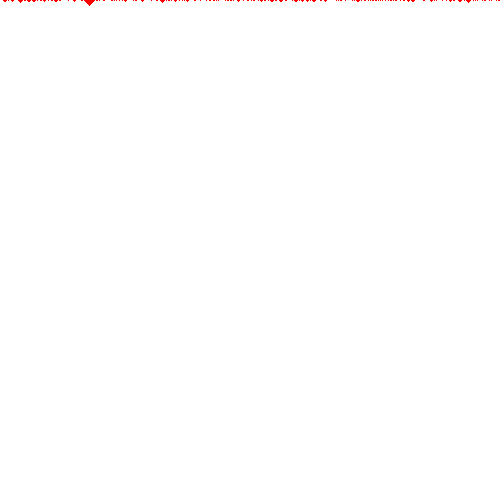} & \includegraphics[width=31mm]{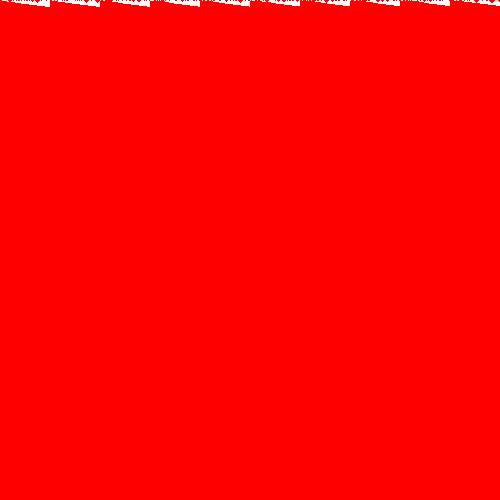} &   \includegraphics[width=31mm]{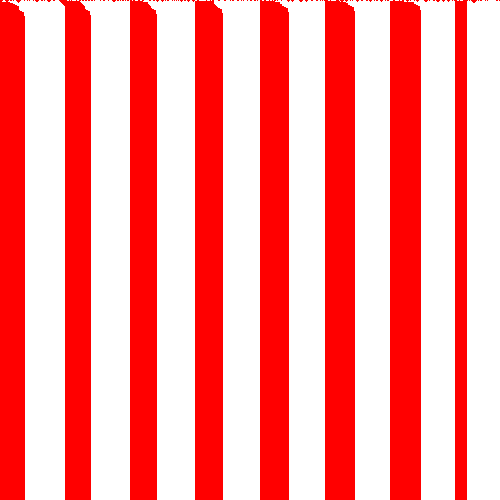} &   \includegraphics[width=31mm]{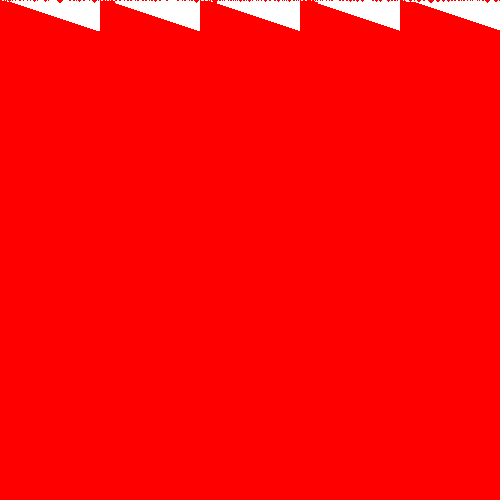}  & \includegraphics[width=31mm]{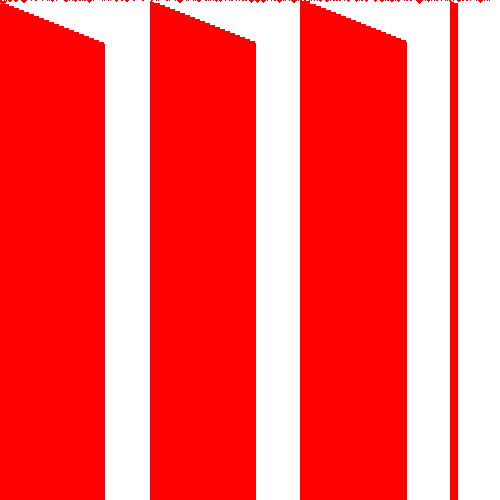}\\		
				
			\end{tabular}
		\end{adjustbox}
		\caption{Space-time diagram of LCA($128,g^b$), for different values of $b$.}
		\label{rule128}
	\end{center}
\end{figure*}

\begin{example}\label{ex1}
	
	Let us consider an LCA with ECA rule $128$. The space-time diagrams of ECA 128 is shown in Figure.~\ref{rule128}(a). In this LCA, ECA $128$ is used with block sizes of $50, 65, 100$ and $150$ respectively. Figure.~\ref{rule128}(b), \ref{rule128}(c), \ref{rule128}(d) and \ref{rule128}(e) show the results of running LCA($128,g^{b}$) on a random initial configuration of size $n=500$, where the color red represents state-1 and white represents state-0.	
	The dynamics of rule 128 gets converged to all-0 configuration. But when applied with a block size, the resultant dynamics converged to all-1 configuration. Such dynamics can be observed only when we select the block size which can divide the cellular space. Figure.~\ref{rule128}(b) and (d) show the dynamics of LCA($128,g^{50}$) and LCA($128,g^{100}$), where $b$ divides the cellular space into equal blocks. But when $b$ is selected such that it does not divide the cellular space, then the dynamics of LCA($128,g^b$) converges to a fixed point configuration.  Figure.~\ref{rule128}(c) and (e) show the dynamics of LCA($128,g^{65}$) and LCA($128,g^{150}$), where $b$ does not divide the cellular space into equal sized blocks.
	 The resulting space-time diagrams show the evolution of the CA over time, where the initial configuration evolves into a complex pattern based on the interaction between the two rules used in different layers.
	 
\end{example}

In summary, this counting scheme is designed to maximize the density of 1s in the current block at time $t+1$ by counting the number of 1s in the left, right, and current blocks at time $t$, and then adding the appropriate number of 1s to the current block at time $t+1$. The concept of averaging can be used in various fields such as in pattern recognition, maximization scheme can be used to identify objects or patterns in an image or video. The maximization scheme can help to enhance the contrast between the object and the background, making it easier to identify the object. LCA with maximization scheme can be used in various machine learning tasks. The maximization scheme can help to identify the most important features or variables that can differentiate between different classes or clusters.

 \subsection{Minimization}

In cellular automata, the concept of minimization involves reducing the density of certain elements in the system. This is typically accomplished by applying a specific rule that governs the transition function of the automaton. The rule $g$ is applied at the block level, which aims to minimize the number of 1s in the current block compared to the number of 1s in the left and right neighbor blocks has been discussed in this third scheme of counting.

To implement the minimization function, we first count the number of 1s present in the left block ($\mathcal{B}^{t}_{i-1}$), current block ($\mathcal{B}^{t}_{i}$), and right block ($\mathcal{B}^{t}_{i+1}$) at time $t$ and find the block with least number of 1s. If the number of 1s in the current block is greater than or equal to number of 1s in the block with least 1s, we remove some 1s from the current block at time $t+1$ to minimize the number of 1s in the system.

As rule $g$ governs the process of minimization in layer 1 considering the blocks of cell, rule $f$ is the classical ECA rules where the neighboring cells are adjacent cells. It operates in layer 0. Rule $g$ consider the elements of entire blocks for identifying the number of 1s that are needed to be dropped whereas rule $f$ only bothers about the individual cells within each block. In this scheme, rule $g$ governs how many number of 1s that should be removed from the current block to perform minimization. This rule is applied to each block in the system, and the resulting pattern is generated by iterating the rule over multiple time steps. We consider $Mn$ as minimization parameter, which is used to calculate the number of 1s that should be dropped or removed from the current block at time $t+1$.

\begin{equation}
	Mn  = |\mathcal{B}^{t}_{i}|_{\#1}-MIN(|\mathcal{B}^{t}_{i-1}|_{\#1},|\mathcal{B}^{t}_{i}|_{\#1},|\mathcal{B}^{t}_{i+1}|_{\#1}) + 1
\end{equation}

$Mn$ number of $1$s will be dropped from the cells in the block $\mathcal{B}^{t+1}_{i}$ by turning the selected state 1 to state 0. Selection of state $1$ cells are chosen from left to right in the block. $Mn$ ensures that the current block has always one cell as state 1 less than the previous least count of 1s. However in order to perform minimization of the current block, $\mathcal{C}$ must be satisfied.


Importantly, $\mathcal{C}$ must be satisfied for rule $g$ to be applied. The conditions ensure that the current block is not already minimized and that removing 1s from it will not cause it to become identical to its left or right neighbor blocks. If the following conditions $\mathcal{C}$ is satisfied, then rule $g$ is applied, and the specified number of 1s are removed from the current block at time $t+1$. 

\begin{figure*}[hbt!]
	\begin{center}
		\begin{adjustbox}{width=\columnwidth,center}
			\begin{tabular}{cccc}
				(a) Rule 30 & (b) ($30,g^{125}$) & (c) ($30,g^{250}$) & (d) ($30,g^{500}$) \\ [6pt]
				\includegraphics[width=31mm]{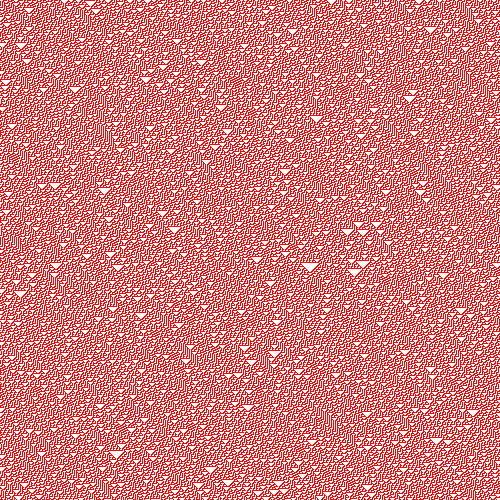} & \includegraphics[width=31mm]{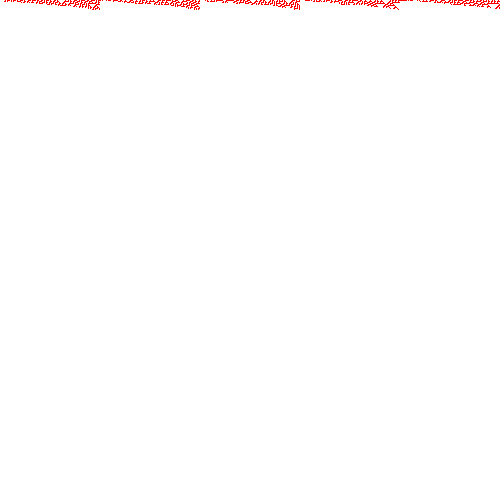} &   \includegraphics[width=31mm]{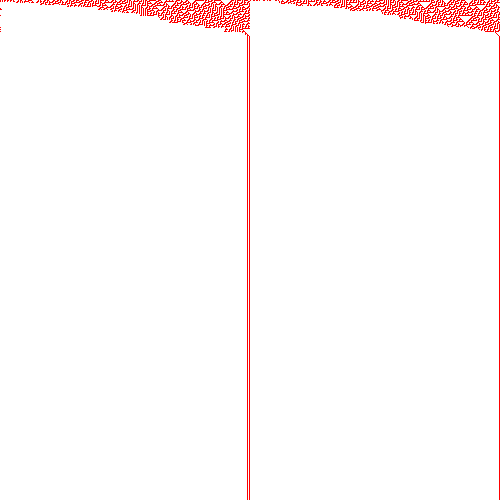} &   \includegraphics[width=31mm]{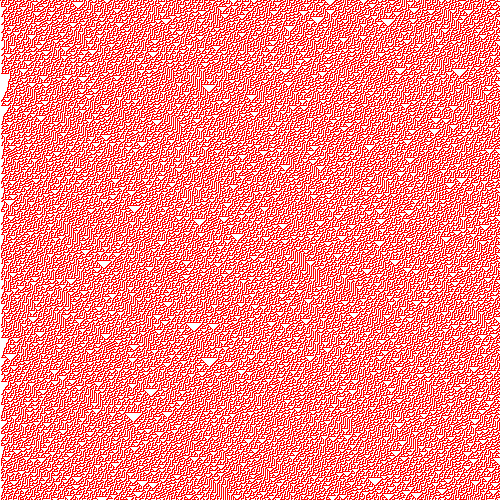}  \\		
				
			\end{tabular}
		\end{adjustbox}
		\caption{Space-time diagram of LCA($30,g^b$) for different $b$.}
		\label{rule30}
	\end{center}
\end{figure*}

\begin{example}\label{ex1}
	
	Let us consider an LCA with rule: ECA $30$. The space-time diagrams of ECA 30 is shown in Figure.~\ref{rule30}(a). In this LCA, ECA $30$ is used with block sizes of $125, 250$ and $500$ respectively. Figure.~\ref{rule30}(b), \ref{rule30}(c) and \ref{rule30}(d) show the results of running LCA($128,g^{b}$) on a random initial configuration of size $n=500$, where the color red represents state-1 and white represents state-0.	
	The dynamics of rule 30 gets represents chaos and randomness. But when applied with a block size 125, the resultant dynamics of LCA($30,g^{125}$) converged to all-0 configuration, see Figure.~\ref{rule30}(b). But in the case of $b=250$, LCA($30,g^{125}$) shows periodic dynamics which gets converged to a fixed point configuration, see Figure.~\ref{rule30}(c). When $b=n$, LCA($30,g^{500}$) shows chaotic dynamics similar to the dynamics of ECA 30.
	
\end{example}

In summary, this counting scheme is designed to minimize the density of 1s in the current block at time $t+1$ by counting the number of 1s in the left, right, and current blocks at time $t$, and then removing the appropriate number of 1s from the current block at time $t+1$. The concept of minimization has applications in various fields, such as data compression, where minimizing the number of 1s in a data sequence can lead to more efficient storage and transmission. In addition, minimization can also be used to model and study biological systems that exhibit a tendency towards minimizing the density of certain components, such as proteins to kill certain viruses like SARS-CoV-2 which includes four proteins as it's fundamental structure named as envelop protein, glyco protein, membrane protein and the nucleocapsid protein which must be reduce to make virus inactive.

\section{LCA based on ECA with modified neighborhood}

In this model, we focus on one-dimensional finite cellular automata with two states, where the cells are arranged in a periodic manner. The set of indices representing the cells is $\mathcal{L}=\mathbb{Z}/n\mathbb{Z}$, where $n$ is the total number of cells. At each time step $t\in\mathbb{N}$, a state from $\mathcal{S}=\{0,1\}$ is assigned to each cell.
In the case of {\em layer $0$}, we consider a neighborhood structure consisting of a cell and its two adjacent cells, and ECA rules are used for $f$. That is, a cell updates its state depending on its own state, the state of its left neighbor, and the state of its right neighbor.
For the block level update (\emph{layer} $1$), we consider a neighborhood structure consisting of a block and its two adjacent blocks. A local transition rule $g$ is applied to update the state of each block. Rule $g$ is applied when the condition $\mathcal{C}$ for a block becomes $\emph{TRUE}$.
It is worth noting here is that at layer 0, we consider three neighborhood structures and ECA rules at the cell level. On the other hand, at layer 1, we consider three neighborhood structures and modified ECA rule with extended neighborhood scheme is applied to update the state each cell of block based on its own state of cells in the block and the state of cells in its left and right neighbor blocks.

Let us discuss the second rule ({\em layer $1$}). Let us consider the lattice is divided into equal size blocks and each block consists of $b$ number of cells, i.e.,$n\%b=0$. We do not consider block sizes that cannot divide the lattice ($n\%b\neq0$) because the dynamical behavior is different as compared to $n\%b=0$. Each block is updated based on self, left block and right block using $g$. The changes are performed synchronously at each time-step in accordance with a local transition rule $g$.

Now a block update occurs by considering each cell update of the block. To update $j^{th}$ cell of the current block, we consider $j^{th}$ cells of the current, left and right blocks. Let us consider a block consists of $b$ number of cells and $c_i^j$ is the state of $j^{th}$ cell at $i^{th}$ block.
Then, the $c_i^j$ is updated by $g$ in the following way,
\begin{equation}
	\nonumber
	c_i^j = g(c_{i-1}^j,c_i^j,c_{i+1}^j), j\in\{0,1,\cdots,b-1\}
\end{equation} 

In the above way, all cells of a block are updated. We consider here $g$ as an ECA rule. Let us assume rule $90$ is used in layer 1. That implies, to update each cell's state $c_i^j$, we consider $c_{i-1}^j,c_{i}^j,c_{i+1}^j$ cell and apply ECA $90$ logic for a block $\mathcal{B}_i=(c_i^j)_{1\leq j\leq b}$.
\begin{equation}
	\nonumber
	c_i^j=c_{i-1}^j\oplus c_{i}^j\oplus c_{i+1}^j
\end{equation}

The above discussion depicts the update by rule $g$ seems like the update occurs at block level i.e. $\mathcal{B}_i=\mathcal{B}_{i-1}\oplus \mathcal{B}_i\oplus \mathcal{B}_{i+1} |_{i\in\mathcal{L}/\mathcal{B}}$.

\begin{figure}[!htpb]
	\centering
		\includegraphics[width=0.8\textwidth]{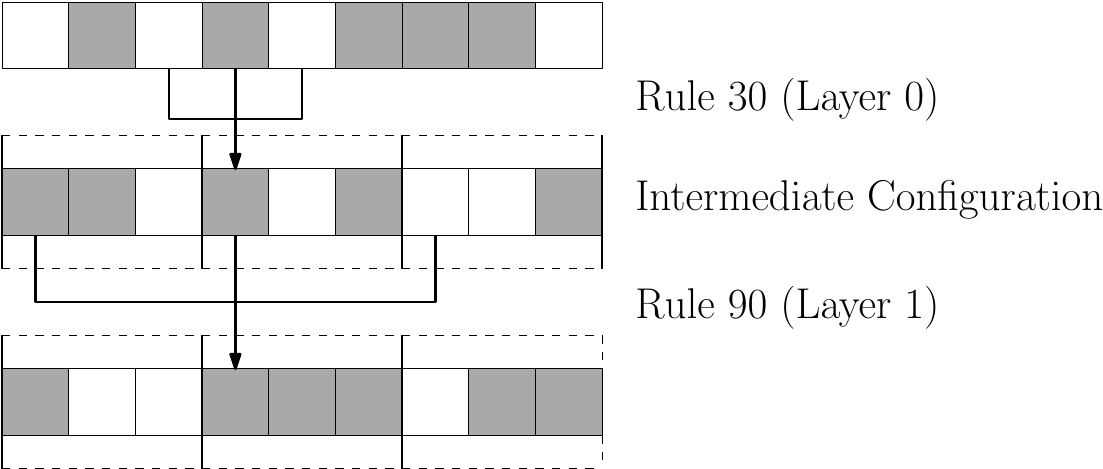}
	
	\caption{Conceptual view of working of LCA($30,90^3$), where black and white cell denotes state-1 and state-0 respectively.}
	\label{2eca}	
\end{figure}

In this model, we assume that the condition $\mathcal{C}$ is always true, which means that for every time step, the rule $f$ is first applied to the initial configuration to generate an intermediate configuration, followed by the application of the rule $g$ on the intermediate configuration to generate the next configuration (as shown in Figure.~\ref{2eca}). This approach is denoted by the notation ($f,g^b$), where $f$ operates on the lower layer (i.e., \emph{layer 0}), $g$ operates on the upper layer (i.e., \emph{layer 1}), and $b$ represents the number of cells in each block (also known as the block size). This notation helps to identify the specific LCA rule used in the model, as different choices of $f$ and $g$ can lead to distinct dynamical behavior and emergent phenomena.

Figure \ref{2eca} illustrates the update procedure of a layered cellular automata (LCA) model, where \emph{layer 0} uses elementary cellular automata (ECA) rule $30$ and \emph{layer 1} uses ECA rule $90$. Initially, the LCA model starts with a given configuration of cells at time $t$. At each time step, the cells are updated according to the rules defined in the LCA model. 
In the first stage, ECA rule $30$ is applied to the initial configuration in \emph{layer 0}. This rule determines the new state of each cell in the next time step based on the states of the cell and its two neighbors. The resulting intermediate configuration is then used as input for \emph{layer 1}.
In the second stage, the intermediate configuration is divided into blocks of size $b$ (as defined in the LCA model), and ECA rule $90$ is applied to each block separately. This rule depends on the states of the current block and its left and right neighbor blocks. The resulting configuration is the updated configuration at time $t+1$. This process of updating is repeated for each time step.
Therefore, in an LCA model with ($f,g^b$) notation, rule $f$ operates on \emph{layer 0} and generates an intermediate configuration, while rule $g$ operates on \emph{layer 1} using block size $b$ and generates the final configuration for the next time step.

\begin{figure*}[hbt!]
	\begin{center}
		\begin{adjustbox}{width=\columnwidth,center}
			\begin{tabular}{cccc}
				(a) Rule 3 & (b) Rule 18 & (c) $b=10$ & (d) $b=50$ \\ [6pt]
				\includegraphics[width=31mm]{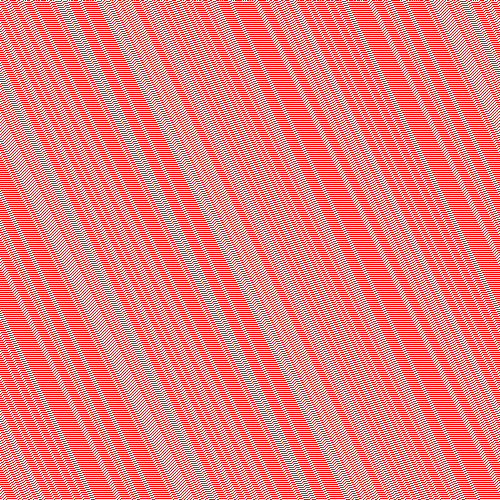} & \includegraphics[width=31mm]{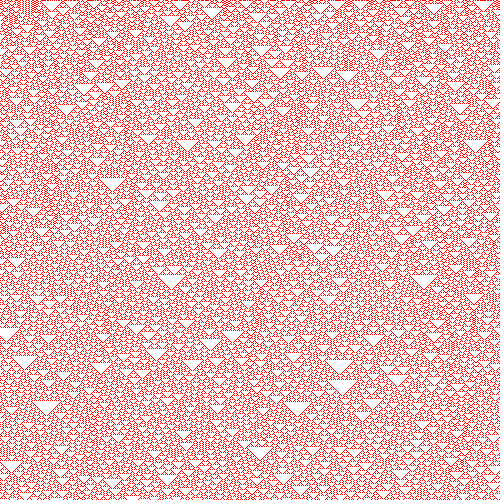} &   \includegraphics[width=31mm]{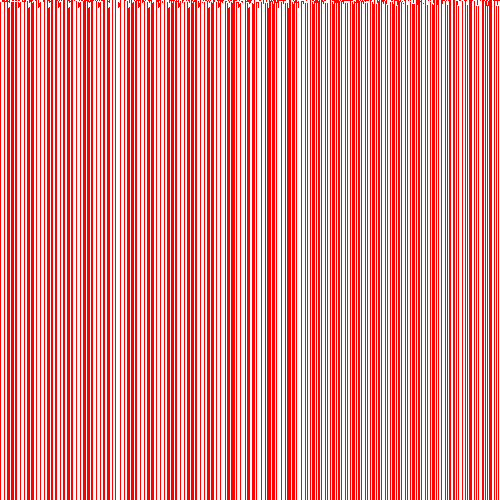} &   \includegraphics[width=31mm]{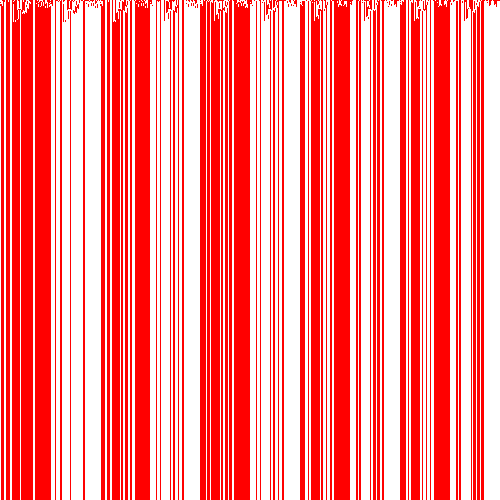}  \\		
				
			\end{tabular}
		\end{adjustbox}
		\caption{Space-time diagram of size 500, starting from a random initial configuration. (a) Space-time diagram of ECA 3. (b) Space-time diagram of ECA 18. (c) Space-time diagram of LCA($18,3^{10}$). (d) Space-time diagram of LCA($18,3^{50}$).}
		\label{18and3}
	\end{center}
\end{figure*}

\begin{example}\label{ex1}
	
	Let us consider an LCA with two rules: ECA $3$ and ECA $18$. The space-time diagrams of the two ECAs are shown in Figure.~\ref{18and3}(a) and~\ref{18and3}(b). In this LCA, ECA $18$ is used in layer $0$ and ECA $3$ is used in layer $1$, with block sizes of $10$ and $50$, respectively. Figure.~\ref{18and3}(c) and~\ref{18and3}(d) show the results of running LCA($18,3^{10}$) and LCA($18,3^{50}$) on a random initial configuration of size $n=500$, where the color red represents state-1 and white represents state-0.
	In LCA($18,3^{10}$), ECA $18$ is applied to the initial configuration to generate an intermediate configuration, which is then divided into blocks of size $10$, on which ECA $3$ is applied as the layer $1$ rule. Similarly, in LCA($18,3^{50}$), ECA $18$ is first applied to the initial configuration, and then the intermediate configuration is divided into blocks of size $50$, on which ECA $3$ is applied as the layer $1$ rule. The resulting space-time diagrams show the evolution of the CA over time, where the initial configuration evolves into a complex pattern based on the interaction between the two rules used in different layers.
	
\end{example}

To summarize, the proposed LCA model has two layers. In the first layer, ECA rules are applied at the cell level to update individual cells. In the second layer, modified ECA rules are applied at the block level to update each cell in the block. This two-layer structure allows for more complex behavior and dynamics to emerge from the simple rules governing individual cells and blocks.

\section{LCA based on Game of Life}

This work basically focuses on \emph{game of life} which is a two dimensional and two-state finite cellular automaton model. The set of indices representing the cells is $\mathcal{L}=\mathbb{Z}/(n\times m)\mathbb{Z}$, where $n\times m$ is the total number of cells. At each time step $t\in\mathbb{N}$, a state from the set $\mathcal{S}=\{0,1\}$ is assigned to each cell.

In the case of layer 0, we consider Moore's 8-neighborhood structures, see Fig~\ref{moor} and apply rules of game of life for the cell update function $f$. This means that a cell is updated based on its current state, as well as the states of its horizontal, vertical and diagonal neighbors.

In contrast, for the block-level update function in layer 1, we follow von Neumann's 4-neighborhood structure, see Figure.~\ref{von}. Here, a block is updated based on its current state, as well as the states of its left, right, top and bottom neighboring blocks, using a local transition rule denoted by $g$. This transition rule $g$ is applied only when the condition $\mathcal{C}$ for a block becomes true. 

As per the definition of LCA, the rule $f$ is applied at the cell level, whereas the rule $g$ is applied at the block level. Rule $f$ is the classical implementation of game of life rules in layer 0. Rule $g$ is the proposed rule which perform the \emph{average} function. The main motivation behind rule $g$ is to balance the density of $1$s in the current block ($\mathcal{B}^{t+1}_{i,j}$) at time $t+1$ by averaging the number of $1$s in the left block ($\mathcal{B}^{t}_{i,j-1}$), current block ($\mathcal{B}^{t}_{i,j}$), right block ($\mathcal{B}^{t}_{i,j+1}$), top block ($\mathcal{B}^{t}_{i+1,j}$) and bottom block ($\mathcal{B}^{t}_{i-1,j}$) at time $t$.

Overall, this setup allows for a hierarchical decision-making process in which the upper layer's decision is influenced by the lower layer's decision. The result is a more complex and sophisticated system that is better suited to modeling real-world phenomena.

Let us discuss about layer 0 rule. The Game of Life is a cellular automaton devised by the mathematician John Conway in 1970. It is a two-dimensional automaton that consists of a grid of cells, where each cell can be in one of two states - alive or dead. The state of each cell at any given time is determined by the state of its eight neighbors (horizontal, vertical, and diagonal). The rules for determining the next state of each cell are as follows:

\begin{itemize}
	\item If a dead cell has exactly three live neighbors, it becomes alive in the next generation.
	\item If a live cell has two or three live neighbors, it remains alive in the next generation. Otherwise, it dies in the next generation.
\end{itemize}

These rules are applied simultaneously to all cells in the grid at each generation, creating a new grid that represents the next generation. The process is then repeated for subsequent generations.

 The changes of states of each cell are performed synchronously at each time step in accordance with a local rule $f:\mathcal{S}^9\rightarrow \mathcal{S}$. Given a set of cells $\mathcal{L}$ and local function $f$, one can define the global transition function for $f$, $\mathcal{G}_f:\mathcal{S}^\mathcal{L}\rightarrow\mathcal{S}^\mathcal{L}$, that is, the image $y=(y_i)_{i\in\mathcal{L}}=\mathcal{G}_f(c)$ of a configuration $c=(c_i)_{i\in\mathcal{L}}\in\mathcal{S}^\mathcal{L}$ is given by,
\begin{align*}\label{eq1}
	\forall i,j \in\mathcal{L},y_i=f(c_{i-1,j-1},c_{i-1,j},c_{i-1,j+1},c_{i,j-1},c_{i,j},c_{i,j+1},c_{i+1,j-1},c_{i+1,j},c_{i+1,j+1})
\end{align*} 





In LCAs, a cell not only gets influenced by its adjacent neighbors but also gets influenced by distant neighbors. Here the second rule $g$ is the external influencer for the model.  Let us discuss the second rule ({\em layer $1$}). Let us consider the lattice is divided into equal size blocks and each block consists of $b$ number of cells in each block. Each block is updated based on self, left block, right block, top block and bottom block using $g$. The changes are performed synchronously at each time-step in accordance with a local transition rule $g:\mathcal{B}^5\rightarrow\mathcal{B}$, given a set of blocks $\mathcal{L}$ and local function $g$, one can define the global transition function $\mathcal{G}_g:(\mathcal{S}^{\mathcal{B}})^{\mathcal{L}/\mathcal{B}}\rightarrow(\mathcal{S}^{\mathcal{B}})^{\mathcal{L}/\mathcal{B}}$.

The main motivation behind rule g is to introduce some influence from distant cells. In order to update a block at \emph{layer 1}, rule $g$ is applied on each block. That is, the image $z = (\mathcal{B}_i)_{i \in \mathcal{L}/\mathcal{B}} = \mathcal{G}_g(c)$ of a configuration $c = (\mathcal{B}_i)_{i \in \mathcal{L}/\mathcal{B}} \in (\mathcal{S}^{\mathcal{B}})^{\mathcal{L}/\mathcal{B}}$ is given by,
\begin{equation}
	\nonumber
	\forall i \in \mathcal{L}/\mathcal{B}, \mathcal{B}_i = g(\mathcal{B}_{i,j-1},\mathcal{B}_{i,j},\mathcal{B}_{i,j+1},\mathcal{B}_{i-1,j},\mathcal{B}_{i+1,j})
\end{equation} 

We have taken block size $(b)=9$. Each block has 9 cells. We assume that the CA is a two-dimensional array of cells and the division of blocks is taken into account starting from the configuration's top-left-most cell (starts from index $(0,0)$). In this direction, we consider the division of CA space as $(n\times m)\% b = 0$, where the whole CA space is divided into equal sized blocks. Here we consider null boundary condition where the blocks present at the end of the CA space is assigned null value. Based on the number of $1$'s present in the left, current, right, top and bottom blocks, modification in the current block is carried out.

Rule $g$ is defined as the number of $1$s that need to be dropped or added in the current block at time $t+1$, which is governed by $K$. Here $K$ corresponds to $\mathcal{C}$, which must be satisfied to apply rule $g$. Essentially, the value of $K$ determines whether the number of $1$s in the current block is greater than or less than the average number of $1$s in the left, current, right, top and bottom blocks. If the number of $1$s in the current block is greater than the average, some $1$s need to be dropped to balance it out. On the other hand, if the number of $1$s in the current block is less than the average, some $1$s need to be added to balance it out. The value of $K$ determines how many $1$s need to be dropped or added to balance the density of $1$s across neighboring blocks.

\begin{equation}
	K  = \floor {|\mathcal{B}^{t}_{i,j}|_{\#1} -\frac{(|\mathcal{B}^{t}_{i,j-1}|_{\#1}+|\mathcal{B}^{t}_{i,j}|_{\#1}+|\mathcal{B}^{t}_{i,j+1}|_{\#1}+|\mathcal{B}^{t}_{i+1,j}|_{\#1}+|\mathcal{B}^{t}_{i-1,j}|_{\#1})}{5}}
\end{equation}

As we have already discussed in section~\ref{averageof1s} if $K<0$, some 1s will re-spawn in place of 0s and if $K>0$, some 1s to be dropped. 
 




\begin{figure}[!htpb]
	\centering
		\includegraphics[width=0.8\textwidth]{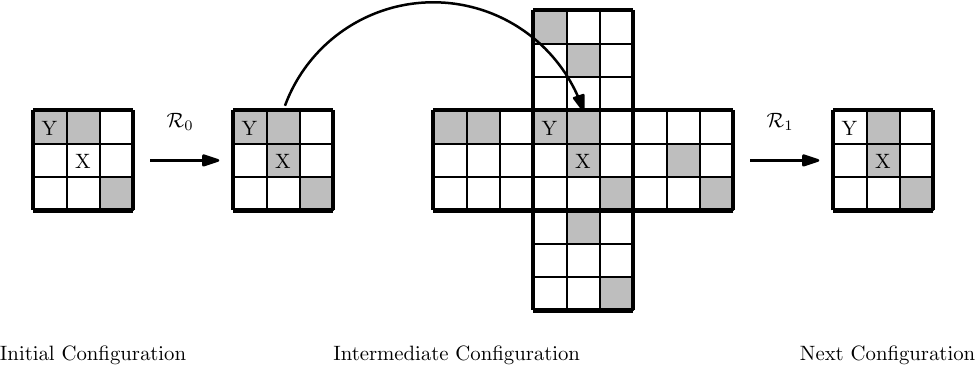}
	
	\caption{Conceptual view of working of LCA on game of life, where $b=9$, $\mathcal{R}_0$ and $\mathcal{R}_1$ are transition function for $layer 0$ and $layer 1$ respectively, black and white cell denotes state-1 and state-0 respectively.}
	\label{lcagol}	
\end{figure}

Figure~\ref{lcagol} shows the implementation of LCA on game of life. Cell marked as X in the initial configuration in dead state but it has 3 neighbor cells in alive state. When rule $f$ is applied to the transition function $\mathcal{R}_0$, the intermediate state is obtained where the cell 'X' becomes alive. In the intermediate state we divide the whole lattice into equal sized blocks where each block contains 9 cells in it. Here we consider von Neumann's neighborhood. Left, current, right, top and bottom blocks are considered for transition function $\mathcal{R}_1$. We calculate the number of 1s present in each block and perform averaging to decide whether to increase or decrease 1s in the current block. Here left, right, current, top and bottom has total twelve 1s. The value of $K$ turns out to be 1. Since $K>0$, we have to reduce one 1 from the configuration. The selection of cells to be modified is done from left to right, horizontally in the current block. Here cell 'Y' is selected to change it's state from $1\rightarrow0$.  

\begin{figure}[hbt!]
	\begin{center}
		\begin{adjustbox}{width=\columnwidth,center}
			\begin{tabular}{cccccc}
				
				\includegraphics[width=31mm]{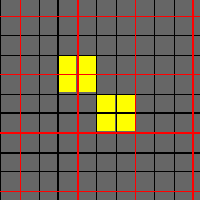} & \includegraphics[width=31mm]{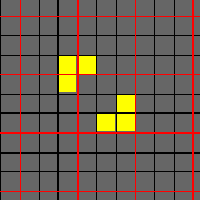} & \includegraphics[width=31mm]{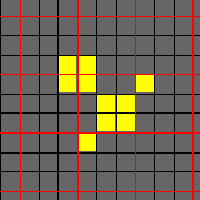} & \includegraphics[width=31mm]{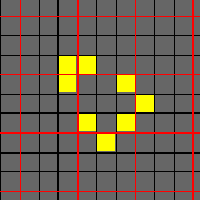} &
				\includegraphics[width=31mm]{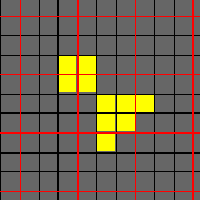}& 
				\includegraphics[width=31mm]{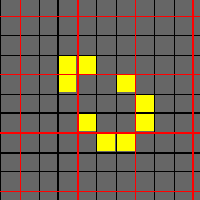} \\
				$t=0$ & $t=1$ & $t=2$ & $t=3$ & $t=4$ & $t=5$\\
				
				
				\includegraphics[width=31mm]{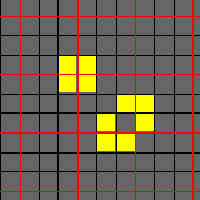} & \includegraphics[width=31mm]{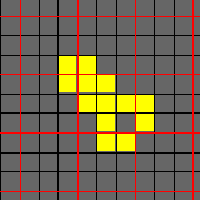} & \includegraphics[width=31mm]{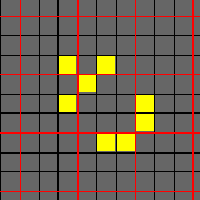} & \includegraphics[width=31mm]{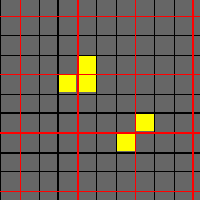} &
				\includegraphics[width=31mm]{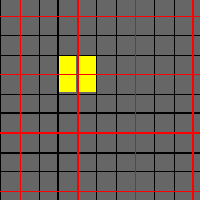}& 
				\\
				$t=6$ & $t=7$ & $t=8$ & $t=9$ & $t=10$ & \\
				
			\end{tabular}
		\end{adjustbox}
		\caption{Dynamics of beacon in LCA}
		\label{beacon}
	\end{center}
\end{figure}

\begin{example}
	Let us discuss the effect of rule $g$ on the dynamics of \emph{beacon}.  Beacon is a well-known oscillator in the Game of Life that consists of two blocks. It is called a beacon because it emits a distinctive ``beacon'' pattern of cell states that alternates between two configurations. The period of the beacon oscillator is 2, meaning it takes two generations for the pattern to complete one cycle.
	
	Figure~\ref{beacon} shows the changes in the dynamics of beacon when rule $g$ influence the configuration. Instead of being an oscillator, after 9 generation, beacon shows the dynamics of still life. Still life are stable patterns remain static and do not exhibit any movement or evolution. The black lines denotes the grid of cells in layer 0 and red line represents grid of blocks which consists of 9 cells in each block. Yellow represents alive state and dark grey represents dead state.
\end{example}

\section{Summary}

In this chapter, we introduced Layered Cellular Automata (LCA), a new model of computation that adds an additional layer of computation to traditional cellular automata. Layer 0 represents an existing model like ECA or the Game of Life, while Layer 1 represents a proposed model. This framework allows for more complex simulations, enabling the study of intricate systems.

We explored different models of LCAs, starting with counting-based LCAs. We discussed concepts such as averaging, which balances the number of 1s in a block based on the density of 1s in its neighboring blocks. We also examined maximization, where the goal is to increase the number of 1s compared to neighboring blocks, and minimization, which aims to decrease the density of 1s.

Additionally, we presented an LCA model based on ECA rules, where ECA is applied to both Layer 0 and Layer 1. In Layer 1, ECA considers left and right neighbor cells that are $b$ cells away from the current cell, leading to a modified neighborhood scheme.

Furthermore, we discussed the implementation of LCAs in the popular Game of Life, applying its rules in Layer 0. In Layer 1, we utilized an averaging concept based on von Neumann's neighborhood for block updates and Moore's neighborhood for individual cell updates. This resulted in intriguing patterns and behaviors.

In summary, Layered Cellular Automata extend traditional CA by introducing additional layers of computation and interlayer rules. This advanced framework enables the modeling of complex systems and phenomena, providing insights into emergent behavior. LCAs have broad applications across various scientific domains, offering a powerful tool for studying intricate systems.

\chapter{Layered Cellular Automata : Classes and Dynamics}
\label{chap4}

\section{Introduction}

In mathematics, dynamics is the study of how things change over time. It is a branch of mathematics that deals with the motion and behavior of objects and systems in motion. It studies the patterns of movement, stability, and chaos in various physical, biological, and social systems. In the context of cellular automata, dynamics refers to the evolution of the system over time, as cells or blocks change their states according to certain rules. The study of dynamics in cellular automata involves analyzing the behavior of the system, such as its stability, periodicity, randomness, and complexity, and identifying patterns and structures that emerge from the interactions between cells or blocks. The dynamics of cellular automata have applications in various fields, such as physics, biology, computer science, and social sciences.

In physics, CA is used to study complex systems and emergent behavior, such as pattern formation and self-organization. It has been used to model physical phenomena, such as fluid dynamics, solid-state physics, and quantum mechanics.

In computer science, CA has been used to design and analyze parallel algorithms and to model and simulate computer networks. CA has also been used in the field of artificial intelligence for developing neural networks and genetic algorithms.

In biology, CA has been used to model and simulate the behavior of biological systems, such as population dynamics, virus interactions and identifying rate of replication, and immune system responses. It has also been used to study the behavior of cells, such as a cell starts to grow and divide itself uncontrollably which eventually turns into cancer.

In social science, CA has been used to study the behavior of complex social systems, such as traffic flow, crowd dynamics, and urban growth. It has also been used to model economic systems, such as identifying market based on population and demand.

The study of dynamical behavior of cellular automata has broad applications in various fields, and its use continues to expand as new applications are discovered.

In the previous chapter, we have discussed a variation of cellular automata (CA) called layered cellular automata (LCA) that incorporate the influence of an upper layer on the lower layer to change its dynamics. In LCA, the lower layer, layer 0, follows the predefined rules of different CA models such as elementary cellular automata (ECA), Game of Life, etc., to update individual cells. However, at layer 1, the blocks of cells are updated based on a local transition rule, denoted as $g$, which in turn affects the dynamics of layer 0. The transition rule for layer 0 is denoted as $f$.

Alan Turing's idea of morphogenesis~\cite{Turing} suggests that a system will select one of the symmetric directions of evolution, indicating a potential way to evolve the system with randomness. This framework of cellular automata is well-suited for studying natural phenomena of this kind.

In LCA, the influence of the upper layer on the lower layer enables the system to exhibit emergent behaviors and patterns, which may not be visible in the lower layer alone. This approach is particularly useful in studying complex systems, such as biological morphogenesis, traffic flow, social dynamics, and pattern formation. LCA can also be used for simulations and modeling, as it allows for the analysis of the behavior of a system over time, given a set of initial conditions and transition rules.

This chapter focuses on studying the impact of layer 1 on layer 0 and classifying the LCAs based on their dynamics. The goal is to understand how the dynamics of the system changes due to this influence.

\section{LCA based on ECA rules in layer 0}

 Elementary cellular automata (ECAs) are one-dimensional arrays of finite automata. These automata are composed of cells that can be in one of two states, and each cell updates its state in discrete time depending on its own state and the states of its two closest neighbors. All cells update their states synchronously, meaning that each cell computes its next state based on the current states of its neighbors simultaneously. ECAs can be thought of as simple models of natural systems that exhibit emergent behavior. They have been studied extensively for their simplicity and their ability to produce complex patterns and behaviors. ECAs are typically represented as a rule table, where each row in the table corresponds to a unique configuration of the cell and its two neighbors, and each entry in the row specifies the new state of the central cell in the next time step. The rule table defines the transition function for the ECA, which governs how the system evolves over time. Based on this, Wolfram \cite{wolfram2002new, Wolfram94} classified the ECA rules into following classes:
\begin{enumerate}
	\item[] \textbf{Class I. Homogeneous behavior}: In this class, the system quickly reaches a homogeneous state, where all cells are in the same state and remain so over time. This is true regardless of the initial configuration of the system.  
	\item[] \textbf{Class II. Periodic behavior}: In this class, the system evolves into a periodic pattern, where the pattern repeats after a fixed number of time steps. The period can be simple, such as a single repeating unit, or more complex, with multiple interacting components. Class II ECAs exhibit regular, repetitive behavior that can be easily predicted and described.
	\item[] \textbf{Class III. Chaotic behavior}: In this class, the system evolves into a complex, aperiodic pattern, where the pattern never repeats and appears to be random or chaotic. Class III ECAs exhibit behavior that is difficult to predict or describe, and small changes in the initial conditions can lead to vastly different outcomes.
	\item[] \textbf{Class IV. Complex behavior}: In this class, the system evolves into a complex, aperiodic pattern that exhibits a high degree of structure and organization. Class IV ECA exhibit behavior that is both complex and predictable, and they are often associated with emergent phenomena and self-organization.
\end{enumerate}	

Li and Packard observed that classification of ECA by Wolfram does not fully distinguish the rules of one class from another, as some rules exhibit two types of behavior, such as chaotic behavior in some regions or two chaotic sections separated by a barrier~\cite{Li90thestructure, Genaro13}. Local chaos is one example of such behavior. Li and Packard revised Wolfram's classification in 1990 and divided the ECA rules into five categories: null, fixed point, periodic, locally chaotic, and chaotic, based on rule space analysis that evaluates the probability of a rule being connected to another rule. Table~\ref{table1} shows the minimum representative rules classified by Stephen Wolfram. According to  Li-Packard's classification Class II is further divided into three class. Row with purple color represents class of fixed point. Row with green color represents class of periodic and row with blue color represnts class of locally chaotic.

	\begin{table}[!htpb]
		\scriptsize
		\begin{center}
			\caption{Wolfram's classification for 88 minimal representative ECAs.}
			\label{table1}
			\begin{adjustbox}{width=\columnwidth,center}
				\begin{tabular}{| c |c c c c c c c c c c c c|} 
					\hline 
					\rowcolor[HTML]{C0C0C0} 
					Class I & 0 & 8 & 32 & 40 & 128 & 136 & 160 & 168 &  &  &  & \\ \hline
					\rowcolor[HTML]{DDA0DD} 
					Class II & 2 & 4 & 10 & 12 & 13 & 24 & 34 & 36 & 42 & 44 & 46 & 56\\ 
					\rowcolor[HTML]{DDA0DD} 
					& 57 & 58 & 72 & 76 & 77 & 78 & 104 & 130 & 132 & 138 & 140 & 152\\
					\rowcolor[HTML]{DDA0DD} 
					& 162 & 164 & 170 & 172 & 184 & 200 & 204 & 232 &&&& \\
					\rowcolor[HTML]{7FFF00} 
					& 1 & 3 & 5  &  6 & 7 & 9 & 11 & 14 & 15 & 19 & 23 & 25\\
					\rowcolor[HTML]{7FFF00} 
					& 27 & 28 & 29  &  33 & 35 & 37 & 38 & 43 & 50 & 51 & 62 & 74\\
					\rowcolor[HTML]{7FFF00} 
					 & 94 & 108 & 134 & 142 & 156 & 178 &&&&&&\\
					\rowcolor[HTML]{87CEFA} 
					& 26 & 73 & 154 &  &  &  &  &  &   &  & & \\ \hline
					\rowcolor[HTML]{FFFF00} 
					Class III & 18 & 22 & 30 & 45 & 60 & 90 & 105 & 122 & 126 & 146 & 150 & \\  \hline
					\rowcolor[HTML]{F08080} 
					Class IV & 41 & 54 & 106 & 110 &  & &  & &  &  &  & \\
					\hline
				\end{tabular}
			\end{adjustbox}
		\end{center}
	\end{table}

On the basis of above classifications, we aim to explore the relationship between the dynamics of two layers of layered cellular automata (LCA) and their resulting classification according to Wolfram \cite{wolfram2002new, Wolfram94} and Li-Packard \cite{Li90thestructure}. Specifically, we investigate whether the dynamics of layer 1 can influence the classification of layer 0. Similar to the classification of the temporally stochastic CAs~\cite{subrata2022}, based on the dynamics, the LCAs are classified as follows:

\textbf{Class A}: This class corresponds to the behavior observed in Wolfram's class I, which is characterized by homogeneity.

\textbf{Class B}: This class corresponds to the behavior observed in Wolfram's class II, which includes fixed-point and periodic behavior but excludes Li-Packard's locally-chaotic rules. 

\textbf{Class C}: This class corresponds to the behavior observed in Wolfram's class III and IV, as well as locally-chaotic rules according to Li-Packard's classification.


\textbf{Phase Transition}: Phase transition is a fascinating phenomenon that has been studied in various non-classical Cellular Automata~\cite{BoureFC12, jca/Fates14, ALONSOSANZ2005383, doi:10.}. It refers to a significant change in the behavior of the system based on a critical value of the non-uniformity rate. The non-uniformity rate represents aspects such as the synchrony rate of a non-classical updating scheme, the mixing rate of different CA rules, or any other non-uniform scheme mixing rate.
In the context of phase transition, two distinct phases are observed: the passive phase and the active phase. In the passive phase, the system converges to a homogeneous fixed point where all cells have the same value, often resulting in a configuration of all 0s. On the other hand, the active phase is characterized by the system exhibiting oscillatory behavior around a fixed non-zero density.
Phase transition has been investigated in various types of non-classical CAs. For instance, the $\alpha$-, $\beta$-, and $\gamma$-synchronous updating schemes have been found to exhibit this phenomenon \cite{BoureFC12, jca/Fates09}. In these cases, the critical non-uniformity rate acts as a threshold that distinguishes between the passive and active phases.
Moreover, the occurrence of phase transition has been explored in Elementary Cellular Automata (ECA) with memory \cite{ALONSOSANZ2005383, doi:10.}. The introduction of memory in ECAs, which allows the cells to retain information about their past states, has been found to affect the transition between the passive and active phases.
More recently, phase transition has been studied in ``Diploid'' cellular automata, which are created by randomly mixing two deterministic ECA rules \cite{Fates17}. The behavior of Diploid CAs exhibits an abrupt change characterized by the emergence of phase transition at a critical non-uniformity rate.
Overall, the investigation of phase transition in non-classical CAs highlights various complex dynamics that can arise from variations in updating schemes, mixing rates, and rule combinations. It provides insights into the emergence of different phases and the critical thresholds that govern their transitions, shedding light on the fundamental behavior of these dynamic systems.


\textbf{Class Transition}: In addition to phase transition, another interesting phenomenon observed in cellular automata is known as ``class transition''. Class transition refers to a block size ($b$)-sensitive dynamic behavior in which the cellular system undergoes a change in its class dynamics at a critical value of $b$.
During the class transition, a critical block size value is identified, and beyond this value, the system's class dynamics shift from one class to another. This implies that the behavior of the cellular automaton becomes fundamentally different depending on the chosen block size.
The occurrence of class transition has been studied in various cellular automata systems. Researchers have explored how changes in certain parameter affect the overall dynamics and behavior of the system. For example, in layered cellular automata, increasing the block size can lead to a transition from homogeneous behavior to periodic behavior or from periodic behavior to chaotic behavior.
The critical value of block size at which class transition occurs can vary depending on the specific cellular automaton rule and the characteristics of the system under study. It represents a threshold at which the system undergoes a qualitative change in its dynamics.
Studying class transition in cellular automata provides insights into the sensitivity of these systems to the block size parameter and highlights the complex relationship between block size and emergent behavior. It contributes to our understanding of the factors that influence the dynamics of cellular automata and can help uncover new insights into the behavior of these systems.

In our study, we aim to explore the qualitative transformations that a cellular automaton (CA) can undergo when the block size is progressively varied. To achieve this, we employ a traditional approach where we visually compare the evolution of configurations, i.e., the space-time diagrams, over a few time steps. This allows us to observe visible changes in the system's behavior as we manipulate the noise rate.
To conduct our experiments, we start with a configuration of a fixed CA size of 500 and let the system evolve for 2000 time steps. However, since it is possible for the system to exhibit different behavior for different runs, we repeat each instance (i.e., for each ($f, g^b$)) 10 times to obtain a more accurate and reliable result. 
Moreover, to ensure the generalizability of our findings, we conduct experiments with various other CA sizes and find that the dynamical behavior of the automaton remains almost the same for different sizes. Therefore, we claim that the cellular system's dynamics are consistent across different CA sizes.

Our research question is whether a change in the class of an ECA occurs when layer 1 influences the dynamics of ECA in layer 0. For example, whether the application of layer 1 can cause an ECA that belongs to a periodic class in layer 0 to exhibit properties closer to those of a chaotic class. We also investigate whether there exists a critical value of $b$ that leads to a phase transition or class transition in the behavior of the ECA.
Through our analysis, we aim to provide numerous examples of these types of LCAs, which may have applications in the study of physical, chemical, and biological systems. Our study highlights the importance of understanding the complex behavior of LCAs and their potential for modeling a variety of natural phenomena.

\section{LCA based on counting}
In his work, Wolfram introduced a classification system for the 256 elementary cellular automata rules \cite{wolfram2002new, Wolfram94}, categorizing them based on their dynamical behavior. The goal of this classification was to provide a systematic framework for understanding and analyzing the diverse dynamics that can emerge from simple local update rules. As discussed in the previous section, these rules were divided into four distinct classes.

In this section, we examine the dynamic properties of the 256 elementary cellular automaton (ECA) rules when applied in a layered cellular automata model. The model consists of two layers: Layer 0, where ECA rules are applied to individual cells, and Layer 1, where different counting schemes are applied to blocks of cells, as discussed in previous chapters.
We denote the ECA rules applied in Layer 0 as Rule $f$, and the different counting schemes applied in Layer 1 as Rule $g$. We explore all 256 ECA rules for various block sizes ($b$). We consider both block sizes that can evenly divide the configuration size ($n$) and those that cannot. Counting scheme used in LCA produce some of the interesting dynamical behavior which has been mapped to different class A, B and C based on the dynamics. Some of LCAs cannot be mapped into these class. Those type of LCAs are said to exhibit \emph{phase transition} and \emph{class transition}.

In the context of our analysis, we examine two rules, denoted as $f$ and $g$, which operate on different layers. We assign the class of $f$ as $\zeta(f)$. The class of the LCA($f,g^b$) is represented by $\zeta(f,g^b)$. By closely studying the space-time diagram, we can extract valuable insights regarding the dynamics of the LCA, which can be summarized as follows:

\begin{itemize}
	\item When the dynamics of rule $f$ and the LCA($f,g^b$) show the same class of dynamics, denoted as $\zeta(f) = \zeta(f,g^b)$, it means that the behavior of the LCA($f,g^b$) is similar to that of rule $f$ for different block sizes. In other words, regardless of the block size $b$, the LCA($f,g^b$) exhibits dynamics that can be attributed to rule $f$ in terms of its qualitative behavior.	
	This observation suggests that the influence of rule $g$ in the LCA($f,g^b$) is not significant enough to alter the class of dynamics exhibited by rule $f$. The dynamics of the LCA($f,g^b$) remain consistent with the dynamics of rule $f$ across different block sizes, indicating that the behavior of rule $f$ dominates in shaping the overall dynamics of the LCA system.
	
	\item  When the dynamics of rule $f$ and the LCA($f,g^b$) differ from each other, denoted as $\zeta(f) \neq \zeta(f,g^b)$, it implies that the behavior of the LCA($f,g^b$) deviates from that of rule $f$ for different block sizes.	
	In this case, the presence of rule $g$ introduces additional dynamics and influences the overall behavior of the LCA system. The LCA($f,g^b$) exhibits dynamics that are distinct from those of rule $f$ alone, indicating that rule $g$ has a significant impact on the system's behavior.
\end{itemize}

Now, let's explore some of the possible scenarios that can arise in the two cases mentioned above:

For first case, where $\zeta(f) = \zeta(g)$ we can find following scenarios:
\begin{itemize}
	\item[1. ] $f$ and LCA($f,g^b$) shows the dynamics of Class A where the system evolves into homogeneous configuration, denoted as $\zeta(f)=\zeta(f,g^b)=\text{Class A}$.
	\item[2. ] $f$ and LCA($f,g^b$) shows the dynamics of Class B, where the system tends to show periodicity, denoted as $\zeta(f)=\zeta(f,g^b)=\text{Class B}$.
	\item[3. ] $f$ and LCA($f,g^b$) shows the dynamics of Class C, where the system tends to show chaos and randomness, denoted as $\zeta(f)=\zeta(f,g^b)=\text{Class C}$.
\end{itemize}

In the second case, where $\zeta(f) \neq \zeta(f,g^b)$, we can observe several scenarios:
\begin{itemize}
	\item[1. ] $\zeta(f)=\text{Class A}$ and $\zeta(f,g^b)=\text{Class B}$.
	\item[2. ] $\zeta(f)=\text{Class A}$ and $\zeta(f,g^b)=\text{Class C}$.
	\item[3. ] $\zeta(f)=\text{Class B}$ and $\zeta(f,g^b)=\text{Class A}$.
	\item[4. ] $\zeta(f)=\text{Class B}$ and $\zeta(f,g^b)=\text{Class C}$.
	\item[5. ] $\zeta(f)=\text{Class C}$ and $\zeta(f,g^b)=\text{Class A}$.
	\item[6. ] $\zeta(f)=\text{Class C}$ and $\zeta(f,g^b)=\text{Class B}$.
\end{itemize}

By combining the dynamics of the ECA rules in Layer 0 with the counting schemes in Layer 1, we investigate the resulting behavior and patterns that emerge in the layered cellular automata model. This analysis allows us to gain insights into the interplay between the local dynamics of the ECA rules and the global behavior induced by the counting schemes in the layered model. Let us discuss the dynamical effects of different counting scheme based on the above cases and scenarios.

\subsection{Averaging}
\label{averagedyn}
When changing the block size do not affect the dynamics of rule $f$ i.e., $f$ is chosen from Class C, the resulting dynamics of LCA remain within Class C. In other words, the overall behavior and patterns exhibited by the CA remain unchanged, denoted as $\zeta(f)=\zeta(f,g^b)$.
To illustrate this, we provide three examples of such behavior: LCA($249,g^{b}$), LCA($1,g^b$) and LCA($107,g^b$). In these cases, when the LCA is defined with the update rules $f$ and $g$ (where $f$ is chosen from Class A, B and C respectively), the resulting behavior is equivalent to the behavior of the CA with the update rule $f$.
This suggests that in certain scenarios where specific combinations of update rules are used, the addition of the second rule $g$ does not significantly alter the behavior of the LCA. 

\begin{figure*}[hbt!]
	\begin{center}
		\scalebox{0.7}{
			\begin{tabular}{ccccc}
				ECA 249 & ($249,g^{2}$) & ($249,g^{50}$) & ($249,g^{200}$) & ($249,g^{400}$) \\[6pt]
				\includegraphics[width=31mm]{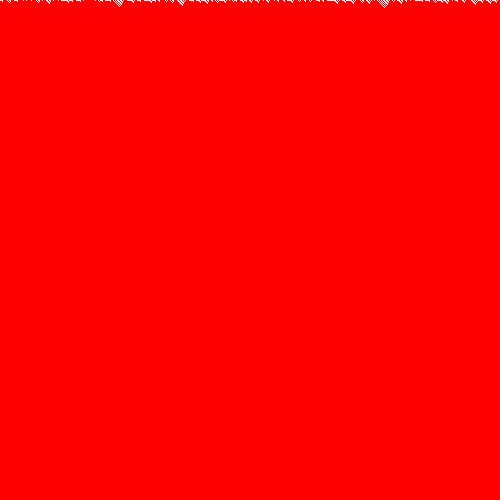} & \includegraphics[width=31mm]{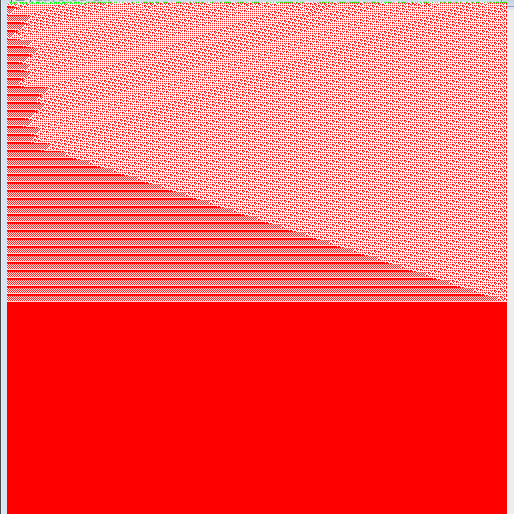} &   \includegraphics[width=31mm]{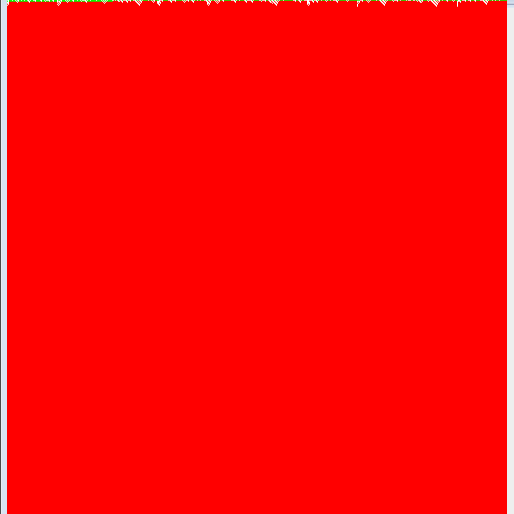} &   \includegraphics[width=31mm]{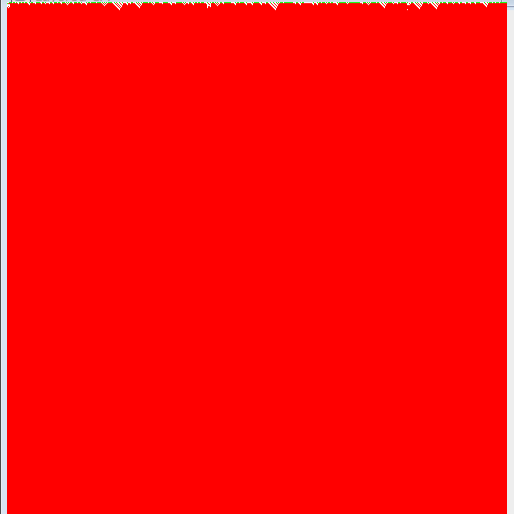} &   \includegraphics[width=31mm]{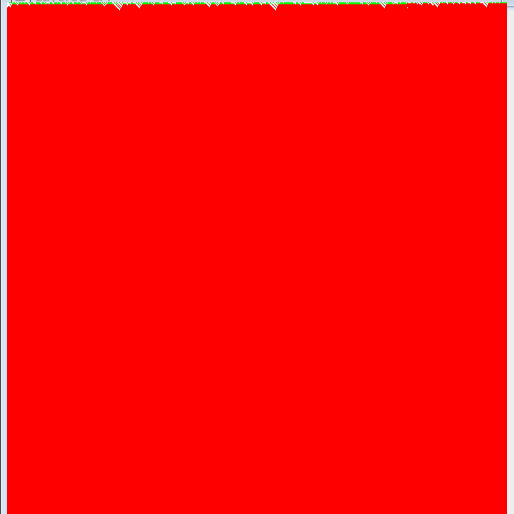} \\
				
				ECA 1 & ($1,g^{50}$) & ($1,g^{150}$) & ($1,g^{250}$) & ($1,g^{350}$) \\[6pt]
				\includegraphics[width=31mm]{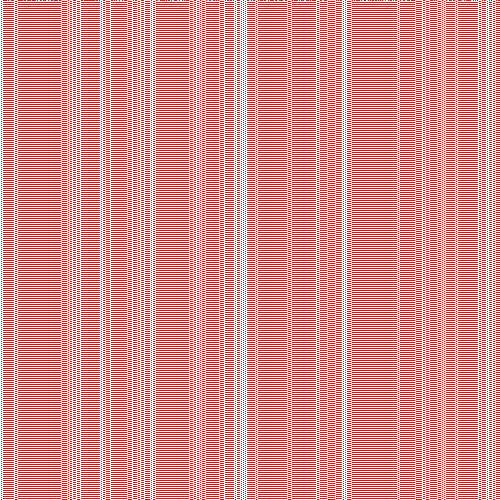} & \includegraphics[width=31mm]{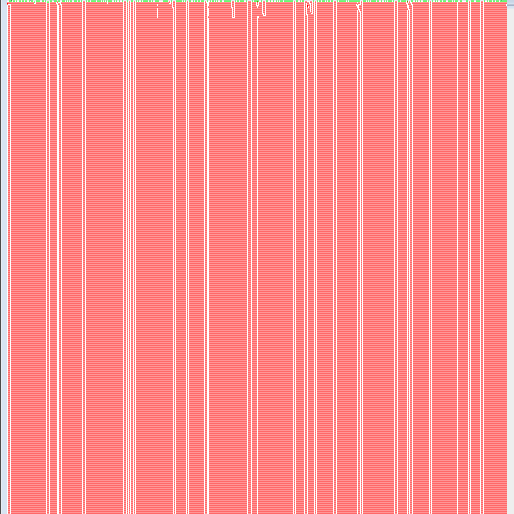} &   \includegraphics[width=31mm]{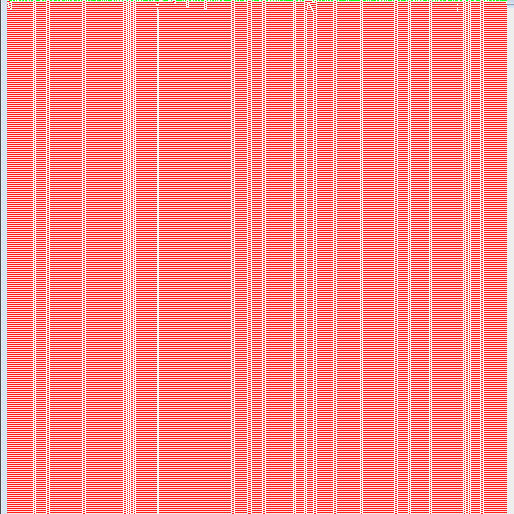} &   \includegraphics[width=31mm]{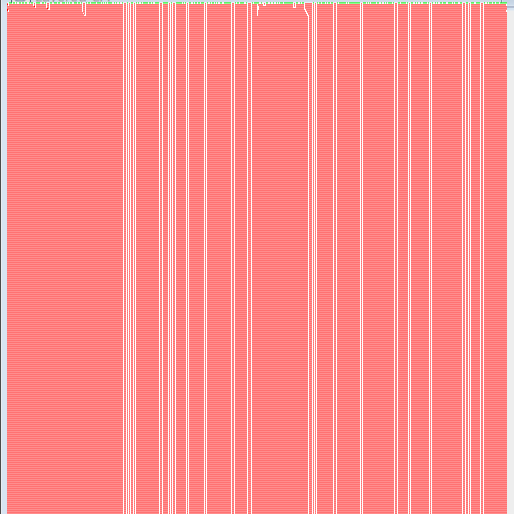} &   \includegraphics[width=31mm]{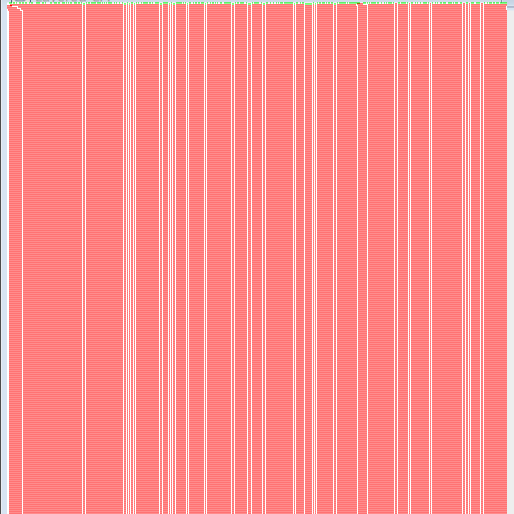} \\
				
				ECA 107 & ($107,g^{4}$) & ($107,g^{25}$) & ($107,g^{150}$) & ($107,g^{300}$) \\[6pt]
				\includegraphics[width=31mm]{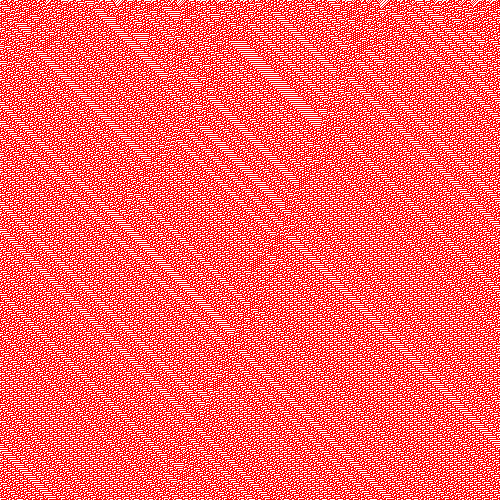} & \includegraphics[width=31mm]{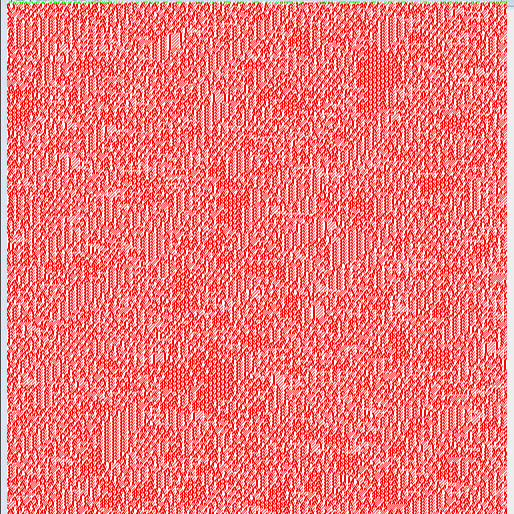} &   \includegraphics[width=31mm]{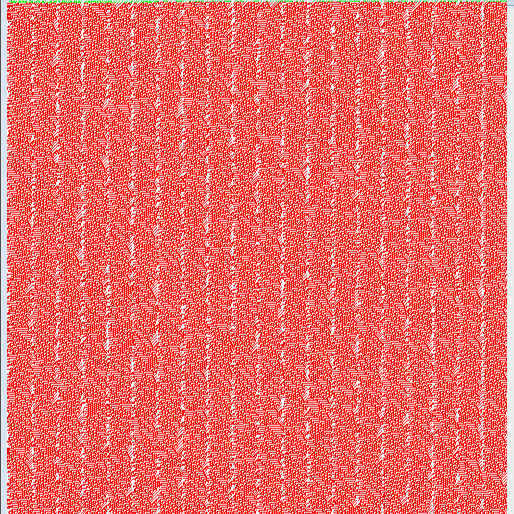} &   \includegraphics[width=31mm]{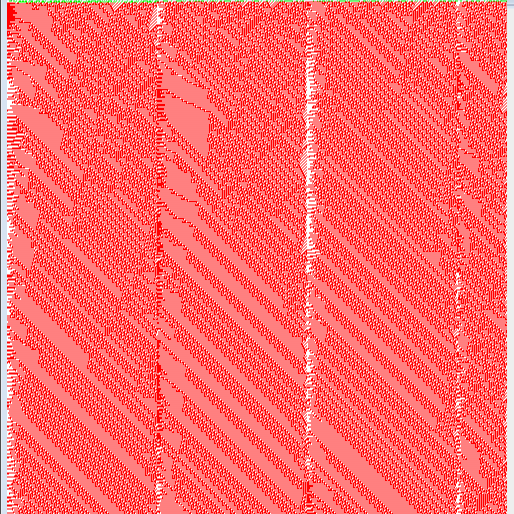} &   \includegraphics[width=31mm]{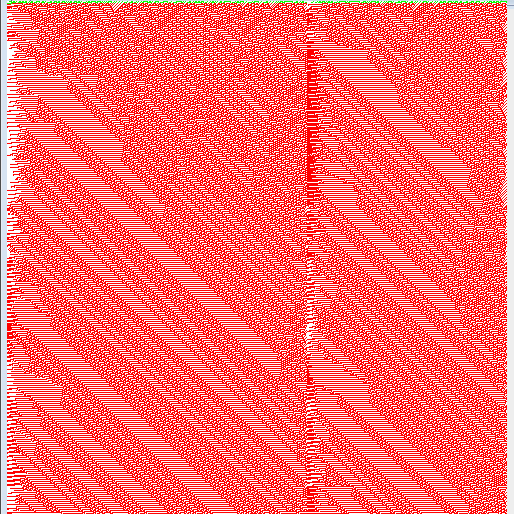} \\
		\end{tabular}}
		\caption{LCA($f,g^{b}$) dynamics when $\zeta$($f,g^{b}$) = $\zeta$($f$) for averaging scheme.}
		\label{avg1}
	\end{center}
\end{figure*}

Figure.~\ref{avg1} depict the space-time diagrams of different LCAs where $\zeta(f)=\zeta(f,g^b)$. In Figure.~\ref{avg1}, ECA $249$, which individually exhibit homogeneous behavior (Wolfram's Class I or Class A dynamics, according to our classification). When ECA $249$ is considered as the default rule ($f$) and noise ($g$) is introduced with varying block sizes, the resulting dynamics remain homogeneous. We provide space-time diagrams for LCA($249,g^b$) with block sizes $b=2,50,200,400$ as examples. Notably, when the block size $b$ is progressively changed, the dynamics of the cellular system remain unaltered. Similarly dynamics for rule $1$ and rule $107$, show the dynamics of Class B and Class C even after $g$ is applied.

\begin{figure*}[hbt!]
	\begin{center}
		\scalebox{0.7}{
			\begin{tabular}{ccccc}
				ECA 234 & ($234,g^{4}$) & ($234,g^{20}$) & ($234,g^{25}$) & ($234,g^{150}$) \\[6pt]
				\includegraphics[width=31mm]{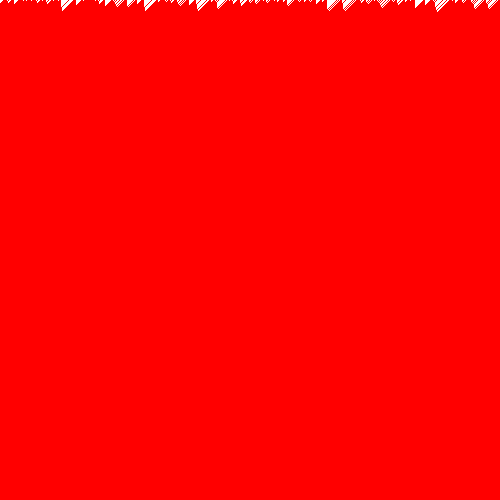} & \includegraphics[width=31mm]{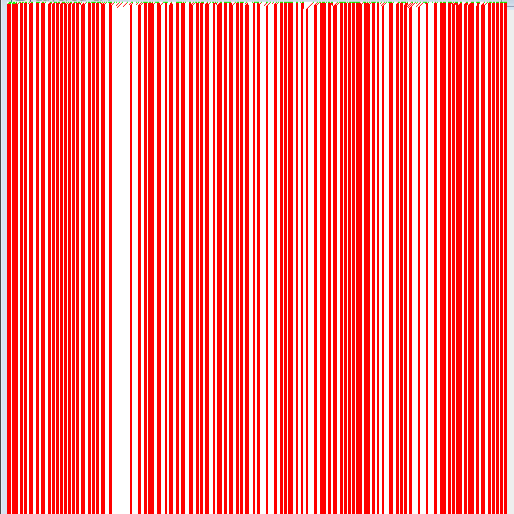} &   \includegraphics[width=31mm]{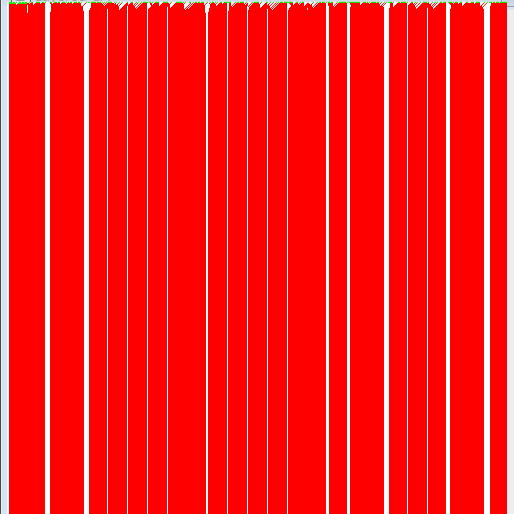} &   \includegraphics[width=31mm]{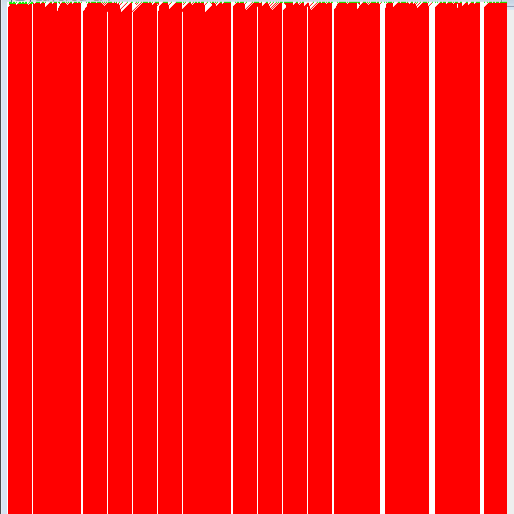} &   \includegraphics[width=31mm]{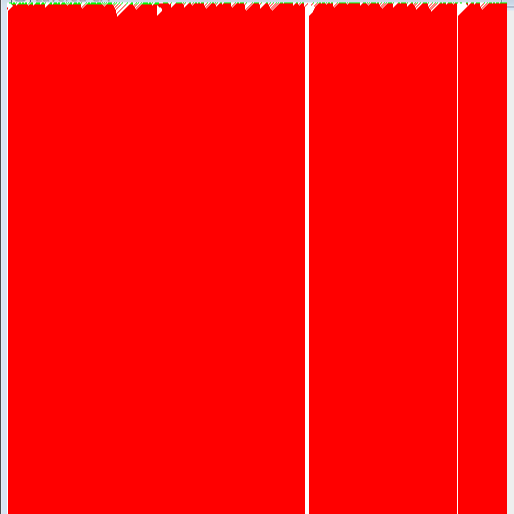} \\
				
				ECA 2 & ($2,g^{100}$) & ($2,g^{125}$) & ($2,g^{300}$) & ($2,g^{400}$) \\[6pt]
				\includegraphics[width=31mm]{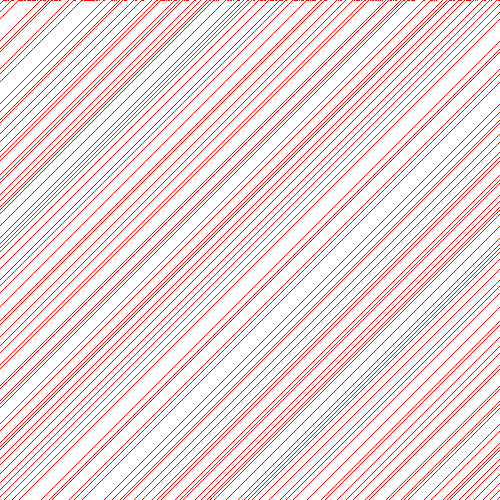} & \includegraphics[width=31mm]{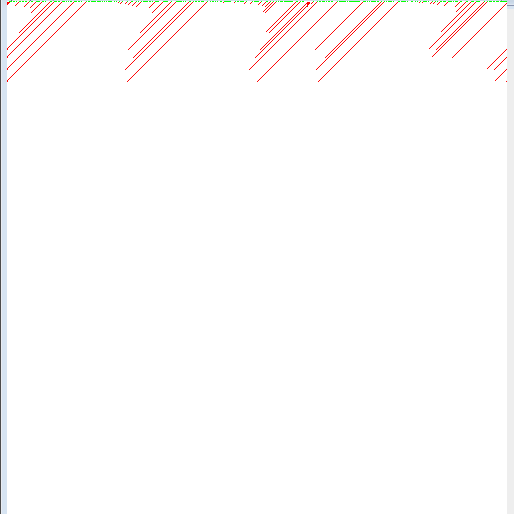} &   \includegraphics[width=31mm]{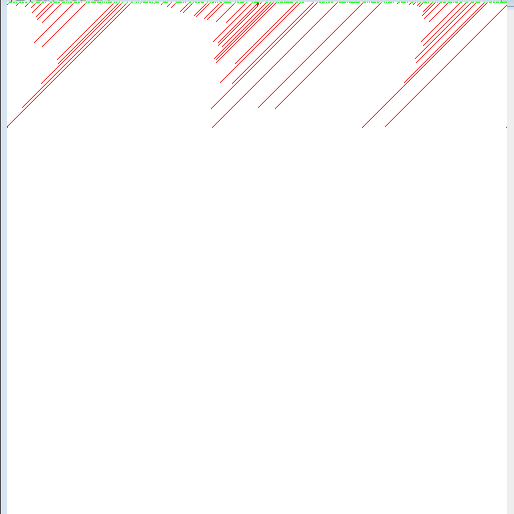} &   \includegraphics[width=31mm]{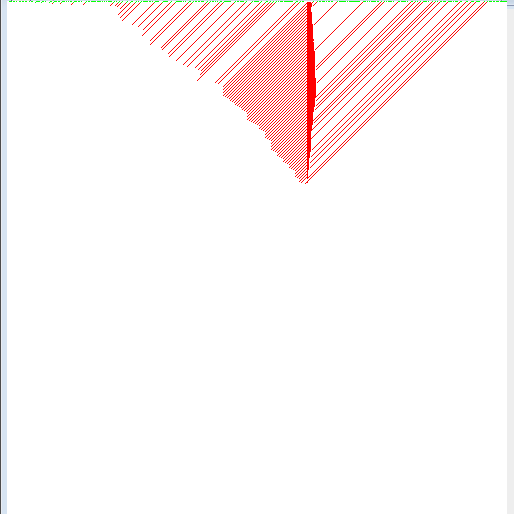} &   \includegraphics[width=31mm]{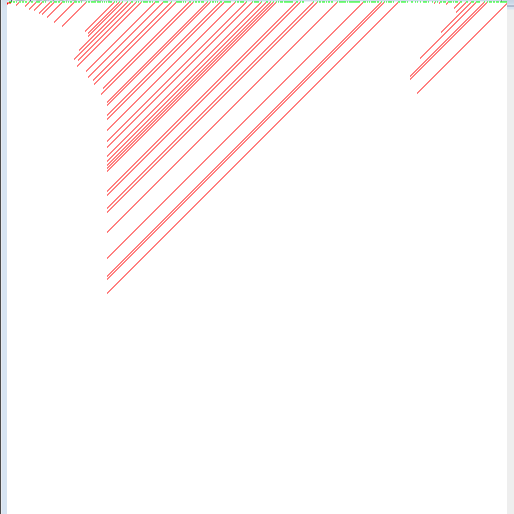} \\
				
				ECA 62 & ($62,g^{20}$) & ($62,g^{125}$) & ($62,g^{150}$) & ($62,g^{200}$) \\[6pt]
				\includegraphics[width=31mm]{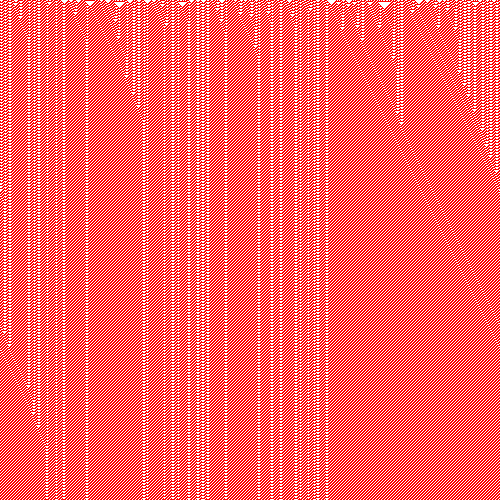} & \includegraphics[width=31mm]{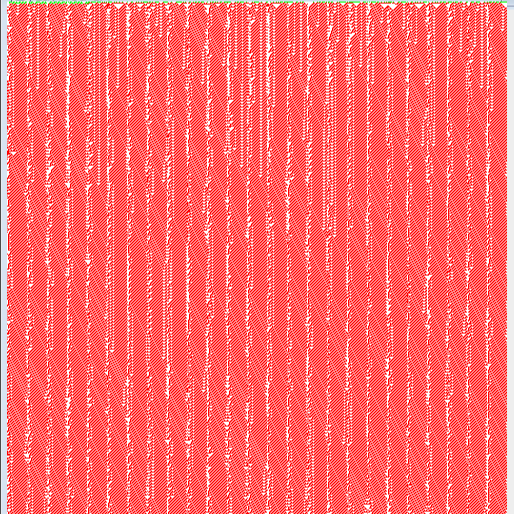} &   \includegraphics[width=31mm]{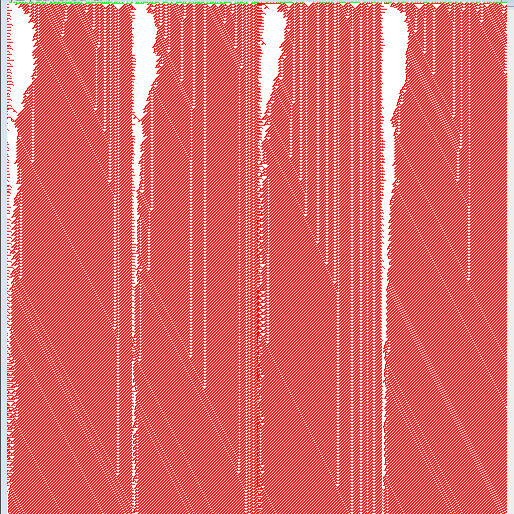} &   \includegraphics[width=31mm]{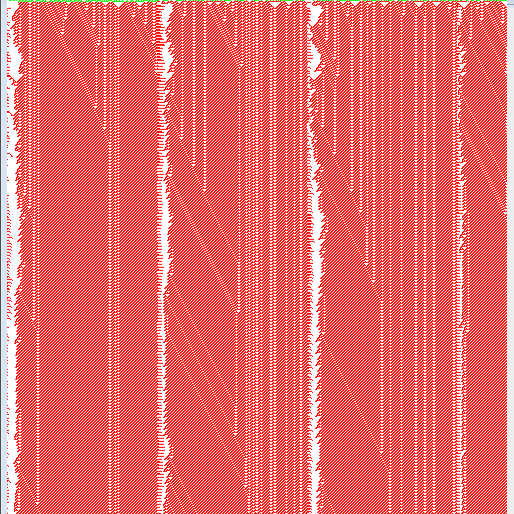} &   \includegraphics[width=31mm]{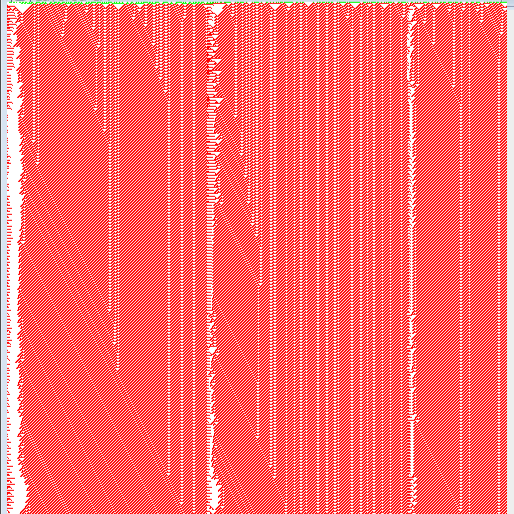} \\
		\end{tabular}}
		\caption{LCA($f,g^{b}$) dynamics when $\zeta$($f,g^{b}$) $\neq$ $\zeta$($f$) for averaging scheme.}
		\label{avg2}
	\end{center}
\end{figure*}

Next observation we made from the analysis of the dynamics is that $g$ having significant influence on the dynamics of $f$. The resulting dynamics of LCA show a change in class. i.e., $f$ is chosen from Class B, the resulting dynamics of LCA changes to Class A. In other words, rule $g$ influenced the behavior and patterns exhibited by the CA show change of class, denoted as $\zeta(f)\neq\zeta(f,g^b)$. One thing is to be noted here that, the change of class is observed for all possible block sizes. 

Figure.~\ref{avg2} depict the space-time diagrams of different LCAs where $\zeta(f)\neq\zeta(f,g^b)$. In Figure.~\ref{avg2}, ECA $234$, which individually exhibit homogeneous behavior (Wolfram's Class I or Class A dynamics, according to our classification). When ECA $234$ is considered as the default rule ($f$) and noise ($g$) is introduced with varying block sizes, the resulting dynamics of LCA shows resemblance towards Class B dynamics. We provide space-time diagrams for LCA($234,g^b$) with block sizes $b=4,20,25,150$ as example for the scenario where $\zeta(f)=$Class A and $\zeta(f,g^b)=$Class B. Notably, when the block size $b$ changed, the dynamics of the cellular system show change in class. Similarly dynamics for rule $2$ and rule $62$ show the dynamics of periodicity, but when $g$ is applied, respective LCAs show the dynamics of Class A and Class C. We didn't find LCAs for rest of the scenario.

\begin{figure*}[hbt!]
	\begin{center}
		\scalebox{0.7}{
			\begin{tabular}{ccccc}
				ECA 240 & ($240,g^{250}$) & ($240,g^{200}$) & ($240,g^{150}$) & ($240,g^{125}$) \\[6pt]
				\includegraphics[width=31mm]{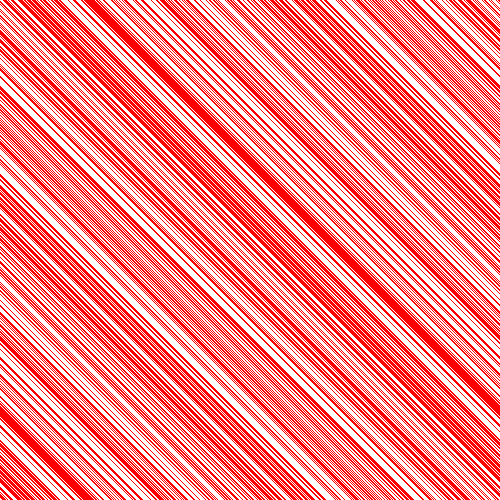} & \includegraphics[width=31mm]{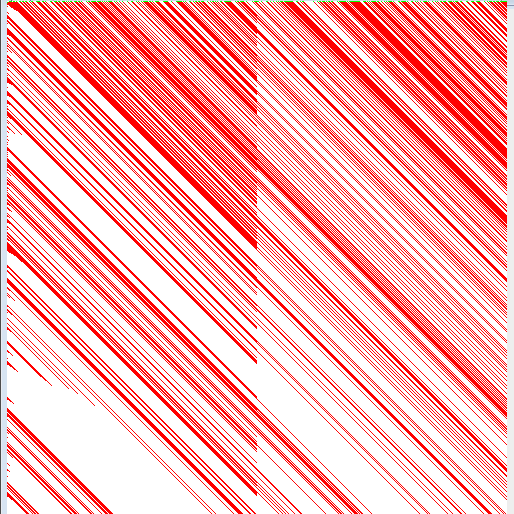} &   \includegraphics[width=31mm]{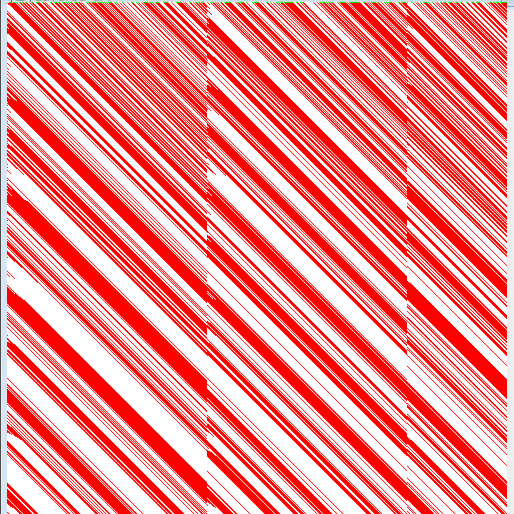} &   \includegraphics[width=31mm]{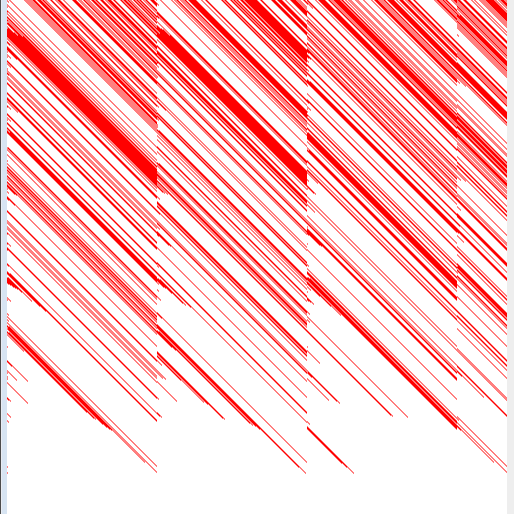} &   \includegraphics[width=31mm]{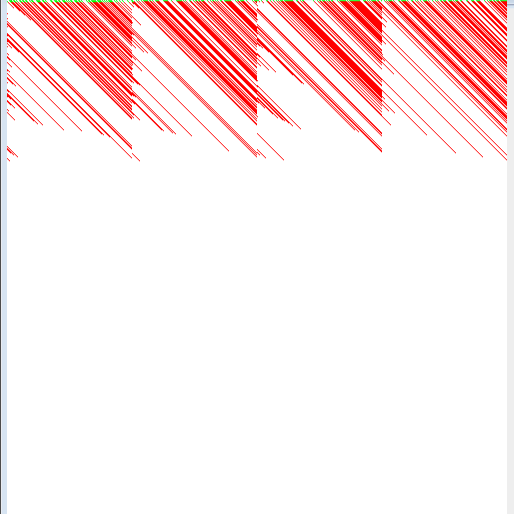} \\
				
				ECA 106 & ($106,g^{150}$) & ($106,g^{128}$) & ($106,g^{127}$) & ($106,g^{100}$) \\[6pt]
				\includegraphics[width=31mm]{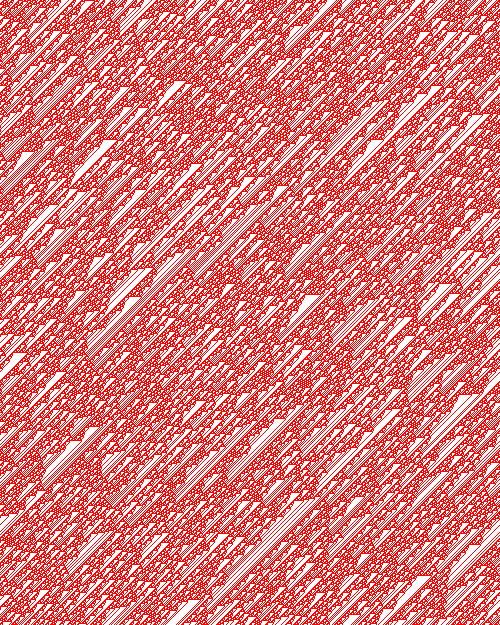} & \includegraphics[width=31mm]{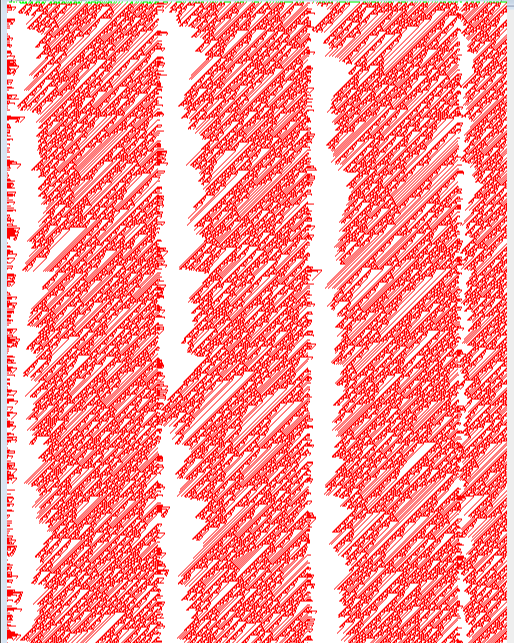} &   \includegraphics[width=31mm]{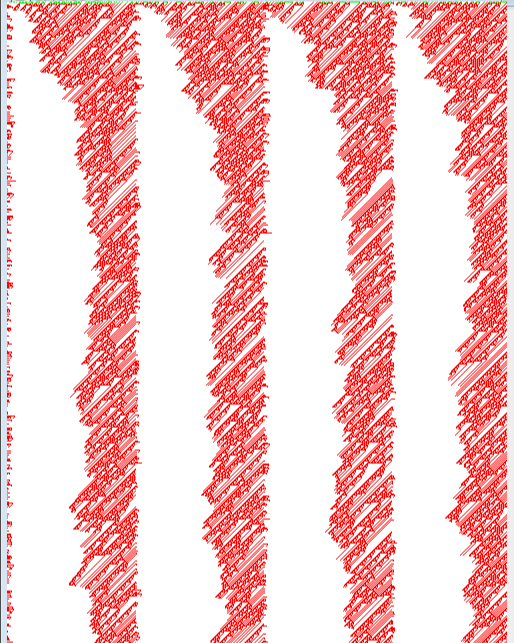} &   \includegraphics[width=31mm]{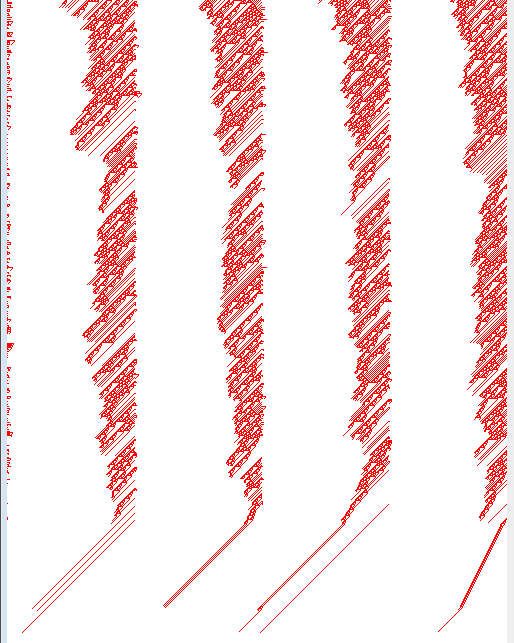} &   \includegraphics[width=31mm]{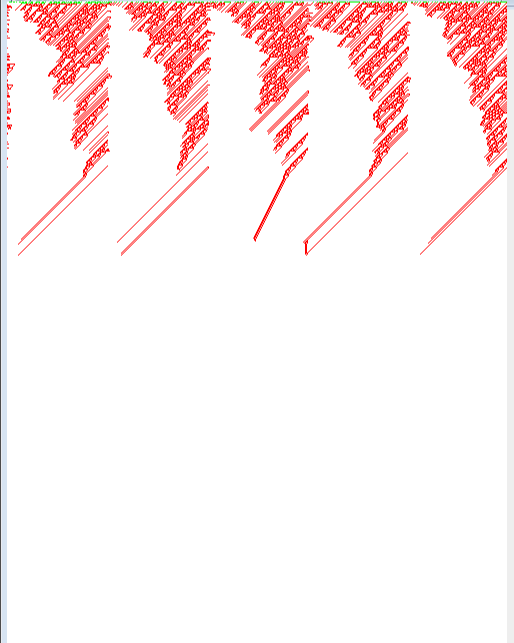} \\
								
				ECA 37 & ($37,g^{400}$) & ($37,g^{250}$) & ($37,g^{200}$) & ($37,g^{25}$) \\[6pt]
				\includegraphics[width=31mm]{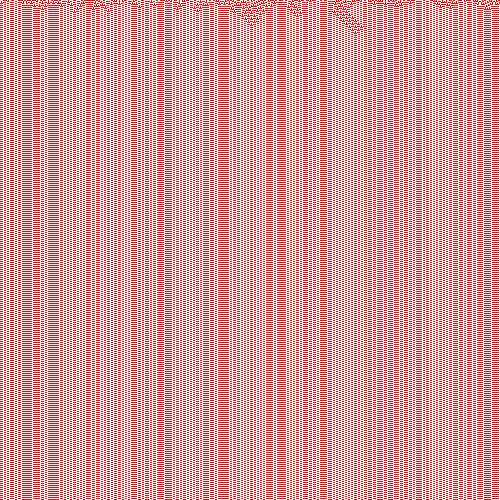} & \includegraphics[width=31mm]{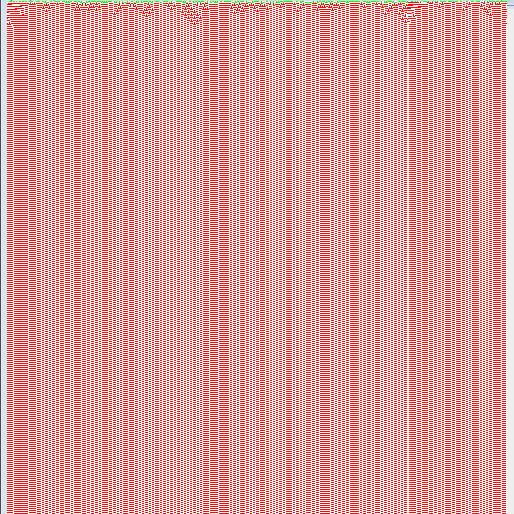} &   \includegraphics[width=31mm]{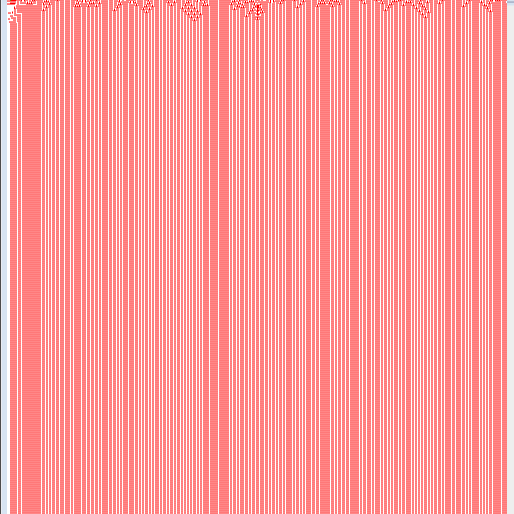} &   \includegraphics[width=31mm]{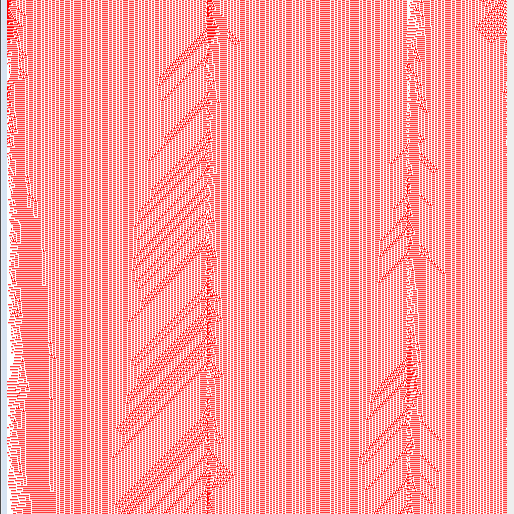} &   \includegraphics[width=31mm]{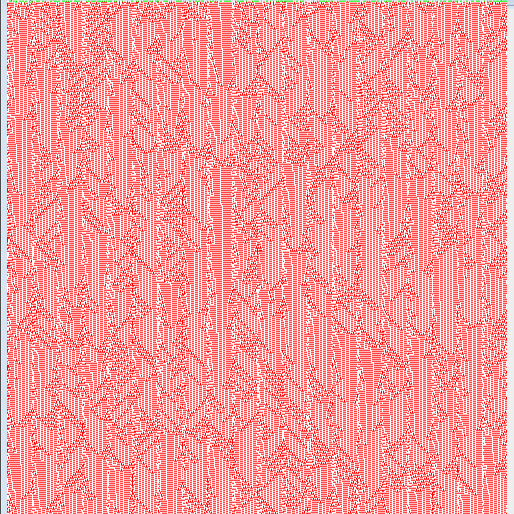} \\
				
				ECA 238 & ($238,g^{150}$) & ($238,g^{50}$) & ($238,g^{25}$) & ($238,g^{10}$) \\[6pt]
				\includegraphics[width=31mm]{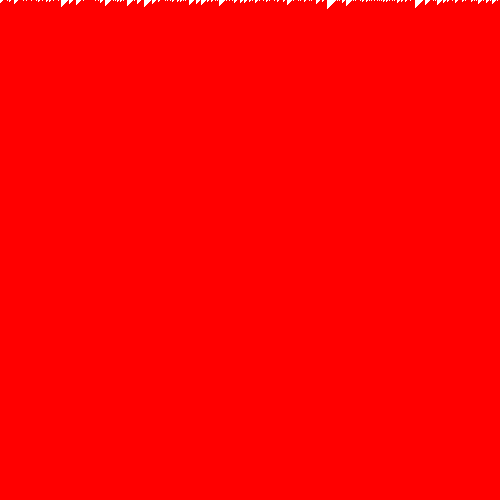} & \includegraphics[width=31mm]{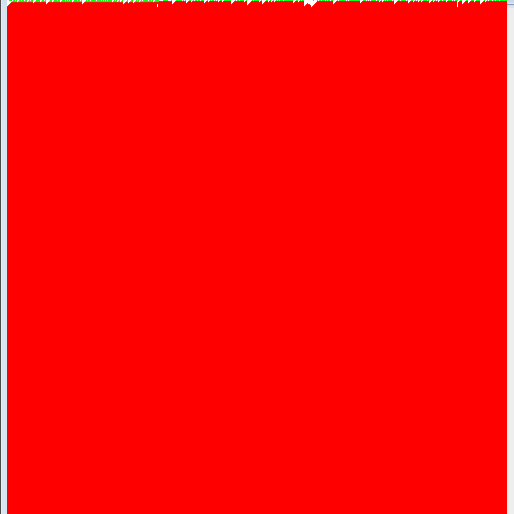} &   \includegraphics[width=31mm]{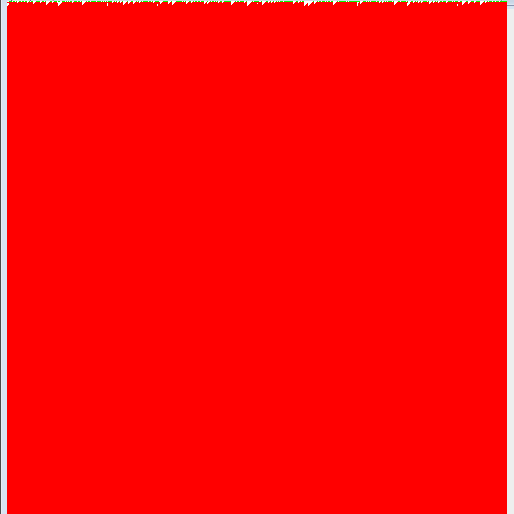} &   \includegraphics[width=31mm]{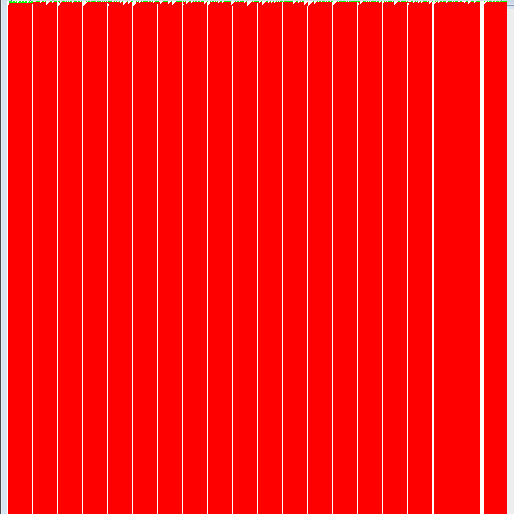} &   \includegraphics[width=31mm]{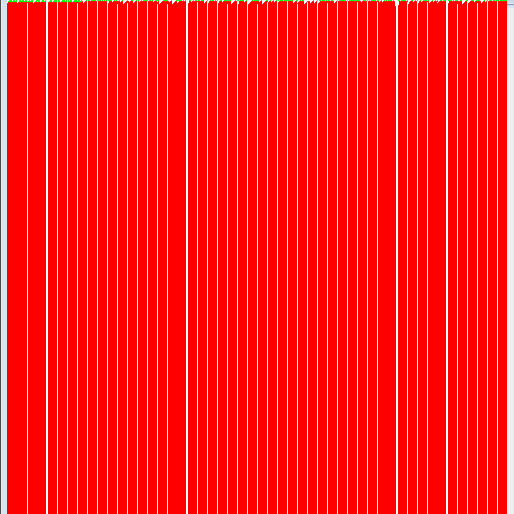} \\
		\end{tabular}}
		\caption{Phase and class transitions in LCA($f,g^{b}$) for averaging scheme.}
		\label{avg3}
	\end{center}
\end{figure*}

Next we investigate the occurrences of phase transition and class transition in the dynamics of LCA. Figure.~\ref{avg3} show examples of LCA showing phase and class transition. ECA 240 and ECA 106 undergoes phase transition for critical block size $b_c=150$ and $b_c=127$. When ECA 240 is used as default rule $f$, on application of noise $g$ by decreasing the block size, the system converges to all-0 configuration when $b$ is 150. Similarly ECA 106 in LCA get converged to all-0 configuration when $b$ is 127. Next, ECA 37 and ECA 238, show class transition for critical block size $b_t=200$ and $b_t=25$. When ECA 37 is used as default rule $f$, on application of noise $g$ by decreasing the block size, the system tends to change its class progressively. At $b=200$, LCA($37,g^b$) changes its state from Class B dynamics to Class C dynamics. Similarly at $b=25$, LCA($238,g^b$) changes its state from Class A dynamics to Class B dynamics, see Figure.~\ref{avg3}.

\subsection{Maximization}

When the dynamics of rule $f$ in a layered cellular automaton (LCA) from Class B remain unaffected by changing the block size, it implies that the resulting dynamics of LCA($f,g^b$) also fall within Class B. This means that the overall behavior and patterns exhibited by the cellular automaton remain unchanged, and we can denote this as $\zeta(f) = \zeta(f,g^b)$.

To illustrate this phenomenon, we can consider three examples: LCA($136,g^b$), LCA($13,g^b$), and LCA($30,g^b$). In these cases, when the LCA is defined with the update rules $f$ and $g$, where $f$ belongs to Class A, B, and C respectively, the resulting behavior is equivalent to that of the CA with the update rule $f$ alone.
This observation suggests that in certain scenarios, when specific combinations of update rules are used, the addition of the second rule $g$ does not significantly alter the behavior of the LCA. The dynamics and patterns remain consistent with the behavior of the original rule $f$, regardless of the block size parameter $b$.

\begin{figure*}[hbt!]
	\begin{center}
		\scalebox{0.7}{
			\begin{tabular}{ccccc}
				ECA 136 & ($136,g^{50}$) & ($136,g^{65}$) & ($136,g^{150}$) & ($136,g^{250}$) \\[6pt]
				\includegraphics[width=31mm]{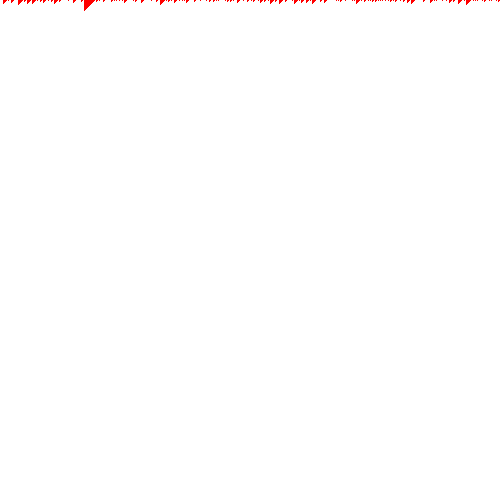} & \includegraphics[width=31mm]{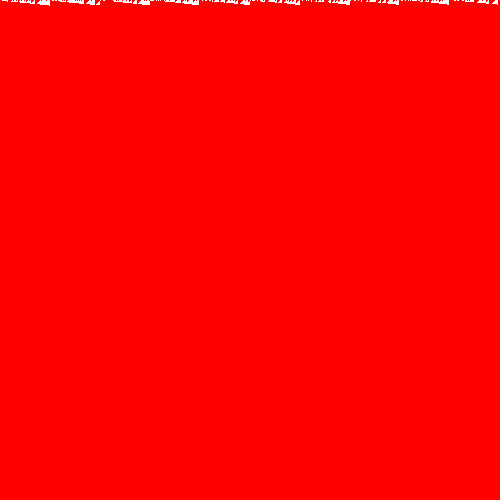} &   \includegraphics[width=31mm]{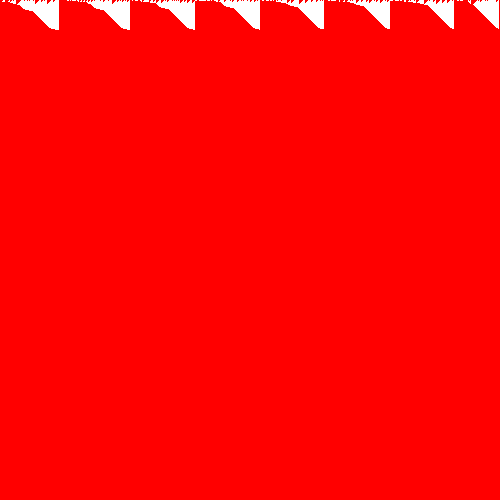} &   \includegraphics[width=31mm]{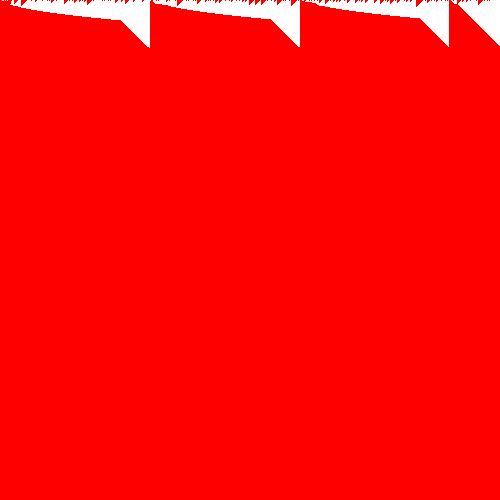} &   \includegraphics[width=31mm]{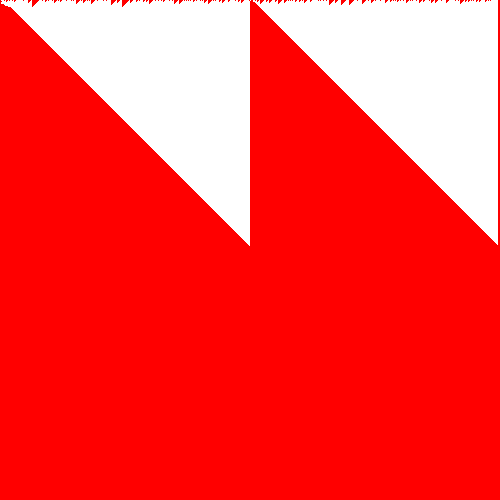} \\
				
				ECA 13 & ($13,g^{25}$) & ($13,g^{65}$) & ($13,g^{100}$) & ($13,g^{300}$) \\[6pt]
				\includegraphics[width=31mm]{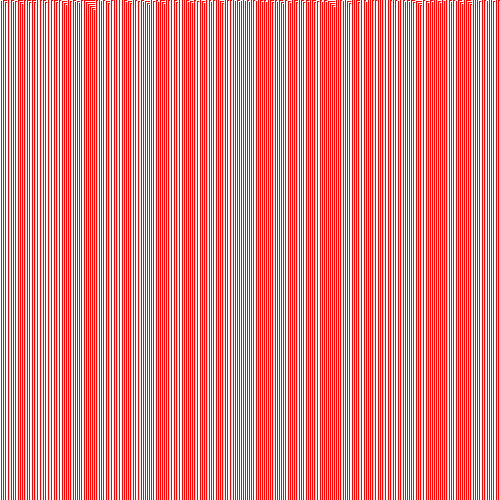} & \includegraphics[width=31mm]{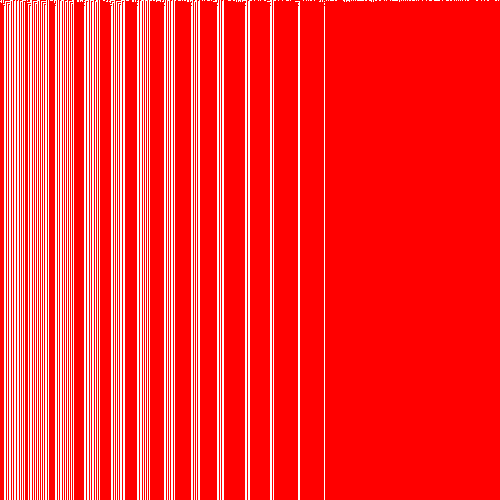} &   \includegraphics[width=31mm]{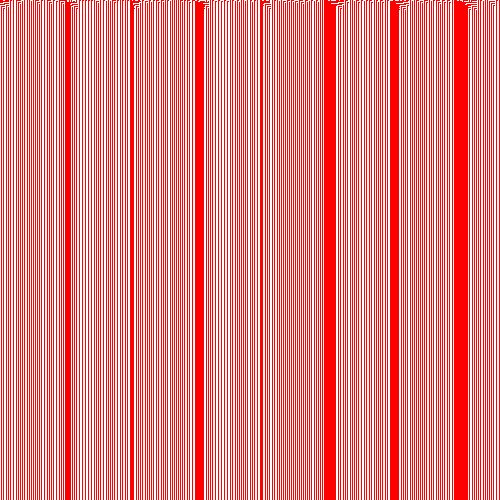} &   \includegraphics[width=31mm]{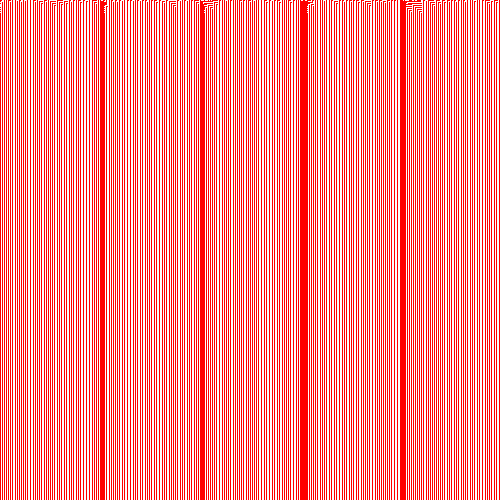} &   \includegraphics[width=31mm]{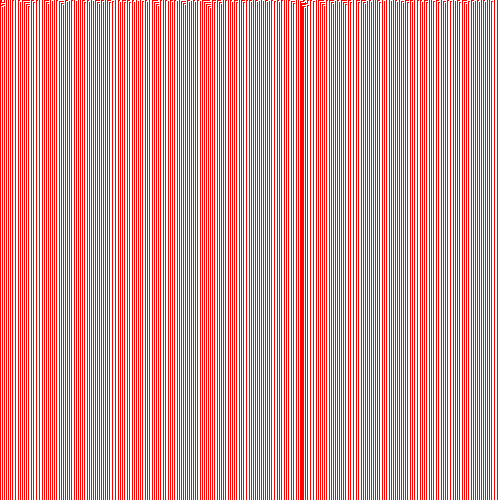} \\
				
				ECA 30 & ($30,g^{100}$) & ($30,g^{150}$) & ($30,g^{250}$) & ($30,g^{350}$) \\[6pt]
				\includegraphics[width=31mm]{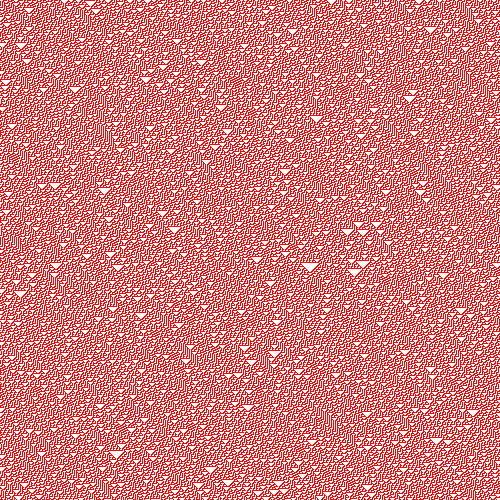} & \includegraphics[width=31mm]{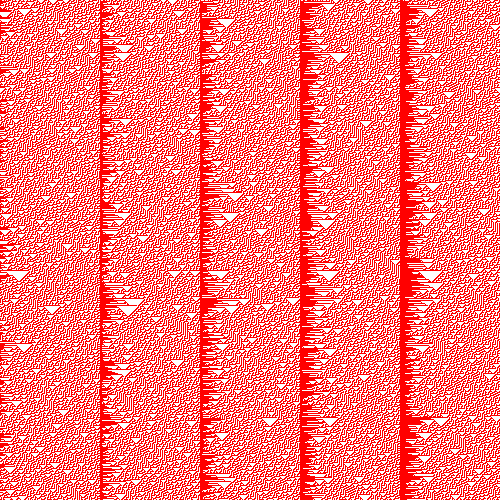} &   \includegraphics[width=31mm]{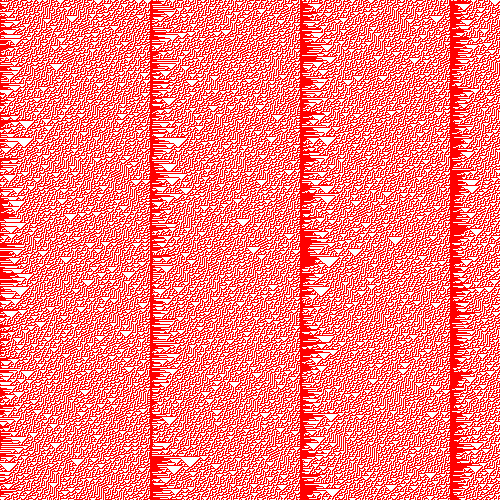} &   \includegraphics[width=31mm]{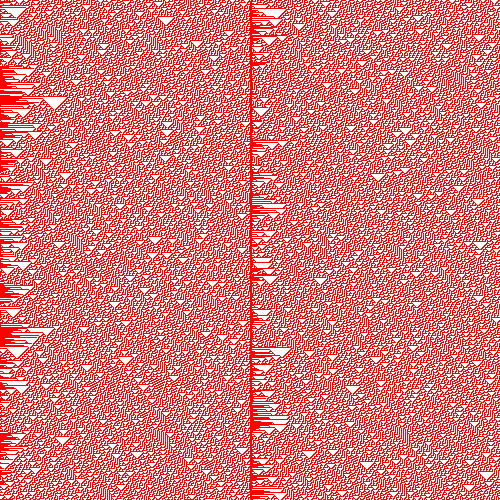} &   \includegraphics[width=31mm]{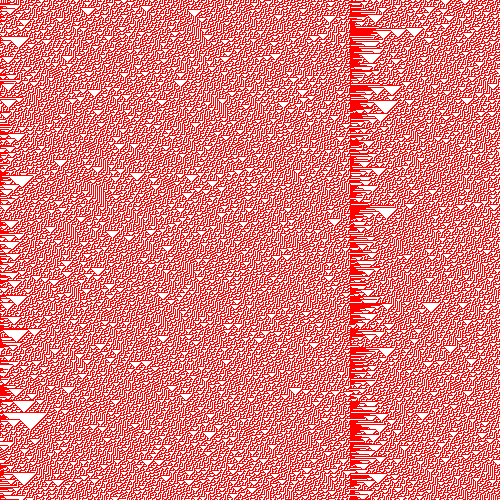} \\
		\end{tabular}}
		\caption{LCA($f,g^{b}$) dynamics when $\zeta$($f,g^{b}$) = $\zeta$($f$) maximization scheme.}
		\label{max1}
	\end{center}
\end{figure*}

In Figure \ref{max1}, we observe the LCA($136,g^b$) with various block sizes. The default rule $f$ is ECA 136, which exhibits homogeneous behavior (belonging to Wolfram's Class I or Class A dynamics, as per our classification) when considered alone. When noise is introduced using different block sizes, the resulting dynamics of LCA($136,g^b$) remain homogeneous. We provide space-time diagrams for LCA($136,g^b$) with block sizes $b=50, 65, 150,$ and $250$ as examples. It is notable that as the block size $b$ is varied, the dynamics of the cellular system remain unchanged.
Similarly, for rule $13$ and rule $30$, which belong to Class B and Class C respectively, applying the noise rule $g$ does not alter their original dynamics. The resulting LCA($13,g^b$) and LCA($30,g^b$) maintain the dynamics of Class B and Class C respectively, even with varying block sizes.

\textbf{Note:} In our analysis of dynamics of LCA based on maximization, we didn't considered $b=1$ as it consider a cell as a block itself. Then if the left, current and right block is in state 1,0,1 respectively. After maximization, it will be 1,1,1. Similarly, 001$\rightarrow$011 and 100$\rightarrow$110. Due to this the whole configuration becomes all-1s.

Next, we observed that the addition of rule $g$ can cause a significant change in the behavior of rule $f$. The resulting dynamics of the LCA may transition to a different class, indicating a distinct pattern of behavior. This highlights the influence of the noise rule on the overall dynamics of the LCA, denoted as $\zeta(f)\neq\zeta(f,g^b)$.

\begin{figure*}[hbt!]
	\begin{center}
		\scalebox{0.7}{
			\begin{tabular}{ccccc}
				ECA 32 & ($32,g^{65}$) & ($32,g^{100}$) & ($32,g^{250}$) & ($32,g^{400}$) \\[6pt]
				\includegraphics[width=31mm]{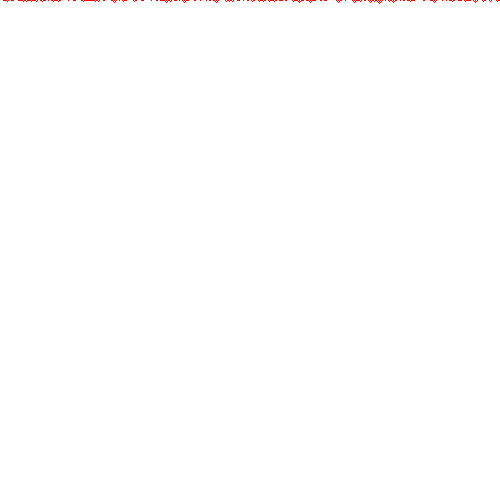} & \includegraphics[width=31mm]{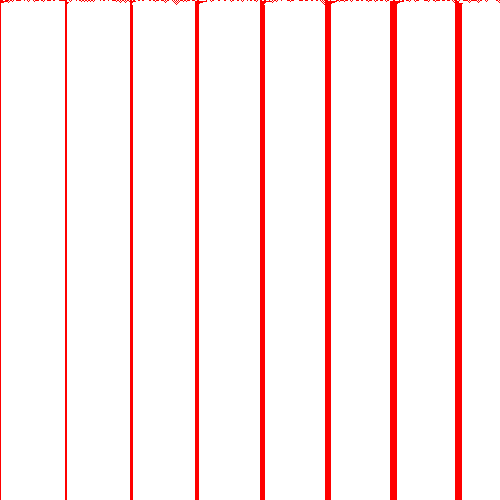} &   \includegraphics[width=31mm]{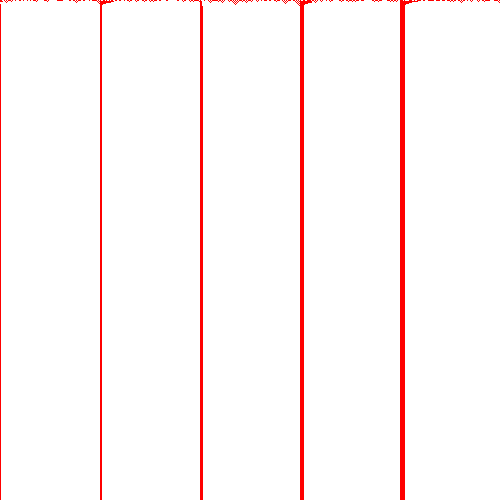} &   \includegraphics[width=31mm]{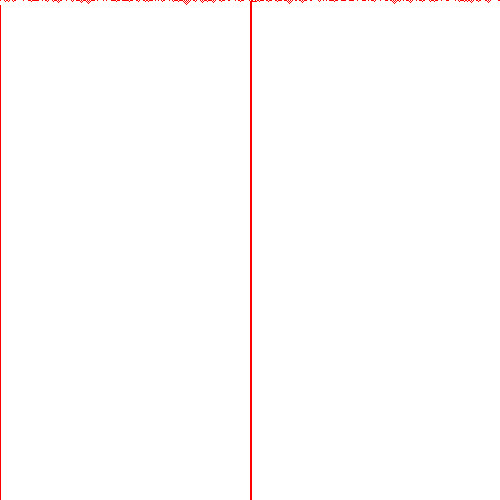} &   \includegraphics[width=31mm]{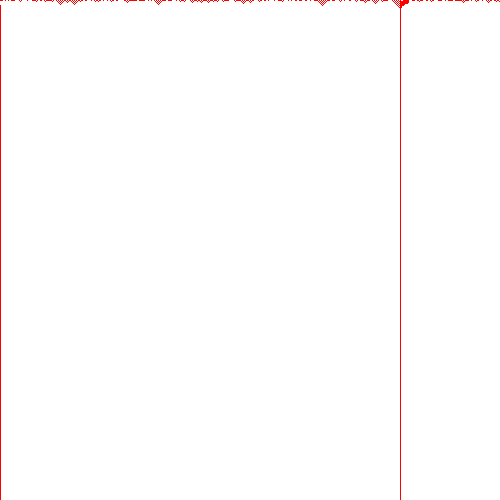} \\
				
				ECA 138 & ($138,g^{50}$) & ($138,g^{65}$) & ($138,g^{250}$) & ($138,g^{400}$) \\[6pt]
				\includegraphics[width=31mm]{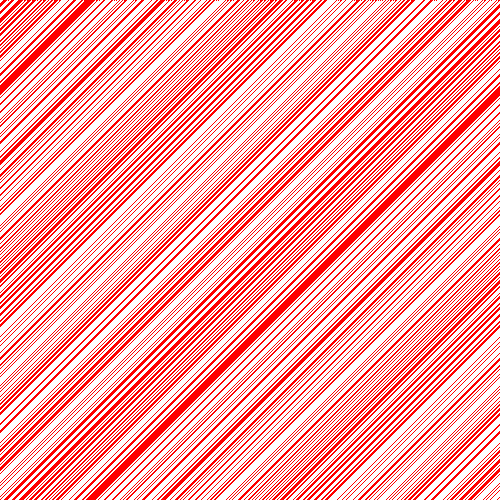} & \includegraphics[width=31mm]{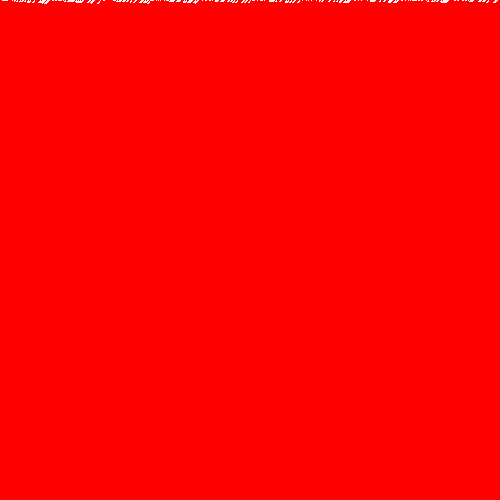} &   \includegraphics[width=31mm]{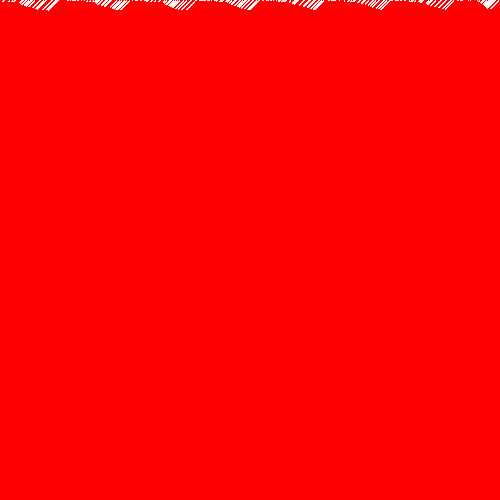} &   \includegraphics[width=31mm]{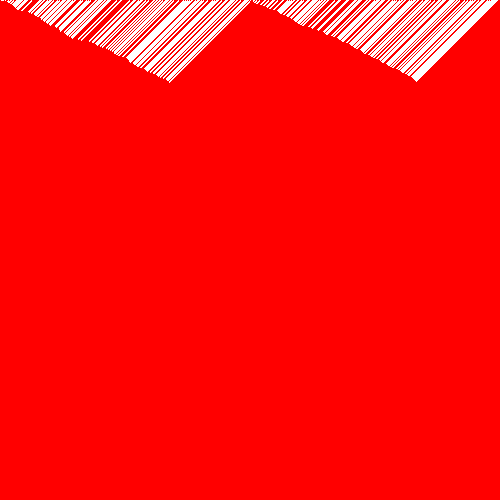} &   \includegraphics[width=31mm]{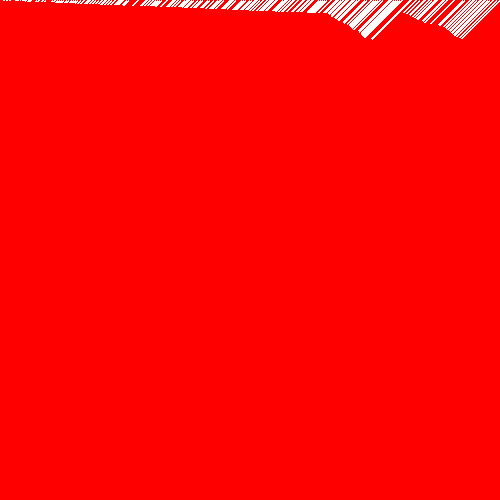} \\
				
				
				ECA 137 & ($137,g^{50}$) & ($137,g^{100}$) & ($137,g^{250}$) & ($137,g^{350}$) \\[6pt]
				\includegraphics[width=31mm]{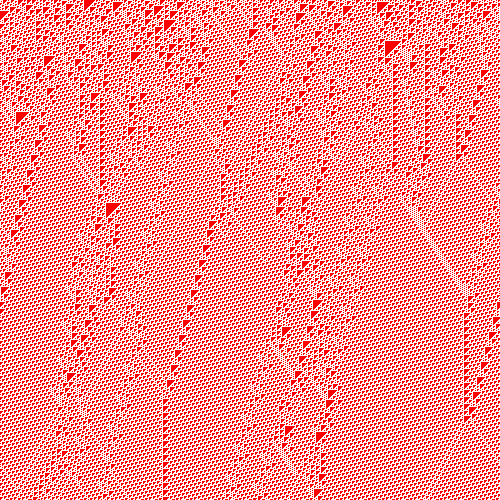} & \includegraphics[width=31mm]{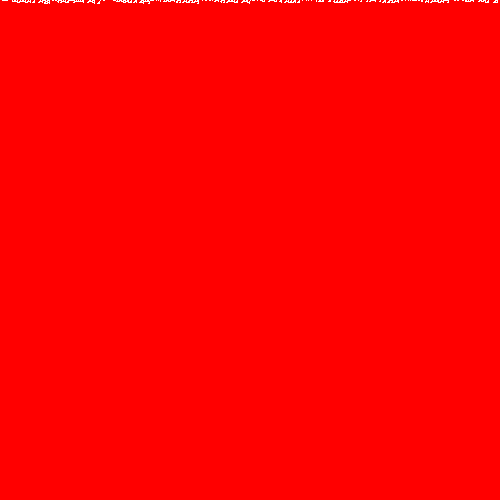} &   \includegraphics[width=31mm]{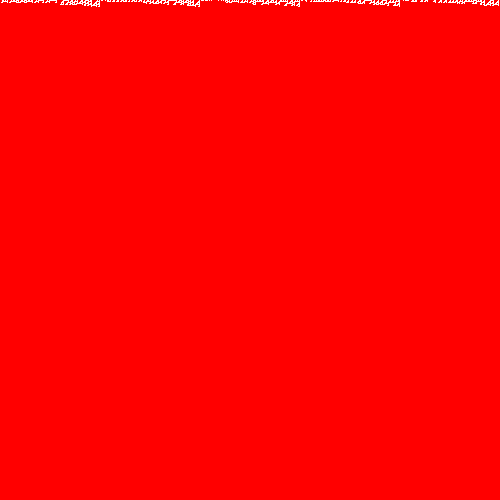} &   \includegraphics[width=31mm]{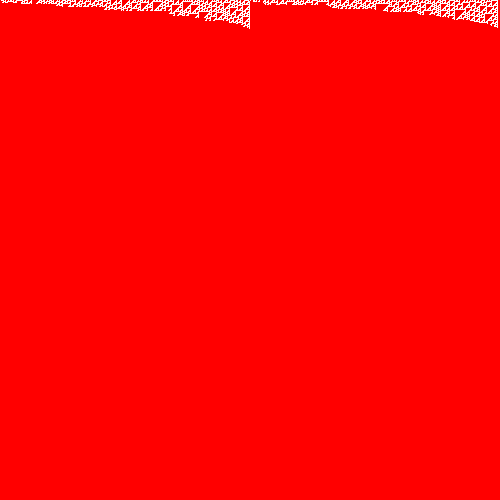} &   \includegraphics[width=31mm]{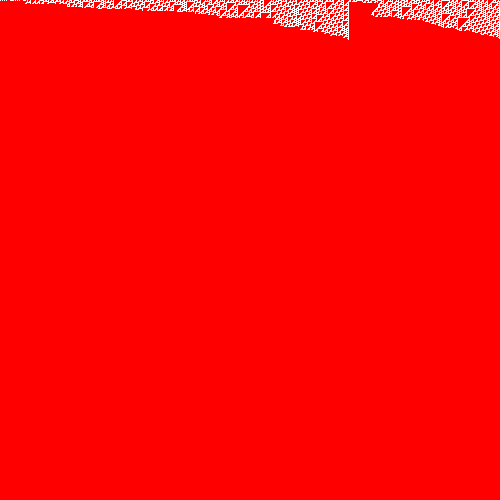} \\
		\end{tabular}}
		\caption{LCA($f,g^{b}$) dynamics when $\zeta$($f,g^{b}$) $\neq$ $\zeta$($f$) for maximization scheme.}
		\label{max2}
	\end{center}
\end{figure*}

Figure~\ref{max2} illustrates the space-time diagrams of various LCAs where the dynamics of $f$ and LCA($f,g^b$) differ, i.e., $\zeta(f)\neq\zeta(f,g^b)$. In this case, we focus on ECA $32$, which exhibits homogeneous behavior (Class A dynamics) when considered individually. However, when noise rule $g$ is introduced with different block sizes, the resulting dynamics of LCA show a resemblance towards Class B dynamics. Space-time diagrams for LCA($32,g^b$) with block sizes $b=65,100,250,400$ are provided as examples of this scenario, where $\zeta(f)=$ Class A and $\zeta(f,g^b)=$ Class B.
Similarly, for ECA $138$ and ECA $137$, which exhibit Class B and Class C dynamics respectively, the introduction of noise rule $g$ leads to the dynamics of Class A in the respective LCAs. We did not find any LCAs that fit the remaining scenarios.

We further examined the occurrence of phase transition and class transition in the dynamics of LCAs. Our investigation did not reveal any instances of phase transition in the studied LCA systems. However, we did observe examples of class transition. 

\begin{figure*}[hbt!]
	\begin{center}
		\scalebox{0.7}{
			\begin{tabular}{ccccc}
				
				ECA 26 & ($26,g^{50}$) & ($26,g^{100}$) & ($26,g^{250}$) & ($26,g^{400}$) \\[6pt]
				\includegraphics[width=31mm]{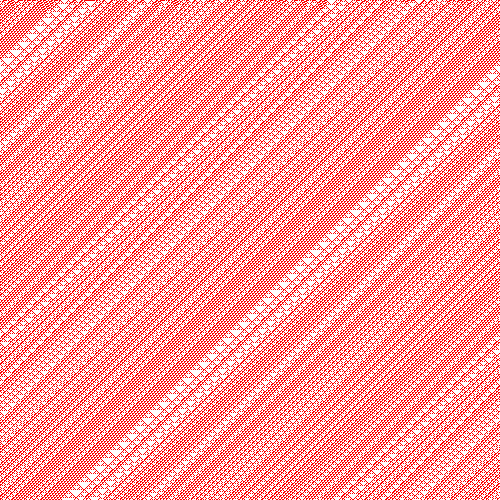} & \includegraphics[width=31mm]{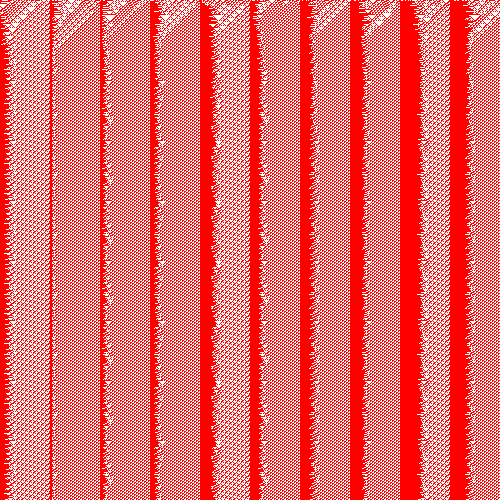} &   \includegraphics[width=31mm]{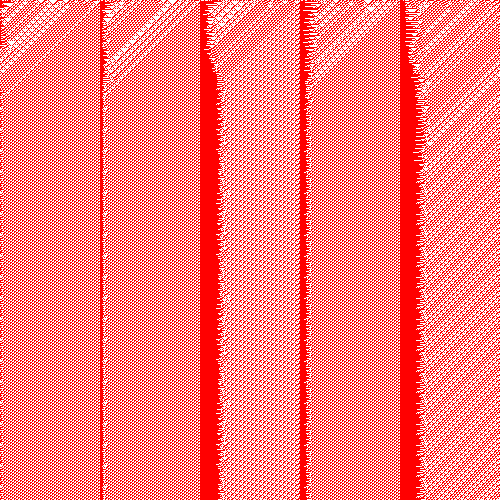} &   \includegraphics[width=31mm]{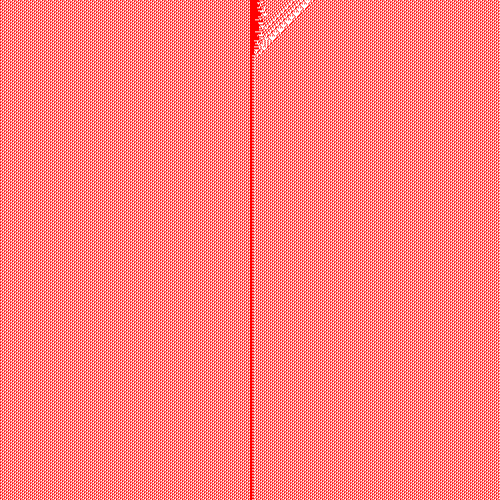} &   \includegraphics[width=31mm]{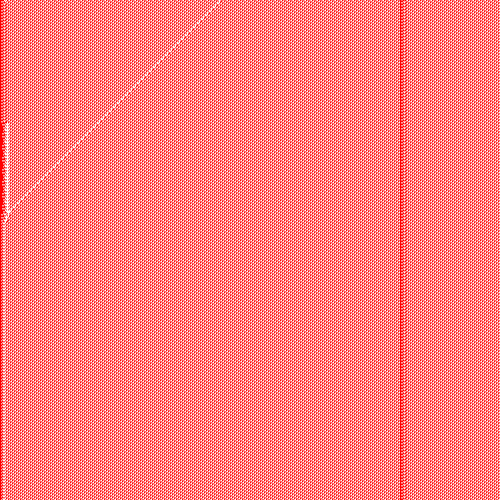} \\
		\end{tabular}}
		\caption{Class transition of LCA($f,g^{b}$) for maximization scheme.}
		\label{max3}
	\end{center}
\end{figure*}

Figure~\ref{max3} illustrates examples of LCAs that exhibit class transition. Specifically, ECA 26 undergoes class transition at a critical block size of $b_t=250$. When ECA 26 is chosen as the default rule $f$, and noise rule $g$ is applied by decreasing the block size, the dynamics of the system progressively change their class. At $b=250$, LCA($26,g^b$) transitions from Class C dynamics to Class B dynamics.

\subsection{Minimization}

When the block size is changed and it does not affect the dynamics of rule $f$, which is chosen from Class C, the resulting dynamics of the LCA also remain within Class C. This means that the overall behavior and patterns exhibited by the cellular automaton (CA) remain unchanged.

To illustrate this observation, we provide three examples: LCA($168,g^{b}$), LCA($57,g^b$), and LCA($105,g^b$). In these cases, when the LCA is defined with the update rules $f$ and $g$ (where $f$ is chosen from Class A, B, and C, respectively), the resulting behavior is equivalent to the behavior of the CA with the update rule $f$. This suggests that in certain scenarios, when specific combinations of update rules are used, the addition of the second rule $g$ does not significantly alter the behavior of the LCA.

\begin{figure*}[hbt!]
	\begin{center}
		\scalebox{0.7}{
			\begin{tabular}{ccccc}
				ECA 168 & ($168,g^{65}$) & ($168,g^{100}$) & ($168,g^{150}$) & ($168,g^{250}$) \\[6pt]
				\includegraphics[width=31mm]{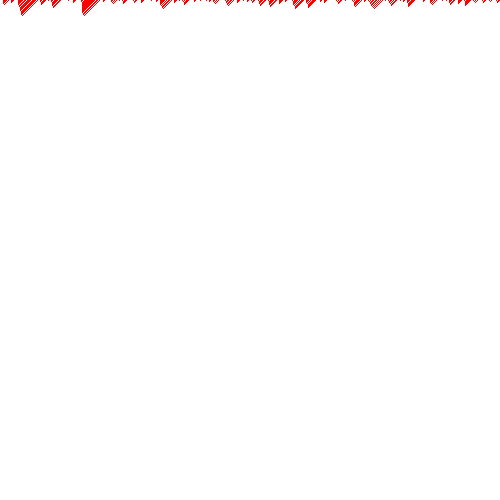} & \includegraphics[width=31mm]{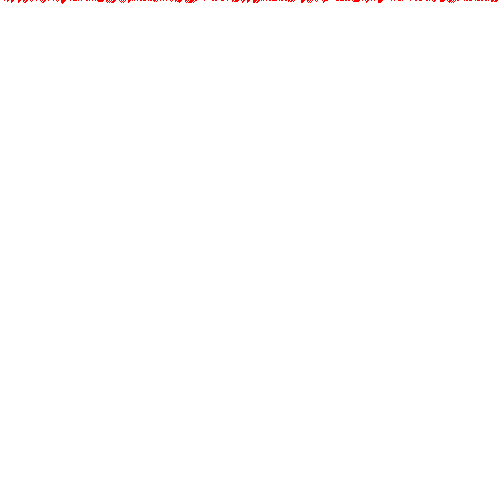} &   \includegraphics[width=31mm]{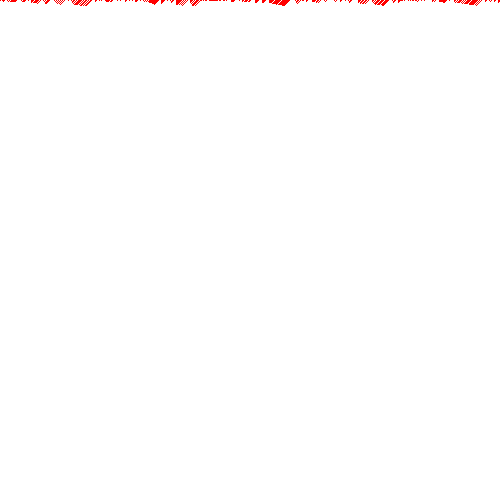} &   \includegraphics[width=31mm]{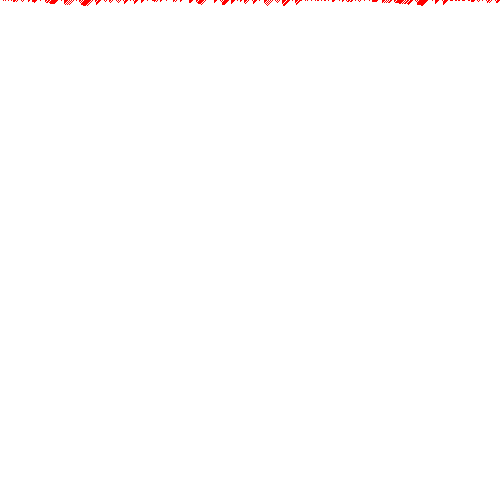} &   \includegraphics[width=31mm]{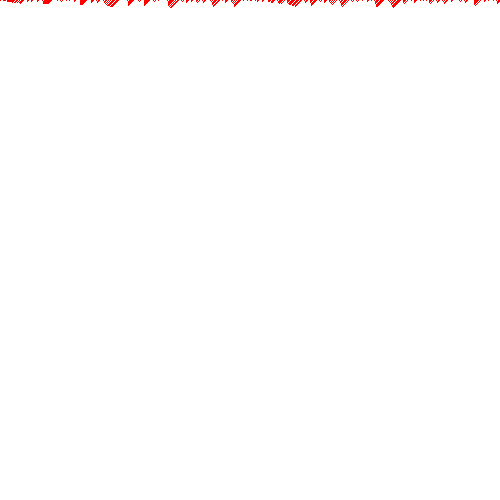} \\
				
				ECA 57 & ($57,g^{50}$) & ($57,g^{65}$) & ($57,g^{150}$) & ($57,g^{250}$) \\[6pt]
				\includegraphics[width=31mm]{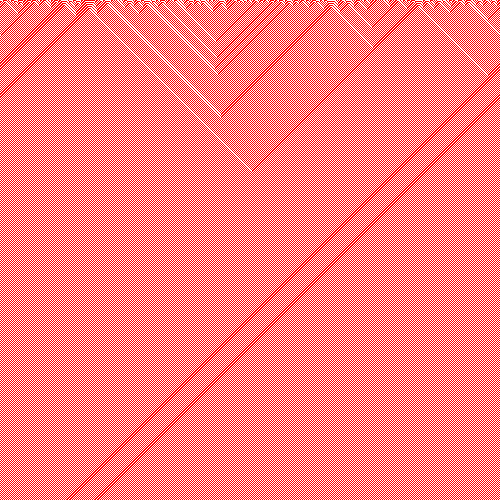} & \includegraphics[width=31mm]{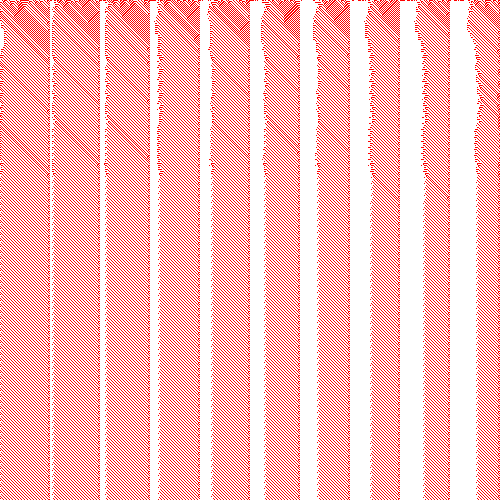} &   \includegraphics[width=31mm]{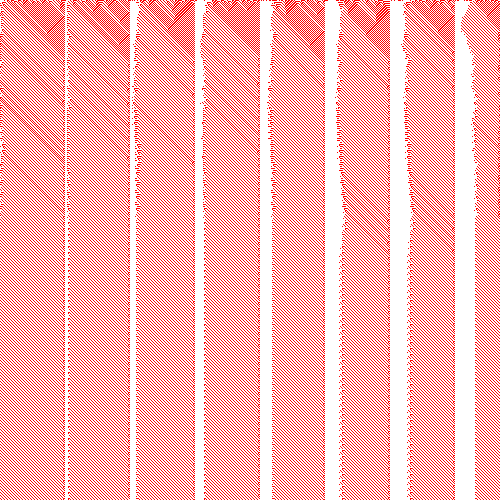} &   \includegraphics[width=31mm]{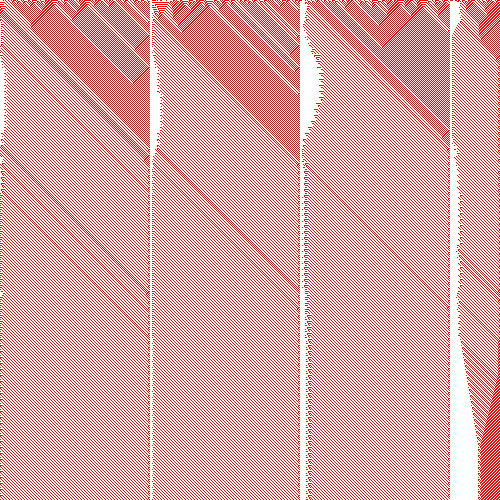} &   \includegraphics[width=31mm]{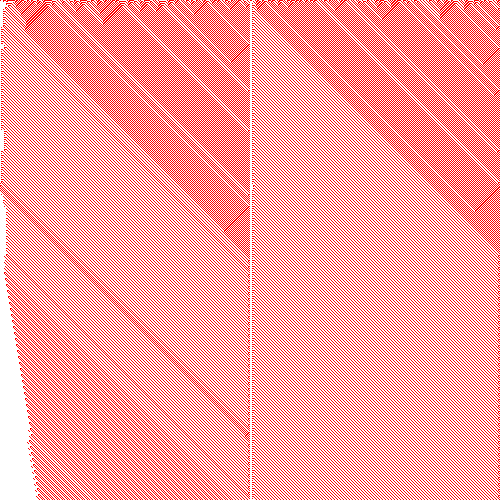} \\
				
				ECA 105 & ($105,g^{50}$) & ($105,g^{65}$) & ($105,g^{150}$) & ($105,g^{250}$) \\[6pt]
				\includegraphics[width=31mm]{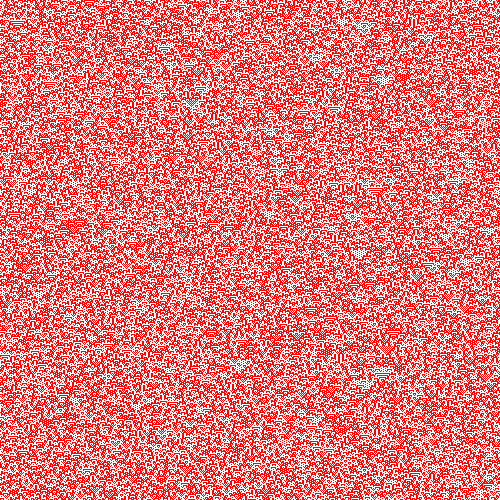} & \includegraphics[width=31mm]{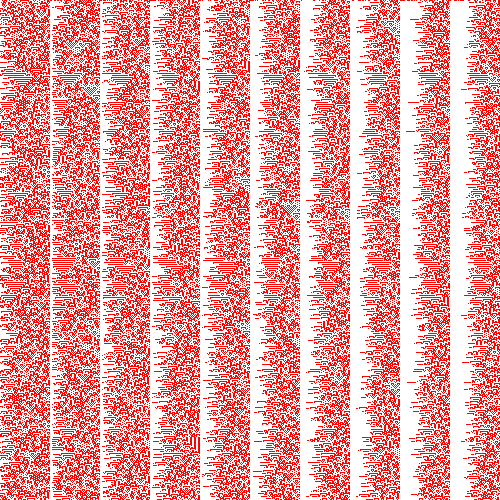} &   \includegraphics[width=31mm]{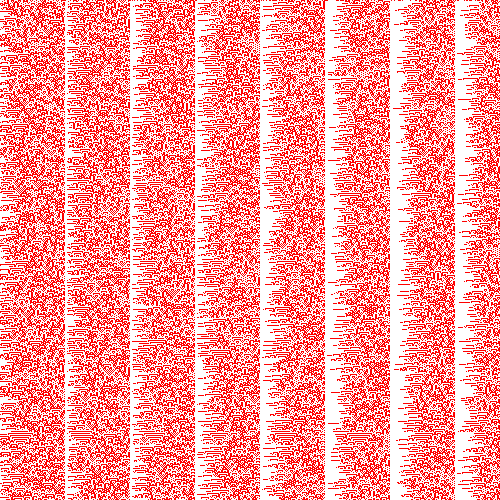} &   \includegraphics[width=31mm]{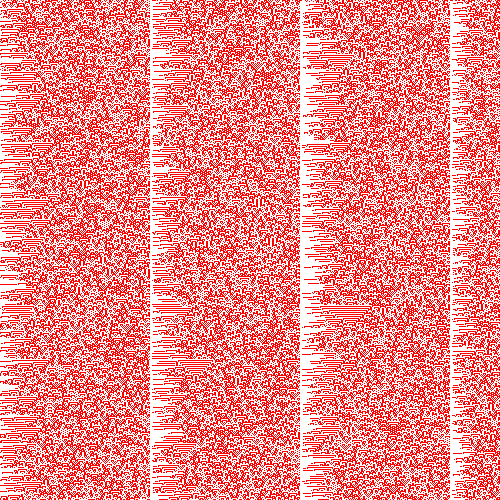} &   \includegraphics[width=31mm]{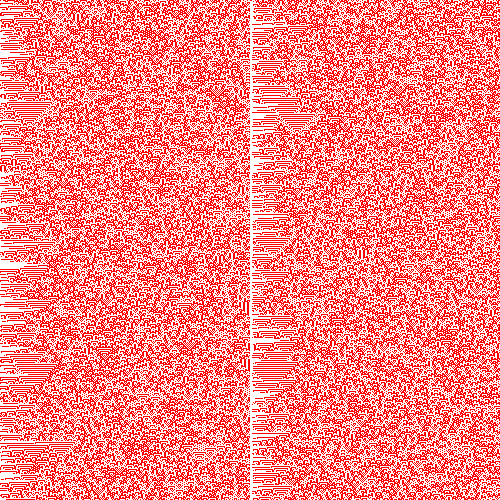} \\
		\end{tabular}}
		\caption{LCA($f,g^{b}$) dynamics when $\zeta$($f,g^{b}$) = $\zeta$($f$) for minimization scheme.}
		\label{min1}
	\end{center}
\end{figure*}

Figure~\ref{min1} presents the space-time diagrams of different LCAs where $\zeta(f)=\zeta(f,g^b)$. In Figure.~\ref{min1}, we consider ECA $168$ as the default rule ($f$), which individually exhibits homogeneous behavior (Wolfram's Class I or Class A dynamics, according to our classification). When noise ($g$) is introduced with varying block sizes, the resulting dynamics of LCA($168,g^b$) remain homogeneous. We provide space-time diagrams for LCA($168,g^b$) with block sizes $b=65,100,150,250$ as examples. Notably, as the block size $b$ is progressively changed, the dynamics of the cellular system remain unaltered, indicating that the behavior and patterns of the CA are not affected by the addition of the noise $g$. Similar observations can be made for LCA($57,g^b$) and LCA($105,g^b$), where the dynamics of Class B and Class C, respectively, are preserved even after the introduction of noise.

\textbf{Note:} Similar to maximization, in our analysis of dynamics of LCA based on minimization, we didn't considered $b=1$. Then if the left, current and right block is in state 1,1,1 respectively. After minimization, it will be 1,0,1. Similarly, 011$\rightarrow$001 and 110$\rightarrow$100. Due to this the whole configuration becomes all-0s.

\begin{figure*}[hbt!]
	\begin{center}
		\scalebox{0.7}{
			\begin{tabular}{ccccc}
				ECA 239 & ($239,g^{25}$) & ($239,g^{65}$) & ($239,g^{125}$) & ($239,g^{200}$) \\[6pt]
				\includegraphics[width=31mm]{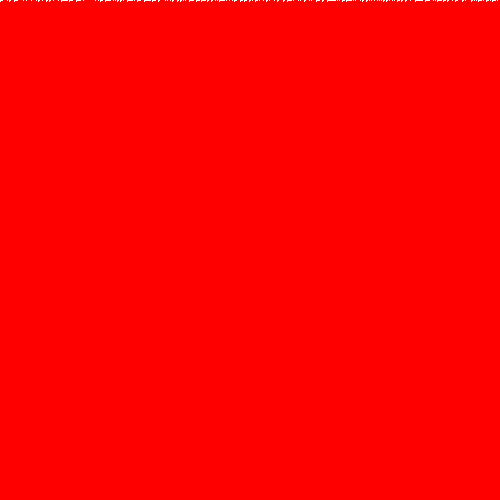} & \includegraphics[width=31mm]{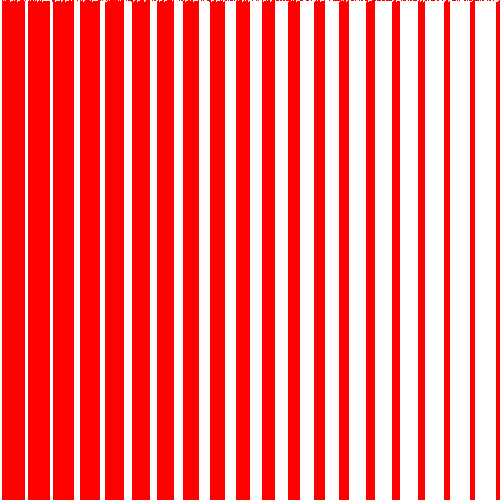} &   \includegraphics[width=31mm]{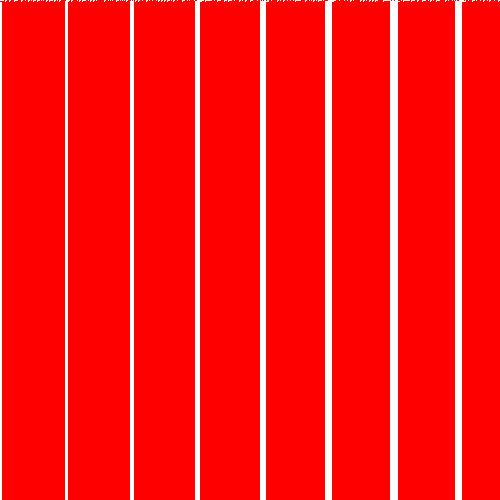} &   \includegraphics[width=31mm]{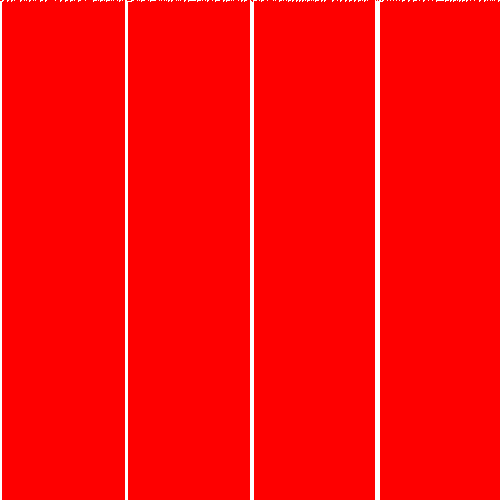} &   \includegraphics[width=31mm]{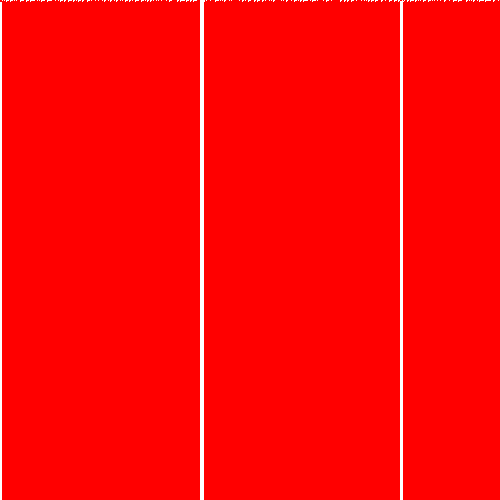} \\
				
				ECA 34 & ($34,g^{50}$) & ($34,g^{65}$) & ($34,g^{150}$) & ($34,g^{250}$) \\[6pt]
				\includegraphics[width=31mm]{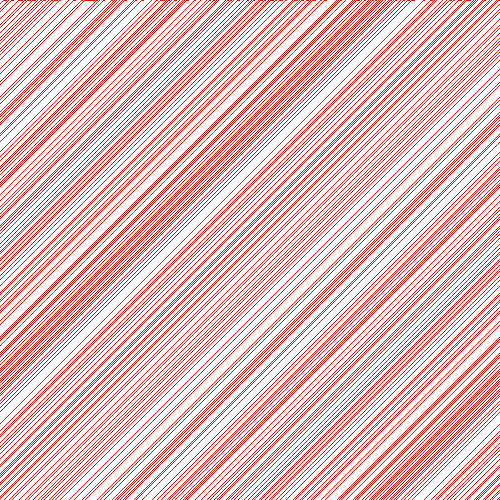} & \includegraphics[width=31mm]{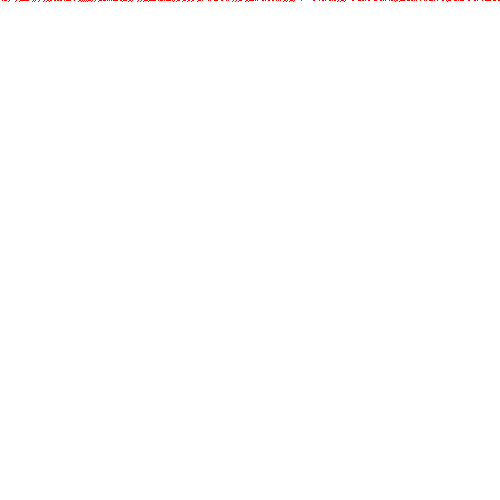} &   \includegraphics[width=31mm]{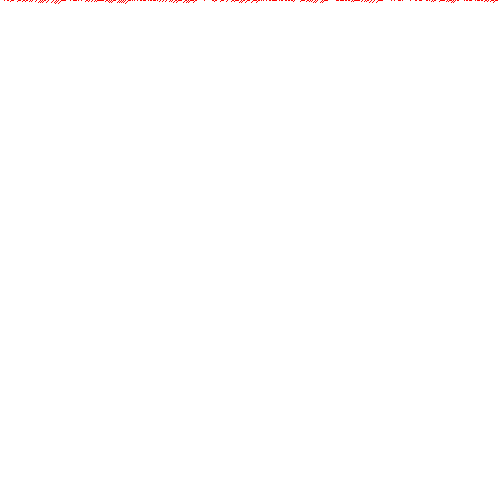} &   \includegraphics[width=31mm]{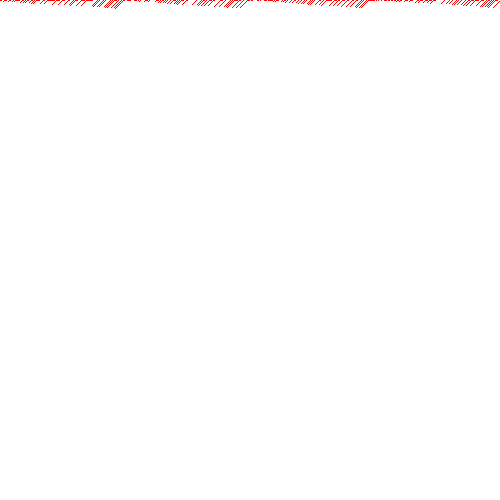} &   \includegraphics[width=31mm]{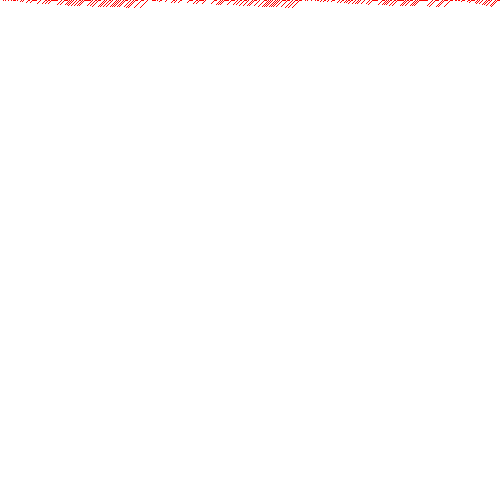} \\
				
				ECA 60 & ($60,g^{125}$) & ($60,g^{250}$) & ($60,g^{300}$) & ($60,g^{400}$) \\[6pt]
				\includegraphics[width=31mm]{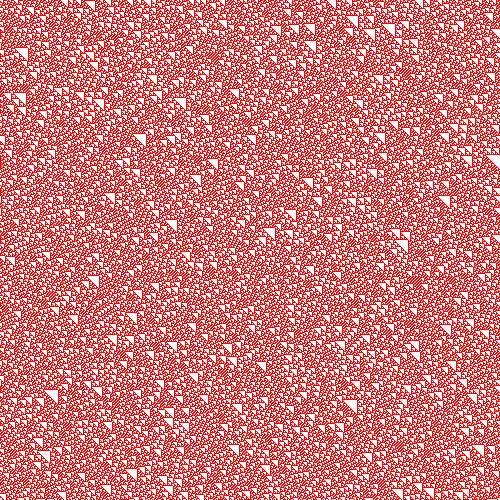} & \includegraphics[width=31mm]{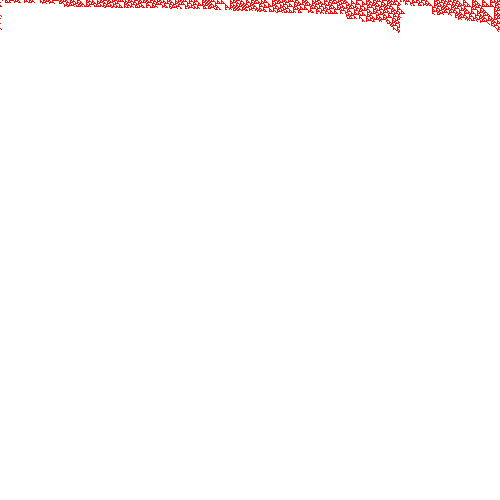} &   \includegraphics[width=31mm]{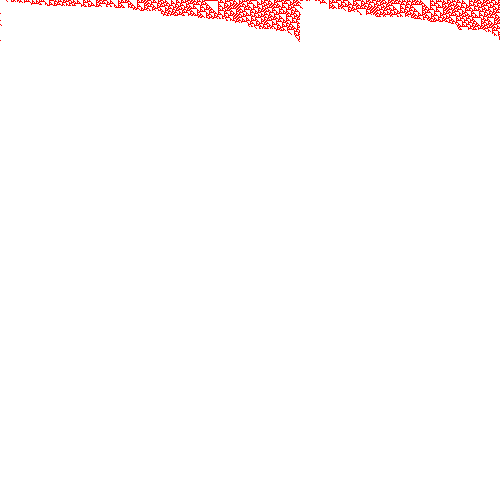} &   \includegraphics[width=31mm]{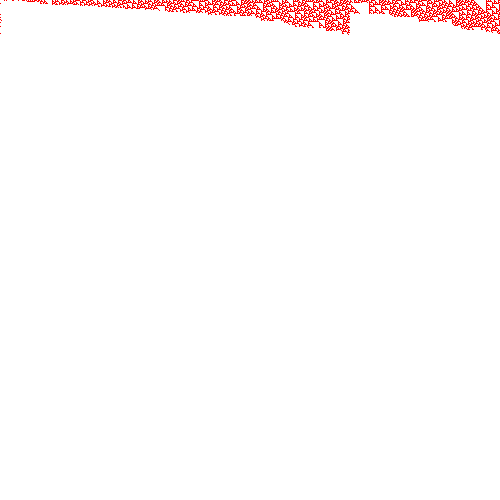} &   \includegraphics[width=31mm]{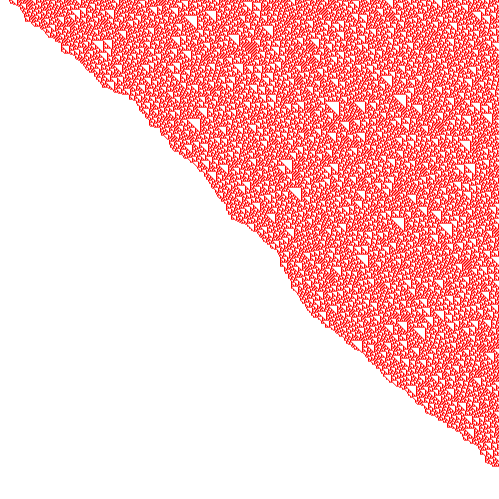} \\
		\end{tabular}}
		\caption{LCA($f,g^{b}$) dynamics when $\zeta$($f,g^{b}$) $\neq$ $\zeta$($f$) for minimization scheme.}
		\label{min2}
	\end{center}
\end{figure*}

Another important observation we made during the analysis of the dynamics is the significant influence of rule $g$ on the dynamics of rule $f$. In this case, the resulting dynamics of LCA show a change in class, specifically when $f$ is chosen from Class B. The introduction of noise $g$ causes the dynamics of LCA to transition from Class B to Class A. This change in class is observed for all possible block sizes, indicating that the behavior and patterns exhibited by the CA undergo a consistent transformation. 

Figure.~\ref{min2} displays the space-time diagrams of different LCAs where the dynamics of $f$ and LCA($f,g^b$) differ, denoted as $\zeta(f)\neq\zeta(f,g^b)$. In this figure, we consider ECA $239$ as the default rule ($f$), which exhibits homogeneous behavior. When noise $g$ is introduced with varying block sizes, the resulting dynamics of LCA show a resemblance to Class B dynamics. We provide space-time diagrams for LCA($239,g^b$) with block sizes $b=25,65,125,200$ as examples of the scenario where $\zeta(f)=$Class A and $\zeta(f,g^b)=$Class B. Notably, as the block size $b$ changes, the dynamics of the cellular system undergo a change in class. Similarly, for rule $34$ and rule $60$, which exhibit Class B and Class C dynamics respectively, the introduction of noise $g$ causes the respective LCAs to exhibit Class A dynamics. However, we didn't find any LCAs that satisfy the rest of the scenarios.

\begin{figure*}[hbt!]
	\begin{center}
		\scalebox{0.7}{
			\begin{tabular}{ccccc}
				
				ECA 41 & ($41,g^{150}$) & ($41,g^{100}$) & ($41,g^{65}$) & ($41,g^{50}$) \\[6pt]
				\includegraphics[width=31mm]{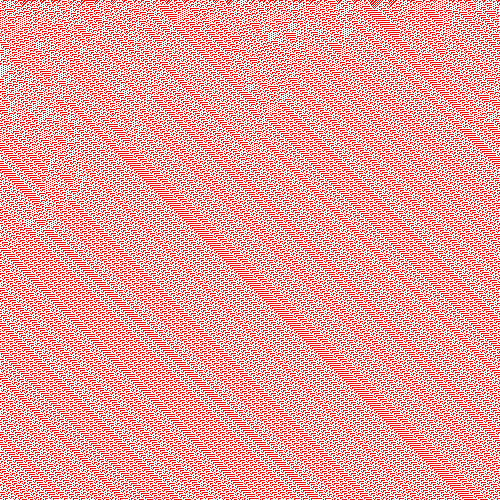} & \includegraphics[width=31mm]{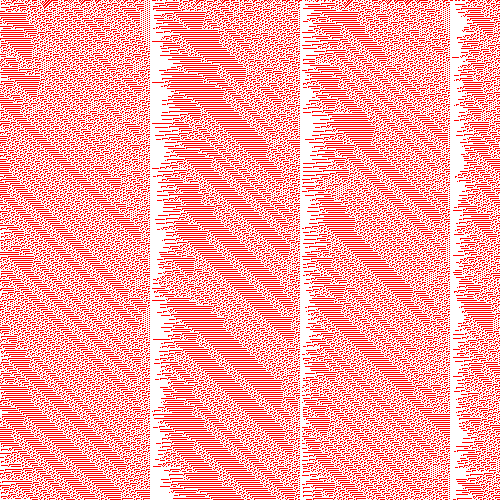} &   \includegraphics[width=31mm]{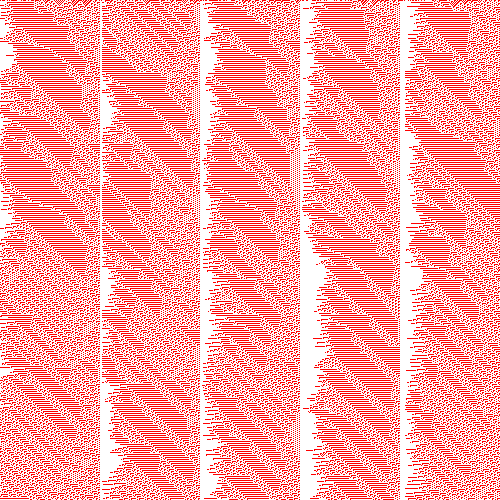} &   \includegraphics[width=31mm]{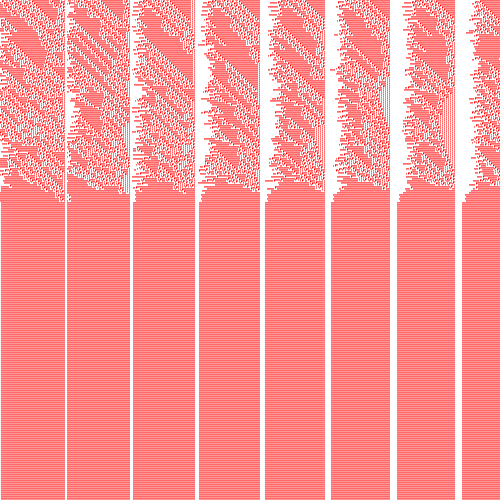} &   \includegraphics[width=31mm]{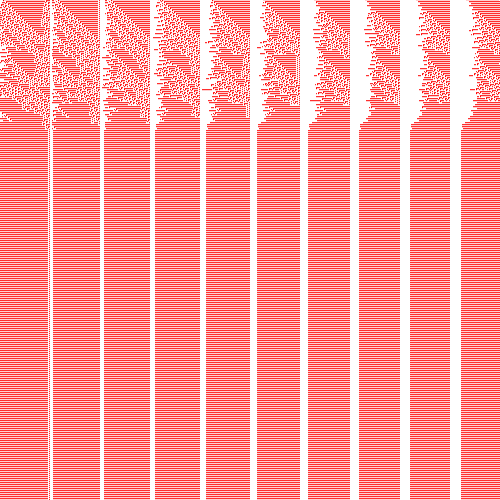} \\
		\end{tabular}}
		\caption{Class transition of LCA($f,g^{b}$) dynamics when counting scheme is minimization.}
		\label{min3}
	\end{center}
\end{figure*}

During our analysis of the dynamics, we did not find any instances of phase transition in the studied LCA systems. However, we did observe examples of class transition. In Figure.~\ref{min3}, we provide illustrations of LCAs that undergo class transition. One such example is ECA 41, which undergoes class transition at a critical block size of $b_t=65$. When Class C rule 41, is chosen as the default rule $f$, and noise rule $g$ is introduced by decreasing the block size, the dynamics of the system progressively transition from Class C to Class B. 

\begin{table}[!htbp] 
	\centering
	\scriptsize
	
	\begin{adjustbox}{width=0.8\columnwidth,center}
		\begin{tabular}{|c|c|c|c|} \hline 
			\textbf{Conditions} & \textbf{Cases} & \textbf{LCAs}& \textbf{$b-$sensitivity} \\ \hline
			&&&\\ \hline
			\multicolumn{4}{|c|}{\textbf{Averaging}} \\ \hline
			&&&\\
			&  $\zeta(f)=\zeta(f,g^b)=$ Class A & ($249,g^{b}$)&\\
			$\zeta$($f$) =  $\zeta$($f,g^b$) &  $\zeta(f)=\zeta(f,g^b)=$ Class B & ($1,g^{b}$)&\\
			&  $\zeta(f)=\zeta(f,g^b)=$ Class C & ($107,g^{b}$)&\\

			&&&\\
			
			&  $\zeta(f)=$ Class A and $\zeta(f,g^b)=$ Class B & ($234,g^{b}$)& Insensitive\\
			&  $\zeta(f)=$ Class B and $\zeta(f,g^b)=$ Class A & ($2,g^{b}$)& \\
			$\zeta(f) \neq \zeta(f,g^b)$ &  $\zeta(f)=$ Class B and $\zeta(f,g^b)=$ Class C & ($62,g^{b}$)& \\
			&  $\zeta(f)=$ Class A and $\zeta(f,g^b)=$ Class C & $-$ & \\
			&  $\zeta(f)=$ Class C and $\zeta(f,g^b)=$ Class A & $-$ & \\
			&  $\zeta(f)=$ Class C and $\zeta(f,g^b)=$ Class B & $-$ & \\
			
			&&&\\
			
			\hline
			&&&\\
			Phase Transition&$\zeta$($f$) = Class B &($240,g^{b}$)&$b_c=150$\\
			& $\zeta$($f$) = Class C &($106,g^{b}$)&$b_c=127$\\
			
			&&&\\
			
			\hline
			&&&\\
			Class Transition&$\zeta$($f$) = Class B and $\zeta$($f,g^b$) = Class C &($37,g^{b}$)&$b_t=200$\\
			& $\zeta$($f$) = Class A and $\zeta$($f,g^b$) = Class B &($238,g^{b}$)&$b_t=25$\\
			&&&\\
			\hline
			\multicolumn{4}{|c|}{\textbf{Maximization}} \\ \hline
			&&&\\
			
			&  $\zeta(f)=\zeta(f,g^b)=$ Class A & ($136,g^{b}$)&\\
			$\zeta$($f$) =  $\zeta$($f,g^b$) &  $\zeta(f)=\zeta(f,g^b)=$ Class B & ($13,g^{b}$)&\\
			&  $\zeta(f)=\zeta(f,g^b)=$ Class C & ($30,g^{b}$)&\\
			
			&&&\\

			&  $\zeta(f)=$ Class A and $\zeta(f,g^b)=$ Class B & ($32,g^{b}$)& Insensitive\\
			&  $\zeta(f)=$ Class B and $\zeta(f,g^b)=$ Class A & ($138,g^{b}$)& \\
			$\zeta(f) \neq \zeta(f,g^b)$ &  $\zeta(f)=$ Class B and $\zeta(f,g^b)=$ Class C & $-$& \\
			&  $\zeta(f)=$ Class A and $\zeta(f,g^b)=$ Class C & $-$ & \\
			&  $\zeta(f)=$ Class C and $\zeta(f,g^b)=$ Class A & ($137,g^{b}$) & \\
			&  $\zeta(f)=$ Class C and $\zeta(f,g^b)=$ Class B & $-$ & \\
			
			&&&\\
			
			\hline
			&&&\\
			Phase Transition&$-$ &$-$&$-$\\
			
			&&&\\
			
			\hline
			&&&\\
			Class Transition&$\zeta$($f$) = Class C and $\zeta$($f,g^b$) = Class B &($26,g^{b}$)&$b_t=250$\\
			&&&\\
			\hline \multicolumn{4}{|c|}{\textbf{Minimization}} \\ \hline
			&&&\\
			
			&  $\zeta(f)=\zeta(f,g^b)=$ Class A & ($168,g^{b}$)&\\
			$\zeta$($f$) =  $\zeta$($f,g^b$) &  $\zeta(f)=\zeta(f,g^b)=$ Class B & ($57,g^{b}$)&\\
			&  $\zeta(f)=\zeta(f,g^b)=$ Class C & ($105,g^{b}$)&\\
			
			&&&\\
			
			&  $\zeta(f)=$ Class A and $\zeta(f,g^b)=$ Class B & ($239,g^{b}$)& Insensitive\\
			&  $\zeta(f)=$ Class B and $\zeta(f,g^b)=$ Class A & ($34,g^{b}$)& \\
			$\zeta(f) \neq \zeta(f,g^b)$ &  $\zeta(f)=$ Class B and $\zeta(f,g^b)=$ Class C & $-$& \\
			&  $\zeta(f)=$ Class A and $\zeta(f,g^b)=$ Class C & $-$ & \\
			&  $\zeta(f)=$ Class C and $\zeta(f,g^b)=$ Class A & ($60,g^{b}$) & \\
			&  $\zeta(f)=$ Class C and $\zeta(f,g^b)=$ Class B & $-$ & \\
			
			&&&\\
			
			\hline
			&&&\\
			Phase Transition&$-$ &$-$&$-$\\
			
			&&&\\
			
			\hline
			&&&\\
			Class Transition&$\zeta$($f$) = Class C and $\zeta$($f,g^b$) = Class B &($41,g^{b}$)&$b_t=65$\\
			&&&\\
			\hline			
			
		\end{tabular}
	\end{adjustbox}	
	\caption{Summary of different dynamics observed for LCAs based on counting ( ``$-$'' indicates, LCA not found for the given condition)}
	\label{Tablesumcont}
\end{table}

\section{LCA based on ECA with modified neighborhood}
\label{S2}

The classification of elementary cellular automata (ECAs) proposed by Wolfram \cite{wolfram2002new, Wolfram94} and Li-Packard \cite{Li90thestructure} is used in this study. The first step is to list down Wolfram's classification for 88 minimal representative ECAs in Table~\ref{table1}. Li-Packard proposed that ECAs belonging to Wolfram's class II show fixed point, periodic, and locally chaotic behavior. These behaviors are marked in Table~\ref{table1} with purple, green, and blue, respectively. Here, $f$ and $g$ are chosen from the set of 88 minimal representative ECAs.

To investigate the dynamical properties of different combinations of rules, we explored all 88$\times$88 = 7744 combinations of $f$ and $g$ for different block sizes ($b$ divides $500$). In other words, we studied the dynamics of cellular automata for each instance of ($f,g^b$) ten times. In this way, we obtained the behavior of the layered cellular automata for different values of block size and could observe any qualitative transformations that occurred as the block size is varied.

Hence, in this study, we analyze the dynamics of $7744$ different combinations of $f$ and $g$ in order to identify similarities in Wolfram's classes using a qualitative and quantitative approach. To do this, we first categorized the $88$ minimal representative ECAs according to Wolfram's classification and recorded the results in Table~\ref{table1}. During this mapping, we merged Wolfram's class III and IV, i.e. chaotic and complex behavior, and Li-Packard's locally chaotic behavior together into a single class, named Class C. While it is true that chaotic and locally chaotic rules have different theoretical dynamics, the space-time diagrams show that locally chaotic rules are more similar to chaotic and complex dynamics than periodic dynamics. Hence, we classified the dynamics of LCA($f,g^b$) into three classes: Class A, which depicts behavior similar to Wolfram's class I, that is homogeneous; Class B, which depicts the behavior of Wolfram's class II (fixed-point and periodic), excluding Li-Packard's locally-chaotic rules; and Class C, which depicts the behavior of Wolfram's class III and IV, and locally chaotic rules.  

In the later stages of this study, we will refer to the individual ECA rules by the class names A, B and C as defined earlier. However, it should be noted that not all couples of $f$ and $g$ can be mapped into these three classes. There are cases where the dynamics of these couples exhibit the phenomenon of {\em phase transition} \cite{ROY2019600} and {\em class transition} (similar to \cite{Martinez12}).

%

We are considering two rules, $f$ and $g$, for different layers, which gives us two possible arrangements: either the classes of $f$ and $g$ are the same, or they are different. We denote the class of $f$, $g$, and LCA($f,g^{b}$) as $\zeta$($f$), $\zeta$($g$), and $\zeta$($f,g^{b}$), respectively.
Based on the space-time diagram, we can derive the observations of the dynamics of layered cellular automata into following ways:


\begin{itemize}
	
	\item[1.] When both $f$ and $g$ belong to the same class of dynamics, denoted as $\zeta(f)=\zeta(g)$ respectively, there are two possible cases. In the first case, if $\zeta(f)=\zeta(g)$, then the LCA($f,g^b$) where $\zeta(f)=\zeta(g)$=Class A, evolves to a homogeneous configuration as the elementary cellular automata $f$ and $g$ do. This means that the dynamics of the LCA become similar to that of ECA $f$ and ECA $g$, and the resulting configuration is completely uniform. 
	
	\item[2.] When both $f$ and $g$ belong to the same class of dynamics, for such condition the case may arise where, $\zeta$($f,g^{b}$)$\neq$$\zeta$($f$). In that case we can say that the interaction rule $g$ with rule $f$ is significantly affecting the dynamics of LCA($f,g^{b}$). On the basis of this case, we observed:
	\begin{itemize}
		\item[-] If the LCA($f,g^{b}$) evolves to a homogeneous configuration (Class A), it suggests that the interaction between the two periodic rules (Class B) has resulted in a mutual synchronization, causing the dynamics to converge to a uniform state. 		
		\item[-] If the LCA($f,g^{b}$) shows a tendency towards chaotic/complex/locally chaotic dynamics (Class C), it means that the interaction between the two periodic rules (Class B) has introduced ``noise'' into the system, causing the dynamics to deviate significantly from the behavior of the individual rules. 
	\end{itemize}
	
	\item[3.] When $\zeta(f)\neq\zeta(g)$, then either $f$ or $g$ tends to exhibits a dominant behavior in the dynamics of LCA($f,g^b$) as the dynamics starts to show its resemblance towards the dynamics of either rule $f$ or rule $g$, resulting in either $\zeta(f,g^b) = \zeta(f)$ or $\zeta(f,g^b) = \zeta(g)$. In the case where $\zeta(f)=\text{Class A}$ and $\zeta(g)=\text{Class C}$, the LCA($f,g^b$) displays a behavior where $\zeta(f,g^b) = \zeta(f)$. Similarly, when $\zeta(f)=\text{Class B}$ and $\zeta(g)=\text{Class C}$, the LCA($f,g^b$) shows a behavior where $\zeta(f,g^b) = \zeta(g)$.

	\item[4.] In contrast, when $\zeta(f)\neq\zeta(g)$, the dynamics of LCA($f,g^b$) are such that neither of the rule classes dominates, meaning that $\zeta(f,g^b)\neq\zeta(f)$ and $\zeta(f,g^b)\neq\zeta(g)$. In this scenario, if $\zeta(f)=\text{Class A}$ and $\zeta(g)=\text{Class C}$, then LCA($f,g^b$) exhibits behavior characterized by $\zeta(f,g^b)=\text{Class B}$. Similarly, if $\zeta(f)=\text{Class B}$ and $\zeta(g)=\text{Class C}$, then the dynamics of LCA($f,g^b$) correspond to $\zeta(f,g^b)=\text{Class A}$.
	
\end{itemize}

Up until now, we have been examining the scenario where the cellular automata's dynamics remained invariant despite varying the block size ($b$). This indicated that the LCA($f,g^{b}$) were not sensitive to changes in the block size. However, there are situations where the dynamics of the cellular automata system are sensitive to $b$.
In such cases, as the block size ($b$) varies, the system's dynamics show variation. This sensitivity to $b$ happens when the cellular automata rules $f$ and $g$ possess a specific relationship. Specifically, when $b$ changes, LCA($f,g^b$) tends to go into a homogeneous state or tends to shift its dynamics from one class to other. Next we discuss about the effects of varying $b$ on the dynamics of an LCA.

\begin{itemize}
	\item[1.] It is noteworthy that LCA($f,g^{b}$) exhibit a significant change in their behavior after a certain critical value of block size ($b$), which is commonly referred to as second-order \emph{phase transition}. This transition is characterized by a sudden change in the system's behavior as $b$ is changed. Specifically, the behavior of the system can be divided into two distinct phases: a passive phase, where the system eventually converges to $0^{\mathcal{L}}$, and an active phase, where the system exhibits a stationary non-zero configuration. The critical value of $b$, denoted as $b_c$, marks the boundary between these two phases. 
	
	\item[2.] In a set of LCA($f,g^{b}$), a critical value of $b$, denoted as $b_t$, marks a transition point where the class dynamics of the system undergoes a significant change. This transition is characterized by the fact that $\zeta$($f,g^{b}$) $\neq$ $\zeta$($f,g^{b^{'}}$), where $b\in[1,b]$ and $b^{'}\in[b,500]$. Specifically, an LCA($f,g^{b}$) with $\zeta$($f$) = Class B and $\zeta$($g$) = Class C shows periodic behavior before $b_t$, but it slowly transforms into chaotic dynamics after reaching the critical value of block size $b_t$. We refer to this phenomenon as a {\em class transition} throughout the remainder of this chapter.
\end{itemize}

This section provides a high-level overview of the different types of dynamics observed in the LCA($f,g^{b}$) system. However, to gain a deeper understanding, it is important to revisit these dynamics with concrete examples. This will enable us to analyze and visualize the behavior of the system and the underlying mechanisms driving these dynamics. Let us explore more in detail for each of the possible dynamics that are explained here.

\subsection{Dynamics when $\zeta$($f$) = $\zeta$($g$)}

Let us start by examining the scenario where $\zeta(f) = \zeta(g)$. Our experiments have revealed two distinct outcomes under this condition: 

\begin{itemize}
	\item $\zeta$($f,g^{b}$) = $\zeta$($f$)
	\item $\zeta$($f,g^{b}$) $\neq$  $\zeta$($f)$
\end{itemize}

When both the update rules, denoted as $f$ and $g$, are chosen from Class A, the resulting dynamics of LCA remain within Class A. In other words, the overall behavior and patterns exhibited by the CA remain unchanged.
On the other hand, when both $f$ and $g$ are selected from either Class B or Class C, only in certain cases does the LCA exhibit similar behavior. This means that the LCA with update rules $f$ and $g$ behaves similarly as dynamics of $f$ and $g$, denoted as $\zeta(f,g^b) = \zeta(f) = \zeta(g)$.
To illustrate this, we provide six examples of such behavior: LCA($1,33^b$), LCA($9,1^b$), LCA($126,30^b$), LCA($105,18^b$), LCA($73,26^b$), and LCA($38,44^b$). In these cases, when the LCA is defined with the update rules $f$ and $g$ (where $f$ and $g$ is chosen from Class B or Class C), the resulting behavior is equivalent to the behavior of the CA with the update rule $f$ and $g$.
This suggests that in certain scenarios where specific combinations of update rules are used, the addition of the second rule $g$ does not significantly alter the behavior of the LCA.

\begin{figure*}[hbt!]
	\begin{center}
		\scalebox{0.7}{
			\begin{tabular}{ccccc}
				ECA 1 & ECA 33 & ($1,33^{4}$) & ($1,33^{10}$) & ($1,33^{25}$) \\[6pt]
				\includegraphics[width=31mm]{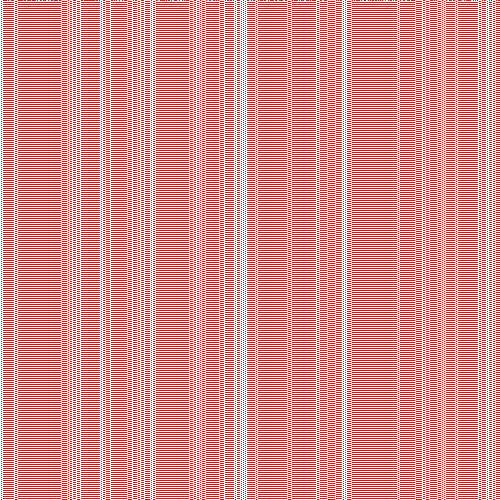} & \includegraphics[width=31mm]{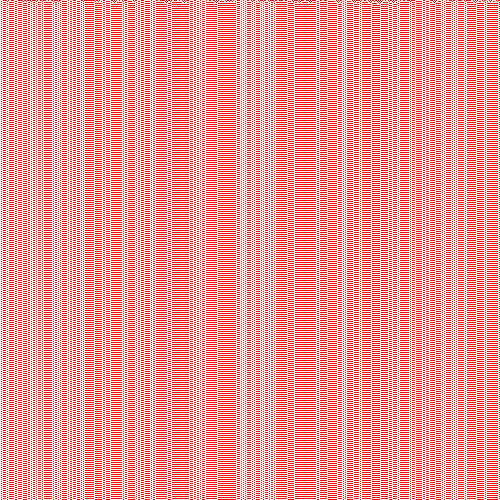} &   \includegraphics[width=31mm]{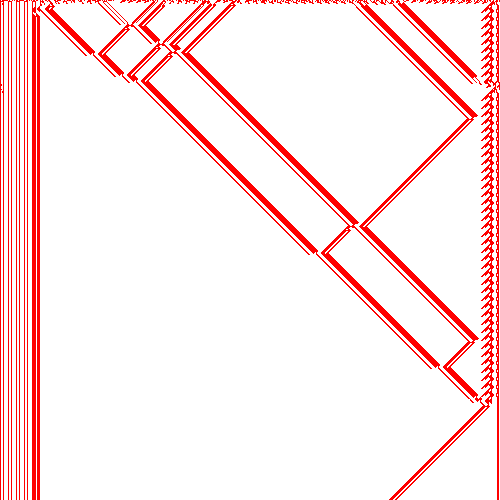} &   \includegraphics[width=31mm]{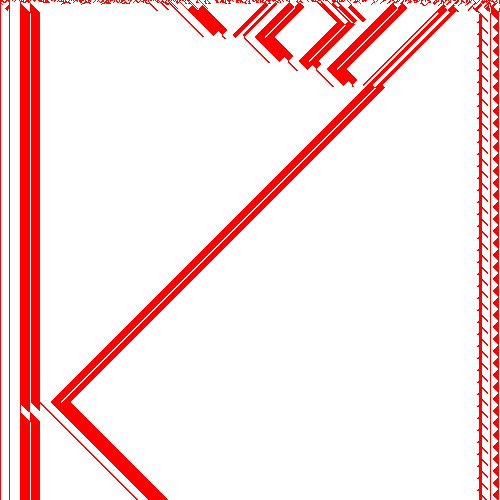} &   \includegraphics[width=31mm]{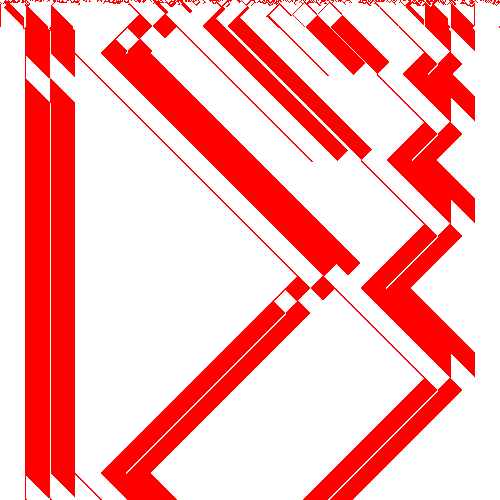} \\
				ECA 9 & ECA 1 & ($9,1^{50}$) & ($9,1^{100}$) & ($9,1^{125}$) \\
				\includegraphics[width=31mm]{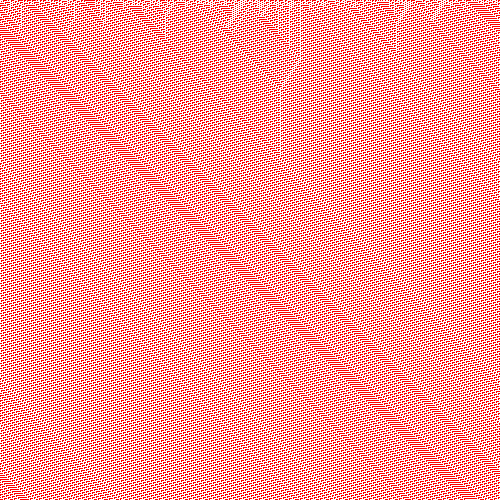} & \includegraphics[width=31mm]{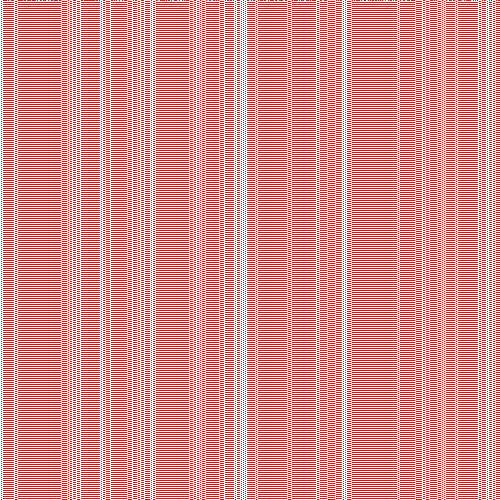} &   \includegraphics[width=31mm]{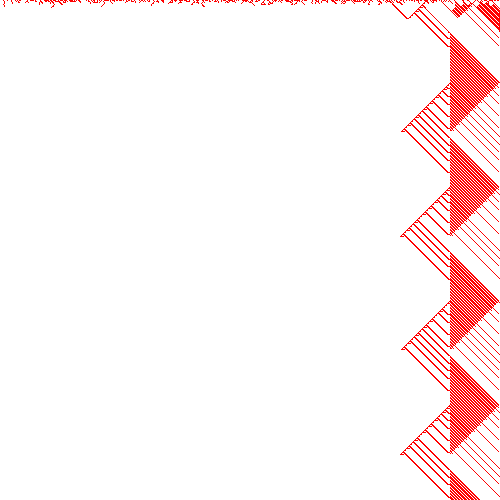} &   \includegraphics[width=31mm]{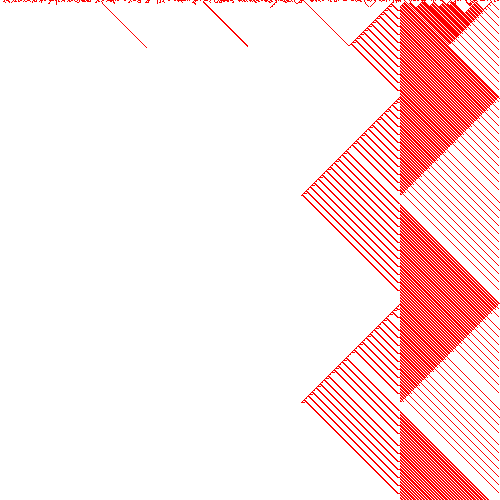}   &   \includegraphics[width=31mm]{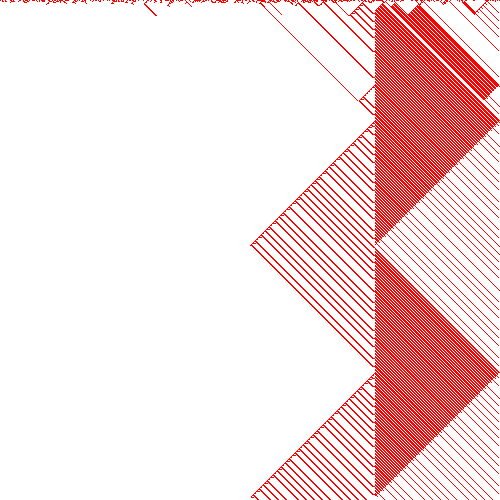} \\
				ECA 126 & ECA 30 & ($126,30^{50}$) & ($126,30^{125}$) & ($126,30^{250}$) \\
				\includegraphics[width=31mm]{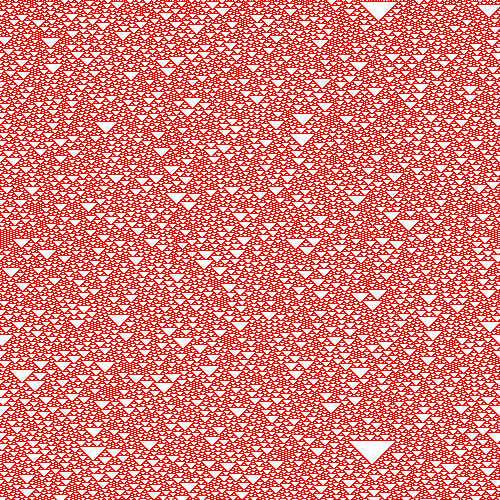} & \includegraphics[width=31mm]{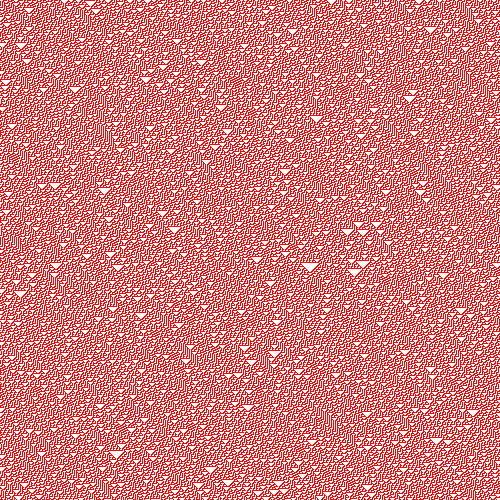} &   \includegraphics[width=31mm]{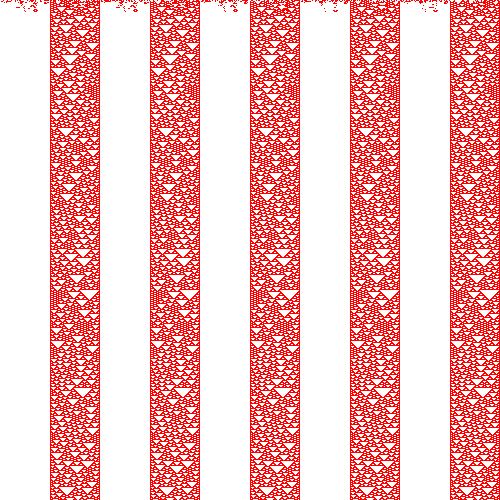} &   \includegraphics[width=31mm]{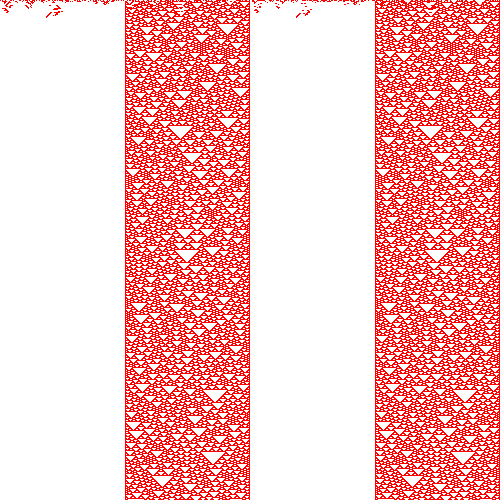}  &   \includegraphics[width=31mm]{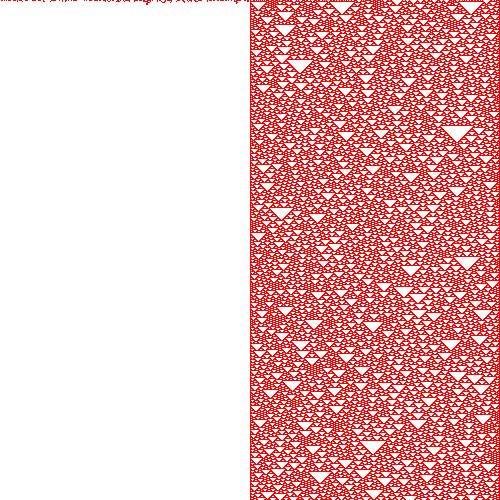} \\
		\end{tabular}}
		\caption{LCA($f,g^{b}$) dynamics when $\zeta$($f,g^{b}$) = $\zeta$($f$) = $\zeta$($g$).}
		\label{Fig2}
	\end{center}
\end{figure*}

\begin{figure*}[hbt!]
	\begin{center}
		\scalebox{0.7}{
			\begin{tabular}{ccccc}
				ECA 105 & ECA 18 & ($105,18^{50}$) & ($105,18^{100}$) & ($105,18^{125}$) \\[6pt]
				\includegraphics[width=31mm]{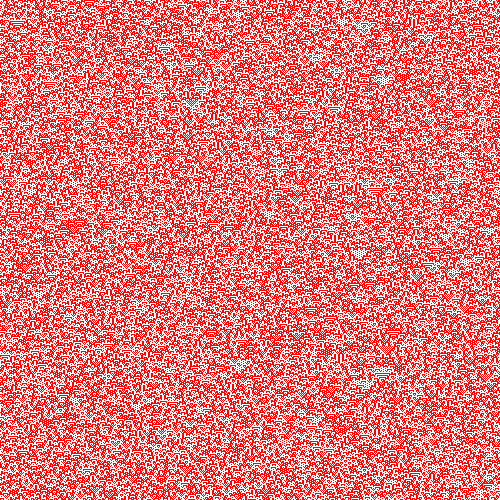} & \includegraphics[width=31mm]{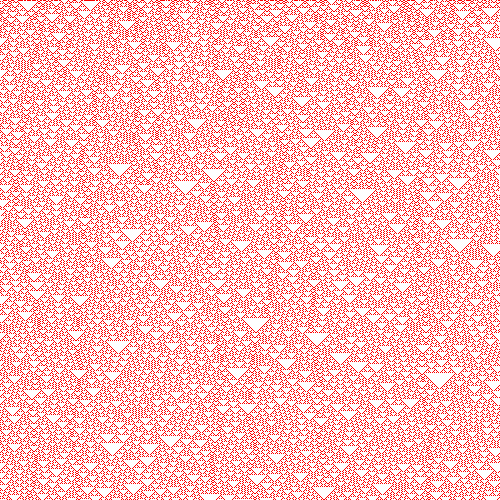} &   \includegraphics[width=31mm]{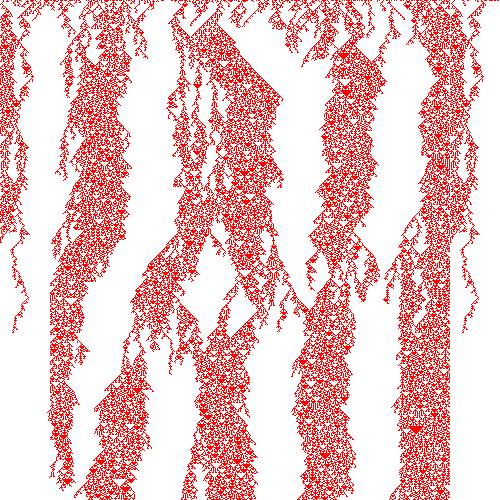} &   \includegraphics[width=31mm]{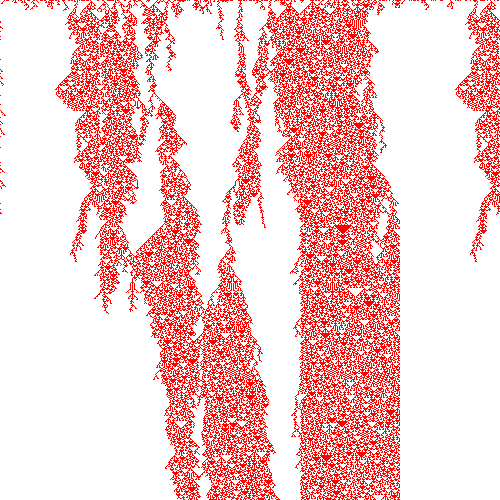} &   \includegraphics[width=31mm]{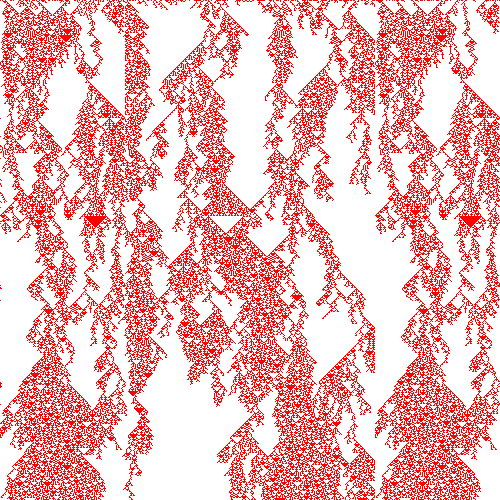} \\
				ECA 73 & ECA 26 & ($73,26^{50}$) & ($73,26^{100}$) & ($73,26^{125}$) \\[6pt]
				\includegraphics[width=31mm]{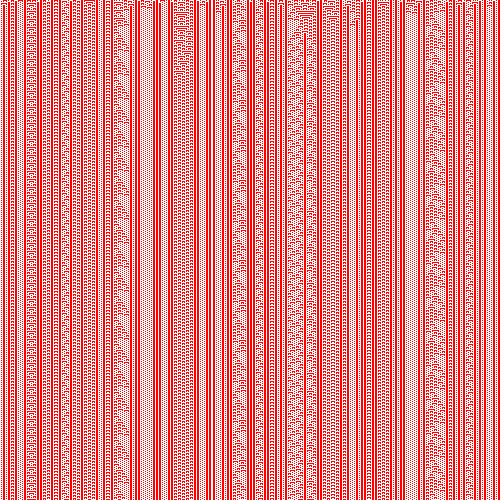} & \includegraphics[width=31mm]{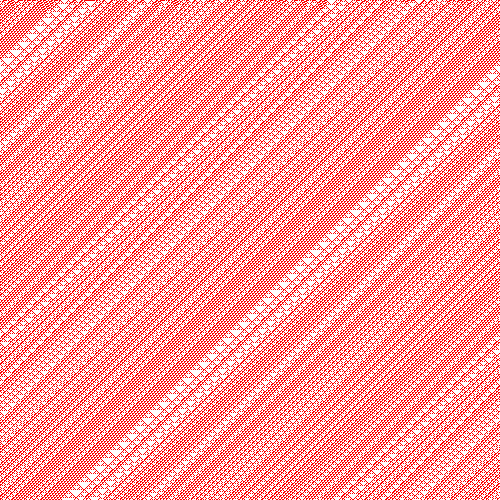} &   \includegraphics[width=31mm]{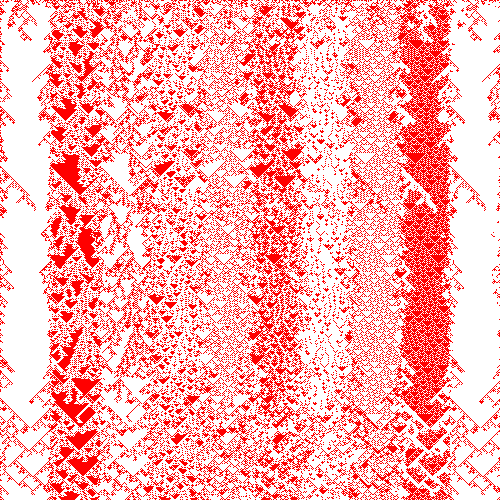} &   \includegraphics[width=31mm]{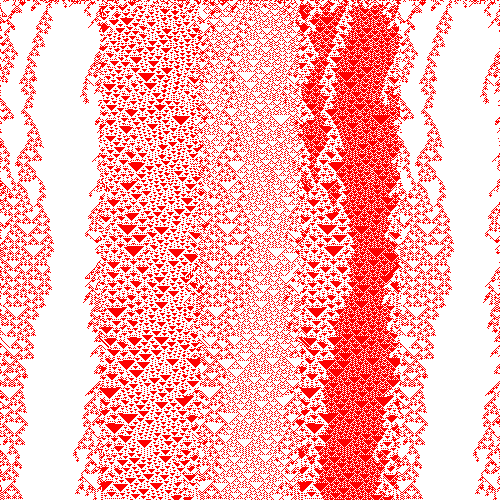} &   \includegraphics[width=31mm]{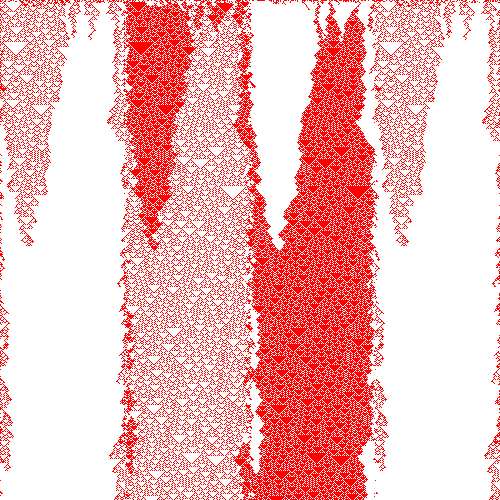} \\
				ECA 38 & ECA 44 & ($38,44^{5}$) & ($38,44^{25}$) & ($38,44^{125}$) \\[6pt]
				\includegraphics[width=31mm]{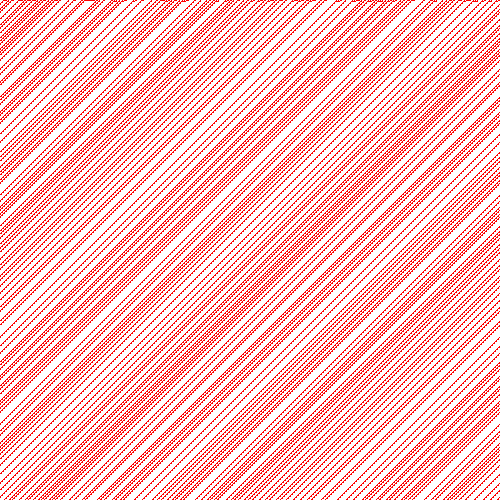} & \includegraphics[width=31mm]{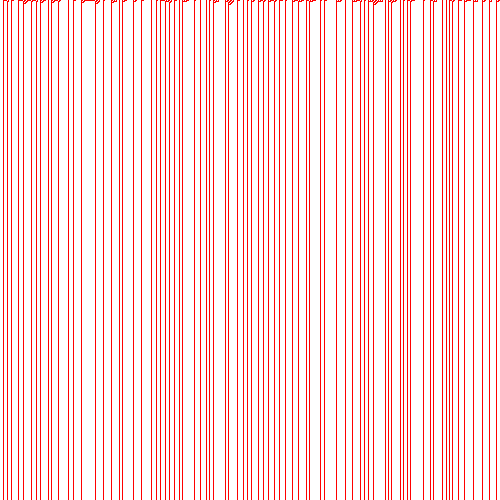} &   \includegraphics[width=31mm]{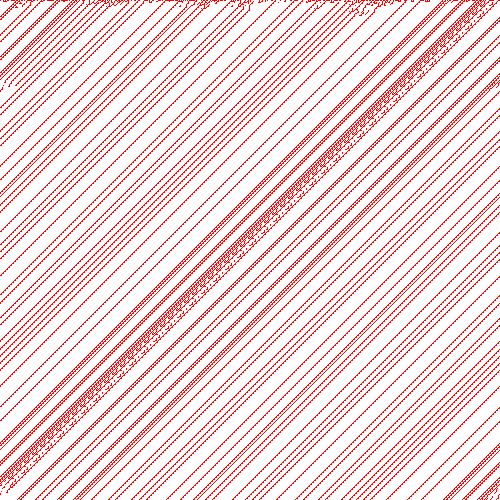} &   \includegraphics[width=31mm]{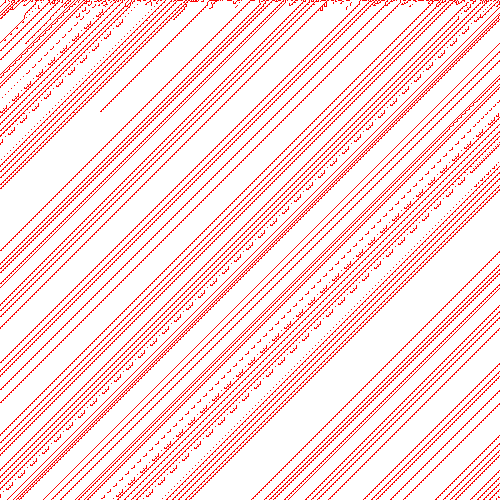} &   \includegraphics[width=31mm]{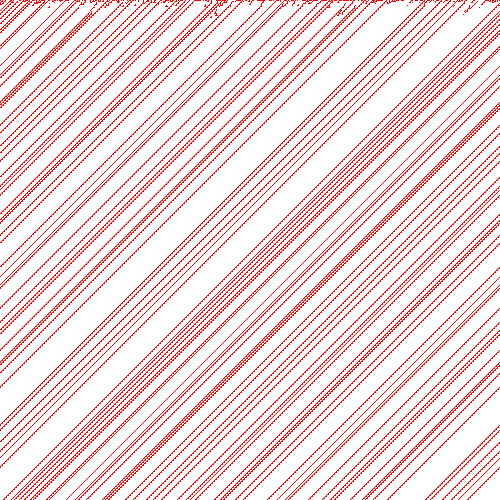} \\
		\end{tabular}}
		\caption{LCA($f,g^{b}$) dynamics when $\zeta$($f,g^{b}$) = $\zeta$($f$) = $\zeta$($g$).}
		\label{TSCA1}
	\end{center}
\end{figure*}

\begin{figure*}[hbt!]
	\begin{center}
		\scalebox{0.7}{
			\begin{tabular}{ccccc}
				ECA 94 & ECA 37 & ($94,37^{125}$) & ($94,37^{250}$) & ($94,37^{500}$) \\[6pt]
				\includegraphics[width=31mm]{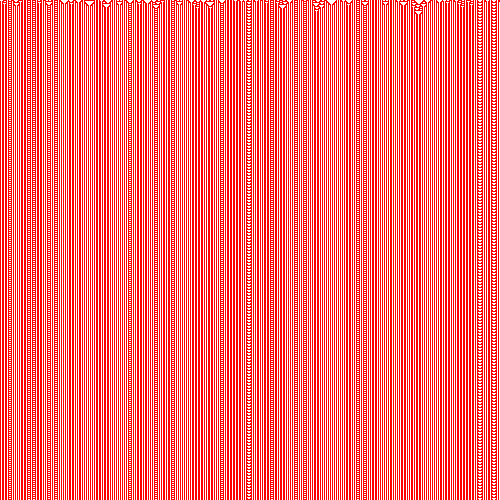} & \includegraphics[width=31mm]{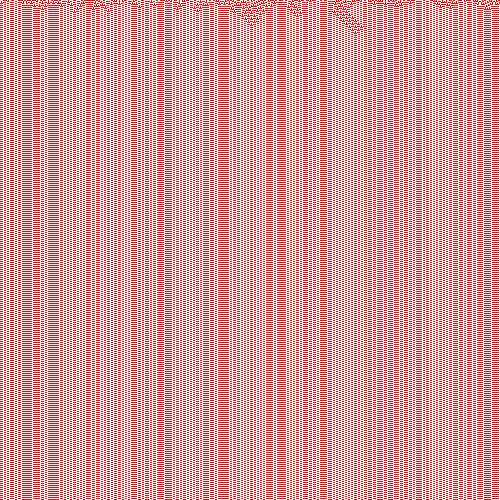} & \includegraphics[width=31mm]{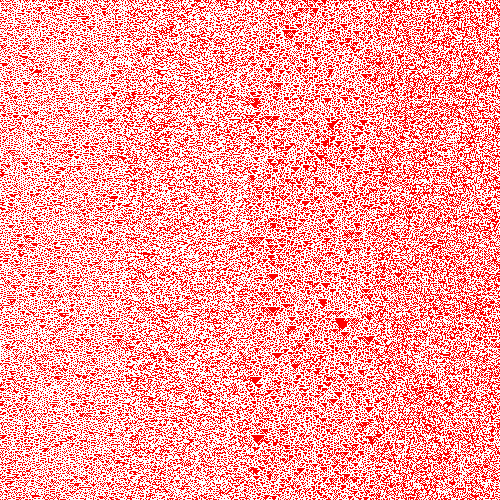} & \includegraphics[width=31mm]{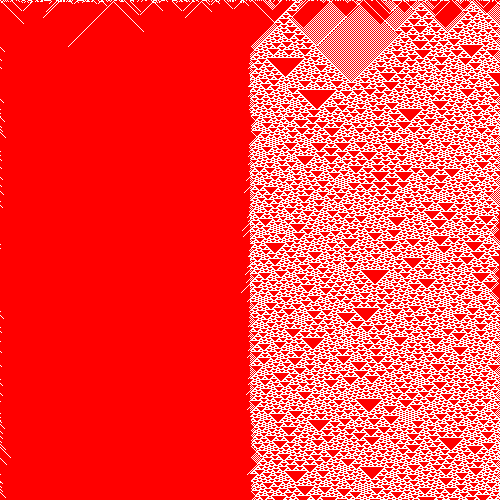} & \includegraphics[width=31mm]{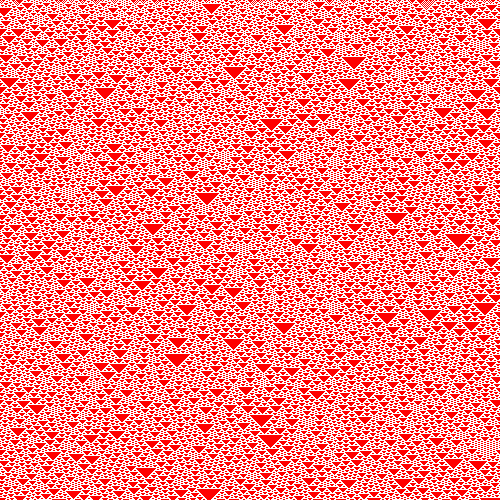} \\
				
				ECA 104 & ECA 51 & ($104,51^{50}$) & ($104,51^{100}$) & ($104,51^{250}$) \\
				
				\includegraphics[width=31mm]{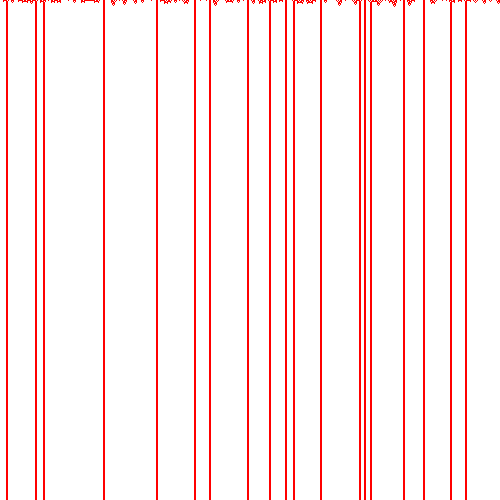}  &  \includegraphics[width=31mm]{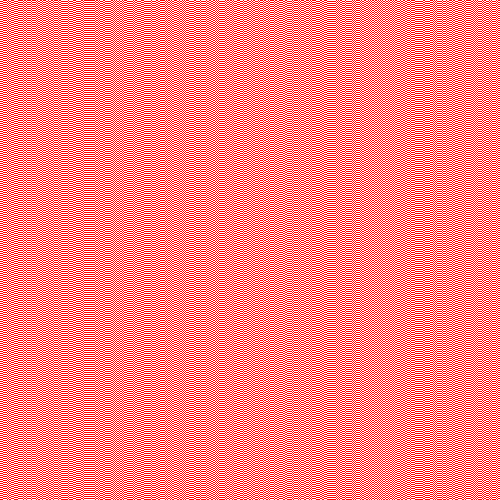}  &   \includegraphics[width=31mm]{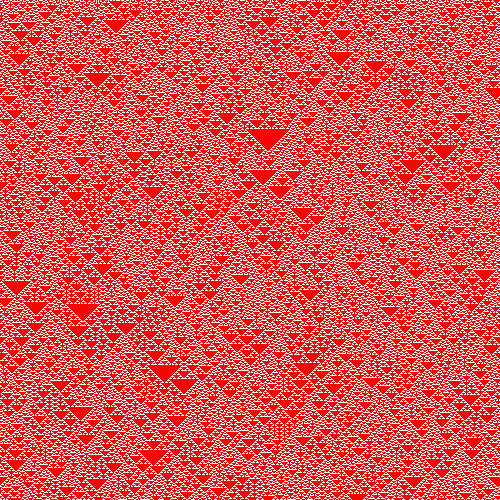}  &  \includegraphics[width=31mm]{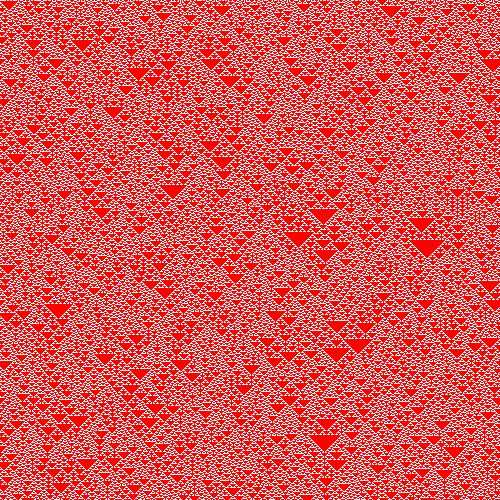}  & \includegraphics[width=31mm]{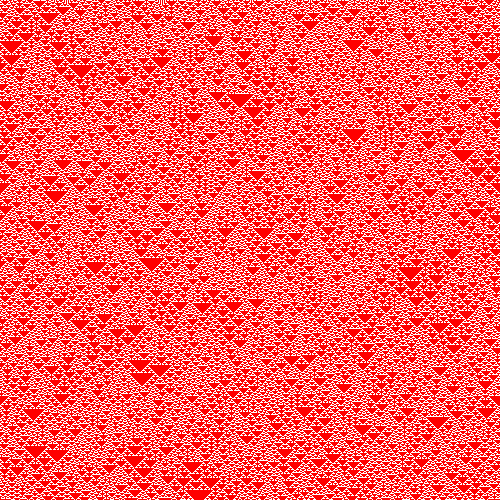} \\
				
				ECA 15 & ECA 19 & ($15,19^{100}$) & ($15,19^{125}$) & ($15,19^{250}$) \\
				
				\includegraphics[width=31mm]{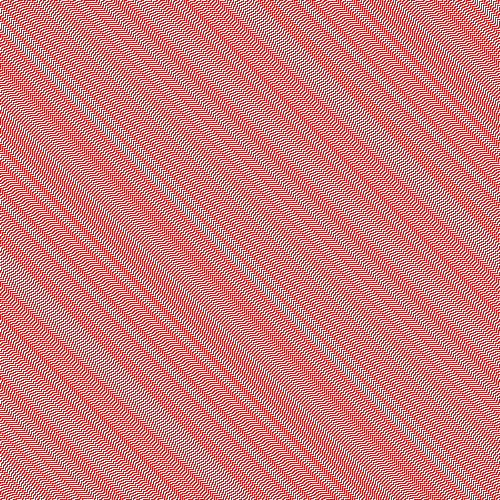} & \includegraphics[width=31mm]{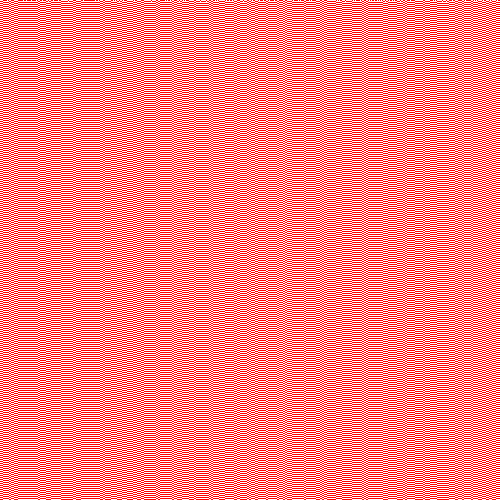}  & \includegraphics[width=31mm]{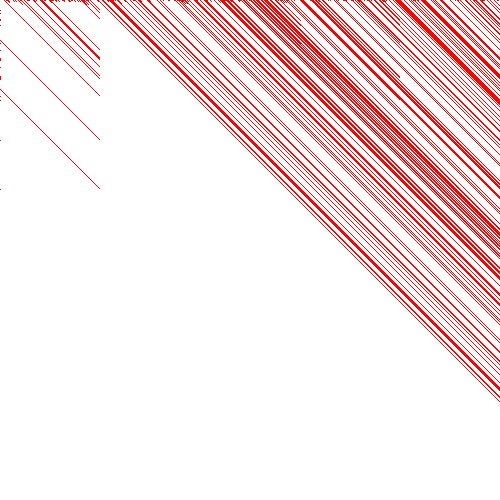} & \includegraphics[width=31mm]{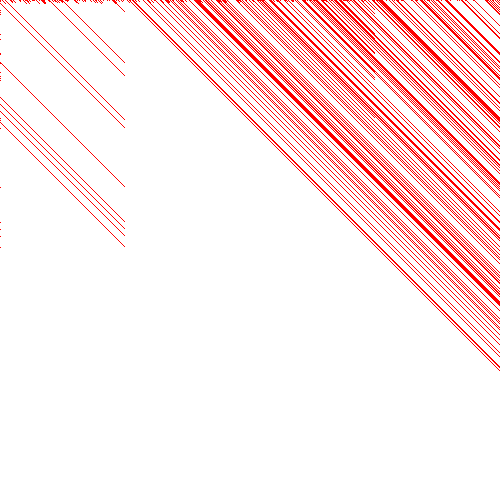}  & \includegraphics[width=31mm]{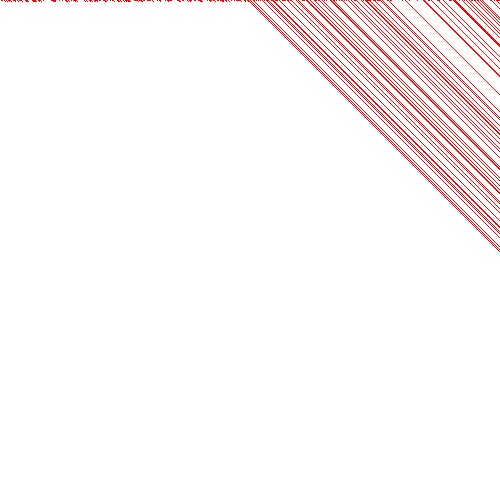} \\
				
					ECA 15 & ECA 1 & ($15,1^{100}$) & ($15,1^{125}$) & ($15,1^{250}$) \\
				
				\includegraphics[width=31mm]{twoeca/15-19/15.png} & \includegraphics[width=31mm]{twoeca/1-33/1.png}  & \includegraphics[width=31mm]{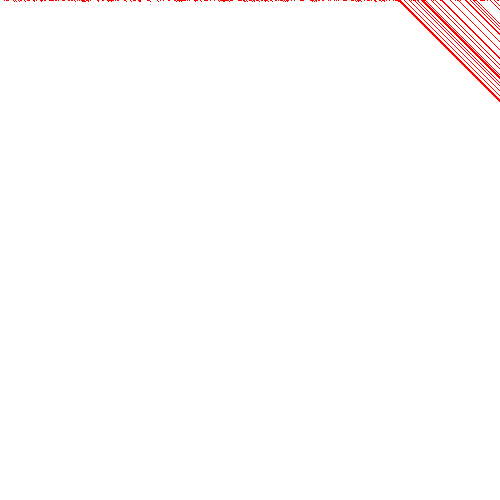} & \includegraphics[width=31mm]{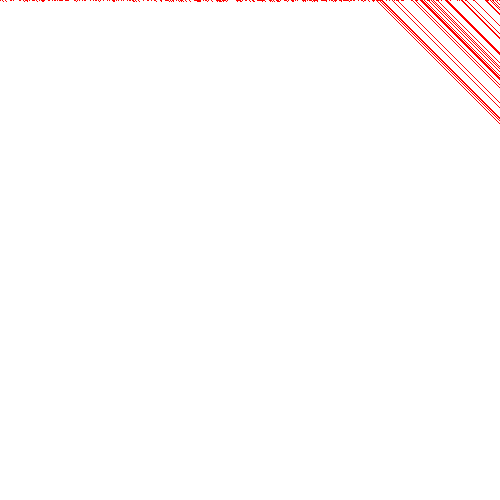}  & \includegraphics[width=31mm]{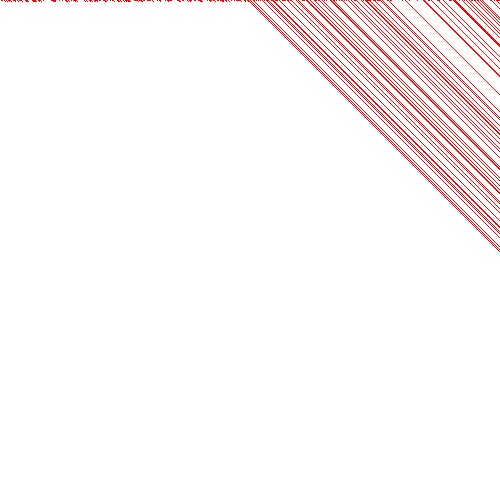} \\
		\end{tabular}}
		\caption{LCA($f,g^{b}$) dynamics when $\zeta$($f,g^{b}$) $\neq$ $\zeta$($f$) and $\zeta$($f$) = $\zeta$($g$).}
		\label{Fig3}
	\end{center}
\end{figure*}

Figure.~\ref{Fig2} and~\ref{TSCA1} depict the space-time diagrams of different LCAs where $\zeta(f,g^b) = \zeta(f) = \zeta(g)$. In Figure.~\ref{Fig2}, ECAs $1$ and $33$, which individually exhibit periodic behavior (Wolfram's Class II or Class B dynamics, according to our classification). When ECA $1$ is considered as the default rule ($f$) and ECA $33$ is introduced as noise ($g$) with varying block sizes, the resulting dynamics remain periodic. We provide space-time diagrams for LCA($1,33^b$) with block sizes $b=4,10$, and $25$ as examples. Notably, when the block size $b$ is progressively changed, the dynamics of the cellular system remain unaltered.
Similarly, for LCA($9,1^b$), where $f=9$ and $g=1$, we observe periodic behavior across different block sizes ($b=50,100,125$), similar to ECA 9 and 1. The dynamics of LCA($126,30^b$) for $b=50,125,250$ (Figure.~\ref{Fig2}) and LCA($105,18^b$) for $b=50,100,125$ (Figure.~\ref{TSCA1}) exhibit chaotic dynamics (Wolfram's class III or Class C dynamics, according to our classification), resembling the dynamics of respective ECAs. Figure.~\ref{TSCA1} showcases LCA($73,26^b$) with ECA 73 and ECA 26, both from Class C. The space-time diagrams for LCA($73,26^b$) with $b=50,100,125$ demonstrate characteristics similar to both ECAs.
Lastly, when ECA $38$ is selected as the default rule ($f$) and ECA $44$ is introduced as noise ($g$), the resulting dynamics remain periodic. Figure.~\ref{TSCA1} provides the space-time diagrams for LCA($38,44^b$) with $b=5,25,125$, illustrating periodic behavior (Wolfram's Class II), which is consistent with the individual dynamics of ECAs $38$ and $44$.

Next, we will discuss the remaining cases where $\zeta(f) = \zeta(g)$ and both $\zeta(f)$ and $\zeta(g)$ belong to either Class B or Class C, but $\zeta(f,g^b) \neq \zeta(f)$. We have identified four such cases: 

\begin{itemize}
	\item[(1)] $\zeta$($f$) = $\zeta$($g$) = Class B and $\zeta$($f,g^{b}$) = Class C
	\item[(2)] $\zeta$($f$) = $\zeta$($g$) = Class B and $\zeta$($f,g^{b}$) = Class A
	\item[(3)] $\zeta$($f$) = $\zeta$($g$) = Class C and $\zeta$($f,g^{b}$) = Class A
	\item[(4)] $\zeta$($f$) = $\zeta$($g$) = Class C and $\zeta$($f,g^{b}$) = Class B
	
\end{itemize}

Let's consider case (1) where $\zeta(f) = \zeta(g)$ belongs to Class B and $\zeta(f,g^b)$ belongs to Class C. We will provide two examples of such behaviors: LCA($94,37^b$) and LCA($104,51^b$).
In Figure.~\ref{Fig3}, we can observe the Class B behavior of ECA $94$ and ECA $37$. However, when we examine LCA($94,37^b$) for different block sizes ($b=125,250,500$), we observe chaotic dynamics instead of the expected periodic behavior. This transition from Class B to Class C dynamics is intriguing and worth studying.
Similarly, in the case of LCA($104,51^b$), we can also observe the transition from Class B to Class C dynamics. Figure.~\ref{Fig3} illustrates the interesting patterns resembling Pascal's triangle in LCA($104,51^b$) for block sizes $b=50,100,250$. 

Moving on to case (2), we focus on the behavior of ECA $15$, ECA $19$, and ECA $1$, which individually exhibit Class B dynamics, as seen in Figure.~\ref{Fig3}. Class B dynamics typically involve the formation of periodic patterns.
However, when we examine the LCAs formed by combining these rules, such as LCA($15,19^b$) and LCA($15,1^b$), with different block sizes ($b=100,125,250$), we observe a distinct change in behavior. Instead of maintaining the periodic dynamics seen in the individual rules, these LCAs exhibit homogeneous behavior, where the patterns converge to all-0 configuration and lack the periodic structure characteristic of Class B.

\begin{figure*}[hbt!]
	\begin{center}
		\scalebox{0.7}{
			\begin{tabular}{ccccc}
				ECA 45 & ECA 18 & ($45,18^{10}$) & ($45,18^{20}$) & ($45,18^{50}$) \\[6pt]
				\includegraphics[width=31mm]{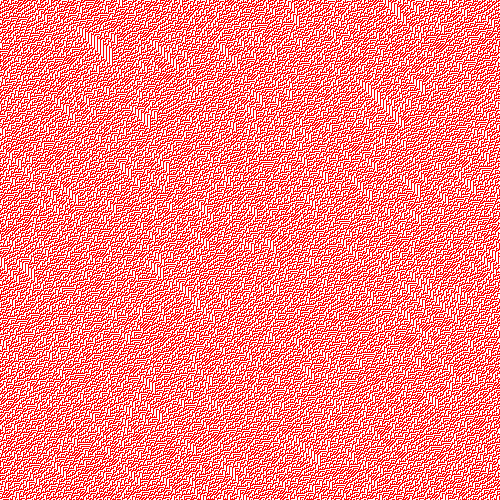} & \includegraphics[width=31mm]{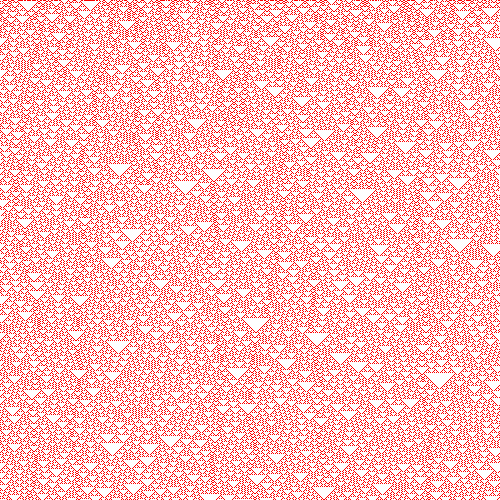} &   \includegraphics[width=31mm]{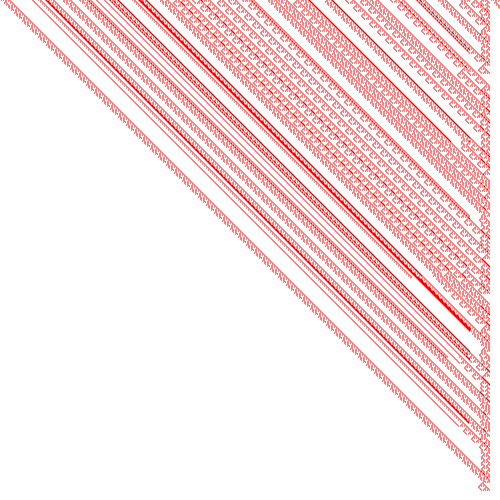} &   \includegraphics[width=31mm]{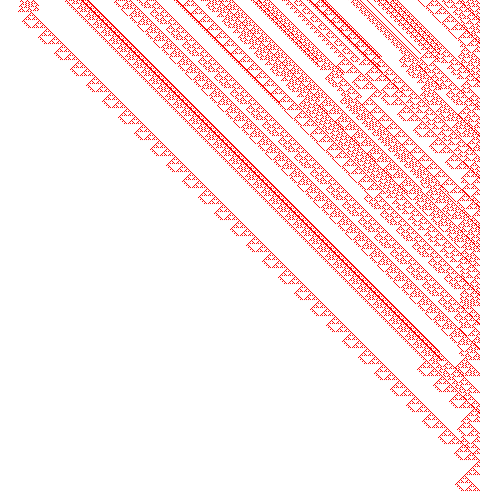} &   \includegraphics[width=31mm]{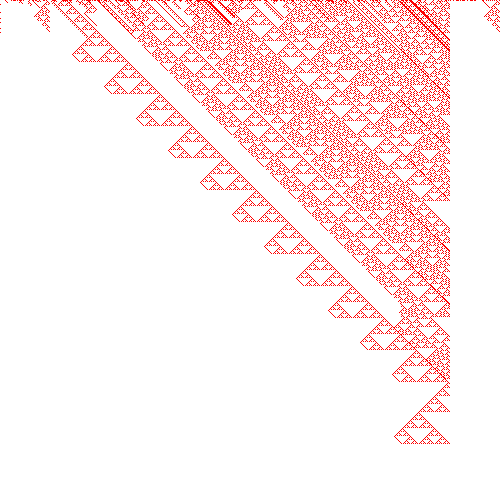} \\
				ECA 110 & ECA 30 & ($110,30^{20}$) & ($110,30^{25}$) & ($110,30^{125}$) \\[6pt]
				\includegraphics[width=31mm]{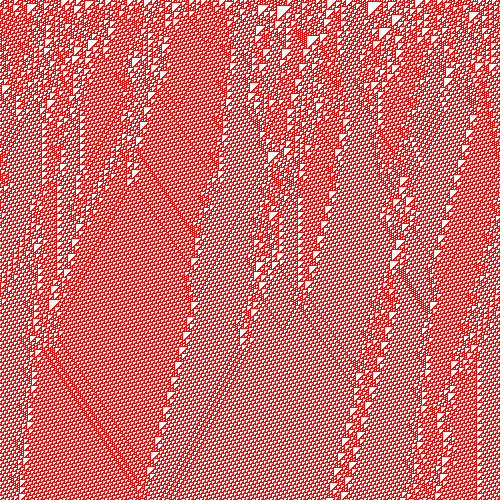} & \includegraphics[width=31mm]{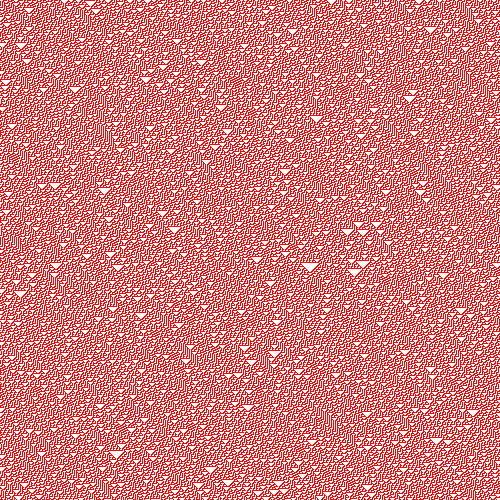} &   \includegraphics[width=31mm]{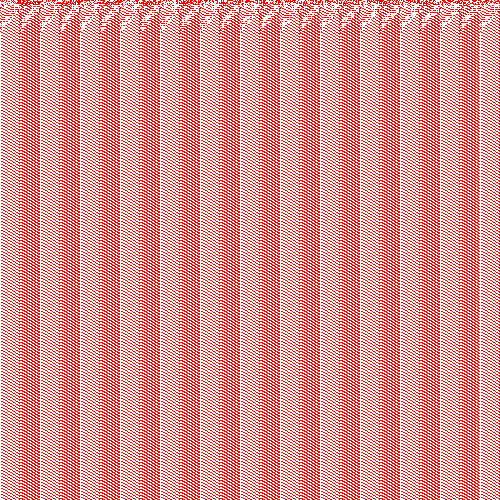} &   \includegraphics[width=31mm]{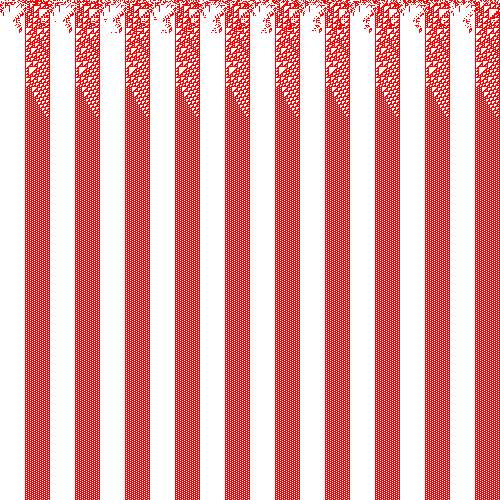} &   \includegraphics[width=31mm]{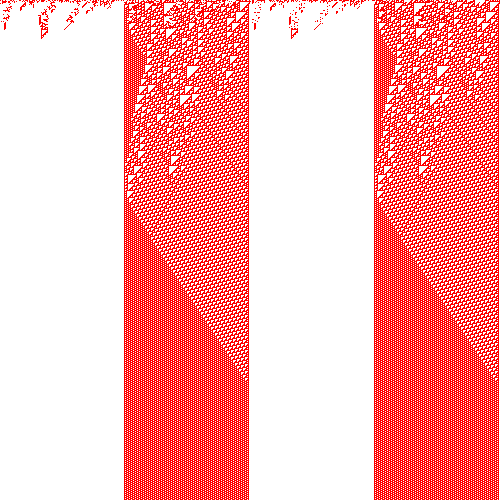} \\
				ECA 126 & ECA 18 & ($126,18^{5}$) & ($126,18^{50}$) & ($126,18^{125}$) \\[6pt]
				\includegraphics[width=31mm]{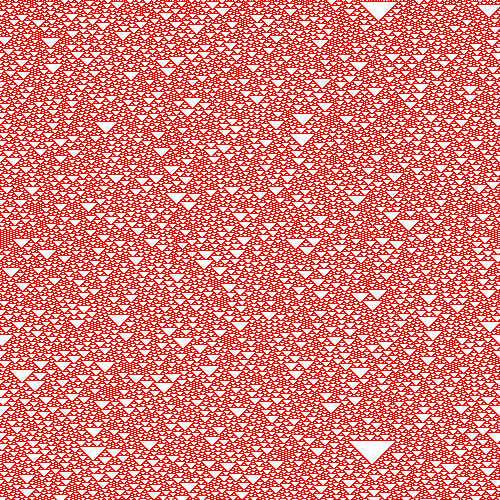} & \includegraphics[width=31mm]{twoeca/45-18/18.png} &   \includegraphics[width=31mm]{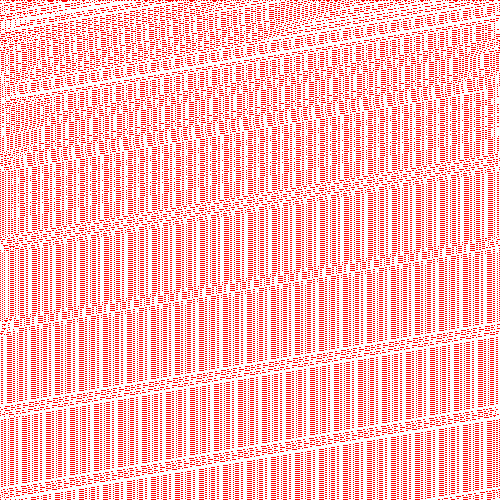} &   \includegraphics[width=31mm]{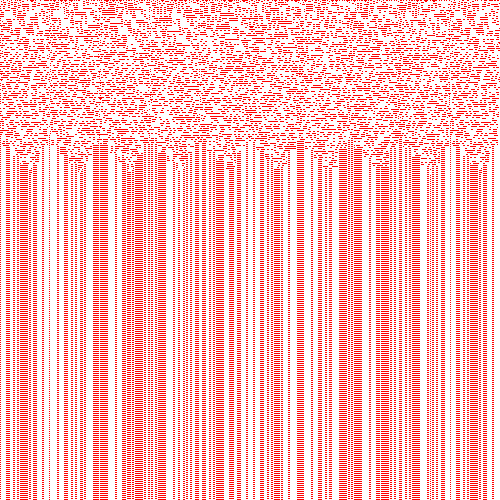} &   \includegraphics[width=31mm]{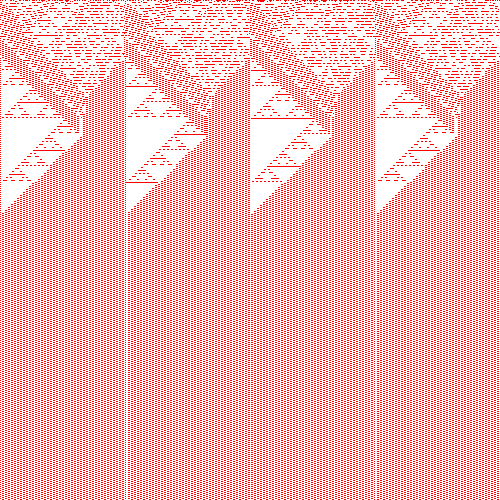} \\

		\end{tabular}}
		\caption{LCA($f,g^{b}$) dynamics when $\zeta$($f,g^{b}$) $\neq$ $\zeta$($f$) and $\zeta$($f$) = $\zeta$($g$).}
		\label{TSCA2}
	\end{center}
\end{figure*}

In case (3), we encounter the scenario where $\zeta(f) = \zeta(g)$ belongs to Class C, but the combination $\zeta(f,g^b)$ results in dynamics classified as Class A. 
For instance, if we consider rule $45$ as the default rule ($f$) and introduce rule $18$ as noise ($g$) with varying block sizes, the resulting dynamics transition towards a homogeneous configuration, characteristic of Class A behavior. It is worth noting that ECA $45$ and ECA $18$ individually exhibit chaotic behavior (Wolfram's Class III).
Figure.~\ref{TSCA2} provides the space-time diagram of LCA($45,18^b$) for three different block sizes: $b=10,20,50$. 

In case (4), we encounter the situation where $\zeta(f) = \zeta(g)$ belongs to Class C, but the combined dynamics $\zeta(f,g^b)$ exhibit behavior classified as Class B.
To illustrate this case, we provide two examples. Firstly, we consider rule 110 as the layer 0 rule ($f$) and rule 30 as the layer 1 rule ($g$). Despite both ECA 110 and ECA 30 individually exhibiting chaotic dynamics (Wolfram's Class III), the resulting dynamics of their combination, LCA($110,30^b$), evolve into a periodic configuration characteristic of Class B. Figure.~\ref{TSCA2} displays the space-time diagrams of LCA($110,30^b$) for different block sizes ($b=20,25,125$).
Similarly, we observe the same phenomenon in LCA($126,18^b$). While ECA 126 and ECA 18 display chaotic dynamics individually, the dynamics of LCA($126,18^b$) exhibit periodic behavior for different block sizes ($b=5,50,125$). 

\subsection{Dynamics when $\zeta$($f$) $\neq$ $\zeta$($g$)}

Next, we investigate the scenario where $\zeta(f) \neq \zeta(g)$, indicating that the default rule $f$ and noise rule $g$ have distinct dynamics. In our experiments, we have observed two types of results in this context. 

\begin{itemize}
	\item $\zeta$($f,g^{b}$) = $\zeta$($f$) or $\zeta$($f,g^{b}$) = $\zeta$($g$)
	\item $\zeta$($f,g^{b}$) $\neq$ $\zeta$($f$) and $\zeta$($f,g^{b}$) $\neq$ $\zeta$($g$)
\end{itemize}

Now let's consider the first possibility, where either $\zeta(f,g^b) = \zeta(f)$ or $\zeta(f,g^b) = \zeta(g)$. In this case, we can identify the following cases:

\begin{itemize}
	\item [(1)]$\zeta$($f$) = Class C, $\zeta$($g$) = Class B and $\zeta$($f,g^b$) = $\zeta$($f$)
	\item [(2)]$\zeta$($f$) = Class C, $\zeta$($g$) = Class B and $\zeta$($f,g^b$) = $\zeta$($g$)
	\item [(3)]$\zeta$($f$) = Class C, $\zeta$($g$) = Class A and $\zeta$($f,g^b$) = $\zeta$($f$)
	\item [(4)]$\zeta$($f$) = Class B, $\zeta$($g$) = Class A and $\zeta$($f,g^b$) = $\zeta$($f$)
	\item [(5)]$\zeta$($f$) = Class B, $\zeta$($g$) = Class C and $\zeta$($f,g^b$) = $\zeta$($f$)
	\item [(6)]$\zeta$($f$) = Class B, $\zeta$($g$) = Class C and $\zeta$($f,g^b$) = $\zeta$($g$)
	\item [(7)]$\zeta$($f$) = Class A, $\zeta$($g$) = Class B and $\zeta$($f,g^b$) = $\zeta$($f$)
	\item [(8)]$\zeta$($f$) = Class A, $\zeta$($g$) = Class B and $\zeta$($f,g^b$) = $\zeta$($g$)
	\item [(9)]$\zeta$($f$) = Class A, $\zeta$($g$) = Class C and $\zeta$($f,g^b$) = $\zeta$($f$)
	\item [(10)]$\zeta$($f$) = Class A, $\zeta$($g$) = Class C and $\zeta$($f,g^b$) = $\zeta$($g$)
	\item [(11)]$\zeta$($f$) = Class B, $\zeta$($g$) = Class A and $\zeta$($f,g^b$) = $\zeta$($g$)
	\item [(12)]$\zeta$($f$) = Class C, $\zeta$($g$) = Class A and $\zeta$($f,g^b$) = $\zeta$($g$)
\end{itemize}

Let's consider case (1), where $\zeta(f)$ is in Class C, $\zeta(g)$ is in Class B, and $\zeta(f,g^b) = \zeta(f)$. In this case, we can observe the chaotic (Class C) behavior of ECA $45$ and the periodic (Class B) behavior of ECA $50$. 

\begin{figure*}[hbt!]
	\begin{center}
		\scalebox{0.7}{
			\begin{tabular}{ccccc}				
				
				ECA 45 & ECA 50 & ($45,50^{5}$) & ($45,50^{20}$) & ($45,50^{50}$) \\
				
				\includegraphics[width=31mm]{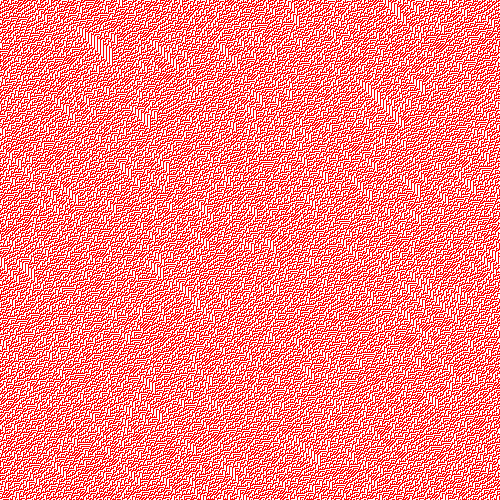}  &  \includegraphics[width=31mm]{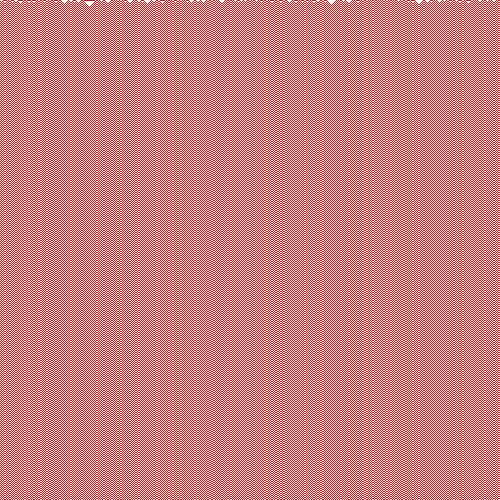}  &   \includegraphics[width=31mm]{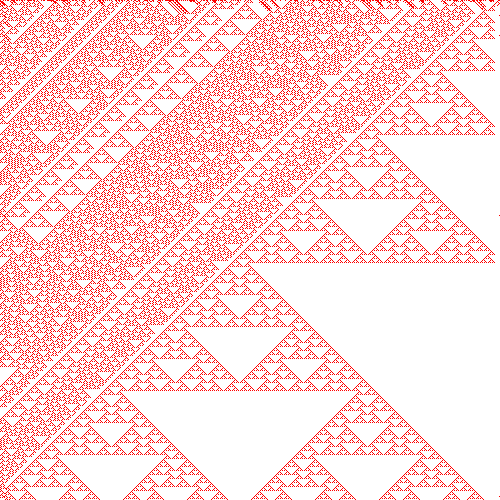}  &  \includegraphics[width=31mm]{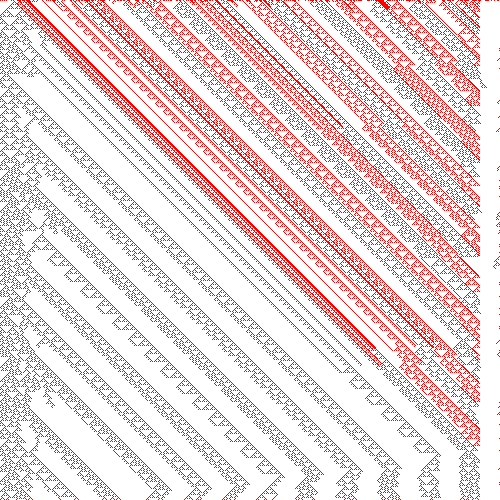}  & \includegraphics[width=31mm]{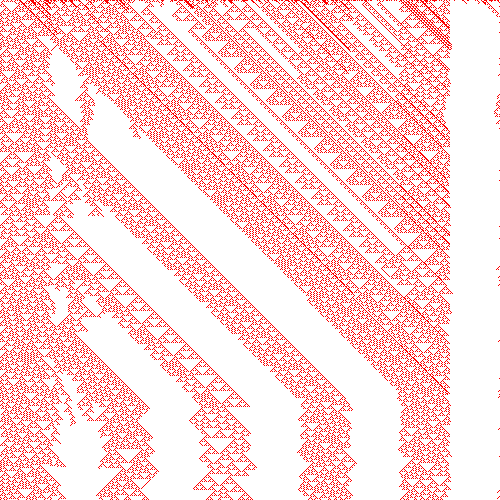} \\
				
				ECA 18 & ECA 3 & ($18,3^{25}$) & ($18,3^{100}$) & ($18,3^{125}$) \\
				
				\includegraphics[width=31mm]{twoeca/45-18/18.png}  &  \includegraphics[width=31mm]{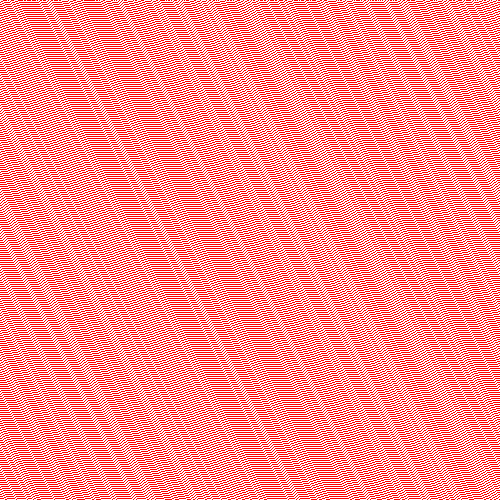}  &   \includegraphics[width=31mm]{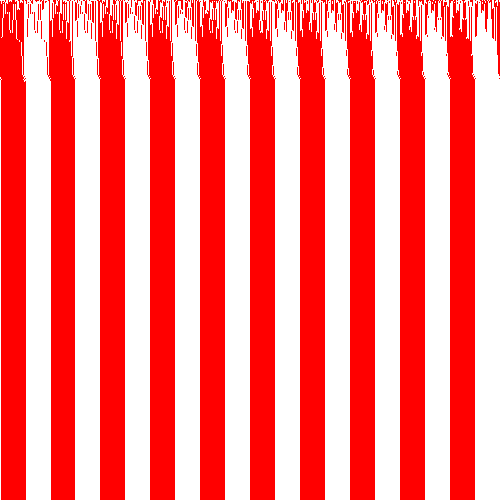}  &  \includegraphics[width=31mm]{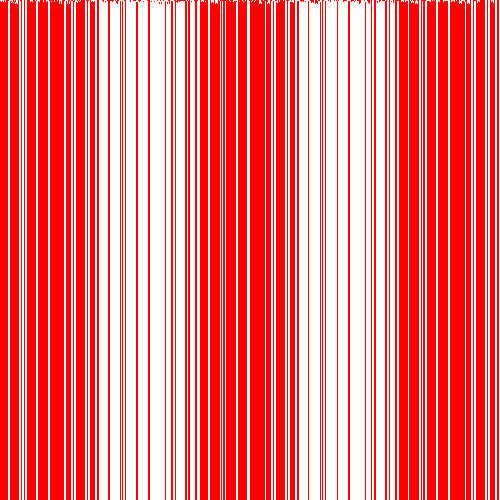}  & \includegraphics[width=31mm]{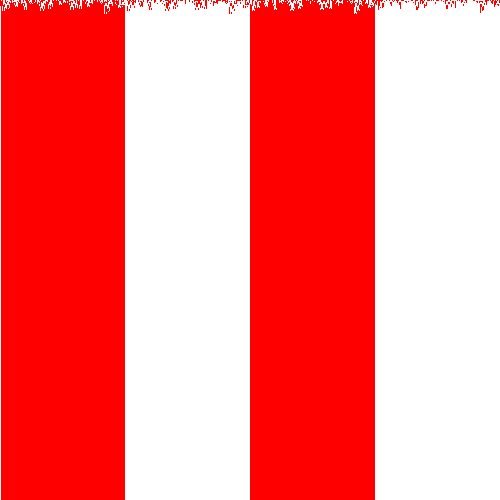} \\
				
				ECA 45 & ECA 160 & ($45,160^{100}$) & ($45,160^{125}$) & ($45,160^{250}$) \\
				
				\includegraphics[width=31mm]{twoeca/45-50/45.png} & \includegraphics[width=31mm]{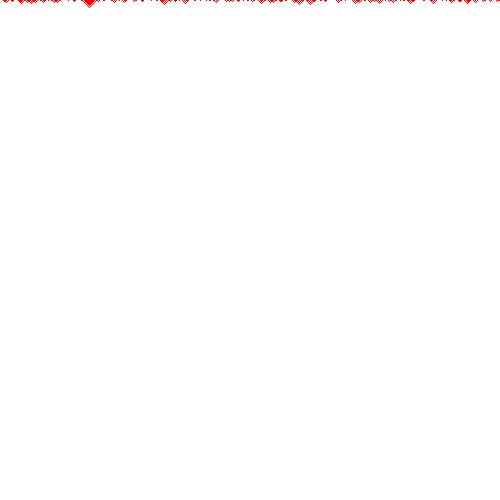}  & \includegraphics[width=31mm]{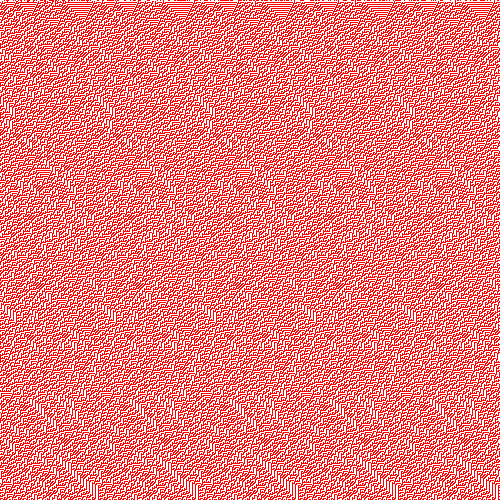} & \includegraphics[width=31mm]{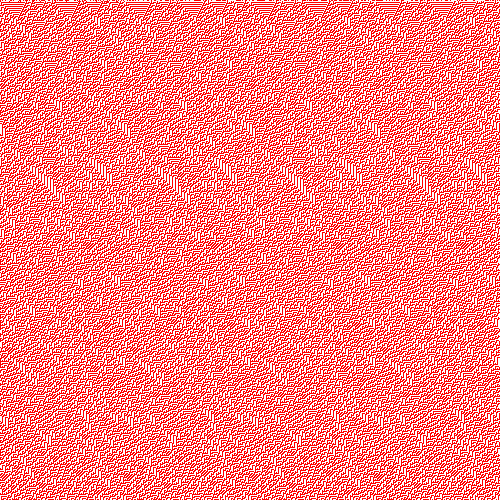}  & \includegraphics[width=31mm]{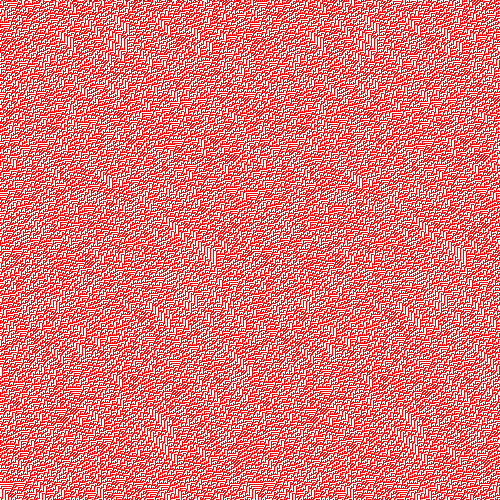} \\
				
				ECA 25 & ECA 128 & ($25,128^{20}$) & ($25,128^{50}$) & ($25,128^{125}$) \\
				
				\includegraphics[width=31mm]{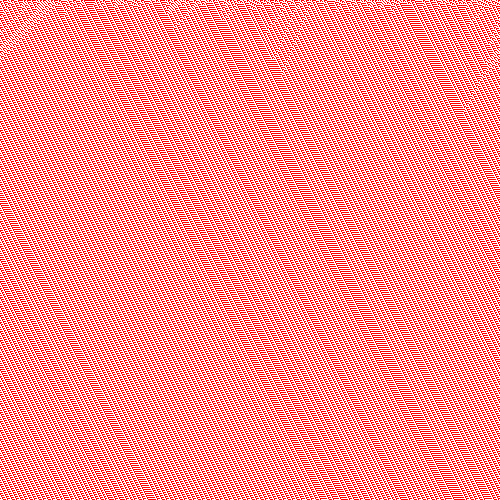} & \includegraphics[width=31mm]{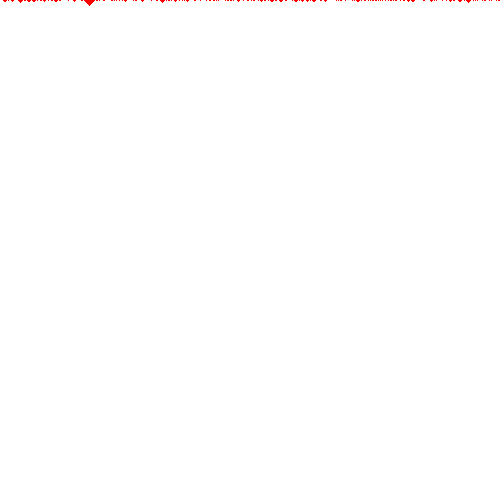}  & \includegraphics[width=31mm]{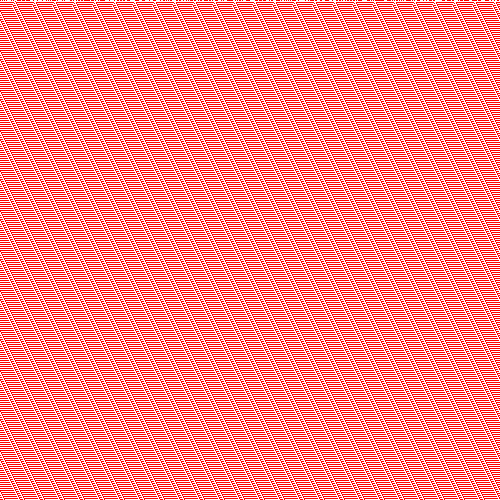} & \includegraphics[width=31mm]{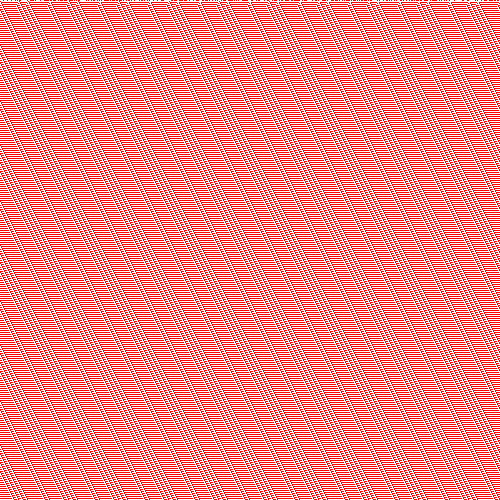}  & \includegraphics[width=31mm]{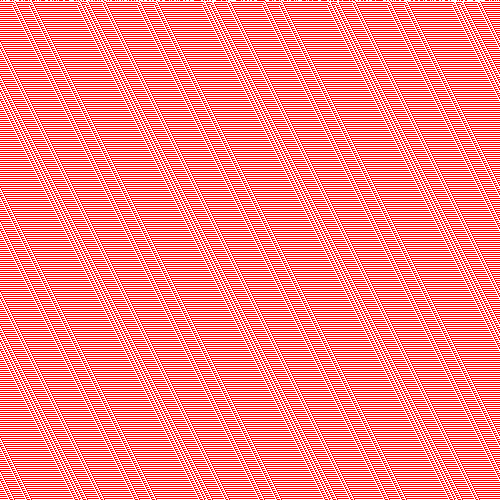} \\
		\end{tabular}}
		\caption{LCA($f,g^{b}$) dynamics where either $\zeta$($f,g^{b}$) = $\zeta$($f$) or $\zeta$($f,g^{b}$) = $\zeta$($g$).}
		\label{Fig4}
	\end{center}
\end{figure*}

However, when we examine the LCA($45,50^b$) for different block sizes ($b=5,20,50$), as shown in Figure.~\ref{Fig4}, we find that the dynamics of the LCA aligns with the dynamics of the default rule $f$, which is chaotic.
This indicates that even though the noise rule $g$ exhibits periodic behavior (Class B), the overall dynamics of the LCA is dominated by the chaotic behavior of the rule $f$. The resulting dynamics of the LCA($45,50^b$) exhibit chaotic behavior, similar to the dynamics of rule $f$. Similarly, Figure~\ref{Fig4} represents the space-time diagram for case 2-4. For LCA($18,3^{b}$), dynamics of rule 3 dominates, similarly for LCA($45,160^{b}$), dynamics of rule 45 dominates and lastly for LCA($25,128^{b}$), dynamics of rule 25 dominates. These LCAs used as example for case 2-4 respectively.

\begin{figure*}[hbt!]
	\begin{center}
		\scalebox{0.7}{
			\begin{tabular}{ccccc}
				ECA 43 & ECA 45 & ($43,45^{50}$) & ($43,45^{100}$) & ($43,45^{250}$) \\
				
				\includegraphics[width=31mm]{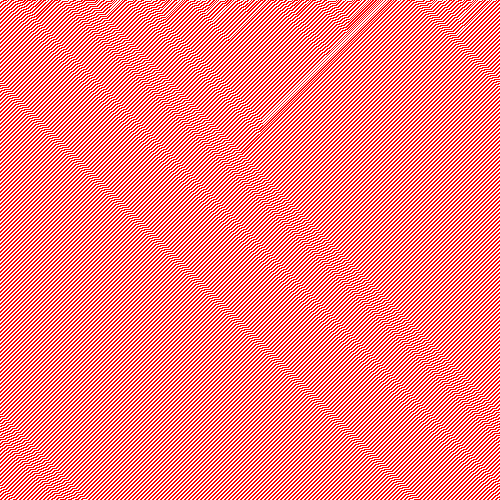}  &  \includegraphics[width=31mm]{twoeca/45-50/45.png}  &   \includegraphics[width=31mm]{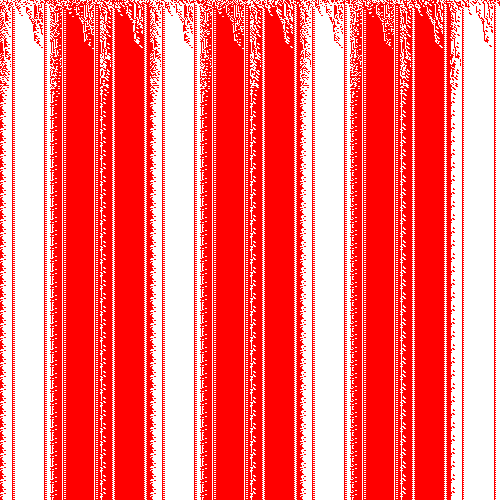}  &  \includegraphics[width=31mm]{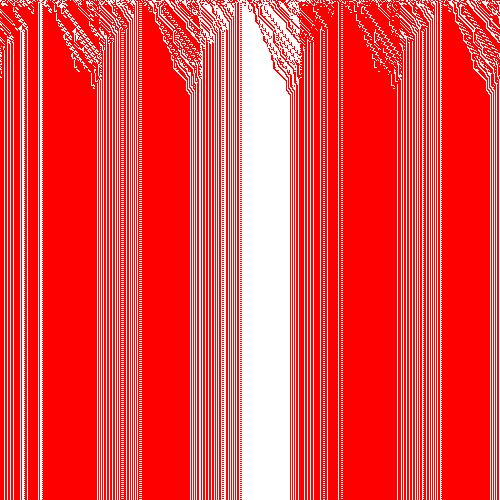}  & \includegraphics[width=31mm]{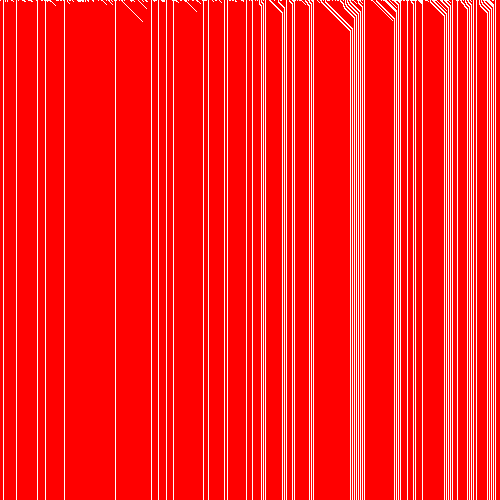} \\
				
				ECA 28 & ECA 60 & ($28,60^{50}$) & ($28,60^{100}$) & ($28,60^{125}$) \\
				
				\includegraphics[width=31mm]{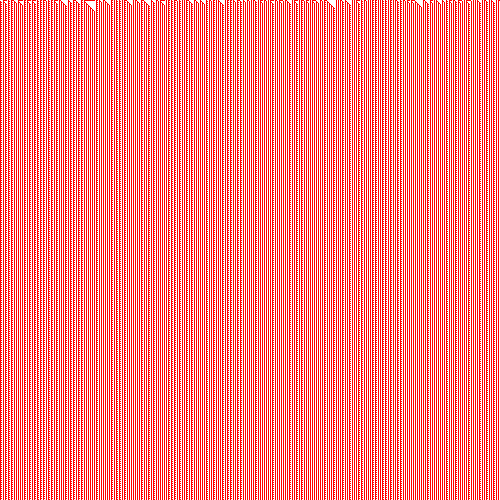}  &  \includegraphics[width=31mm]{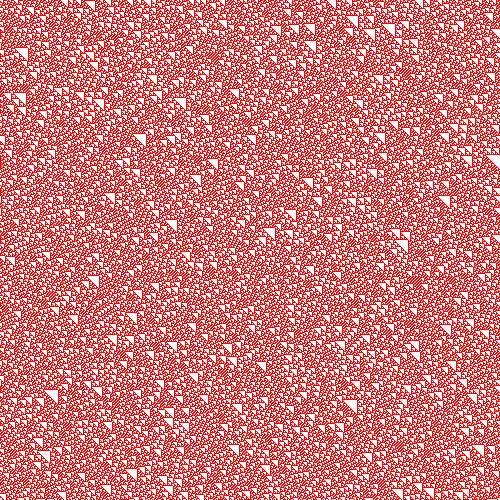}  &   \includegraphics[width=31mm]{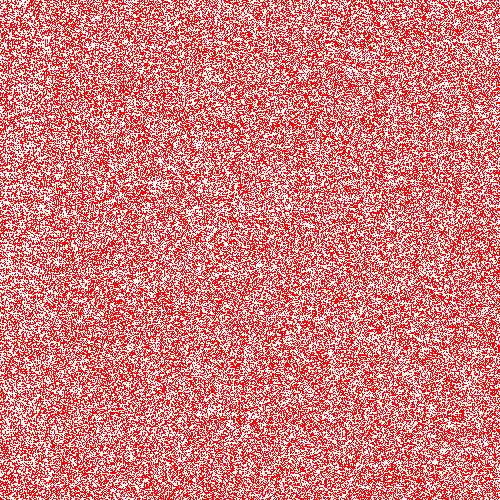}  &  \includegraphics[width=31mm]{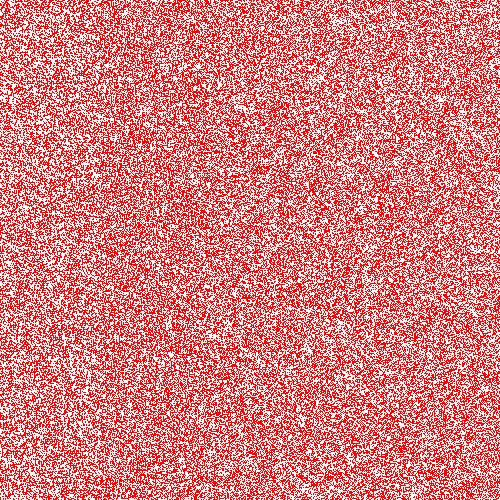}  & \includegraphics[width=31mm]{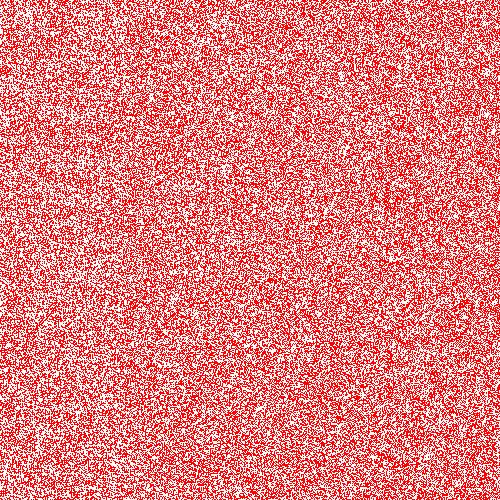} \\
				
				ECA 168 & ECA 28 & ($168,28^{10}$) & ($168,28^{25}$) & ($168,28^{50}$) \\
				
				\includegraphics[width=31mm]{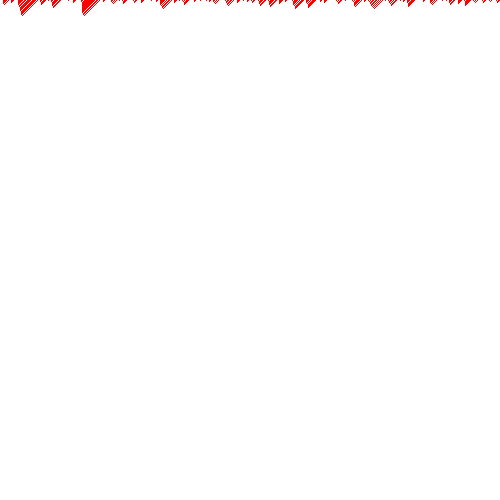} & \includegraphics[width=31mm]{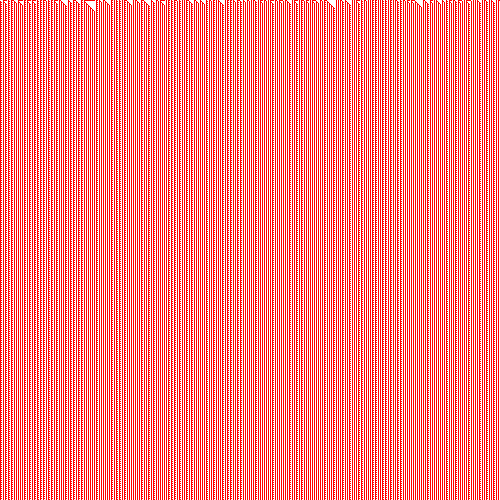}  & \includegraphics[width=31mm]{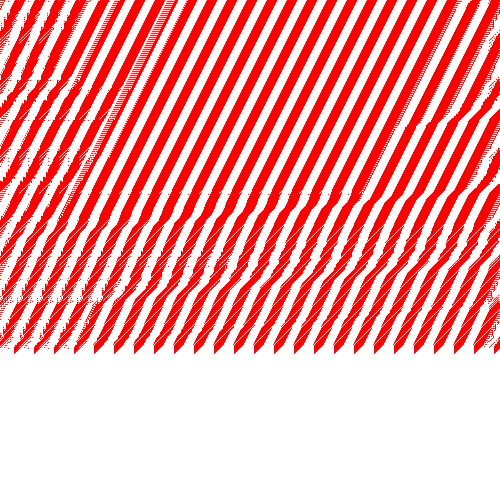} & \includegraphics[width=31mm]{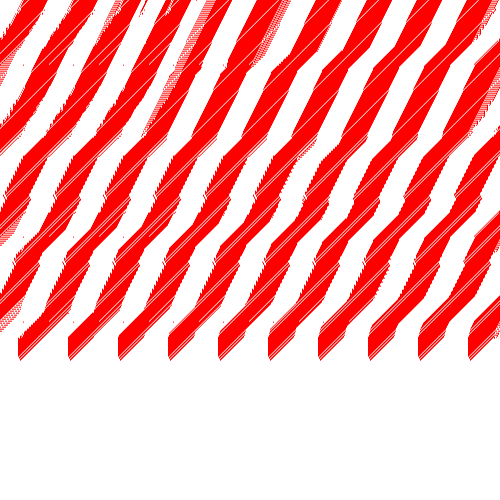}  & \includegraphics[width=31mm]{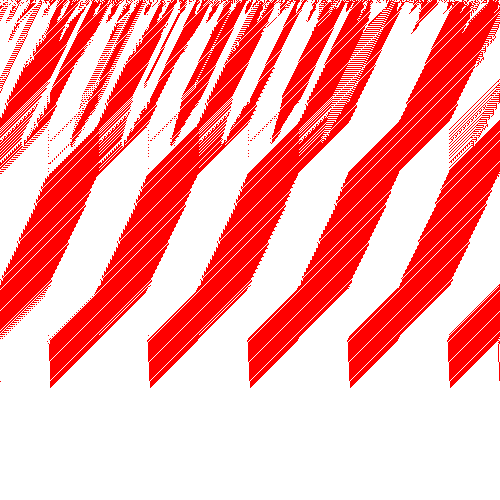} \\
				
				ECA 136 & ECA 29 & ($136,29^{50}$) & ($136,29^{100}$) & ($136,29^{125}$) \\
				
				\includegraphics[width=31mm]{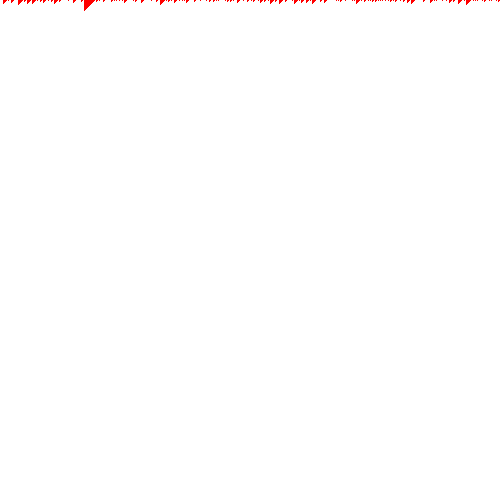} & \includegraphics[width=31mm]{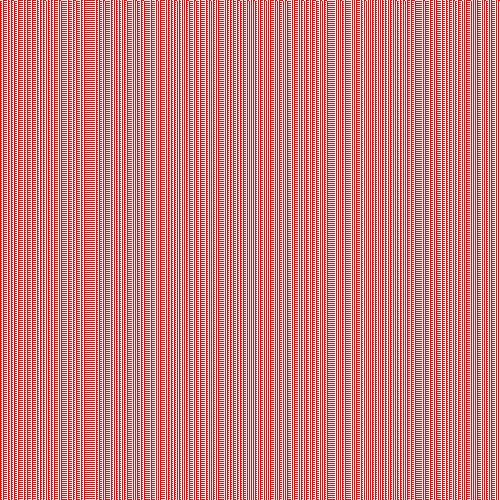}  & \includegraphics[width=31mm]{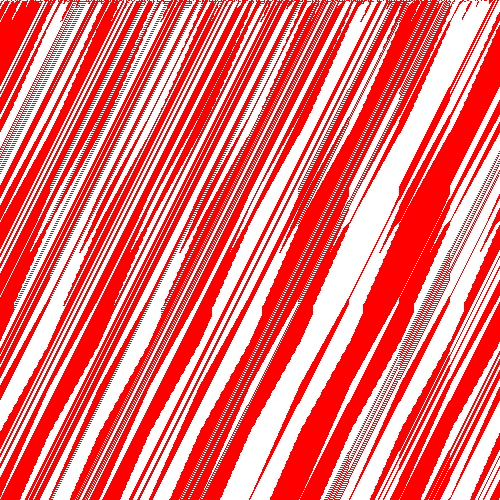} & \includegraphics[width=31mm]{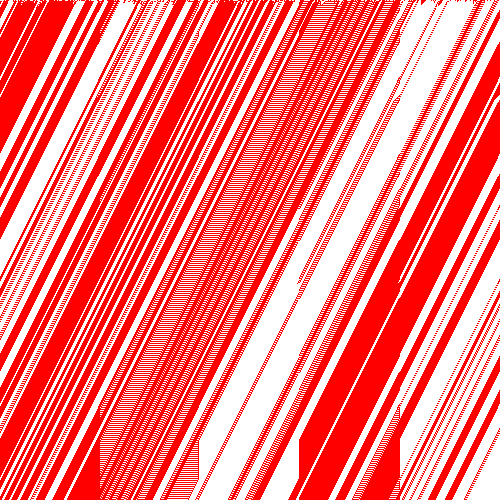}  & \includegraphics[width=31mm]{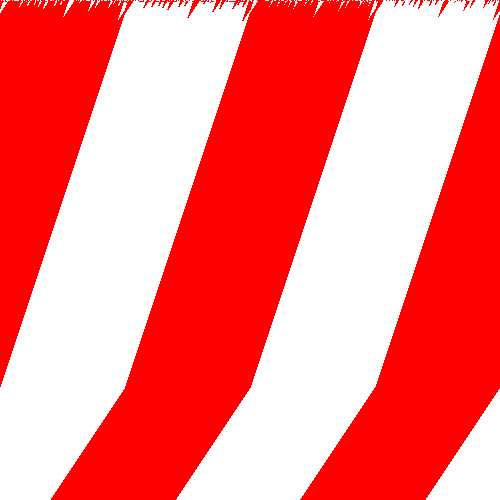} \\
		\end{tabular}}
		\caption{LCA($f,g^{b}$) dynamics where either $\zeta$($f,g^{b}$) = $\zeta$($f$) or $\zeta$($f,g^{b}$) = $\zeta$($g$). Here, $\zeta$($f$) $\neq$ $\zeta$($g$).}
		\label{TSCA4}
	\end{center}
\end{figure*}

Next, let's consider case (5), where $\zeta(f)$ is in Class B, $\zeta(g)$ is in Class C, and $\zeta(f,g^b) = \zeta(f)$. In this case, we have ECA $43$ exhibiting periodic dynamics (Class B) and ECA $45$ displaying chaotic dynamics (Class C) individually.
To demonstrate this case, we consider rule $43$ as the default rule and rule $45$ as the noise rule for different block sizes ($b=50,100,250$) as shown in Figure.~\ref{TSCA4}. The resulting dynamics of the LCA($43,45^b$) exhibit periodic behavior, similar to the dynamics of rule $43$. Even though the noise rule $g$ displays chaotic behavior (Class C), the resulting dynamics of the LCA align with the periodic behavior of the default rule $f$. 
In addition, we explore case 6-8, where the dynamics of the LCA is dominated by either the default rule or the noise rule.
In the case of LCA($28,60^b$), as shown in Figure.~\ref{TSCA4}, the dynamics are predominantly influenced by rule 60, with the resulting space-time diagram exhibiting patterns similar to the dynamics of rule 60.
Similarly, in the case of LCA($168,28^b$), the dynamics are primarily driven by rule 168, as observed in Figure.~\ref{TSCA4}. The space-time diagram showcases patterns that align with the dynamics of rule 168.
Lastly, for LCA($136,29^b$), the dominant influence comes from rule 29, as depicted in Figure.~\ref{TSCA4}. The resulting dynamics of the LCA resemble the behavior of rule 29.
These specific LCAs, chosen as examples for case 6-8, demonstrate the prevalence of either the default rule or the noise rule in determining the overall dynamics.

\begin{figure*}[hbt!]
	\begin{center}
		\scalebox{0.7}{
			\begin{tabular}{ccccc}
				ECA 168 & ECA 126 & ($168,126^{50}$) & ($168,126^{100}$) & ($168,126^{125}$) \\
				
				\includegraphics[width=31mm]{twoeca/168-28/168.png} & \includegraphics[width=31mm]{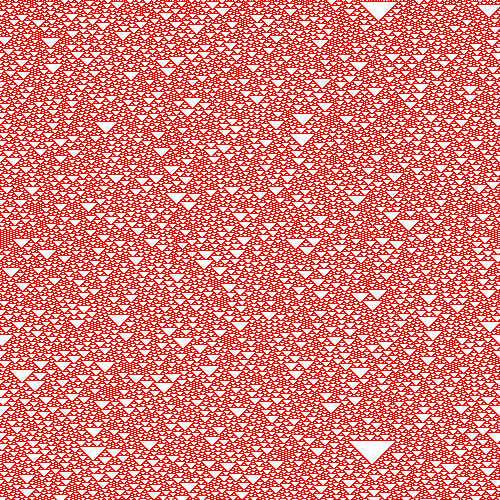}  & \includegraphics[width=31mm]{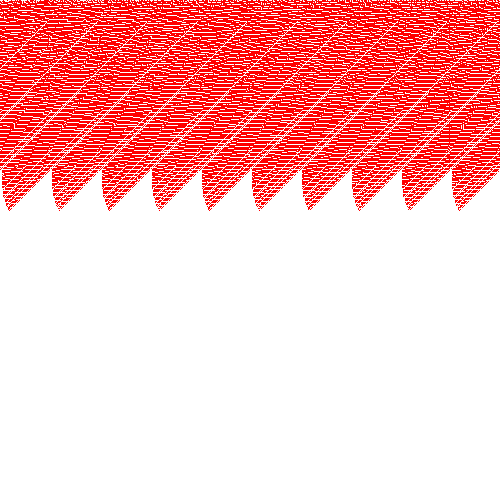} & \includegraphics[width=31mm]{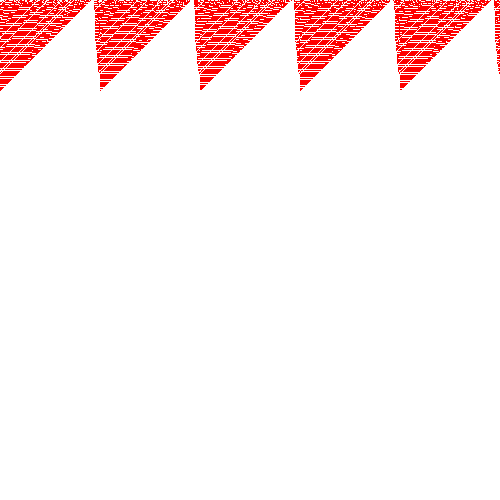}  & \includegraphics[width=31mm]{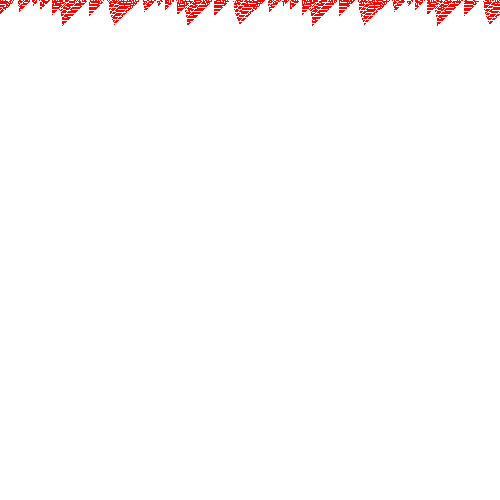} \\
				
				ECA 40 & ECA 45 & ($40,45^{50}$) & ($40,45^{100}$) & ($40,45^{125}$) \\
				
				\includegraphics[width=31mm]{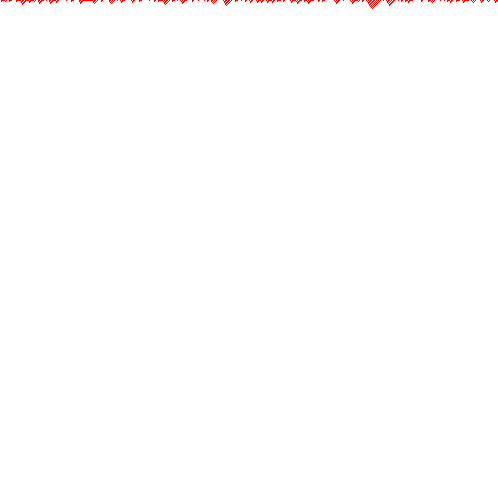} & \includegraphics[width=31mm]{twoeca/45-18/45.png}  & \includegraphics[width=31mm]{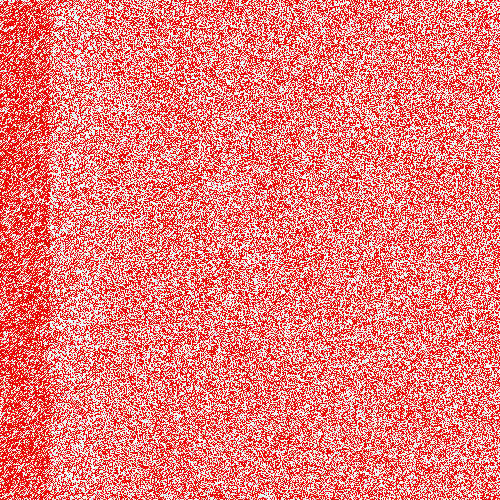} & \includegraphics[width=31mm]{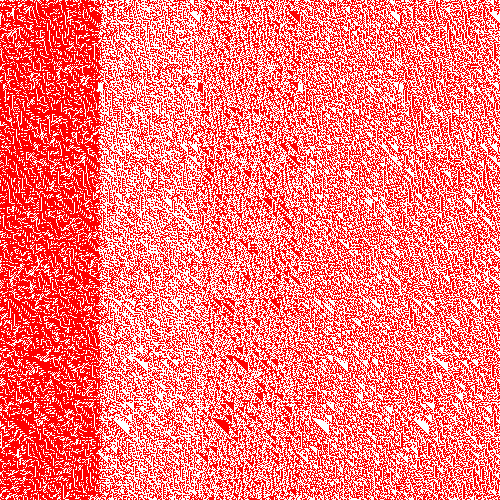}  & \includegraphics[width=31mm]{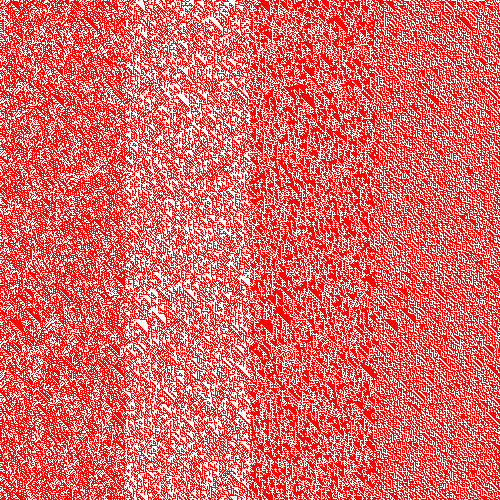} \\
				
				ECA 140 & ECA 40 & ($140,40^{50}$) & ($140,40^{100}$) & ($140,40^{125}$) \\
				
				\includegraphics[width=31mm]{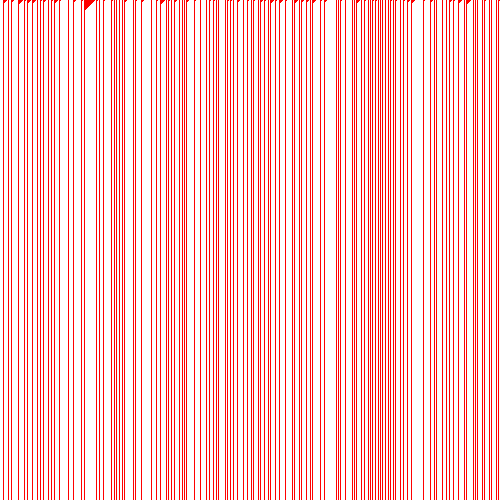} & \includegraphics[width=31mm]{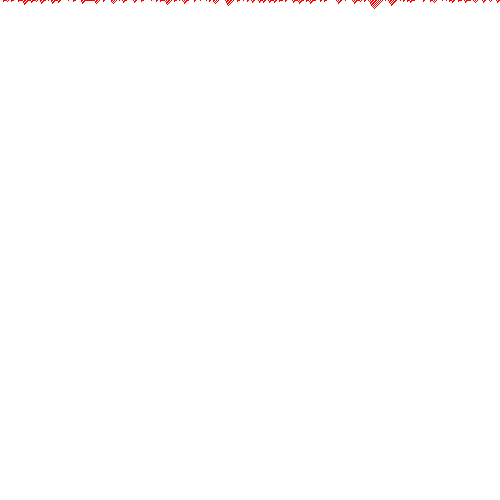}  & \includegraphics[width=31mm]{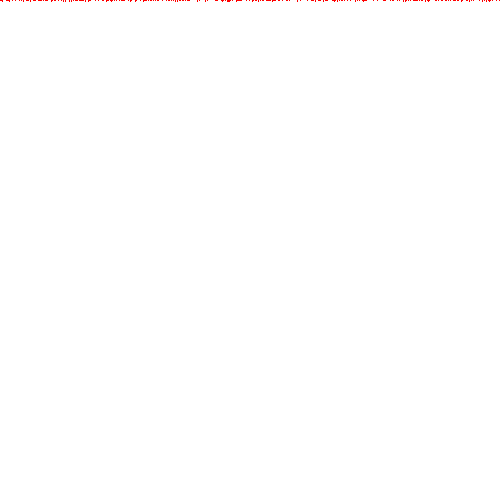} & \includegraphics[width=31mm]{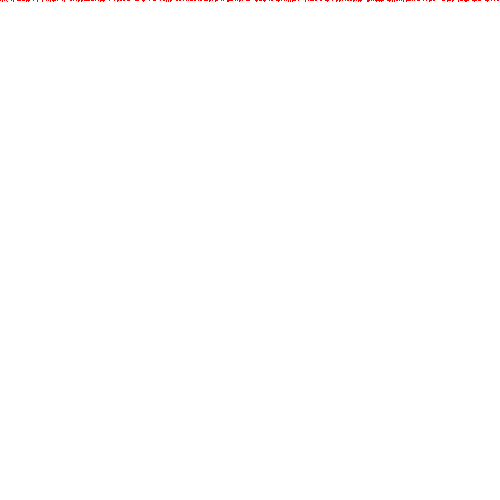}  & \includegraphics[width=31mm]{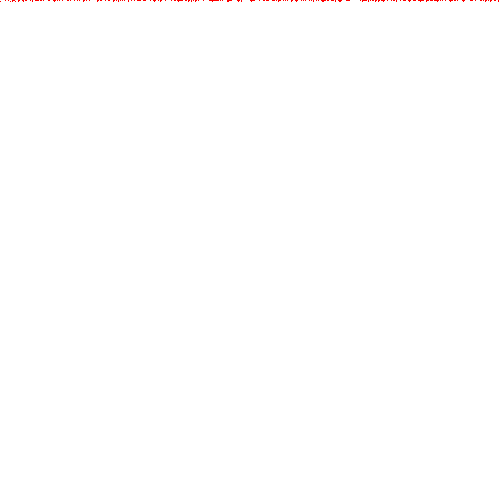} \\
				
				ECA 126 & ECA 136 & ($126,136^{100}$) & ($126,136^{125}$) & ($126,136^{250}$) \\
				
				\includegraphics[width=31mm]{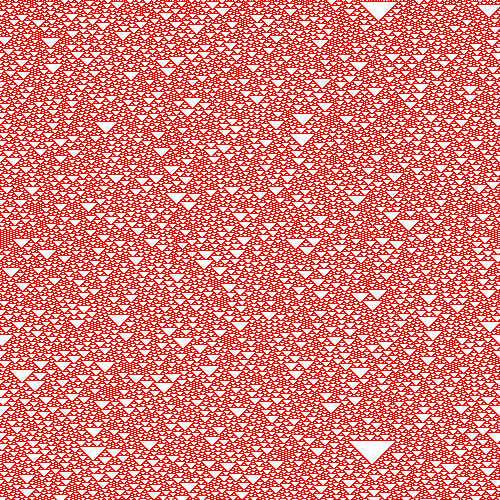} & \includegraphics[width=31mm]{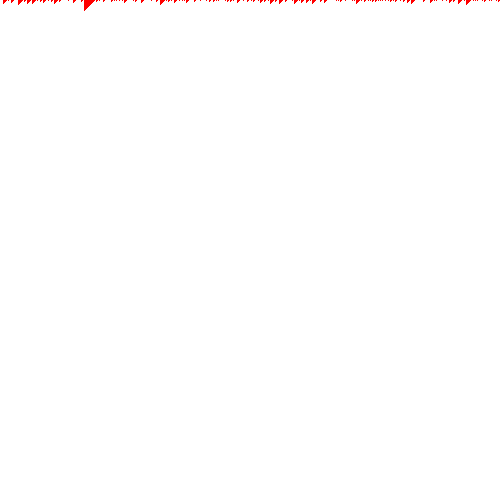}  & \includegraphics[width=31mm]{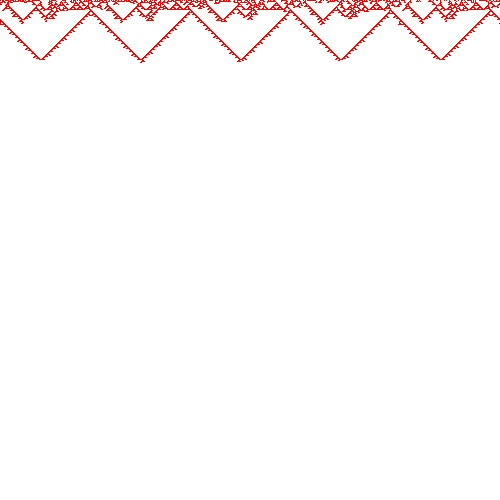} & \includegraphics[width=31mm]{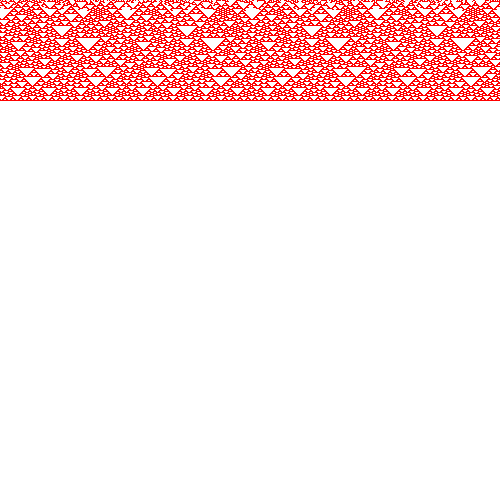}  & \includegraphics[width=31mm]{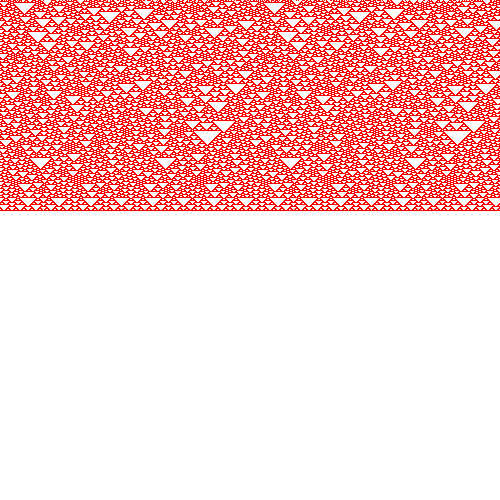} \\
		\end{tabular}}
		\caption{LCA($f,g^{b}$) dynamics where either $\zeta$($f,g^{b}$) = $\zeta$($f$) or $\zeta$($f,g^{b}$) = $\zeta$($g$). Here, $\zeta$($f$) $\neq$ $\zeta$($g$).}
		\label{TSCA5}
	\end{center}
\end{figure*}

Let's now consider case (9), we explore the scenario where $\zeta(f)$ is in Class A, $\zeta(g)$ is in Class C, and $\zeta(f,g^b) = \zeta(f)$. In this case, we have ECA $168$ exhibiting periodic dynamics (Class A), while ECA $126$ displays chaotic dynamics (Class C) individually.

To demonstrate this case, we consider rule $168$ as the default rule and rule $126$ as the noise rule for different block sizes ($b=50,100,125$), as shown in Figure.~\ref{TSCA5}.
Remarkably, the resulting dynamics of the LCA($168,126^b$) exhibit homogeneous behavior, similar to the dynamics of rule $168$.
Despite the chaotic behavior of the noise rule $g$ (Class C), the overall dynamics of the LCA align with the homogeneous behavior of the default rule $f$ (Class A). This suggests that the dominant influence of the default rule $f$.
Likewise, Figure.~\ref{TSCA5} illustrates the space-time diagrams for cases 10-12. In LCA($40,45^b$), the dynamics are primarily influenced by rule 45, as observed in the space-time diagram. Similarly, for LCA($140,40^b$), the dynamics are predominantly driven by rule 40. Lastly, in the case of LCA($126,136^b$), the dominant influence comes from rule 136.
These specific LCAs, selected as examples for cases 10-12, demonstrate the prevalence of either the default rule or the noise rule in shaping the overall dynamics. 

\begin{figure*}[!h]
	\begin{center}
		\scalebox{0.7}{
			\begin{tabular}{ccccc}
				ECA 37 & ECA 40 & ($37,40^{10}$) & ($37,40^{20}$) & ($37,40^{25}$) \\[6pt]
				\includegraphics[width=31mm]{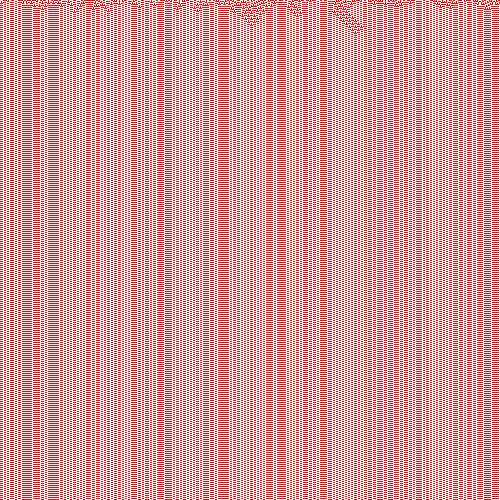} & \includegraphics[width=31mm]{twoeca/37-40/40.png} & \includegraphics[width=31mm]{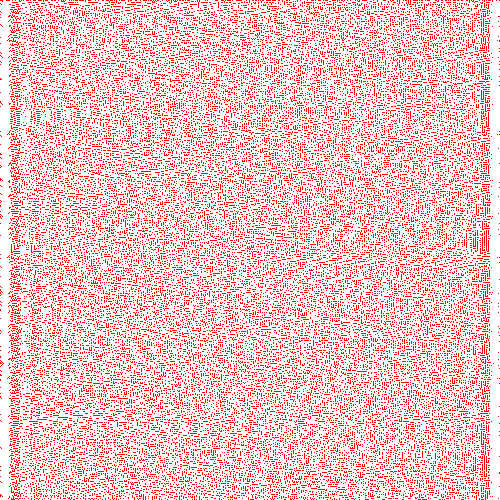} & \includegraphics[width=31mm]{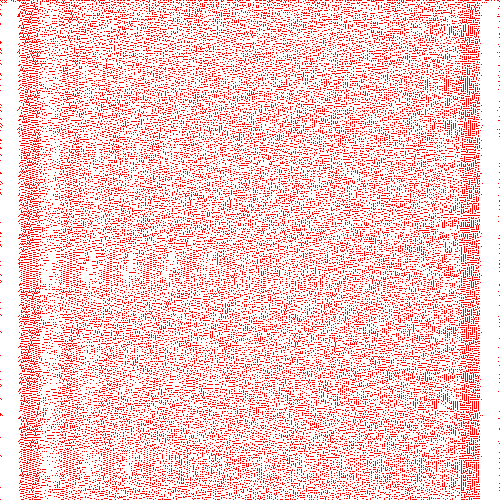} & \includegraphics[width=31mm]{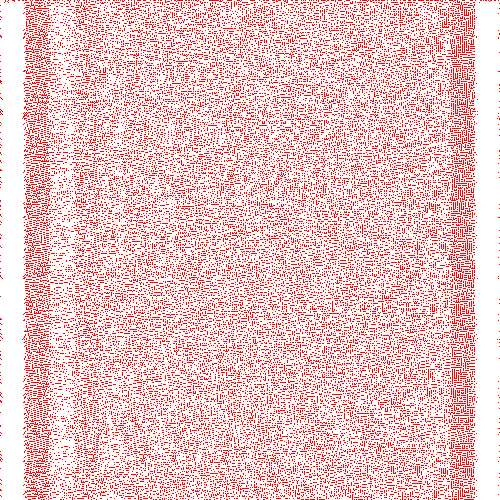} \\
				
				ECA 90 & ECA 56 & ($90,56^{50}$) & ($90,56^{100}$) & ($90,56^{125}$) \\[6pt]
				\includegraphics[width=31mm]{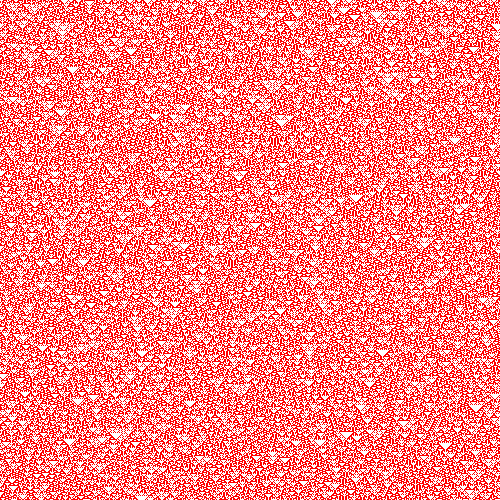} & \includegraphics[width=31mm]{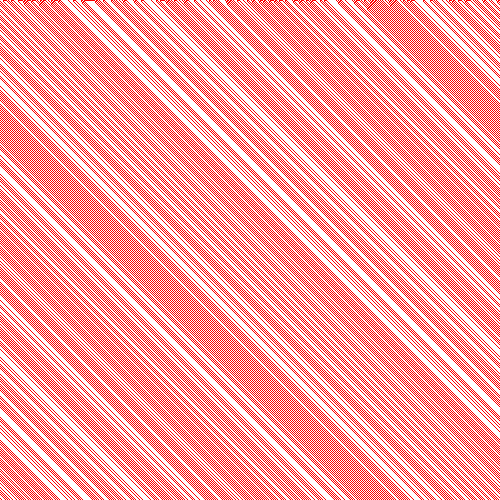} & \includegraphics[width=31mm]{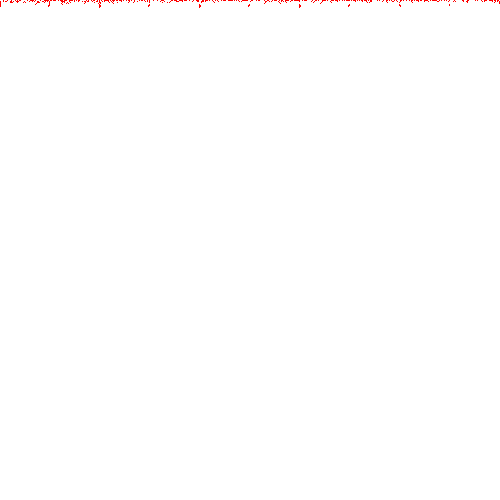} & \includegraphics[width=31mm]{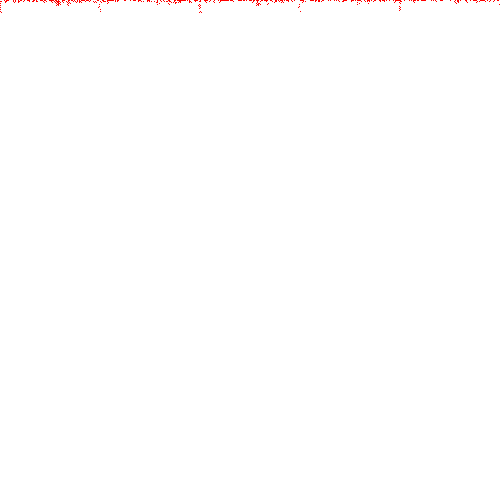} & \includegraphics[width=31mm]{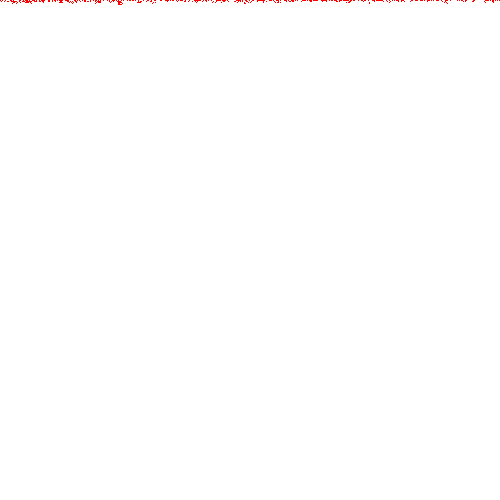} \\
				
				ECA 41 & ECA 32 & ($41,32^{100}$) & ($41,32^{125}$) & ($41,32^{250}$) \\[6pt]
				\includegraphics[width=31mm]{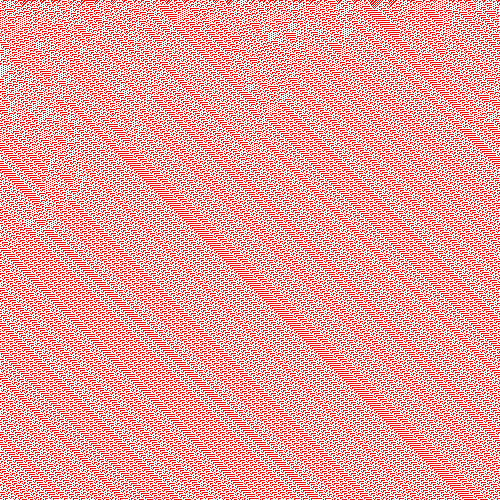} & \includegraphics[width=31mm]{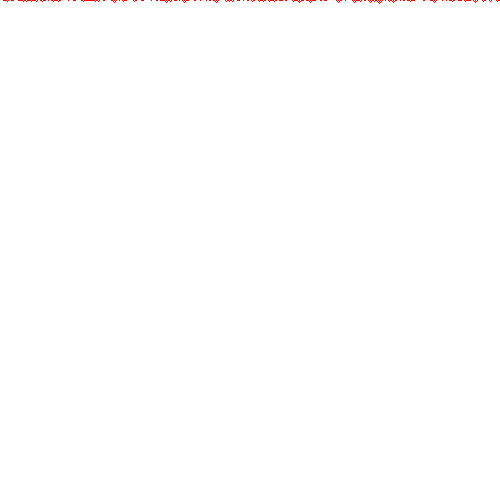} & \includegraphics[width=31mm]{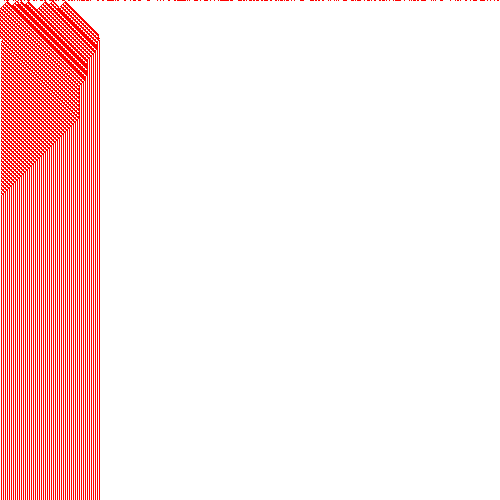} & \includegraphics[width=31mm]{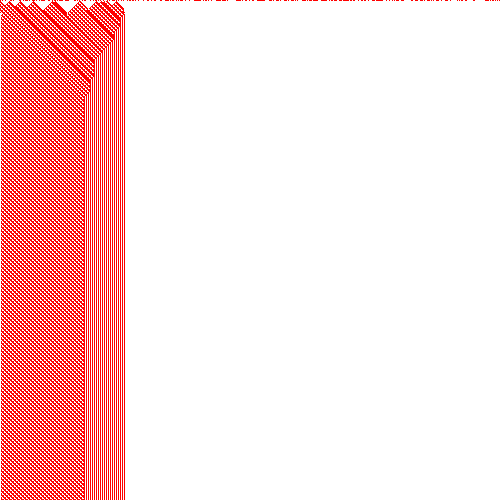} & \includegraphics[width=31mm]{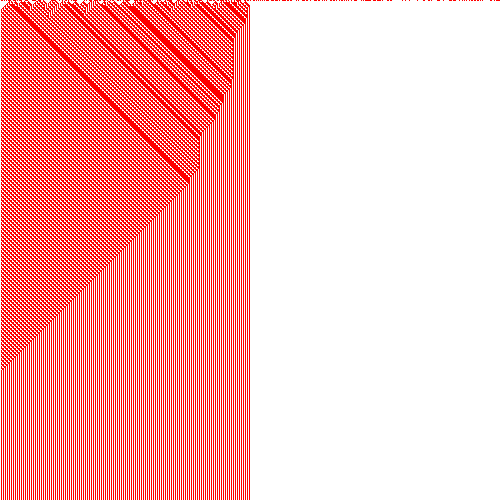} \\

		\end{tabular}}

		\caption{LCA($f,g^{b}$) dynamics where $\zeta$($f,g^{b}$) $\neq$ $\zeta$($f$) and $\zeta$($f,g^{b}$) $\neq$ $\zeta$($g$). Here, $\zeta$($f$) $\neq$ $\zeta$($g$).}
		\label{Fig4A}
	\end{center}
\end{figure*}

Now, we move on to the remaining scenarios where $\zeta(f) \neq \zeta(g)$, $\zeta(f,g^b) \neq \zeta(f)$, and $\zeta(f,g^b) \neq \zeta(g)$. In these cases, we have the following six scenarios to consider. Let's explore each of them in detail. 

\begin{itemize}
	\item [(1)] $\zeta$($f$) = Class B, $\zeta$($g$) = Class A and $\zeta$($f,g^b$) = Class C
	\item [(2)] $\zeta$($f$) = Class C, $\zeta$($g$) = Class B and $\zeta$($f,g^b$) = Class A
	\item [(3)] $\zeta$($f$) = Class C, $\zeta$($g$) = Class A and $\zeta$($f,g^b$) = Class B
	\item [(4)] $\zeta$($f$) = Class A, $\zeta$($g$) = Class B and $\zeta$($f,g^b$) = Class C
	\item [(5)] $\zeta$($f$) = Class B, $\zeta$($g$) = Class C and $\zeta$($f,g^b$) = Class A
	\item [(6)] $\zeta$($f$) = Class A, $\zeta$($g$) = Class C and $\zeta$($f,g^b$) = Class B
\end{itemize}

\begin{figure*}[hbt!]
	\begin{center}
		\scalebox{0.7}{
			\begin{tabular}{ccccc}
				ECA 136 & ECA 37 & ($136,37^{20}$) & ($136,37^{25}$) & ($136,37^{50}$) \\[6pt]
				\includegraphics[width=31mm]{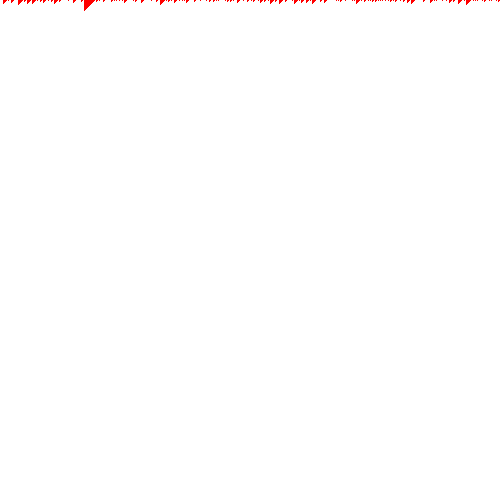} & \includegraphics[width=31mm]{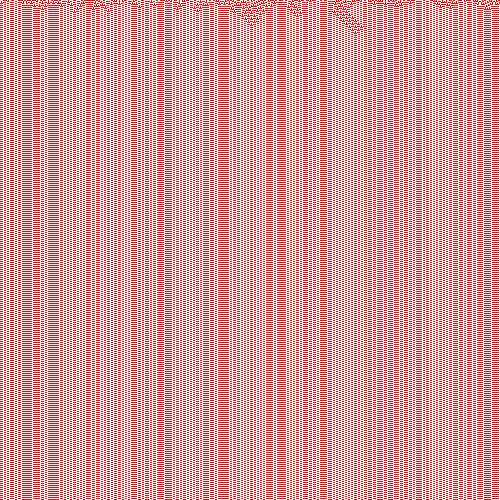} & \includegraphics[width=31mm]{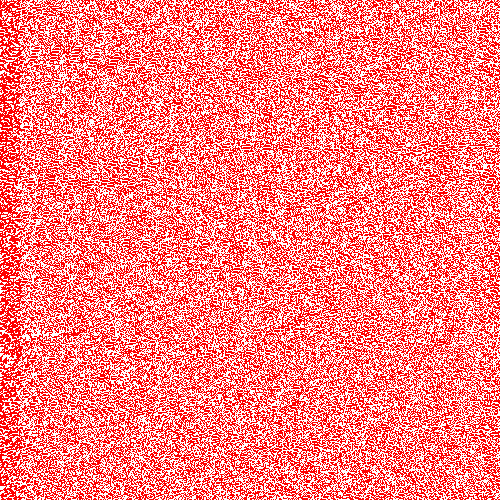} & \includegraphics[width=31mm]{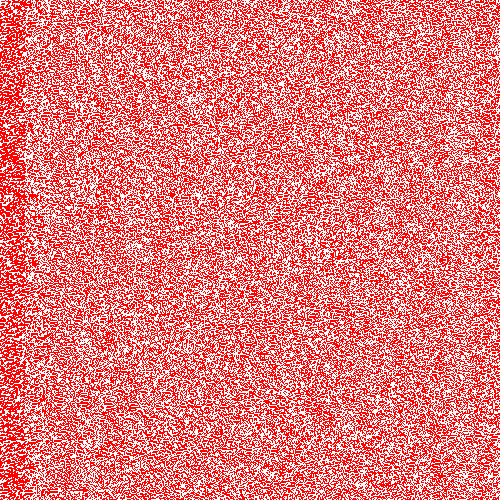} & \includegraphics[width=31mm]{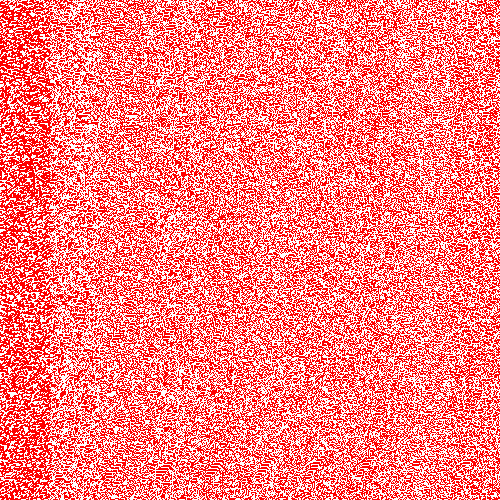} \\
				
				ECA 15 & ECA 18 & ($15,18^{50}$) & ($15,18^{100}$) & ($15,18^{125}$) \\[6pt]
				\includegraphics[width=31mm]{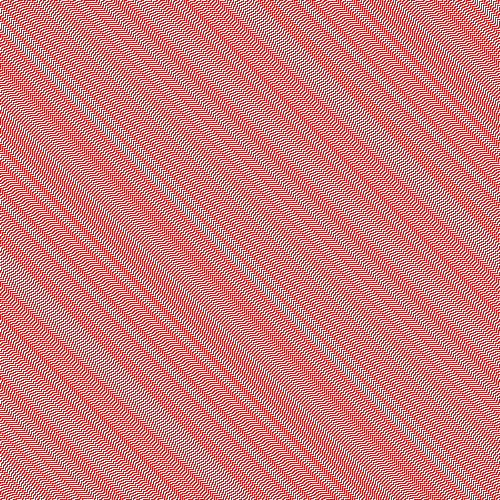} & \includegraphics[width=31mm]{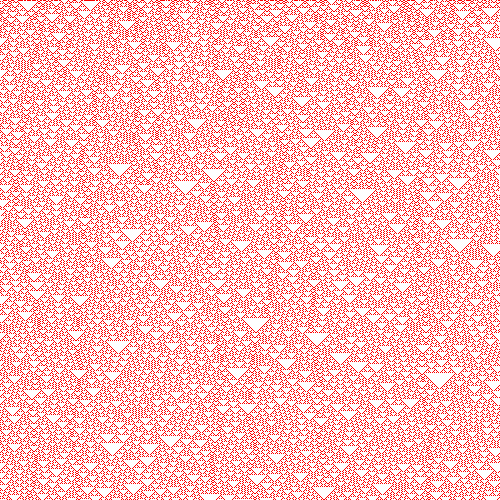} & \includegraphics[width=31mm]{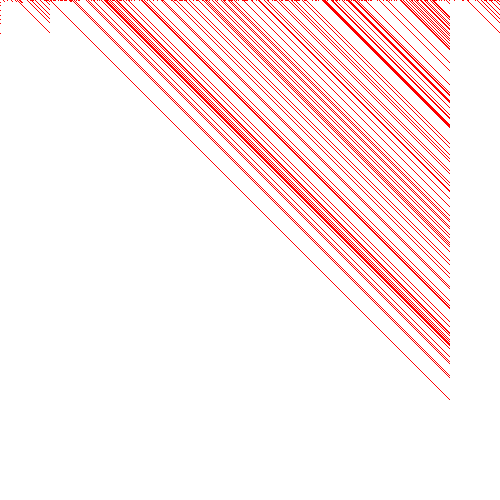} & \includegraphics[width=31mm]{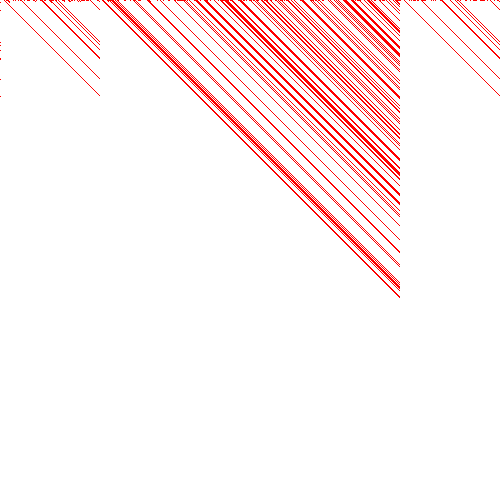} & \includegraphics[width=31mm]{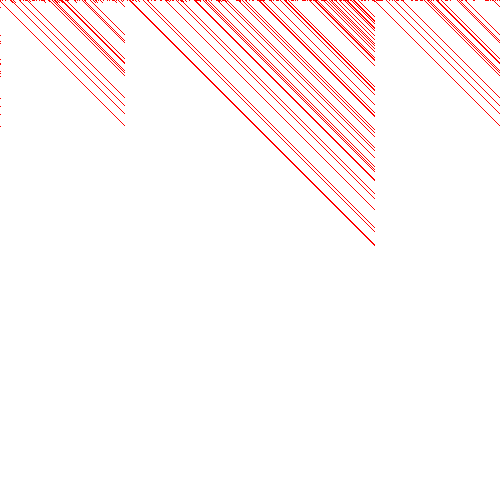} \\		
				
				ECA 40 & ECA 105 & ($40,105^{125}$) & ($40,105^{250}$) & ($40,105^{500}$) \\[6pt]
				\includegraphics[width=31mm]{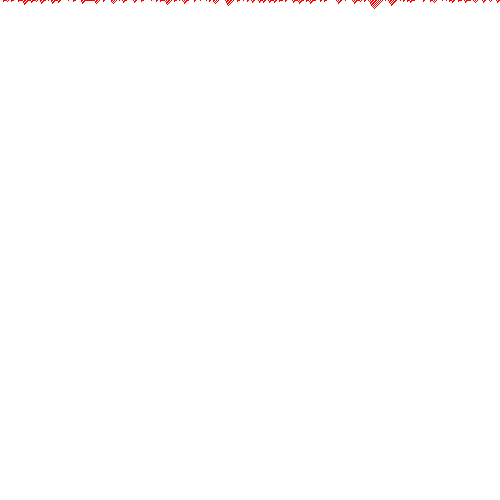} & \includegraphics[width=31mm]{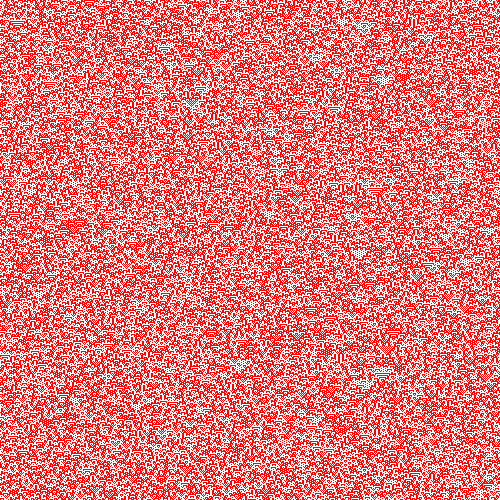} & \includegraphics[width=31mm]{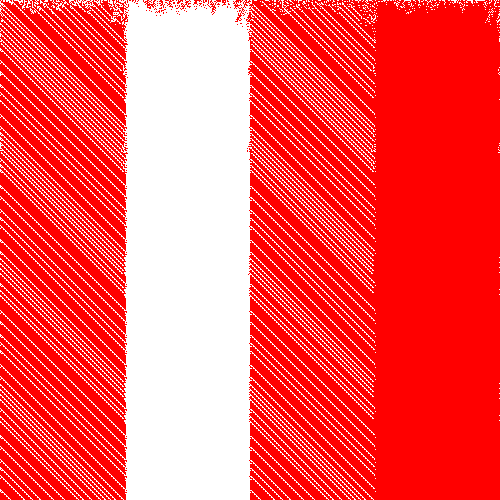} & \includegraphics[width=31mm]{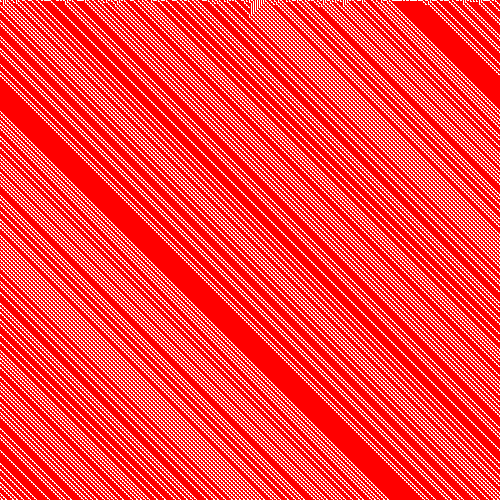} & \includegraphics[width=31mm]{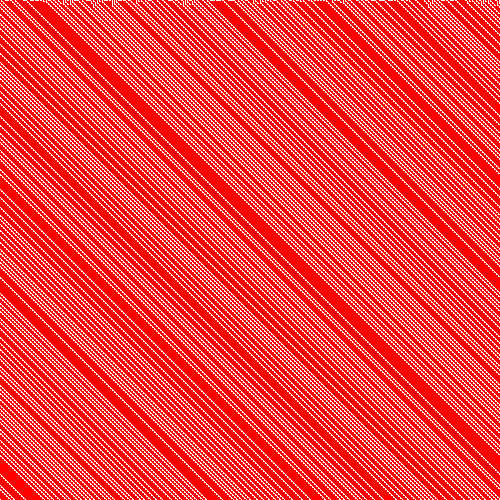} \\

			\end{tabular}}
		
		\caption{LCA($f,g^{b}$) dynamics where $\zeta$($f,g^{b}$) $\neq$ $\zeta$($f$) and $\zeta$($f,g^{b}$) $\neq$ $\zeta$($g$). Here, $\zeta$($f$) $\neq$ $\zeta$($g$).}
		\label{Fig4B}
	\end{center}
\end{figure*}

In case (1), where $\zeta(f)$ is in Class B and $\zeta(g)$ is in Class A, and $\zeta(f,g^b) \neq \zeta(f)$ as well as $\zeta(f,g^b) \neq \zeta(g)$, interesting dynamics can be observed. Specifically, when considering Elementary Cellular Automaton (ECA) rule 37 as the default rule and ECA rule 40 as the noise rule for different block sizes ($b=10,20,25$). The resulting dynamics of LCA($37,40^b$) show chaotic behavior (Class C) instead of the expected behavior based on the individual rules. This implies that the combination of these specific rules in an LCA leads to the emergence of chaotic dynamics, which is a unique characteristic that does not match the dynamics of either Rule 37 or Rule 40 alone. In Figure.~\ref{Fig4A}, we can observe the space-time diagrams for cases 1-3. LCA($90,56^b$) and LCA($41,32^b$) exhibit homogeneous and periodic behavior, respectively, which is distinct from the dynamics of the individual rules $f$ and $g$. The resulting behavior in the LCA is completely different from what would be expected based on the dynamics of the constituent rules.
Similarly, Figure~\ref{Fig4B} showcases the dynamics for cases 4-6. LCA($136,37^b$), LCA($15,18^b$), and LCA($40,105^b$) are taken as examples. LCA($136,37^b$) exhibits chaotic behavior, LCA($15,18^b$) displays homogeneous behavior, and LCA($40,105^b$) demonstrates periodic behavior. These dynamics differ from the individual behaviors of rules $f$ and $g$.
These observations highlight that in layered cellular automata, the combination of rules in an LCA can lead to emergent behaviors that deviate from the dynamics of the constituent rules. The resulting dynamics in the LCA can be chaotic, homogeneous, or periodic, indicating the complex and unpredictable nature of layered cellular automata dynamics.

\begin{figure*}[hbt!]
	\begin{center}
		\scalebox{0.8}{
		\begin{tabular}{cccc}
			($30,152^{125}$) & ($30,152^{100}$) & ($30,152^{50}$) & ($30,152^{25}$) \\[6pt]
			\includegraphics[width=33mm]{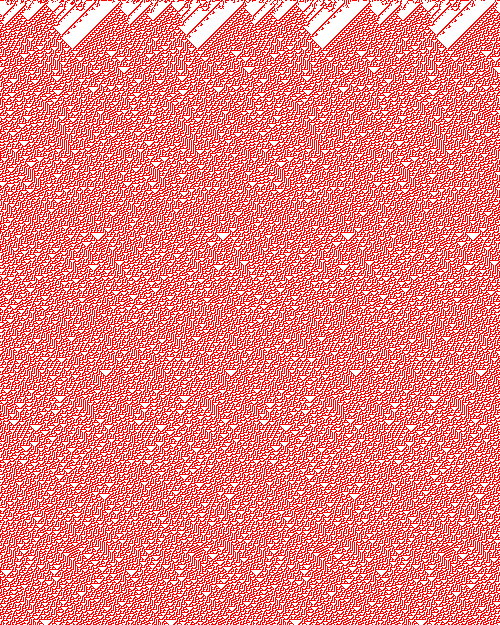} & \includegraphics[width=33mm]{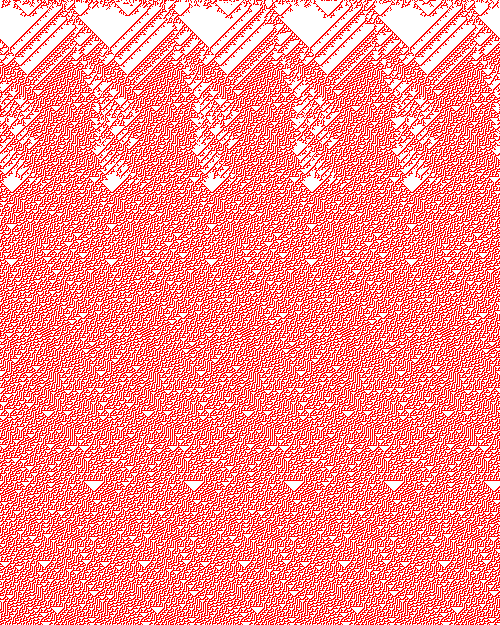}  &   \includegraphics[width=33mm]{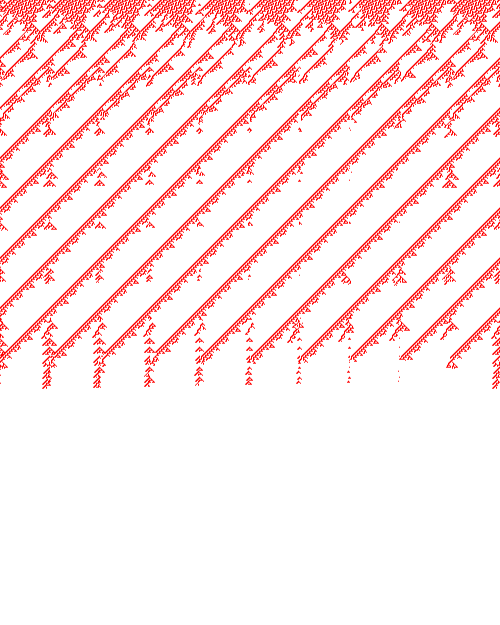}  &   \includegraphics[width=33mm]{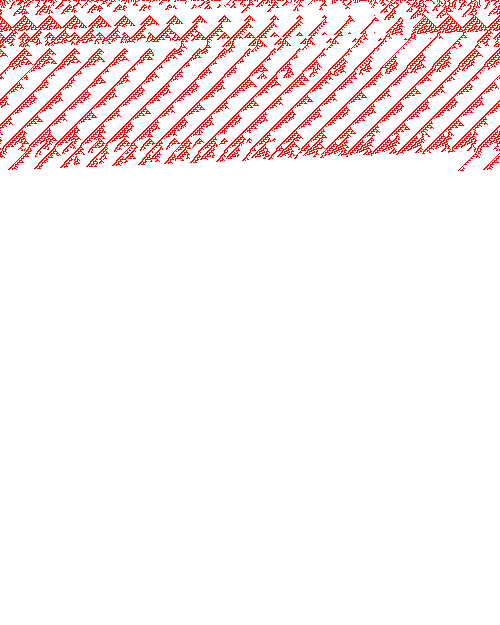}\\
			
			($168,122^{10}$) & ($168,122^{20}$) & ($168,122^{25}$) & ($168,122^{50}$) \\[6pt]
			\includegraphics[width=33mm]{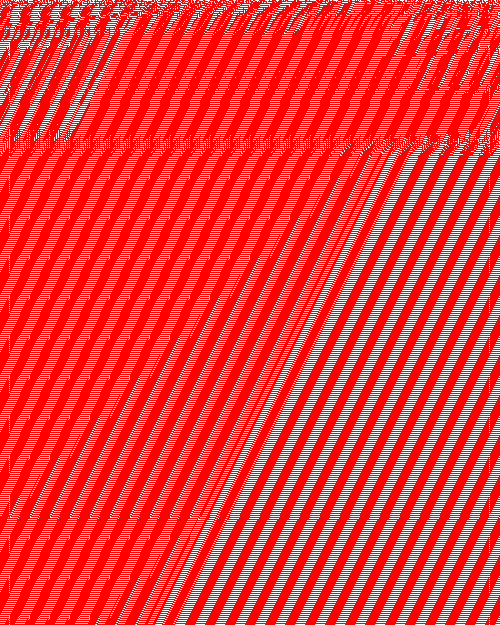} & \includegraphics[width=33mm]{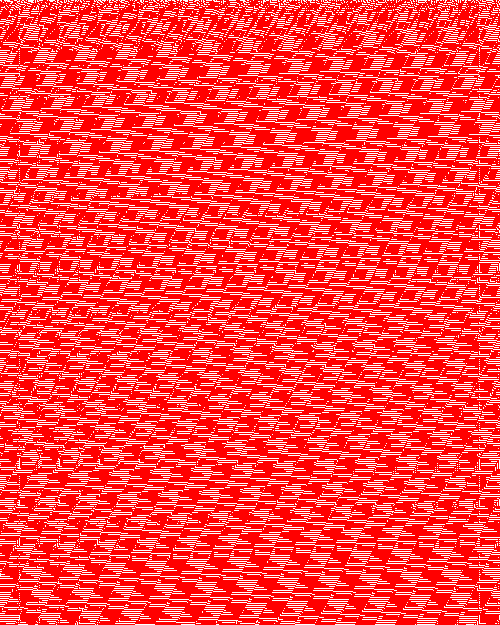}  &   \includegraphics[width=33mm]{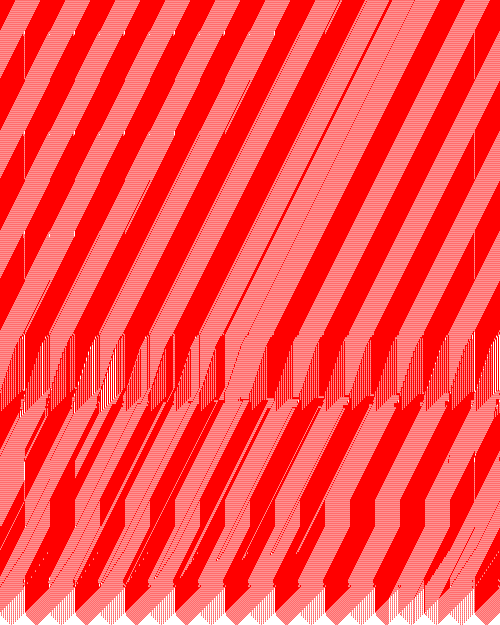}  &   \includegraphics[width=33mm]{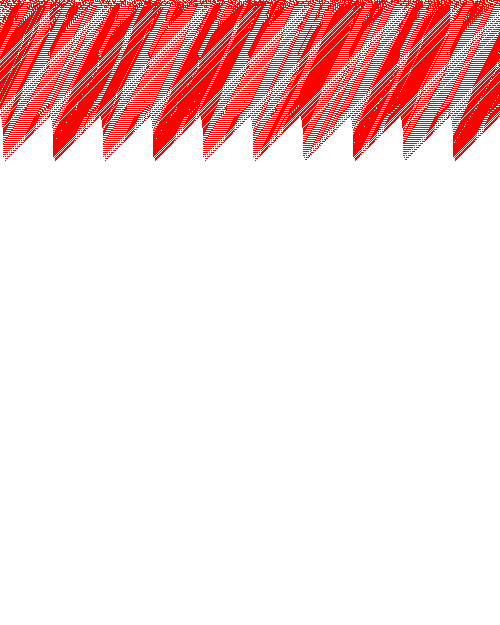}\\
		
			($168,94^{25}$) & ($168,94^{50}$) & ($168,94^{100}$) & ($168,94^{125}$) \\[6pt]
			\includegraphics[width=33mm]{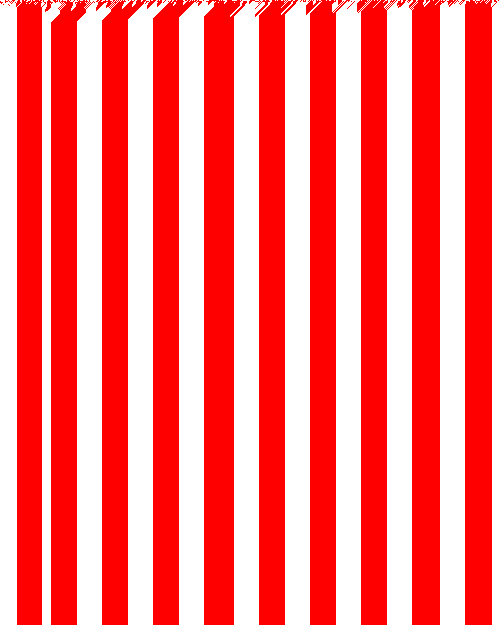} & \includegraphics[width=33mm]{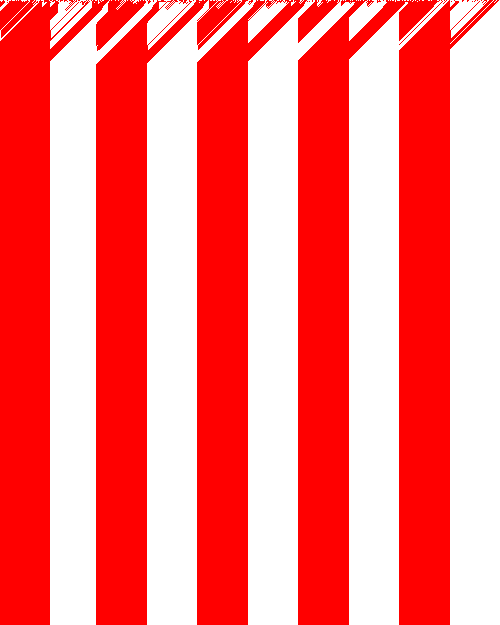}  &   \includegraphics[width=33mm]{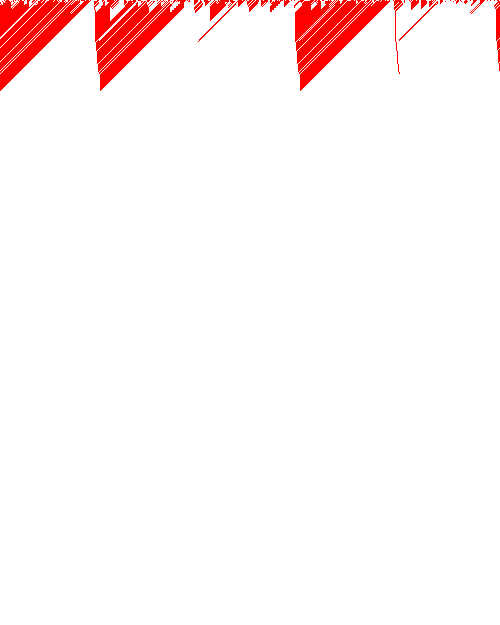}  &   \includegraphics[width=33mm]{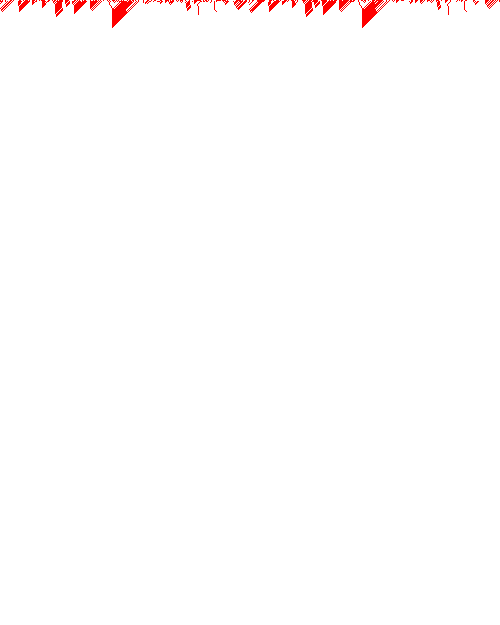}\\
			
			($26,136^{250}$) & ($26,136^{125}$) & ($26,136^{100}$) & ($26,136^{50}$) \\[6pt]
			\includegraphics[width=33mm]{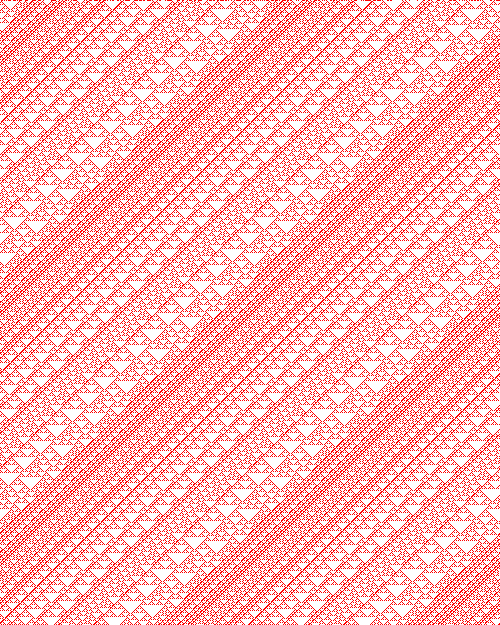} & \includegraphics[width=33mm]{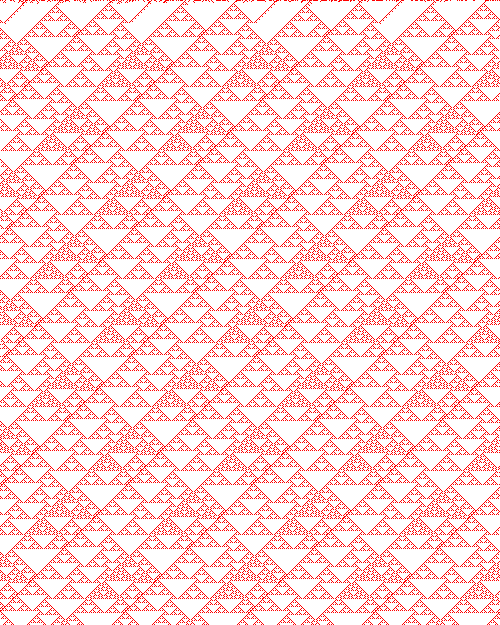}  &   \includegraphics[width=33mm]{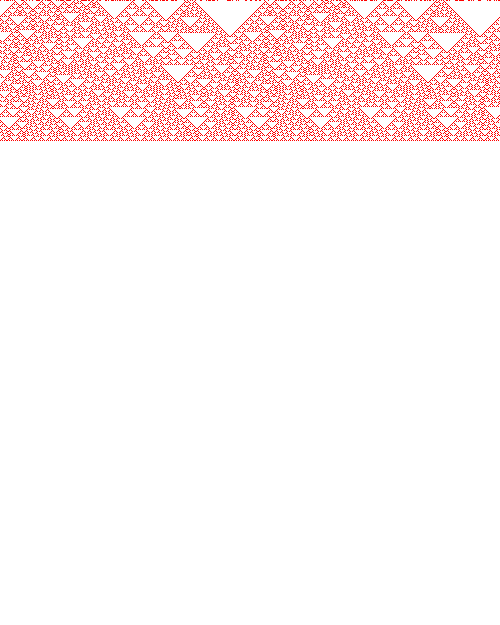}  &   \includegraphics[width=33mm]{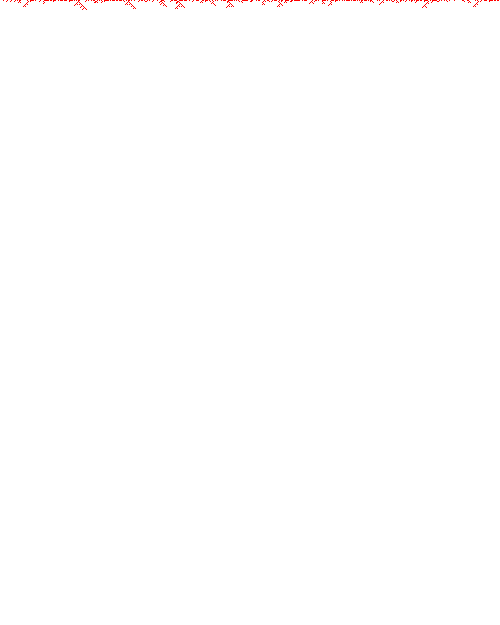}\\
		\end{tabular}}
		\caption{Phase transition behavior of LCA($30,152^{b}$), LCA($168,122^{b}$), LCA($168,94^{b}$), LCA($26,136^{b}$).}
		\label{Fig5}
	\end{center}
\end{figure*}

\subsection{Phase transition dynamics} 

Next, we investigate phase transition for the following model where for a critical block size denoted $b_{c}$ the whole system converge to all-0 configuration. The experiments involve examining the behavior of the system as the block size varies. When the block size is below the critical value, the system exhibits dynamics that are different from when the block size exceeds the critical value. Specifically, below $b_c$, the system displays active dynamics, whereas above $b_c$, the system undergoes a transition and converges to a passive state with an all-0 configuration. To highlight the effect of phase transition, we consider four specific LCAs in our experiments. These examples are chosen to showcase the diverse behaviors observed during the transition. 

In Figure~\ref{Fig5}, we present various examples of phase transition to visually demonstrate this phenomenon. The LCAs chosen for illustration are LCA($30,152^b$), LCA($168,122^b$), LCA($168,94^b$), and LCA($26,136^b$). These examples exhibit convergence to an all-0 configuration as the block size changes. For LCA($30,152^b$), the critical block size ($b_c$) is found to be 50. This means that when the block size is larger than 50, the system displays active dynamics, but as the block size decreases below this threshold, the system undergoes a transition and converges to an all-0 configuration in a passive phase. Similarly, for LCA($168,122^b$), the critical block size is determined to be 25, indicating that the transition from active to passive behavior occurs when the block size is 25. for LCA($168,94^b$) and LCA($26,136^b$), $b_c$ is 100.

\begin{figure*}[hbt!]
	\begin{center}
		\scalebox{0.7}{
		\begin{tabular}{ccccc}
			
			ECA 41 & ECA 122 & ($41,122^{100}$) & ($41,122^{125}$) & ($41,122^{250}$) \\
			
			\includegraphics[width=31mm]{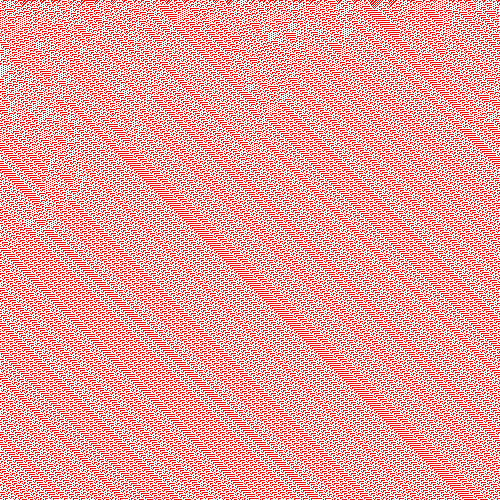} & \includegraphics[width=31mm]{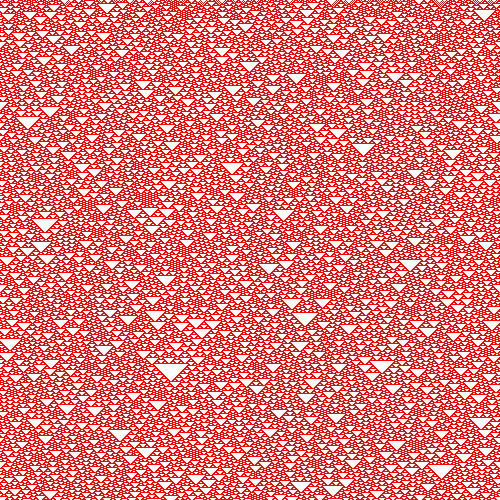}  & \includegraphics[width=31mm]{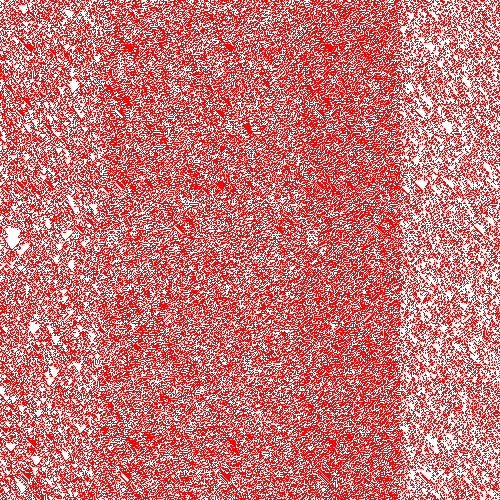} & \includegraphics[width=31mm]{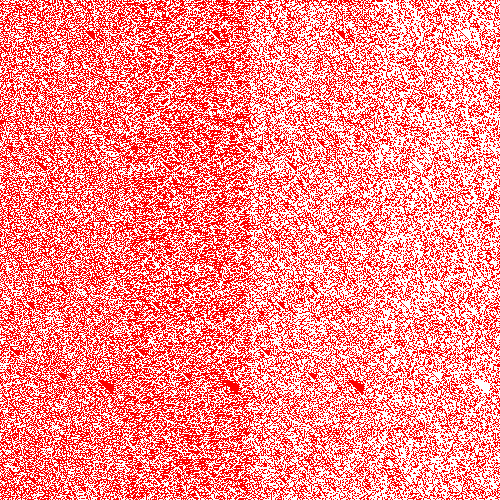}  & \includegraphics[width=31mm]{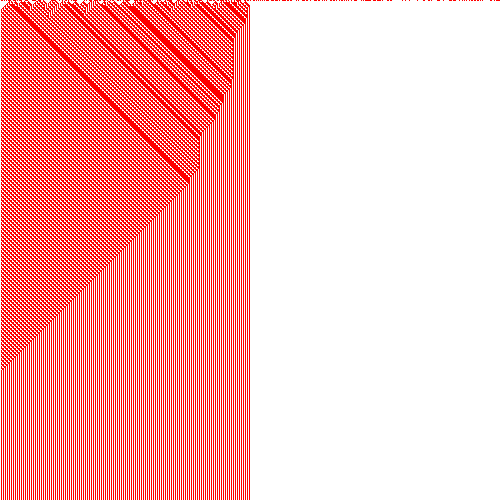} \\
			
			ECA 146 & ECA 18 & ($146,18^{50}$) & ($146,18^{100}$) & ($146,18^{125}$) \\
			
			\includegraphics[width=31mm]{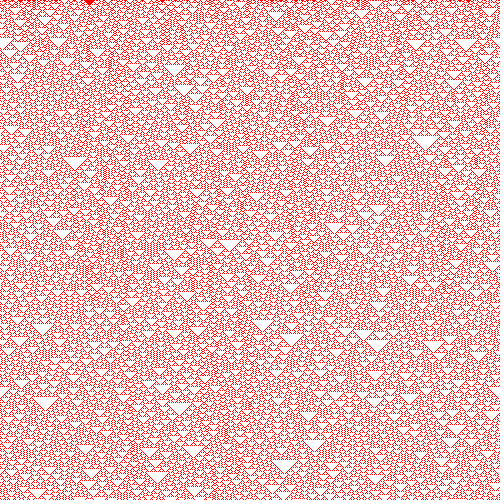} & \includegraphics[width=31mm]{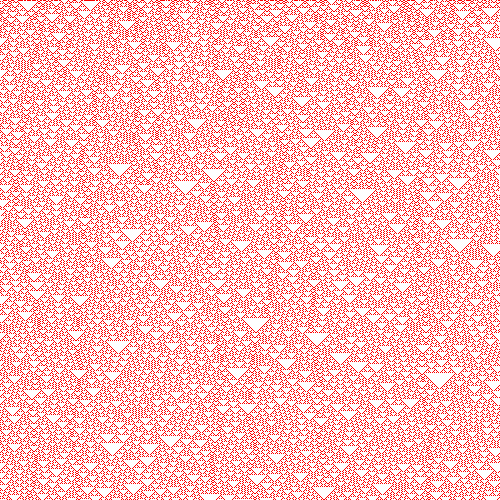}  & \includegraphics[width=31mm]{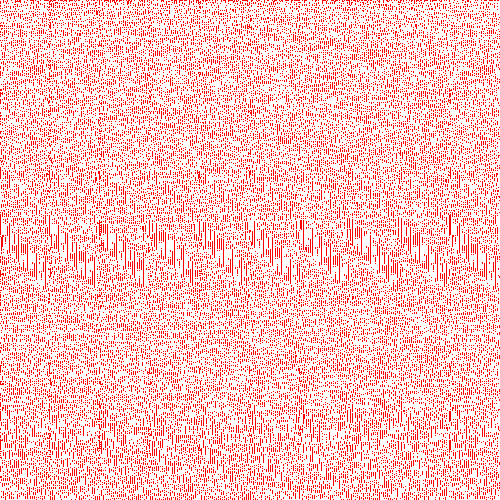} & \includegraphics[width=31mm]{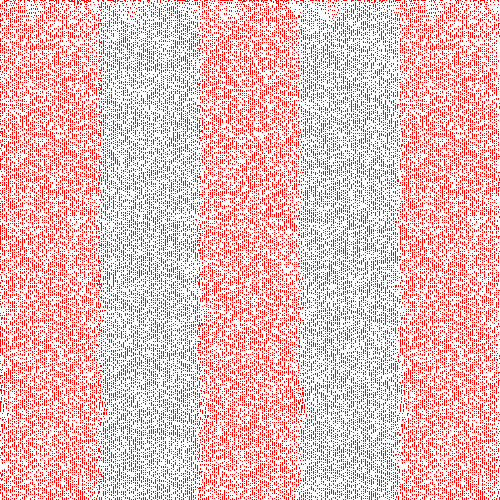}  & \includegraphics[width=31mm]{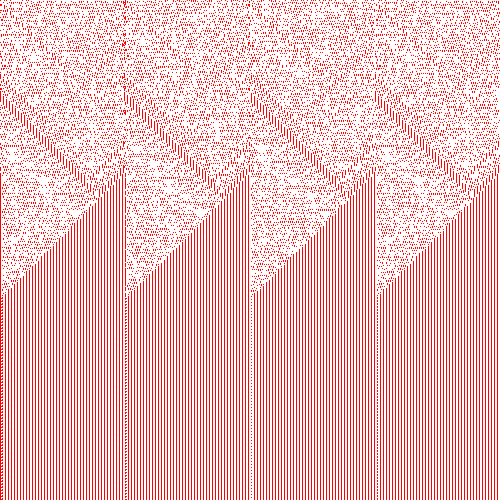} \\
			
		ECA 37 & ECA 19 & ($37,19^{100}$) & ($37,19^{125}$) & ($37,19^{250}$) \\		
		\includegraphics[width=31mm]{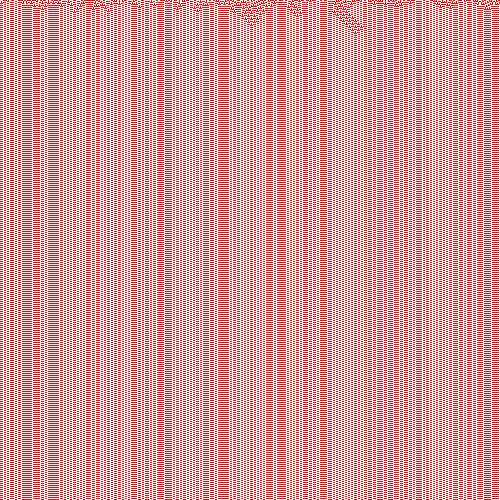} & \includegraphics[width=31mm]{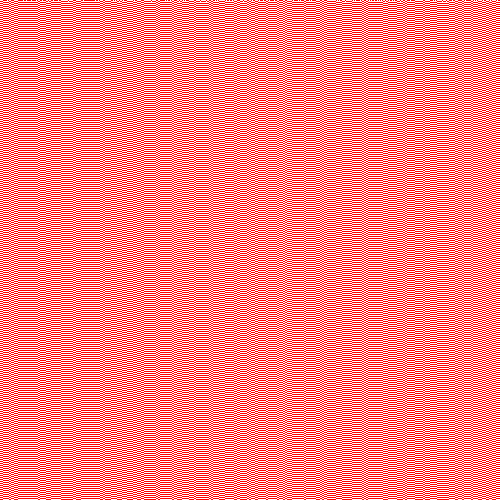}  & \includegraphics[width=31mm]{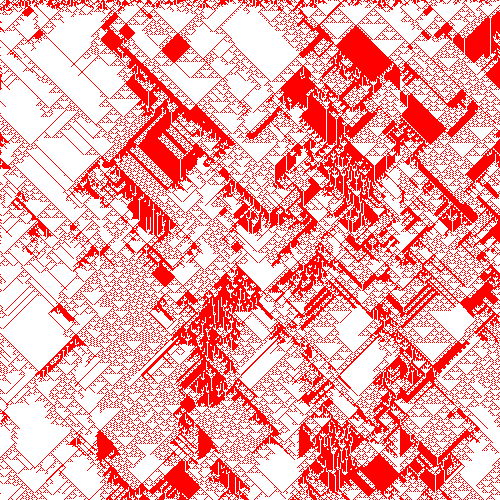} & \includegraphics[width=31mm]{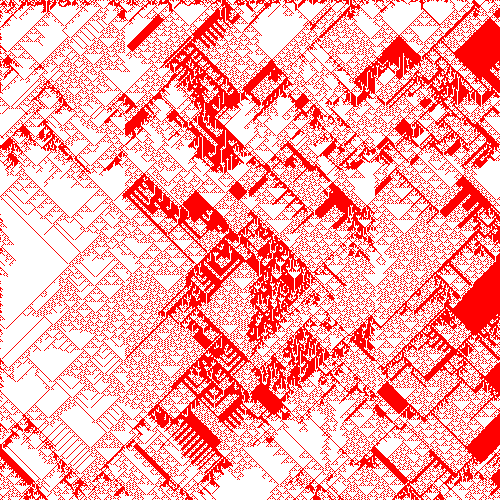}  & \includegraphics[width=31mm]{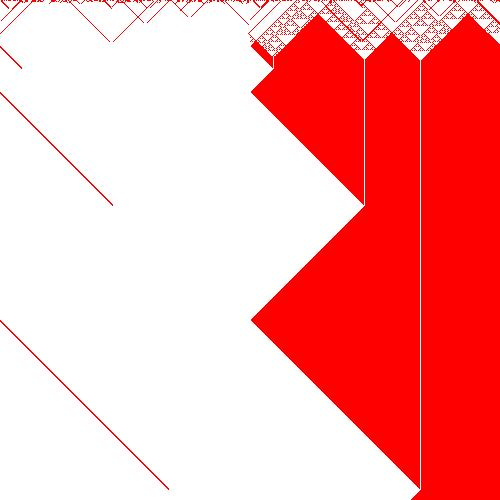} \\
		
		ECA 38 & ECA 1 & ($38,1^{100}$) & ($38,1^{125}$) & ($38,1^{250}$) \\
		
		\includegraphics[width=31mm]{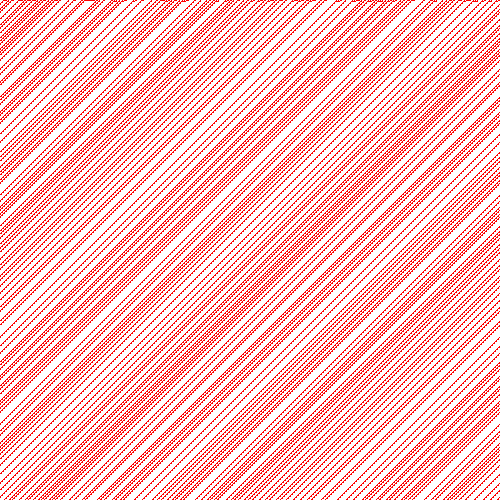} & \includegraphics[width=31mm]{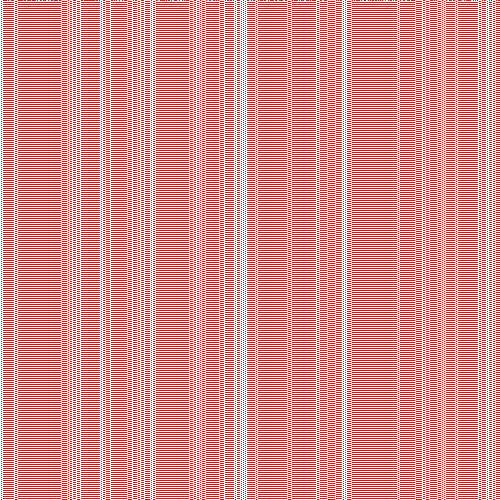}  & \includegraphics[width=31mm]{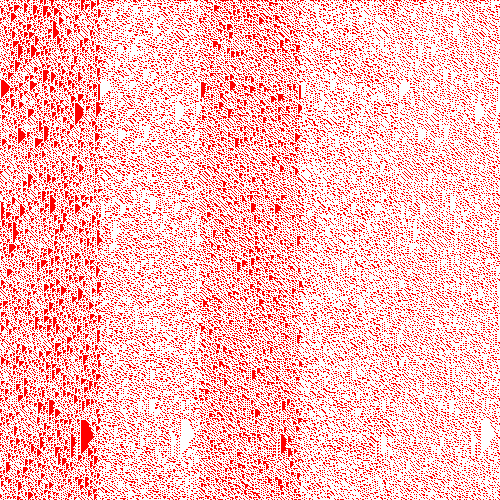} & \includegraphics[width=31mm]{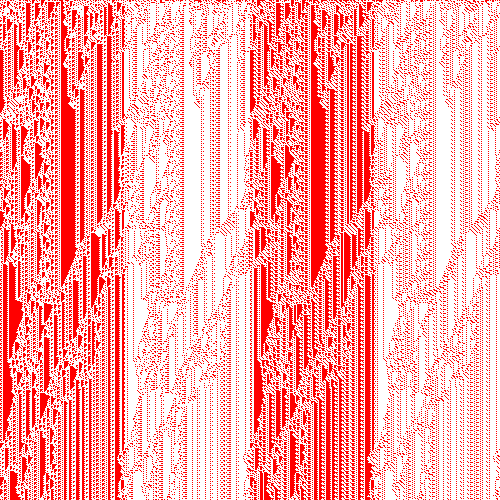}  & \includegraphics[width=31mm]{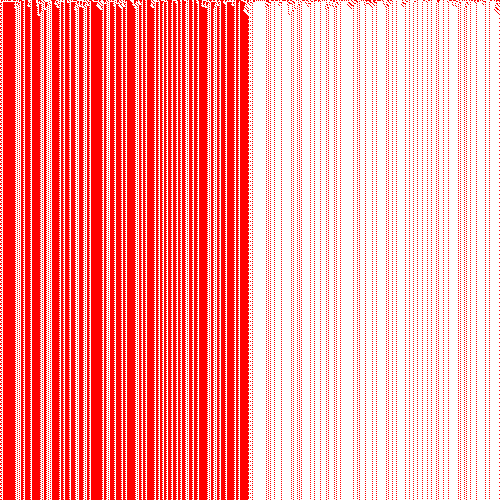} \\
			
		\end{tabular}}
		
		\caption{Class transition dynamics of LCA($41,122^{b}$), LCA($146,18^{b}$), LCA($37,19^{b}$), ($38,1^{b}$) where $\zeta$($f$) = $\zeta$($g$).}
		\label{Fig7}
	\end{center}
\end{figure*}

\subsection{Class transition dynamics}
\label{ctd}


Our next investigation focuses on class transition, which refers to a block size ($b$)-sensitive dynamic behavior in cellular systems. In this case, the class dynamics of the system change for a critical value of $b$, denoted as $b_t$. Mathematically, this can be expressed as $\zeta(f,g^b) \neq \zeta(f,g^{b'})$, where $b$ ranges from 1 to $b$ and $b'$ ranges from $b$ to 500.
During our experiments, we observe two types of results under the scenario of class transition.

\begin{itemize}
	\item $\zeta$($f$) = $\zeta$($g$)
	\item $\zeta$($f$) $\neq$ $\zeta$($g$)
\end{itemize}
In the first case, we consider situations where both $f$ and $g$ are chosen from the same class. This leads to two distinct scenarios:
\begin{itemize}
	\item[(1)] $\zeta$($f$) = $\zeta$($g$) = Class C
	\item[(2)] $\zeta$($f$) = $\zeta$($g$) = Class B
\end{itemize}

Let's consider case (1) where $\zeta(f) = \zeta(g) = \text{Class C}$. We provide two specific examples, LCA($41,122^b$) and LCA($146,18^b$), to illustrate this phenomenon. Figure~\ref{Fig7} visualizes the dynamics of these LCAs, where both chaotic or complex rules are involved.
In the case of LCA($41,122^b$), we observe the chaotic behavior of ECA $41$ and ECA $122$ in Figure~\ref{Fig7}. When $b$ is set to 100 and 125, LCA($41,122^b$) exhibits chaotic dynamics belonging to Class C. However, for $b=250$, LCA($41,122^b$) undergoes a transition and displays periodic dynamics belonging to Class B.
Similarly, for LCA($146,18^b$), the system shows chaotic dynamics for $b=50$ and $b=100$. However, at $b=125$, the system undergoes a transition and its dynamics become more periodic in nature, resembling Class B behavior.
The critical block size $b_t$ for LCA($41,122^b$) and LCA($146,18^b$) is found to be 250 and 125 respectively.

Moving on to case (2) with $\zeta(f) = \zeta(g) = \text{Class B}$, we consider two examples: LCA($37,19^b$) and LCA($38,1^b$). Figure~\ref{Fig7} visualizes the dynamics of these LCAs, where both periodic rules are involved.
In the case of LCA($37,19^b$), we observe the periodic behavior of ECA $37$ and ECA $19$ in Figure~\ref{Fig7}. For $b=100$ and $b=125$, LCA($37,19^b$) exhibits chaotic dynamics belonging to Class C. However, at $b=250$, LCA($37,19^b$) transitions to periodic dynamics belonging to Class B.
Similarly, for LCA($38,1^b$), the system displays chaotic dynamics for $b=100$ and $b=125$. At $b=250$, LCA($38,1^b$) undergoes a transition and exhibits periodic dynamics.
The critical block size $b_t$ for LCA($37,19^b$) and LCA($38,1^b$) is determined to be 250, representing the point at which the transition from chaotic (Class C) to periodic (Class B) dynamics occurs. 

Next we discuss the class transformation dynamics for $\zeta$($f$) $\neq$ $\zeta$($g$). So we have following six situations:
\begin{itemize}
	\item[(1)] $\zeta$($f$) = Class B and $\zeta$($g$) = Class C
	\item[(2)] $\zeta$($f$) = Class C and $\zeta$($g$) = Class B	
	\item[(3)] $\zeta$($f$) = Class C and $\zeta$($g$) = Class A
	\item[(4)] $\zeta$($f$) = Class A and $\zeta$($g$) = Class C	
	\item[(5)] $\zeta$($f$) = Class B and $\zeta$($g$) = Class A
	\item[(6)] $\zeta$($f$) = Class A and $\zeta$($g$) = Class B
\end{itemize}

\begin{figure*}[hbt!]
	\begin{center}
		\scalebox{0.8}{
		\begin{tabular}{ccccc}
			
			ECA 37 & ECA 18 & ($37,18^{10}$) & ($37,18^{20}$) & ($37,18^{50}$) \\			
			\includegraphics[width=31mm]{twoeca/37-19/37.png} & \includegraphics[width=31mm]{twoeca/15-18/18.png}  & \includegraphics[width=31mm]{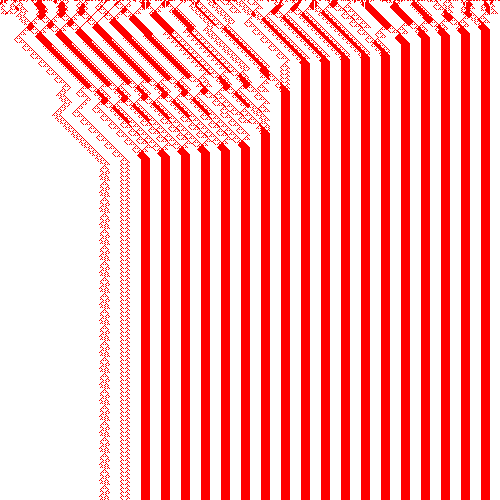} & \includegraphics[width=31mm]{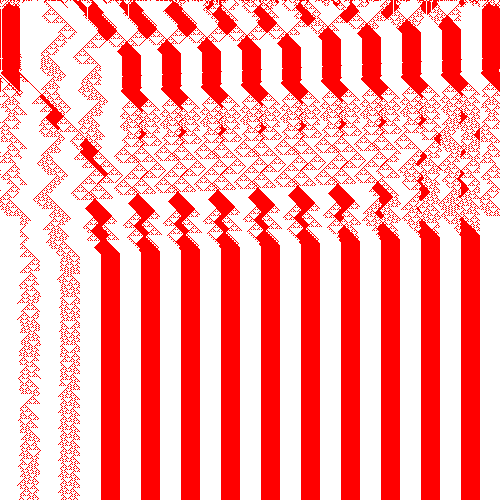}  & \includegraphics[width=31mm]{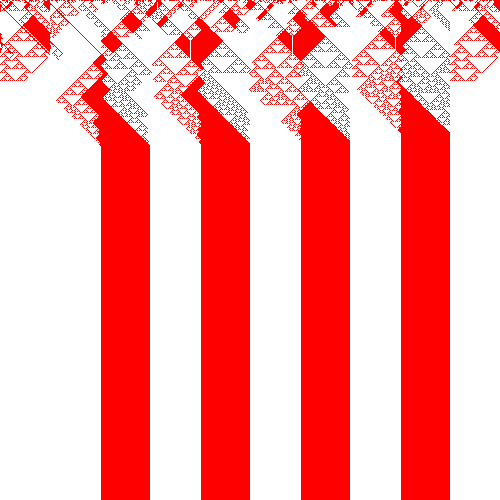} \\
			
			ECA 122 & ECA 44 & ($122,44^{25}$) & ($122,44^{50}$) & ($122,44^{100}$) \\			
			\includegraphics[width=31mm]{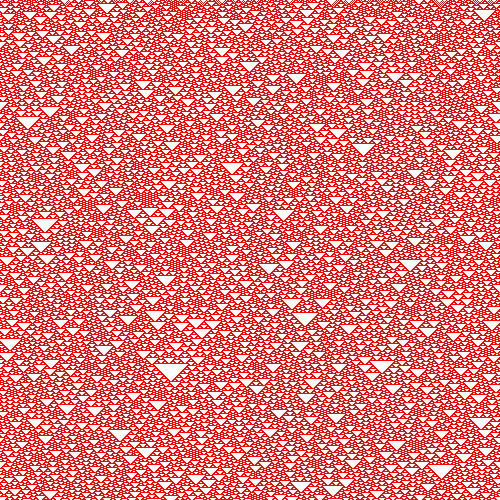} & \includegraphics[width=31mm]{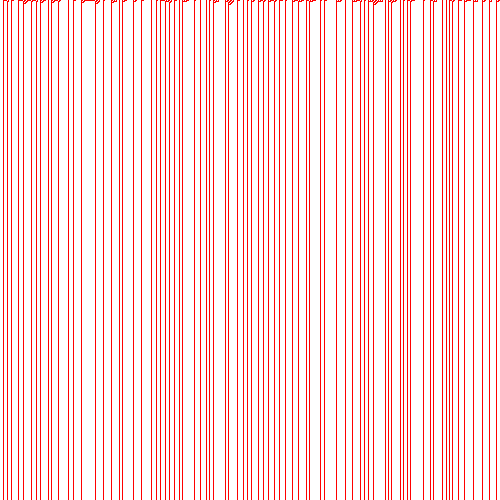}  & \includegraphics[width=31mm]{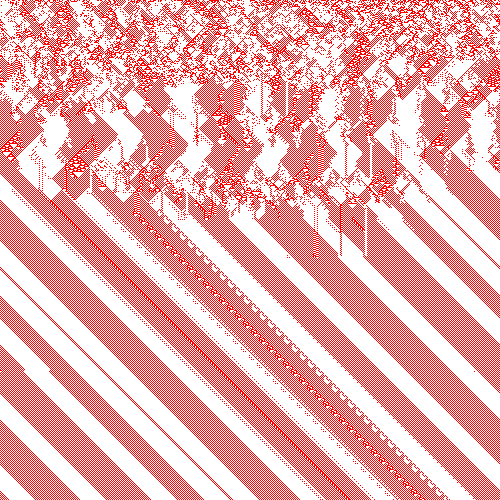} & \includegraphics[width=31mm]{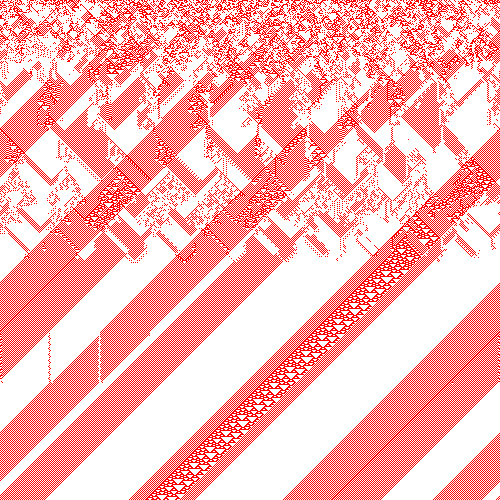}  & \includegraphics[width=31mm]{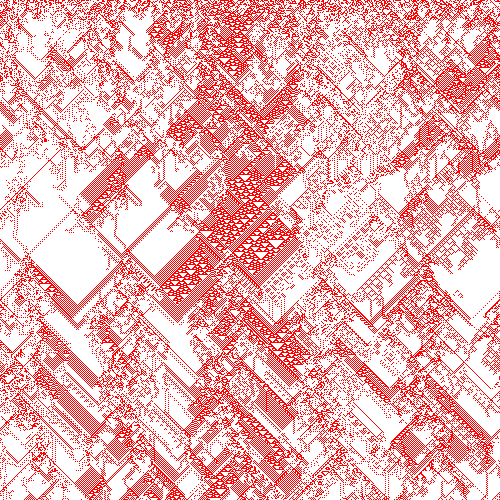} \\
			
		ECA 22 & ECA 168 & ($22,168^{100}$) & ($22,168^{125}$) & ($22,168^{250}$) \\			
		\includegraphics[width=31mm]{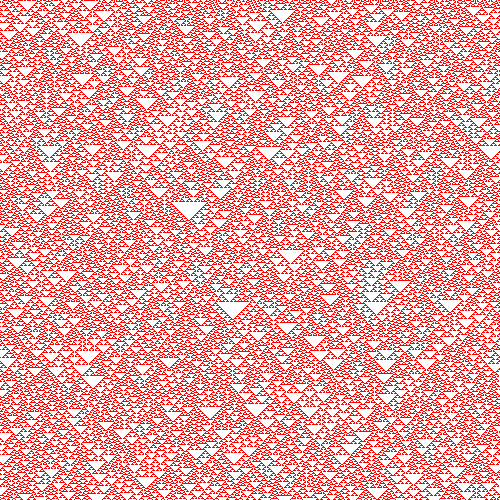} & \includegraphics[width=31mm]{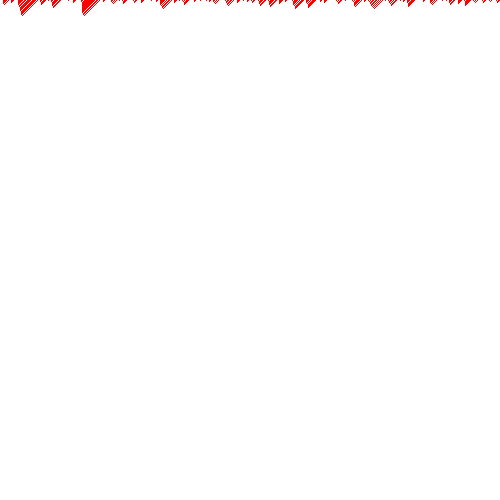}  & \includegraphics[width=31mm]{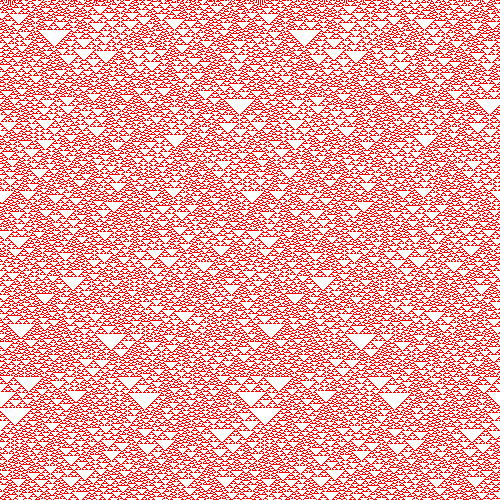} & \includegraphics[width=31mm]{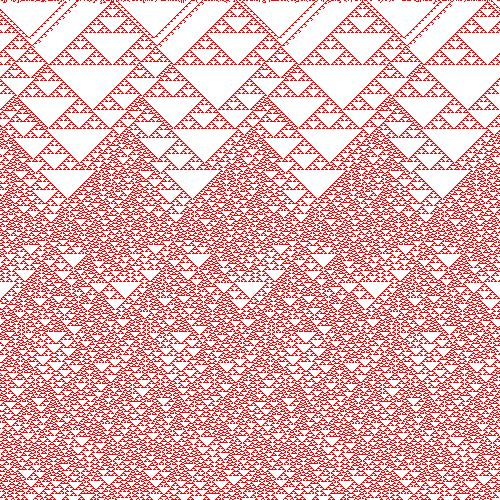}  & \includegraphics[width=31mm]{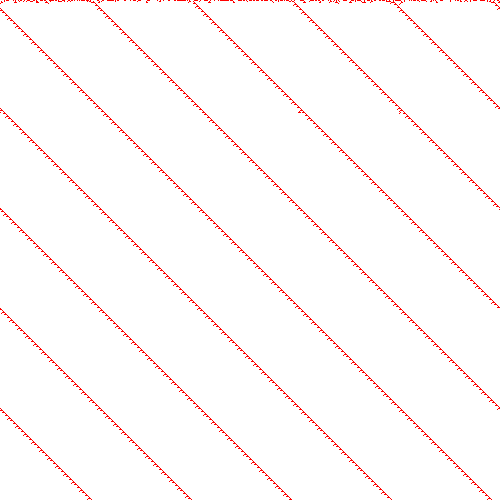} \\
		
		ECA 128 & ECA 73 & ($128,73^{50}$) & ($128,73^{100}$) & ($128,73^{125}$) \\			
		\includegraphics[width=31mm]{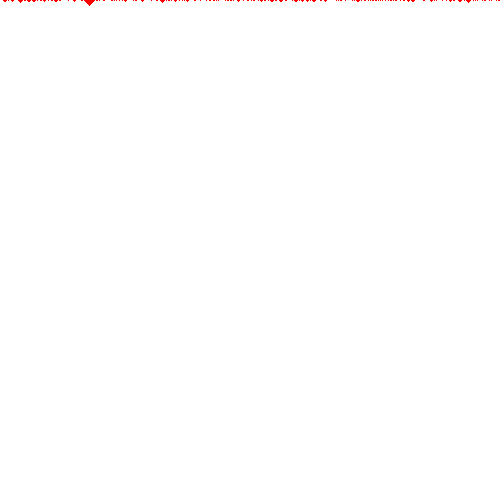} & \includegraphics[width=31mm]{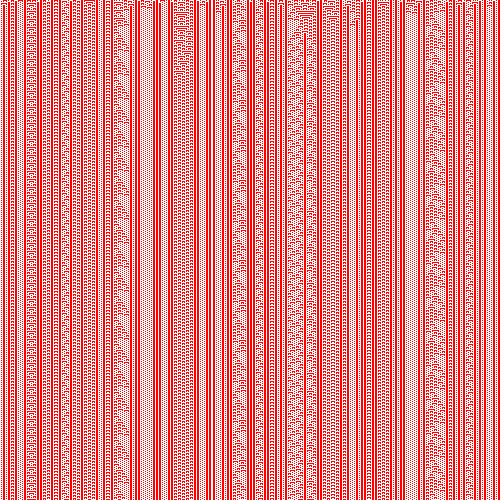}  & \includegraphics[width=31mm]{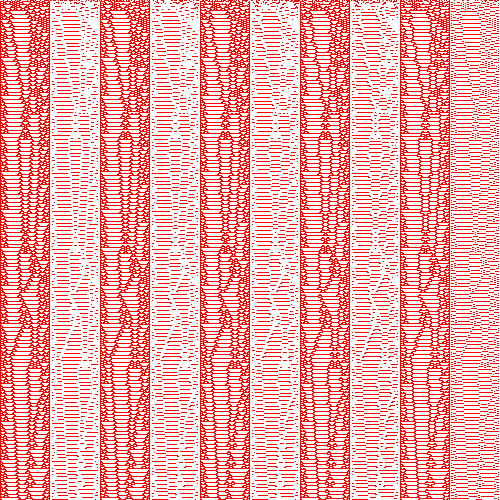} & \includegraphics[width=31mm]{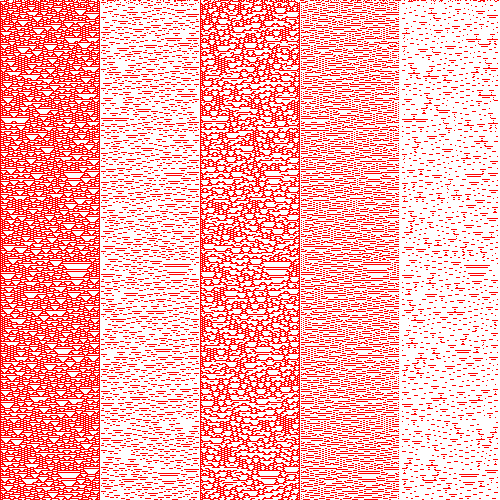}  & \includegraphics[width=31mm]{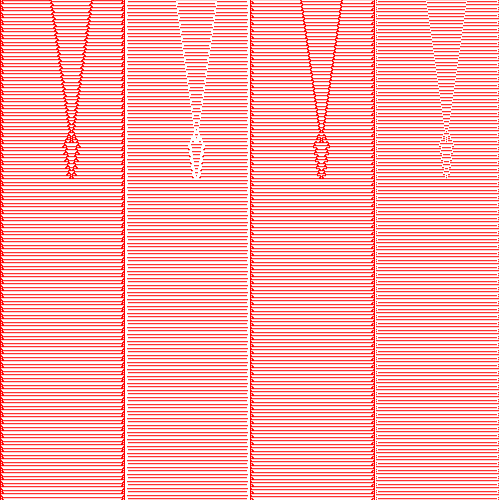} \\
		
		ECA 62 & ECA 168 & ($62,168^{125}$) & ($62,168^{100}$) & ($62,168^{50}$) \\			
		\includegraphics[width=31mm]{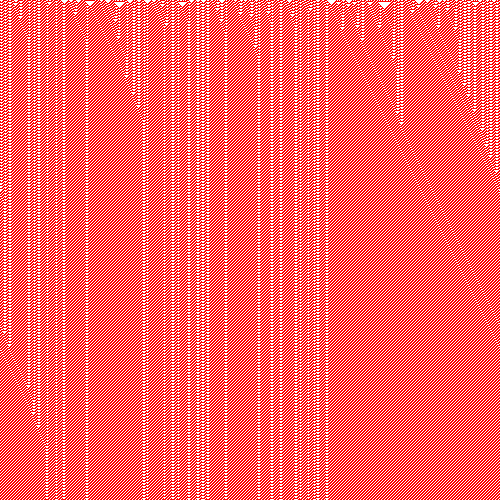} & \includegraphics[width=31mm]{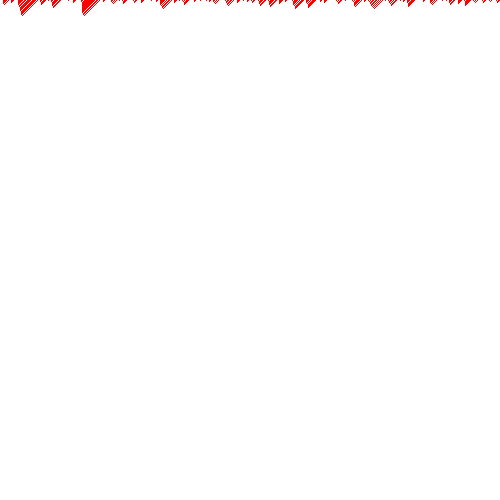}  & \includegraphics[width=31mm]{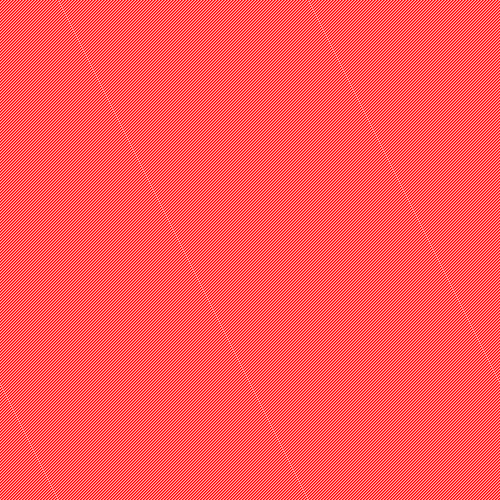} & \includegraphics[width=31mm]{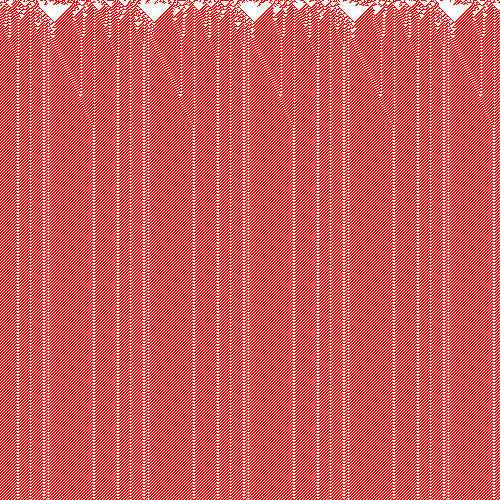}  & \includegraphics[width=31mm]{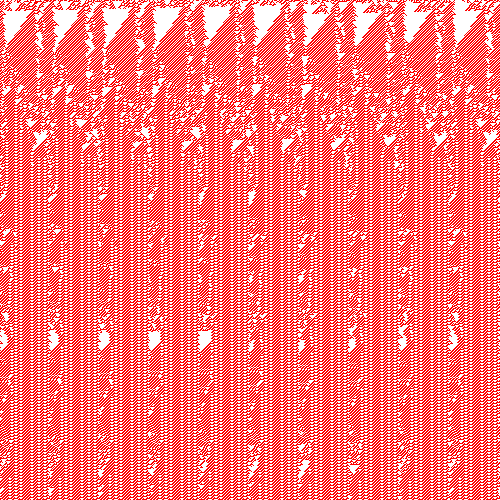} \\
			
			ECA 128 & ECA 37 & ($128,37^{100}$) & ($128,37^{125}$) & ($128,27^{250}$) \\			
			\includegraphics[width=31mm]{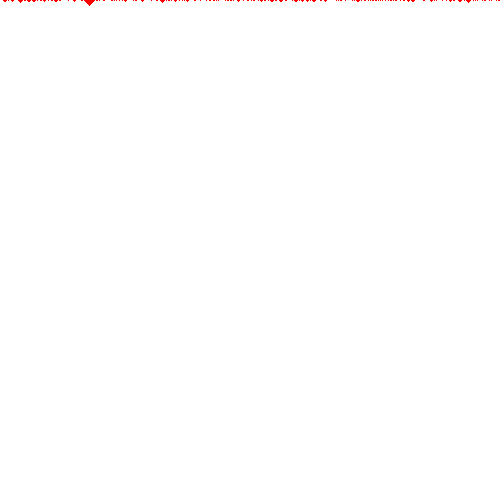} & \includegraphics[width=31mm]{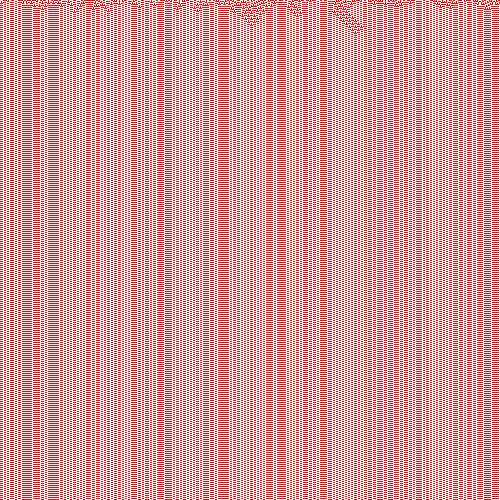}  & \includegraphics[width=31mm]{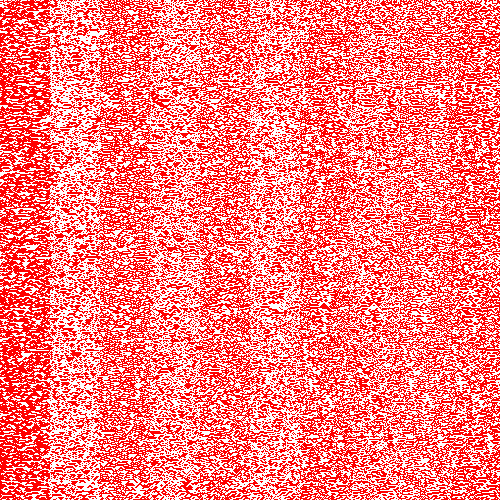} & \includegraphics[width=31mm]{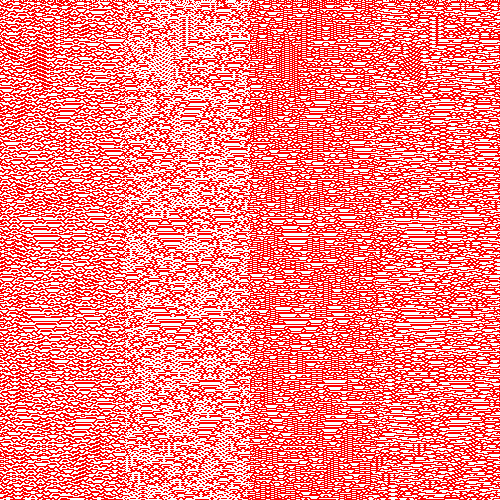}  & \includegraphics[width=31mm]{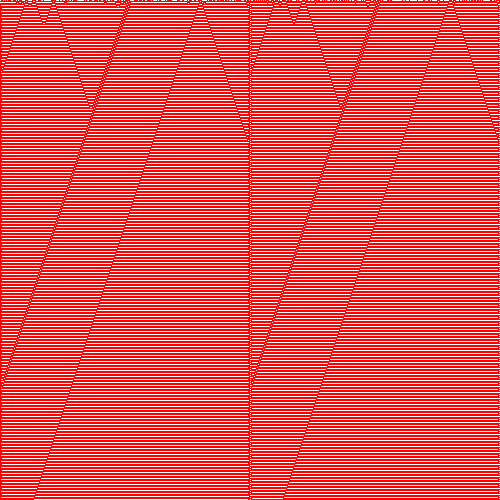} \\
			
		\end{tabular}}
		
		\caption{Class transition dynamics of LCA($37,18^{b}$), LCA($122,44^{b}$), LCA($22,168^{b}$), LCA($128,73^{b}$), LCA($62,168^{b}$), LCA($128,37^{b}$) where $\zeta$($f$) $\neq$ $\zeta$($g$).}
		\label{Fig8}
	\end{center}
\end{figure*}

Let's begin with case (1), where $\zeta(f) = \text{Class B}$ and $\zeta(g) = \text{Class C}$. We examine the dynamics of LCA($37,18^b$) to illustrate this case. In Figure~\ref{Fig8}, we observe that ECA $37$ exhibits chaotic dynamics belonging to Class B, while ECA $18$ displays Class C dynamics.
LCA($37,18^b$) demonstrates chaotic dynamics for $b=10$ and $b=20$ (see Figure~\ref{Fig8}). However, as we progressively increase the block size $b$, LCA($37,18^b$) undergoes a transition and exhibits periodic dynamics (see Figure~\ref{Fig8}) at a critical block size $b_t=50$.

Moving on to case (2), we consider $\zeta(f) = \text{Class C}$ and $\zeta(g) = \text{Class B}$. In Figure~\ref{Fig8}, ECA $122$ shows chaotic dynamics belonging to Class C, while ECA $44$ exhibits homogeneous dynamics belonging to Class A. We examine the dynamics of LCA($122,44^b$) to investigate this case.
LCA($122,44^b$) displays periodic dynamics for $b=25$ and $b=50$ (see Figure~\ref{Fig8}). However, as the block size $b$ increases beyond $b_t=100$, the cellular system undergoes a transition and exhibits chaotic dynamics. The space-time diagrams of LCA($122,44^b$) are visualized in Figure~\ref{Fig8}. 

For case 3-6, Figure~\ref{Fig8} shows the dynamics of LCA($22,168^b$), LCA($128,73^b$), LCA($62,168^b$) and LCA($128,37^b$). Respective examples of LCAs are used to describe the dynamics for case 3-6. For LCA($22,168^b$), LCA($128,73^b$), LCA($62,168^b$) and LCA($128,37^b$), their respective critical block size is $b_t=250,125,50,250$.
The examples provided and the corresponding analysis highlight the class transformation dynamics observed in layered cellular automata. These cases demonstrate how the combination of rules from different classes can lead to diverse dynamics and transitions based on the block size parameter.

\pagebreak
\begin{table}[!htbp] 
	\centering
	\scriptsize
	
	\begin{adjustbox}{width=\columnwidth,center}
		\begin{tabular}{|c|c|c|c|} \hline 
			\textbf{Conditions} & \textbf{Cases} & \textbf{LCAs}& \textbf{$b-$sensitivity} \\ \hline
			&&&\\
			& \textbf{\underline{$\zeta(f,g^b)$=$\zeta(f)$}} & &\\
			&  $\zeta(f,g^b)=\zeta(f)$ = Class A & ($8,40^{b}$) ($128,32^{b}$) ($136,40^{b}$)&\\
			&  $\zeta(f,g^b)=\zeta(f)$ = Class B & ($1,33^{b}$) ($9,1^{b}$) ($38,44^{b}$)&\\
			$\zeta$($f$) =  $\zeta$($g$) &  $\zeta(f,g^b)=\zeta(f)$ = Class C & ($126,30^{b}$) ($105,18^{b}$) ($73,26^{b}$)&\\
			&&&\\
			& \textbf{\underline{$\zeta(f,g^b)\neq\zeta(f)$}} & &\\
			
			& $\zeta$($f$) = $\zeta$($g$) = Class B and $\zeta$($f,g^{b}$) = Class C & ($94,37^{b}$) ($104,51^{b}$)&\\
			
			& $\zeta$($f$) = $\zeta$($g$) = Class B and $\zeta$($f,g^{b}$) = Class A & ($15,19^{b}$) ($15,1^{b}$) &\\
			& $\zeta$($f$) = $\zeta$($g$) = Class C and $\zeta$($f,g^{b}$) = Class A & ($45,18^{b}$)&\\
			& $\zeta$($f$) = $\zeta$($g$) = Class C and $\zeta$($f,g^{b}$) = Class B & ($110,30^{b}$) ($126,18^{b}$)&\\
			&&&\\
			&&&\\
			&&&\\
			& \textbf{\underline{$\zeta$($f,g^{b}$) = $\zeta$($f$) or $\zeta$($f,g^{b}$) = $\zeta$($g$)}} & &\\
			&  $\zeta$($f$) = Class C, $\zeta$($g$) = Class B and $\zeta$($f,g^b$) = $\zeta$($f$) & ($45,50^{b}$)& \\	
			&  $\zeta$($f$) = Class C, $\zeta$($g$) = Class B and $\zeta$($f,g^b$) = $\zeta$($g$) & ($18,3^{b}$)& \\	
			&  $\zeta$($f$) = Class C, $\zeta$($g$) = Class A and $\zeta$($f,g^b$) = $\zeta$($f$) & ($45,160^{b}$)& \\	
			&  $\zeta$($f$) = Class B, $\zeta$($g$) = Class A and $\zeta$($f,g^b$) = $\zeta$($f$) & ($25,128^{b}$)& Insensitive\\
	    	&  $\zeta$($f$) = Class B, $\zeta$($g$) = Class C and $\zeta$($f,g^b$) = $\zeta$($f$) & ($43,45^{b}$)& \\
			&  $\zeta$($f$) = Class B, $\zeta$($g$) = Class C and $\zeta$($f,g^b$) = $\zeta$($g$) & ($28,60^{b}$)& \\
			&  $\zeta$($f$) = Class A, $\zeta$($g$) = Class B and $\zeta$($f,g^b$) = $\zeta$($f$) & ($168,28^{b}$)& \\
			&  $\zeta$($f$) = Class A, $\zeta$($g$) = Class B and $\zeta$($f,g^b$) = $\zeta$($g$) & ($136,29^{b}$)& \\
			$\zeta(f)\neq\zeta(g)$&  $\zeta$($f$) = Class A, $\zeta$($g$) = Class C and $\zeta$($f,g^b$) = $\zeta$($f$) & ($168,126^{b}$)& \\
			&  $\zeta$($f$) = Class A, $\zeta$($g$) = Class C and $\zeta$($f,g^b$) = $\zeta$($g$) & ($40,45^{b}$)& \\
			&  $\zeta$($f$) = Class B, $\zeta$($g$) = Class A and $\zeta$($f,g^b$) = $\zeta$($g$) & ($140,40^{b}$)& \\
			&  $\zeta$($f$) = Class C, $\zeta$($g$) = Class A and $\zeta$($f,g^b$) = $\zeta$($g$) & ($126,136^{b}$)& \\
			&&&\\
			& \textbf{\underline{$\zeta$($f,g^{b}$) $\neq$ $\zeta$($f$) and $\zeta$($f,g^{b}$) $\neq$ $\zeta$($g$)}} & &\\
			&  $\zeta$($f$) = Class B, $\zeta$($g$) = Class A and $\zeta$($f,g^b$) = Class C & ($37,40^{b}$)& \\
			&  $\zeta$($f$) = Class C, $\zeta$($g$) = Class B and $\zeta$($f,g^b$) = Class A & ($90,56^{b}$)& \\
			&  $\zeta$($f$) = Class C, $\zeta$($g$) = Class A and $\zeta$($f,g^b$) = Class B & ($41,32^{b}$)& \\
			&  $\zeta$($f$) = Class A, $\zeta$($g$) = Class B and $\zeta$($f,g^b$) = Class C & ($136,37^{b}$)& \\
			&  $\zeta$($f$) = Class B, $\zeta$($g$) = Class C and $\zeta$($f,g^b$) = Class A & ($15,18^{b}$)& \\
			&  $\zeta$($f$) = Class A, $\zeta$($g$) = Class C and $\zeta$($f,g^b$) = Class B & ($40,105^{b}$)& \\
			
			&&&\\
			
			\hline
			&&&\\
			&$\zeta$($f$) = Class C and $\zeta$($g$) = Class B &($30,152^{b}$)&$b_c=25$\\
			Phase Transition& $\zeta$($f$) = Class A and $\zeta$($g$) = Class C &($168,122^{b}$)&$b_c=50$\\
			& $\zeta$($f$) = Class A and $\zeta$($g$) = Class B &($168,94^{b}$)&$b_c=125$\\
			& $\zeta$($f$) = Class C and $\zeta$($g$) = Class A &($26,136^{b}$)&$b_c=50$\\
			&&&\\
			
			\hline
			&&&\\
			& \textbf{\underline{$\zeta(f)$=$\zeta(g)$}} & &\\
			&$\zeta$($f$) = $\zeta$($g$) = Class C &($41,122^{b}$) ($146,18^{b}$)& $b_t=250$ $b_t=125$\\
			&$\zeta$($f$) = $\zeta$($g$) = Class B &($37,19^{b}$) ($38,1^{b}$)& $b_t=250$ $b_t=250$\\
			&&&\\
			& \textbf{\underline{$\zeta$($f$) $\neq$ $\zeta$($g$)}} & &\\
			Class Transition& $\zeta$($f$) = Class B and $\zeta$($g$) = Class C &($37,18^{b}$)&$b_t=50$\\
			& $\zeta$($f$) = Class C and $\zeta$($g$) = Class B &($122,44^{b}$)&$b_t=100$\\
			& $\zeta$($f$) = Class C and $\zeta$($g$) = Class A &($22,168^{b}$)&$b_t=250$\\
			& $\zeta$($f$) = Class A and $\zeta$($g$) = Class C &($128,73^{b}$)&$b_t=125$\\
			& $\zeta$($f$) = Class B and $\zeta$($g$) = Class A &($62,168^{b}$)&$b_t=50$\\
			& $\zeta$($f$) = Class A and $\zeta$($g$) = Class B &($128,37^{b}$)&$b_t=250$\\
			&&&\\
			\hline
			
		\end{tabular}
	\end{adjustbox}	
	\caption{Summary of different dynamics observed for LCAs based on modified neighborhood}
	\label{Tablesumeca}
\end{table}

\section{LCA based on Game of Life}

LCA (Layered Cellular Automaton) refers to a model where the outcome of lower layer is influenced by upper layer imitating the hierarchical structure observed in nature with the aim to capture more complex dynamics. One application of LCA is the extension of the Game of Life, a well-known cellular automaton, by introducing additional layer.

In the LCA based on Game of Life, each layer represents a distinct rule set or modification to the original Game of Life rules where the upper layer influences the behavior of the lower layers. This hierarchical structure allows for more intricate and varied dynamics to emerge.
Layer 0 follows the standard Game of Life rules ($f$), where the state of a cell is determined solely by its 8 neighbor cells. Layer 1, on the other hand, introduces averaging as rule $g$ to influence the behavior of the cells in layer 0, as discussed in section~\ref{averagedyn}. These rules could be based on different neighborhood configurations, as rule $f$ follows Moore's neighborhood scheme and rule $g$ follows von Neumann's neighborhood scheme.

Game of life exhibits a wide range of dynamics. These dynamics refer to the various patterns and behaviors that emerge as the game progresses through different generations. Here are some of the different dynamics observed in the Game of Life:
\begin{itemize}
	\item \textbf{Still Life:} Still life patterns, are configurations that remain unchanged over time. They are stable and do not evolve or move. Examples of still life patterns include blocks, beehives, and boats.
	\item \textbf{Oscillators:} Oscillators are patterns that repeat their configuration after a certain number of generations. They oscillate between two or more distinct states. Common types of oscillators include blinkers (period 2), toads (period 2), and pulsars (period 3).
	\item \textbf{Spaceships:} Spaceships are patterns that move across the grid while maintaining their overall shape. They exhibit periodic motion, repeating their configuration after a certain number of generations. Common examples of spaceships in the Game of Life are gliders, lightweight spaceships, middleweight spaceships, and heavyweight spaceships.
	\item \textbf{Chaos:} Chaos refers to patterns that exhibit unpredictable and highly complex behavior. These patterns rapidly evolve and interact in a way that makes it difficult to determine long-term outcomes. Chaos arises when a pattern becomes highly unstable and generates rapid and unpredictable changes.
	\item \textbf{Infinite Growth:} In some cases, certain patterns can lead to unbounded growth, where the number of live cells continually increases with each generation. Examples include the Gosper glider gun, which emits gliders indefinitely, or the R-pentomino, which generates various structures as it expands.
	\item \textbf{Stabilization:} Some initial configurations may exhibit transient behavior before stabilizing into stable patterns, oscillators, or spaceships. The Game of Life allows for complex interactions and transformations that eventually reach a stable state. One such example is Penta-decathlon. Initially, the dynamic seems to keep growing. However, after a 15 number of generations, the pattern converges and stabilizes into a repeating cycle.
	\item \textbf{Chaotic Decay:} Certain patterns or interactions can lead to chaotic decay, where complex structures break down into smaller components or dissipate entirely over time. This decay can result from collisions, overcrowding, or instability within the pattern. Diehard pattern is one such example for chaotic decay where the pattern exhibits chaotic behavior as the live cells interact with their neighbors and generate new patterns. However, as time progresses, the number of live cells in the pattern steadily decreases until eventually, no live cells remain after 130 generation.
\end{itemize}
These different dynamics make the Game of Life a fascinating and complex system to explore. They showcase the emergence of complex behavior from simple rules, highlighting the richness and diversity of patterns that can arise in cellular automata.

\begin{figure*}[hbt!]
	\begin{center}
		\begin{adjustbox}{width=\columnwidth,center}
			\begin{tabular}{cccc}
				
				\includegraphics[width=31mm]{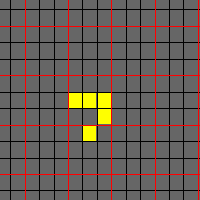} & \includegraphics[width=31mm]{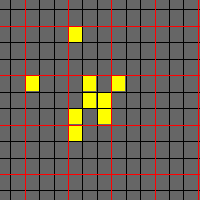} & \includegraphics[width=31mm]{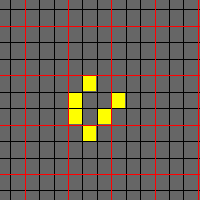} & \includegraphics[width=31mm]{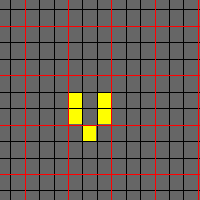} \\
				$t=0$ & $t=1$ & $t=2$ & $t=3$ \\
				
				\includegraphics[width=31mm]{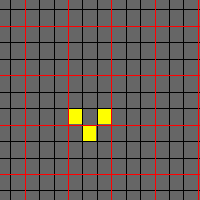} & \includegraphics[width=31mm]{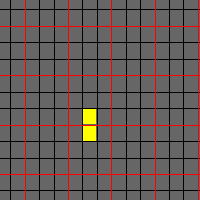} & \includegraphics[width=31mm]{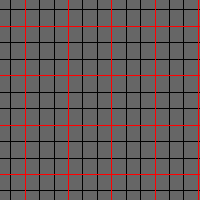} &  \\
				$t=4$ & $t=5$ & $t=6$ &  \\
				
			\end{tabular}
		\end{adjustbox}
		\caption{Dynamics of glider in LCA}
		\label{glider}
	\end{center}
\end{figure*}

Through our analysis of different initial configurations, we have observed distinct outcomes that deviate from the original dynamics of the Game of Life. One such example is the glider pattern, which typically repeats its movement diagonally across the grid while maintaining its shape. Gliders have the ability to travel indefinitely unless they encounter obstacles or reach the boundaries of the grid.
However, when the dynamics of the glider are influenced by rule $g$, a notable change occurs. Instead of continuing its diagonal movement across the grid, the glider pattern gradually diminishes and eventually dies out after a certain number of generations. This alteration in behavior is demonstrated in Figure~\ref{glider}, where the dynamics of the glider pattern under the influence of rule $g$ are depicted. After just five generations, the glider pattern ceases to exist.
This observation highlights how introducing variations to the original dynamics of the Game of Life, such as through modified rules, can lead to different outcomes and ultimately impact the behavior of well-known patterns like the glider.

\begin{figure*}[hbt!]
	\begin{center}
		\begin{adjustbox}{width=\columnwidth,center}
			\begin{tabular}{cccc}
				
				\includegraphics[width=31mm]{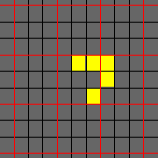} & \includegraphics[width=31mm]{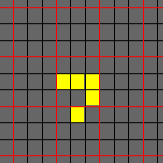} & \includegraphics[width=31mm]{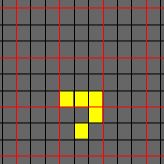} &  \\
				(a) & (b) & (c) &  \\
				
			\end{tabular}
		\end{adjustbox}
		\caption{Change in dynamics due to position of initial configuration in lattice.}
		\label{glider1}
	\end{center}
\end{figure*}

During our experiments, we made a critical observation regarding the position of the initial configuration in the lattice, which significantly influences
future outcomes. Figure.~\ref{glider1} shows different position of glider in the lattice. Black grids represent cells and red grid represent blocks. If the initial configuration of glider is similar to Figure.~\ref{glider1}(a), then glider dies out after 15 generation. On the other hand, if the initial configuration resembles Figure~\ref{glider1}(b), the glider pattern will cease to exist after only 5 generations as shown in Figure.~\ref{glider}, exhibiting a different outcome. However, when the glider pattern is positioned according to Figure~\ref{glider1}(c), it exhibits its true dynamics by continuing its diagonal movement across the grid. In this case, the glider pattern maintains its shape and traverses the lattice without dying out. Hence we can say that glider in LCA is position sensitive.

\begin{figure*}[hbt!]
	\begin{center}
		\begin{adjustbox}{width=\columnwidth,center}
			\begin{tabular}{cccccc}
				
				\includegraphics[width=31mm]{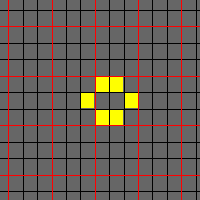} & \includegraphics[width=31mm]{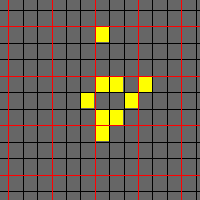} & \includegraphics[width=31mm]{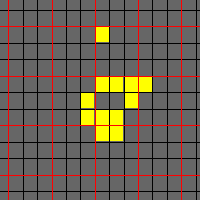} & \includegraphics[width=31mm]{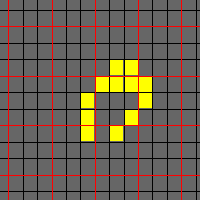} &
				\includegraphics[width=31mm]{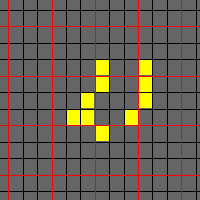}& 
				\includegraphics[width=31mm]{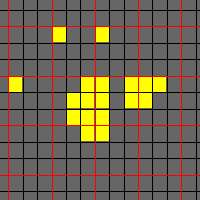} \\
				$t=0$ & $t=1$ & $t=2$ & $t=3$ & $t=4$ & $t=5$\\
				
				
				\includegraphics[width=31mm]{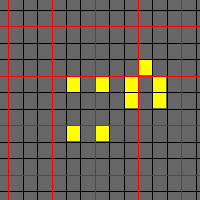} & \includegraphics[width=31mm]{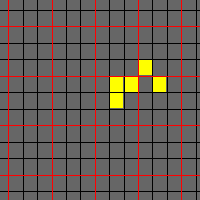} & \includegraphics[width=31mm]{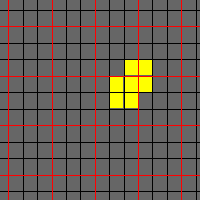} & \includegraphics[width=31mm]{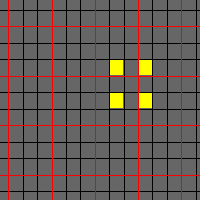} &
				\includegraphics[width=31mm]{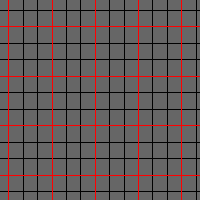}& 
				\\
				$t=6$ & $t=7$ & $t=8$ & $t=9$ & $t=10$ & \\
				
			\end{tabular}
		\end{adjustbox}
		\caption{Dynamics of bee-hive in LCA}
		\label{bee}
	\end{center}
\end{figure*}

Next, we observe the change in dynamics of bee-hive pattern in LCA. Bee-hive is an example of a still life pattern in the Game of Life, representing a static and non-changing structure. Each cell is surrounded by other neighboring cells, creating a tightly packed structure. The cells form a stable arrangement where each cell has exactly two or three neighboring cells. The cells remain in their original positions, and the pattern retains its shape indefinitely. But due to influence of rule $g$, the bee-hive pattern dies out after several iteration. Figure.~\ref{bee} shows the effect of $g$ on the dynamics. After 9 generation, the pattern cease to exist. 
Similar to glider, we found out that bee-hive is also a position sensitive pattern in LCA. It give three different outcomes based on the position which are:
\begin{itemize}
	\item Still life, which is similar to the outcome of bee-hive pattern in Game of Life.
	\item Dies out after several generation as shown in Figure.~\ref{bee}.
	\item Chaos, as the pattern grows rapidly and unpredictably.
\end{itemize}

Next we observed some of the still life pattern for our experiment. We focused on four different still life patterns: block, loaf, tub, and boat. For the block, loaf, and tub patterns, we found that their dynamics remained unaltered even after applying the noise rule $g$. These patterns retained their original structure and exhibited no evolution or alteration. They acted as stable entities in the LCA system, maintaining their static nature regardless of the noise influence.
However, when we considered the boat pattern as the initial configuration and introduced the noise rule $g$, we observed two distinct types of behavior. In some cases, the boat pattern displayed chaotic dynamics, undergoing complex and unpredictable changes over time. On the other hand, in other instances, the boat pattern exhibited still life behavior. It remained unchanged and stable throughout the evolution of the LCA system, even in the presence of the noise rule. Hence boat pattern is a position sensitive configuration.

Similarly, we experimented with oscillators. We considered three oscillators - blinker, beacon and toad. For the blinker oscillator, we observed that the dynamics remained unchanged when the noise rule $g$ was applied. The blinker continued to oscillate between its two configurations without any alteration. In the case of the beacon oscillator, we made an intriguing observation of position sensitivity. Upon the application of rule $g$, the beacon pattern exhibited three distinct types of dynamical behavior. Firstly, it could undergo chaotic dynamics, characterized by unpredictable and complex changes in its configuration. Secondly, it could display oscillation, where it alternated between a limited set of configurations. Lastly, it could exhibit still life behavior, maintaining a stable pattern without any evolution. Figure.~\ref{beacon} shows the dynamical change of beacon from being oscillator to still life. Similarly, we found that the toad oscillator also displayed position sensitivity. It manifested three different types of dynamical behavior when subjected to the noise rule $g$. These included chaotic dynamics, where the pattern exhibited complex and unpredictable changes, the pattern dying out completely, and oscillation, where it alternated between a set of configurations. 

When the rule $g$ is applied to the diehard pattern in the game of life, we observed some intriguing and diverse dynamical behaviors. Initially, the diehard pattern exhibits its characteristic behavior of evolving over generations, with some cells dying out while others continue to survive. After approximately 130 generations, the pattern eventually ceases to exist, which is the expected behavior in the standard game of life rules.
However, when we introduced the influence of rule $g$ on the diehard pattern, we observed five distinct types of dynamical behavior, highlighting the pattern's position sensitivity. These behaviors include still life, where the pattern remains unchanged over generations; dies out, where the pattern quickly disappears; chaos, characterized by rapid and unpredictable changes; glider, where the pattern moves across the grid in a specific direction; and oscillation, where the pattern undergoes periodic oscillations.
The presence of these different behaviors in the diehard pattern under the influence of rule $g$ demonstrates the pattern's sensitivity to its initial position

\begin{table}[htbp]	
	\begin{center}			
		\begin{adjustbox}{width=\columnwidth,center}
			\begin{tabular}{|c|c|c|}
				\hline
				\textbf{Pattern} & \textbf{Dynamics observed in Game of Life} & \textbf{Dynamics observed in LCA} \\
				\hline
				Blinker & Oscillation & Oscillation \\
				\hline
				Beacon & Oscillation & Chaos, still life, oscillation\\
				\hline
				Toad & Oscillation & Dies out, chaos, oscillation\\
				\hline
				Block & Still life & Still life\\
				\hline
				Loaf & Still life & Still life\\
				\hline
				Boat & Still life & Chaos, still life\\
				\hline
				Tub & Still life & Still life\\
				\hline
				Glider & Diagonal movement across the grid & Dies out, Glider\\
				\hline
				R-pentomino & Oscillation, still life, Dies out & Still life, Dies out, Chaos\\
				\hline
				Diehard & Dies out & Still life, Dies out, Chaos, Glider, oscillation\\
				\hline
				
			\end{tabular}
		\end{adjustbox}
		\caption{Dynamics of different patterns in LCA}
		\label{table2}
	\end{center}
\end{table}

In Table~\ref{table2}, we have summarized the different outcomes of patterns in LCA based on the influence of rule $g$ where the block size ($b$) is 9. We observed that for certain patterns, the application of $g$ did not result in any significant change in their behavior. These patterns include the blinker, loaf, block, and tub. Regardless of the application of $g$, these patterns maintained their original dynamics.

However, we have also found that some patterns were highly sensitive to the influence of $g$ and exhibited different dynamics depending on their initial position in the grid. These patterns include beacon, toad, boat, glider, r-pentomino, and diehard. The behavior of these patterns varied across different runs of the LCA, based on their position within the grid. For example, the glider pattern, which typically travels diagonally across the grid, showed different termination points or decay patterns depending on its initial position and the influence of $g$.

This position sensitivity highlights the intricate relationship between the initial configuration, the applied rules, and the resulting dynamics in LCA. It showcases how small variations in the starting conditions can lead to diverse and sometimes unpredictable outcomes.

\section{Summary}
\label{S5}

In this chapter, we have explored the dynamics of Layered Cellular Automata (LCA) and their relationship with the classes of Elementary Cellular Automata (ECA) and the influence of rule $g$ on the default rule $f$.

We have begun by examining the different classes of ECAs, namely Class A, Class B, and Class C, which represent homogeneous, periodic, and chaotic dynamics, respectively. We have then investigated how these classes of dynamics are manifested in LCAs.

Throughout the chapter, we have presented various cases based on the relationship between the classes of $f$, $g$, and the combined dynamics of $f$ and $g^b$ in LCA. We discussed cases where the dynamics of the LCA aligned with the dynamics of either $f$ or $g$, as well as cases where the LCA exhibited different dynamics compared to the individual rules.

Additionally, we have explored specific examples of LCA dynamics, such as oscillators, still life configurations, and the Game of Life, to illustrate the diverse behaviors that can emerge in different LCA systems.

Furthermore, we have examined the impact of rule $g$ on specific patterns, such as gliders and the R-pentomino, and observed how the position of these patterns within the grid and the influence of $g$ could lead to different outcomes and dynamics.

Overall, this chapter has provided a comprehensive analysis of the classes and dynamics of LCA. It highlights the intricate interplay between the default rule, the noise rule, and the initial configurations in determining the behavior.

\chapter{Pattern Classification with Layered Cellular Automata}
\label{chap5}

\section{Introduction}

Pattern classification with cellular automata (CA)~\cite{maji2003theory,NiloyIV} is a field that explores the use of CA models as classifiers for different types of patterns. 
In pattern classification, the goal is to train a CA model to recognize and classify input patterns into different classes or categories. This involves selecting appropriate rules and configurations for the CA that allow it to accurately differentiate between patterns belonging to different classes.

The process typically involves a training phase and a testing phase. In the training phase, the CA model is trained using a set of labeled patterns from different classes. The model is updated and its parameters are adjusted iteratively until it achieves satisfactory classification performance. The training phase aims to identify the rules and configurations that optimize the CA's ability to accurately classify patterns.
In the testing phase, the trained CA model is evaluated using a separate set of patterns that were not used in the training phase. The model's classification accuracy is measured by comparing its predicted class labels with the known labels of the test patterns.


Cellular automata which exhibit convergence to fixed points regardless of the initial seed, have found extensive applications in pattern classification tasks ~\cite{DasMNS09,Sethi2016,RAGHAVAN1993145,Maji05,Sethi6641432,CPLX:CPLX21749,Sethi2015}.
In this chapter, we discuss the application of pattern classification on proposed layered cellular automata models. We have discussed pattern classification on different models of LCA with the aim to explore and develop novel approaches to enhance the performance and effectiveness of CA models as pattern classifiers. These efforts aim to harness the inherent capabilities of CA to tackle various classification problems in domains such as image recognition, data mining, and bioinformatics.
The performance of the proposed classifier is assessed using widely-used standard datasets obtained from \url{http://www.ics.uci.edu/~mlearn/MLRepository.html}. 
These datasets helped us to evaluate the effectiveness of pattern classification on different LCA models and generate a benchmark for comparing the performance of the proposed classifier against existing methods. 

\subsection{Convergence}\label{convergence}
During the process of evolution, a cellular automaton (CA) tends to reach a specific set of configurations known as an attractor. If this set contains only one configuration, it is called a fixed point attractor. When all the attractors of a CA are fixed points, we refer to the CA as convergent. A fixed point attractor occurs when the state of each cell in the CA's grid remains the same in the next iteration, given the current states of its neighboring cells and the CA's update rules. This means that the CA has reached a stable state where further evolution does not lead to any changes.
\begin{definition}\label{def0}
	An LCA ($f$, $g^b$) is classified as a convergent LCA if, regardless of the initial configuration and the values of $b$ (where $1\leq b \leq n$, and $n$ represents the number of cells in the LCA), the CA reaches a fixed point and remains in that state indefinitely.
\end{definition}
In simpler terms, for a convergent LCA, no matter how we start the CA and the size of the blocks ($b$), the CA will eventually reach a stable configuration and remain in that state permanently. Based on the number of fixed point attractors, an LCA can be classified based on the number of fixed point attractors it possesses. If all possible configurations of the LCA converge to a single configuration, it is referred to as a single attractor LCA. On the other hand, if each possible configuration converges to more than one fixed point configuration, the LCA is termed a multiple attractor LCA. In a single attractor LCA, regardless of the initial configuration, the LCA evolves and settles into a single fixed point attractor. This means that all possible configurations of the LCA eventually converge to the same configuration. In contrast, a multiple attractor LCA exhibits a different behavior. For each possible configuration, the LCA converges to more than one fixed point attractor. This implies that different initial configurations lead to different stable states. The LCA's evolution results in a diverse set of attractors, each corresponding to a specific initial configuration.

The distinction between single attractor and multiple attractor LCAs is important for understanding the dynamics and behavior of LCAs. It provides insights into the range of stable states that an LCA can exhibit and how different initial conditions can lead to different outcomes. The study of multiple attractor LCAs is particularly interesting as it allows for the exploration of complex and diverse patterns of behavior and offers potential applications in various fields, including pattern classification, optimization, and simulation of complex systems.

Following a large number of experiments, we identify the set of LCAs that converge to fixed points. From these LCAs, we eliminate two types of LCAs: those associated with a single fixed point attractor and those with a large number of attractors.

LCAs with a single fixed point attractor are excluded because they do not align with the concept of a two-class pattern classifier. When all patterns converge to a single class, it contradicts the objective of distinguishing between two distinct classes.

On the other hand, LCAs with a large number of attractors, specifically those using the Elementary Cellular Automaton (ECA) rule 204, are also excluded. ECA 204 generates attractors that correspond to every possible initial configuration. For an LCA with $n$ cells, this results in $2^n$ attractors. Consequently, as $n$ grows larger, storing and managing such a large number of attractors becomes impractical and memory-intensive. Moreover, this type of LCA operates primarily as a pattern-matching mechanism rather than an effective classifier.

Hence, both LCAs with a single fixed point attractor and those generating a large number of attractors, particularly using ECA 204, are not suitable for our classification purposes. Our focus lies on LCAs that exhibit stable convergence behavior and can effectively distinguish between two classes of patterns.

Based on experimentation, we can conclude that the convergence behavior of different LCAs can change depending on the values of $n$ (size) and $b$ (block size). For instance, in Fig~\ref{Fig5}, the dynamics of LCA$(168,94^b)$ exhibit a transition in convergence behavior. After reaching a critical value of $b$ (e.g., $b = 100$), it converges to an all-$0$ configuration. However, for smaller values of $b$, such as $b = 25$ and $b = 50$, the LCA oscillates around a fixed non-zero density. This significant change in behavior is referred to as a second-order phase transition~\cite{Sethi2016}.
Similarly, in Fig~\ref{Fig8}, LCA$(131,136^b)$ demonstrates a class transition phenomenon. For $b = 50$, it converges to a fixed point configuration. However, for smaller values of $b$, such as $b = 10$ and $b = 20$, it exhibits chaotic dynamics. In this case, the dynamics of the system change, resulting in a transition between different classes. It is worth noting that for the current study, these LCAs with phase transition and class transition are excluded.

\section{Design of a Pattern classifier}\label{pattern_classifier}

An LCA with $n$ cells and $k$ fixed points can be employed as a $k$-class classifier. Each class corresponds to a distinct set of configurations that all converge to the same fixed point. Hence, the fixed point itself can be considered as a representative symbol for that class. To create a two-class classifier, a subset of fixed points out of the total $k$ fixed points is assigned to represent one class, while the remaining fixed points represent the other class.
From an implementation perspective, all the fixed points, along with their corresponding class information, are stored in memory. When the class of an input pattern ($P$) needs to be determined, the LCA is initiated with the pattern as the initial seed. The LCA evolves until it settles into a fixed point, and based on which fixed point it converges to, the class of $P$ is determined.

In summary, the fixed points of the LCA serve as representatives of classes in the classification task. By running the LCA with an input pattern and observing which fixed point it converges to, the class of the pattern can be determined. This approach allows the LCA to function as a classifier capable of distinguishing between classes based on the convergence behavior of the system.

To assess the performance of a classifier, it is important that the patterns representing different classes are evenly distributed within the attractor basins. However, in real-world datasets, the attractor basins may overlap or mix patterns from two or more classes, making the classification task more challenging.
To measure the effectiveness of the classifier in this context, we typically evaluate its performance in terms of classification accuracy. Classification accuracy is defined as the ratio of the number of patterns correctly classified by the classifier to the total number of patterns in the dataset. The formula for calculating accuracy is as follows:
\begin{align}
\text{Accuracy} = \frac{\text{No of patterns which are properly classified}}{\text{Total no. of patterns}} \times 100\%
\end{align}

By calculating the classification accuracy, we can determine how well the classifier is able to correctly assign patterns to their respective classes despite the potential mixing of attractor basins. A higher classification accuracy indicates that the classifier is performing well in distinguishing and assigning patterns to their appropriate classes, while a lower accuracy suggests that the classifier may struggle to correctly classify patterns.

The evaluation of classification accuracy provides a quantitative measure of the classifier's performance, enabling comparisons between different classifiers or assessing the impact of various factors on the classification task. It serves as a valuable metric in assessing the effectiveness of the classifier in handling the complexities of real-world datasets with potentially overlapping attractor basins.

\begin{figure}
	\centering
	\includegraphics[width=0.5\textwidth]{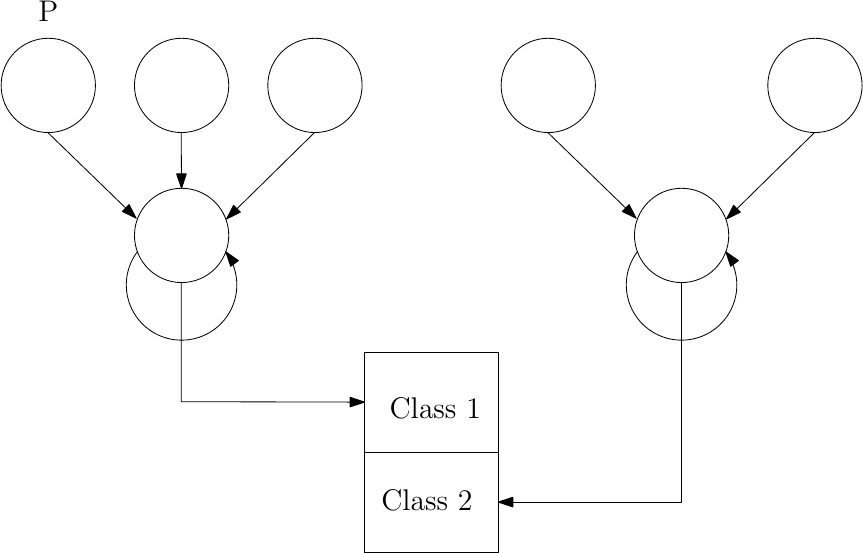}
	\caption{Classification scheme using multiple fixed point attractors.}
	\label{multiple-fixed-point}
\end{figure}

While a multiple-attractor LCA may not necessarily be an optimal pattern classifier, its performance and effectiveness can still be measured through the use of the \emph{training phase} and \emph{testing phase}. This approach allows us to evaluate how well the LCA can generalize its learned knowledge to correctly classify patterns beyond the training set and provides a measure of its capability in handling unseen data.

\subsection{Training phase}\label{training}
Multiple attractor LCAs can be considered as potential candidates for pattern classification. To select the best classifier among these candidates, a training process is conducted using two separate datasets, namely $P_1$ and $P_2$.
In the training phase, an LCA is selected from the set of candidate LCAs and initialized with patterns from both $P_1$ and $P_2$. The LCA is then iteratively updated until it converges to a fixed point. During this process, the number of patterns from each pattern set that converge to each attractor is recorded.
Based on these convergence counts, the attractors are categorized into two sets: attractorset-1 and attractorset-2. If more patterns from $P_1$ converge to an attractor than patterns from $P_2$, the attractor is considered to belong to Class 1 and is stored in attractorset-1. Conversely, if more patterns from $P_2$ converge to an attractor, it is classified as Class 2 and placed in attractorset-2. By categorizing the attractors based on the convergence behavior of patterns from $P_1$ and $P_2$, the training phase determines the associations between attractors and their corresponding classes. This process allows for the selection of the best classifier among the multiple attractor LCAs by identifying the attractor sets that provide the most accurate and reliable classification of patterns from the two disjoint datasets. The following formula is used to determine the accuracy of an LCA:
\begin{align}
	Accuracy &= \frac{\sum_{i=1}^{m} max(n_1^i, n_2^i)}{\big|P_1\big|+\big|P_2\big|}
\end{align}

Here, we consider $n^i_1$ and $n^i_2$ as the maximum numbers of patterns that converge to the $i^{th}$ fixed point attractor of an LCA from dataset $P_1$ and $P_2$ respectively. $\big|P_1\big|$ and $\big|P_2\big|$ represent the total number of patterns in the two datasets used for pattern classification. During this phase, the goal is to generate an LCA with optimal efficiency. The outputs of the training phase are attractorset-$1$, attractorset-$2$ and an LCA with maximum accuracy.


The outputs obtained during the training phase, including the LCA with the highest accuracy, as well as the attractor sets labeled as ``attractorset-1'' and ``attractorset-2'', play a crucial role in the subsequent testing phase. These attractor sets serve as inputs and form the foundation for pattern classification, enabling the evaluation of the trained LCA's performance and effectiveness. During the testing phase, the classifier's capability to accurately classify patterns from new and unseen data is examined. This phase builds upon the knowledge and insights gained from the training phase, where the LCA was trained to recognize and classify patterns. 

\subsection{Testing phase}\label{testing}
In the testing phase, a new set of patterns is used to assess the effectiveness of the designed classifier. This phase takes as input the attractor sets ``attractorset-1'' and ``attractorset-2'' obtained from the training phase, along with the pattern sets $P_1$ and $P_2$.
During the testing phase, the LCA obtained from the training phase is loaded with the patterns from $P_1$ and $P_2$. The LCA is then iteratively updated until all the patterns converge to one of the fixed point attractors. The main objective is to measure the efficiency of the LCA as a classifier based on the number of patterns successfully identified.
To evaluate the LCA's efficiency, the following approach is employed: For each attractor present in ``attractorset-1'', only the number of patterns from dataset $P_1$ that converge to that particular attractor are counted as correctly identified patterns. Similarly, for each attractor in ``attractorset-2'', only the number of patterns from dataset $P_2$ that converge to that attractor are considered as correctly identified patterns.

By distinguishing the attractors based on the corresponding datasets, the testing phase provides a quantitative measure of the LCA's efficiency in correctly identifying patterns from the respective datasets. This evaluation process enables an assessment of the classifier's performance in terms of its ability to accurately classify patterns from unseen data and further validates its effectiveness as a pattern classifier.

Next, we will explore the pattern classification capabilities of various LCA models discussed in the previous chapter. We will evaluate their effectiveness and compare them to other efficient classifiers. By conducting this evaluation, we aim to determine the strengths and weaknesses of the LCA models as pattern classifiers and analyze how they stack up against other established classification methods. This comparative analysis will provide valuable insights into the potential applications and performance of LCA models in the field of pattern classification.

\section{LCA based on averaging}

The idea behind this model is to maintain the balance of the number of 1s in a block. To achieve that we perform averaging function, where we consider the number of 1s from left, right neigbors. If average is greater than number of 1s in the current block, then some 1s are added to maintain balance, similarly if 1s is greater in current block as compared to average, then some 1s are dropped. 

\begin{figure*}[hbt!]
	\begin{center}
		\scalebox{0.8}{
			\begin{tabular}{ccccc}				
				ECA 52 & ($52,g^{25}$) & ($52,g^{50}$) & ($52,g^{125}$) & ($52,g^{150}$) \\[6pt]
				\includegraphics[width=31mm]{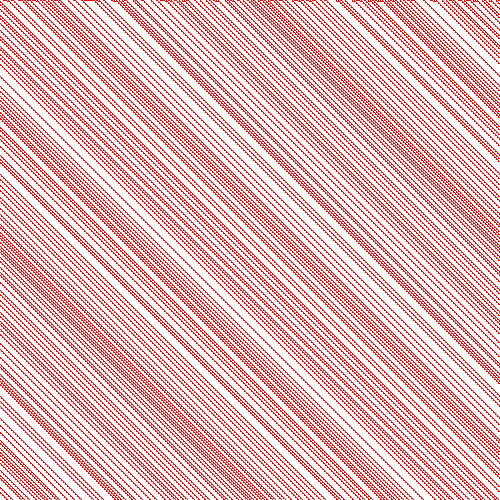} & \includegraphics[width=31mm]{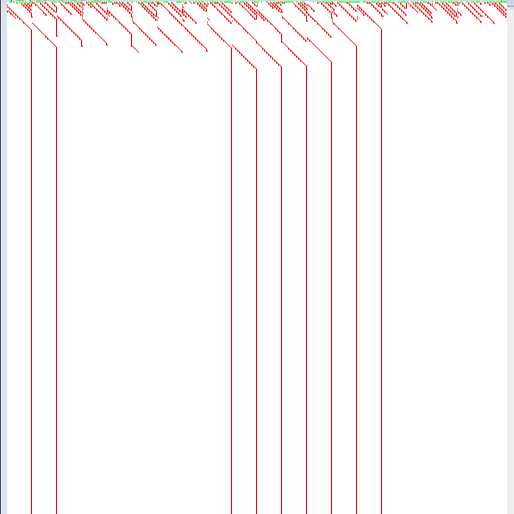} &   \includegraphics[width=31mm]{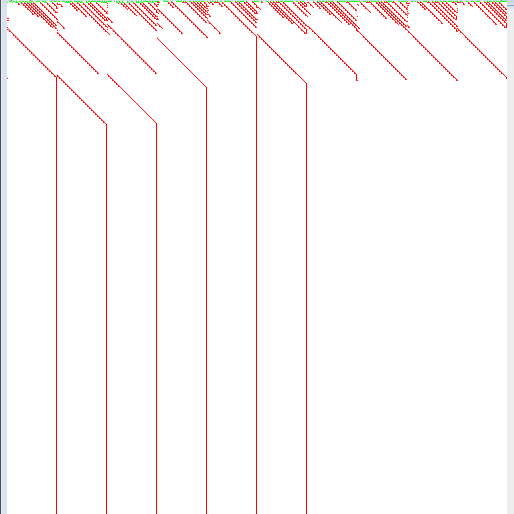} &   \includegraphics[width=31mm]{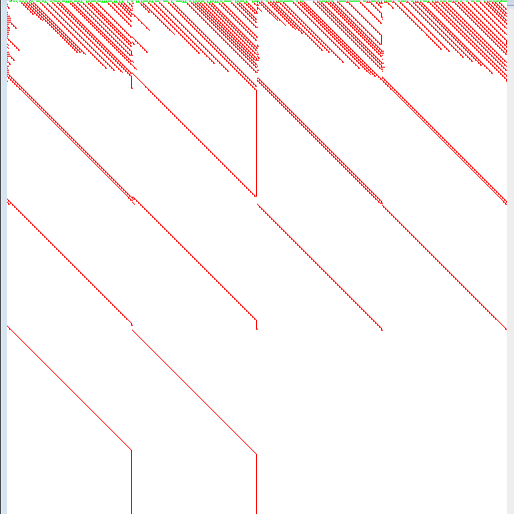} &   \includegraphics[width=31mm]{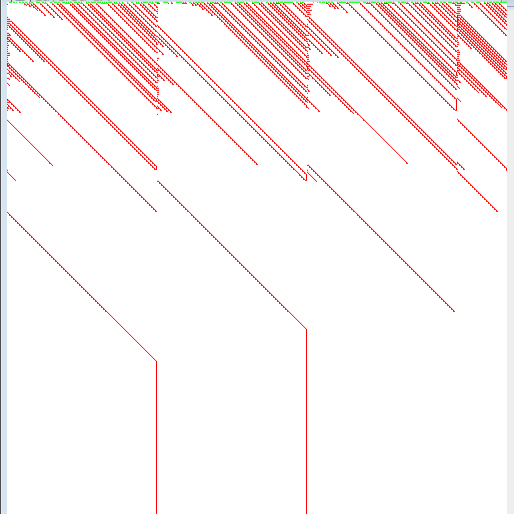} \\
		\end{tabular}}
		\caption{Convergent LCA($52,g^{b}$) dynamics for averaging scheme}
		\label{con-avg}
	\end{center}
\end{figure*}

Through experimentation, we have identified a set of LCAs that consistently exhibit convergence behavior. Regardless of the initial configuration and the value of $b$, these LCAs tend to converge to fixed-point configurations. 
An example of such a convergent LCA is illustrated in Figure.~\ref{con-avg}, specifically LCA($52,g^b$). In this case, Rule 52 is selected as the rule $f$, which is a periodic rule. However, when the dynamics of rule $f$ are influenced by rule $g$, the resulting behavior of the LCA demonstrates convergence to a fixed-point configuration. Figure.~\ref{con-avg} provides visual representations of the fixed-point configurations obtained from LCA($52,g^b$) for different values of $b$, such as $b=25, 50, 125, 150$.

These types of convergent LCAs are particularly relevant for the design of pattern classifiers. However, it is important to note that the demand for pattern classifiers typically leans towards multiple attractor LCAs. While the convergent LCAs offer stability and convergence to fixed-point configurations, the ability to capture multiple attractors is desired for effective pattern classification tasks.

Previously, we have discussed that to design a pattern classifier, we exclude certain types of LCAs. Specifically, we avoid LCAs that are characterized by a single fixed point and those with a large number of attractors specifically, LCAs with ECA $204$.

After applying these exclusion criteria, we narrow down the pool of potential LCAs to a smaller subset. Specifically, from a total of $190$ convergent LCAs identified for $n=11$, we focus on a subset of $148$ LCAs (refer to Table~\ref{tablemonk2}). These $148$ LCAs emerge as strong candidates for the proposed pattern classifier.

We have considered all $256$ ECA rules for all possible block sizes $b$, $1\leq b \leq n$ where $b$ is the block size and $n$ is the CA size, to identify convergent LCAs. The list of LCAs (CA size $n=11$) that converge to Fixed Points is shown in Table.~\ref{convergent LCA}. However, to show a large number of attractors, we exclude some trivial rules, such as rule $204$. The LCAs designated as {\em Sing} and {\em Mult} define single attractor LCAs and multiple attractor LCAs respectively. The first column in this Table.~\ref{convergent LCA} displays the size of each blocks in rule $g$ block, the second column shows the rule $f$, and the last column defines attractor.

\begin{table}[h]
	\scriptsize
	\begin{center}
		\caption{Convergent LCAs and numbers of attractors for different block size $b$. }
		\label{convergent LCA}
		\begin{adjustbox}{width=0.95\columnwidth,center}
			\begin{tabular}{|c|c|c|} \hline 
					$\#b$&Rule $f$&Attr.\\\hline
					1&\makecell{2,  8,  10,  16,  18,  24,  26,  32,  34,  40,  42,  48, , 50,  56,  58,  64,  66,  72,  74,  80,\\  82,  96,  98,  104, 106,  112,  114,  129,  131,  137,  139,  145,  147,  149,  151,  153,\\ , 155,  159,  161,  163,  169,   171,  177,  179,  181,  183,  185,  187, , 191,  193,  195,  201,\\  203,  209,  211,  213,  215,  219,  223,  225,  , 227,  233,  235,  241,  243,  245,  247,  251,  255}& Sing\\\hline
					2&\makecell{ 2,  8,  16, 18,  24,  26,  32,  34,  40,  42,  48,  56, 64,  66,  80, 88,  90,  96,  98,   120,  151,  191,\\  209,  227,  253,  255,}&Sing \\\hline
					3&\makecell{2,  8,  16,  24,  32,  34,  40,  48,  56,  64,  66,  80,  88,  96,  98,  112,  120,  129,  137,  151,  161,\\  253,  255, }&Sing \\\hline
					4&\makecell{
					2,  8,  10,  16,  18,  24,  26,  32,  34,  40,  42,  48,  56,  64,  66,  80,  82,  88,  90,  96,  98,  112,\\  120,  137,  169,  251,  253,  255,
					}& Sing\\\hline
				
					5&\makecell{  2,  8,  10,  16,  24,  32,  40,  48,  56,  64,  66,  80,  88,  96,  112,  249,  251,  253,  255, 
					}&Sing\\\hline
					6&\makecell{ 2,  8,  10,  16,  20,  24,  32,  34,  40,  42,  48,  52,  56,  64,  66,  80,  88,  96,  98,  112,  120,  249,\\  251,  253,  255,
					}& Sing\\\hline
					7&\makecell{2,  8,  10,  16,  20,  24,  32,  34,  40,  42,  48,  52,  56,  64,  66,  80,  88,  96,  98,  112,  120,  195,\\  249,  251,  253,  255,
					}& Sing\\\hline
					8&\makecell{2,  8,  10,  16,  20,  24,  32,  34,  40,  42,  48,  52,  56,  64,  66,  80,  88,  96,  112,  120,  249,\\  251,  253,  255,
					}&Sing\\\hline
					9&\makecell{2,  8,  10,  16,  20,  24,  32,  34,  40,  42,  48,  52,  56,  64,  66,  80,  88,  96,  98,  112,  249,  251,\\  253,  255, 
					}&Sing\\\hline
					10&\makecell{2,  8,  10,  16,  20,  24,  32,  34,  40,  42,  48,  52,  56,  64,  66,  80,  96,  98,  112,  120,  249,\\  251,  253,  255, }& Sing\\\hline&&\\\hline
					1&\makecell{  4,  6,  12,  14,  20,  22,  28,  30,  36,  38,  44,  46,  52,  54,  60,  62,  68,  70,  76,  78,  84,  86,  88,\\  90,  92,  94,  100,  102,  108,  110,  116,  118,  120,  122,  124,  126,  128,  130,  132,  133,\\  134,  135,  136,  138,  140,  141,  142,  143,  144,  146,  148,  150,  152,\ 154,  156,  158,  160, \\ 162,  164,  165,  166,  167,  168,  170,  172,  173,  174,  175,  176,  178,}& Mult\\\hline
					2&\makecell{4,  10,  12,  14,  20,  22,  28,  30,  36,  44,  52,  54,  68,  72,  74,  76,  78,  84,  86,  92,  94,  100,\\  104,  106,  116,  118,  124,  }&Mult \\\hline
					3&\makecell{4,  6,  10,  12,  14,  22,  36,  38,  42,  44,  62,  68,  72,  74,  76,  78,  92,  100,  104, 106,  124,\\  126,  128,  130, }&Mult\\\hline
					4&\makecell{4,  12,  36,  44,  68,  72,  74,  76,  78,  86,  100,  104,  128,  130,  132,  135,  136,  138, 140,\\  144,  146,  152,  154,  160,  162,  164,  168,  170,  172,}&Mult\\\hline					
					5&\makecell{4,  12,  20,  36,  38,  52,  68,  72,  74,  76,  78,  100,  104,  124,  128,  130,  132,  136,  138,  140, 
					}&Mult\\\hline
					6&\makecell{4,  6,  12,  14,  36,  38,  44,  46,  68,  72,  74,  76,  78,  100,  104,  106,  128,  130,  132,  134,\\  136,  138,  140,  142,  144,  148,
					}& Mult\\\hline
					7&\makecell{4,  6,  12,  14,  36,  38,  44,  46,  68,  72,  74,  76,  78,  100,  104,  128,  130,  132,  134,  136,\\  138,  140,  142,  144,  148,  152,  160, 
					}& Mult\\\hline
					8&\makecell{4,  6,  12,  14,  36,  38,  44,  68,  72,  74,  76,  78,  100,  104,  128,  130,  132,  136,  138,  140,\\  142,  144,  148,  152,  158,
					}&Mult\\\hline
					9&\makecell{4,  6,  12,  14,  36,  38,  44,  46,  68,  72,  74,  76,  78,  100,  104,  106,  128,  130,  132,  136,\\  138,  140,  141,  142,  144,
					}&Mult\\\hline
					10&\makecell{4,  6,  12,  14,  36,  38,  44,  46,  68,  72,  74,  76,  78,  100,  104,  106,  128,  130,  132,  134,\\  136,  138,  140,  142,  144, 
					}& Mult\\\hline
			\end{tabular}
		\end{adjustbox}
	\end{center}
\vspace{-2 em}
\end{table}

\subsection{Design of pattern classifier}


	Let us assume for a real-time data set {\em Monk-2}, where the size of the data is $11$. We have considered a $11$-cell convergent LCA($238,g^{10}$) which has $8$ fixed point attractors. The attractors are $00000000000$, $00001111110$, $01111000000$, $01111100000$, $01111110000$, $11111111111$, $00000111110$ and $00111111110$.
	The fixed points $00000000000$, $00001111110$, $01111000000$, $01111100000$, $01111110000$, and $11111111111$ represent Class 1 because more patterns from the pattern set $P_1$ to converge to these fixed point attractors and the rest fixed points $00000111110$ and $00111111110$ represents Class 2. A pattern, say $00100111110$ is given, and the LCA runs with $00100111110$ as seed. After some time, the CA reaches a fixed point $00111111110$. Since $00111111110$ represents Class 2, the class of $00100111110$ is declared as Class 2. Hence this multiple attractor LCA can act as a two-class pattern classifier, see Figure.~\ref{multiple-fixed-point}.

\subsection{Training phase}
To select the best classifier among the candidate LCAs (listed in Table~\ref{convergent LCA}), we train them using patterns from two separate datasets, $P_1$ and $P_2$. 
As an example, let us consider the Monk-$2$ dataset ($11$-bit data) for classification. Let us take the  LCA $(76, g^5)$ as a two-class pattern classifier (similar to Fig~\ref{multiple-fixed-point}), with two patterns set $P_1$ and $P_2$ loaded to the  LCA as Class 1 and Class 2, respectively. $P_1$ and $P_2$ contain a total of $169$ patterns, out of which $15$ patterns of $P_2$ and $31$ patterns of $P_1$ are wrongly identified as in Class 1 and Class 2, respectively. Hence, $123$ patterns are properly classified, which gives training accuracy as $72.781\%$. 

To get the best candidate  LCA, we train all the multiple attractor LCAs of Table~\ref{tablemonk2} by Monk-$2$ dataset. The result of the training accuracy is noted in Table~\ref{tablemonk2}. We find that the   LCA($76,g^{7}$) with training accuracy $82.248\%$, has the highest training accuracy.

\subsection{Testing phase}
The effectiveness of the LCA is gauged by how many patterns the classifier was able to identify. In this phase, the attractor sets \emph{attractorset-I} and \emph{attractorset-II} with an LCA (getting from the training phase) and the (new) pattern sets ($P_1$ and $P_2$) are taken as input in this phase. The patterns of $P_1$ and $P_2$ are put into the LCA, which is then updated until all of the patterns converge to any fixed point attractor.
Table~\ref{tableavg} shows LCAs along with their training and testing accuracy for various datasets.

\begin{table}[!ht]
	\scriptsize
	\centering
	\caption{Accuracies in training utilizing candidate LCAs on the Monk-2 dataset.}
	\label{tablemonk2}
	\begin{adjustbox}{width=1\columnwidth,center}
		\begin{tabular}{|cccccccccccc|} \hline 
			LCAs & \makecell{ Accuracy\\ (in \%)} & \makecell{ Number of\\ Attractor} & LCAs & \makecell{ Accuracy\\ (in \%)} & \makecell{ Number of\\ Attractor}  & LCAs & \makecell{ Accuracy\\ (in \%)} & \makecell{ Number of\\ Attractor}& LCAs & \makecell{ Accuracy\\ (in \%)} & \makecell{ Number of\\ Attractor}\\\hline
			$(4,g^1)$  & 62.13 & 2 & $(4,g^2)$  & 62.13 & 2 & $(4,g^5)$  & 62.721 & 4 & $(4,g^6)$  & 62.13 & 5 \\ $(4,g^7)$  & 62.13 & 5 & $(4,g^8)$  & 62.13 & 4 & $(4,g^9)$  & 62.721 & 6 & $(4,g^{10})$  & 63.313 & 8 \\ $(12,g^1)$  & 62.13 & 2 & $(12,g^2)$  & 62.13 & 2 & $(12,g^3)$  & 63.905 & 3 & $(12,g^4)$  & 64.497 & 8 \\ $(12,g^5)$  & 63.313 & 9 & $(12,g^6)$  & 68.639 & 16 & $(12,g^7)$  & 71.597 & 23 & $(12,g^8)$  & 65.68 & 10 \\ $(12,g^9)$  & 62.13 & 6 & $(12,g^{10})$  & 66.863 & 24 &  $(68,g^{1})$ & 62.13 & 2 & $(68,g^{2})$ & 62.13 & 2 \\ $(68,g^{4})$ & 66.864 & 11 & $(68,g^{5})$ & 69.231 & 15 & $(68,g^{6})$ & 66.272 & 14 & $(68,g^{7})$ & 66.864 & 19 \\ $(68,g^{8})$ & 67.456 & 19 & $(68,g^{9})$ & 62.722 & 11 & ($(68,g^{10})$ & 66.864 & 24 & $(76,g^{1})$ & 62.13 & 2 \\ $(76,g^{2})$ & 62.722 & 8 & $(76,g^{3})$ & 68.639 & 24 & $(76,g^{4})$ & 71.598 & 34 & $(76,g^{5})$ & 72.781 & 50 \\ $(76,g^{6})$ & 80.473 & 65 & $(76,g^{7})$ & 82.248 & 59 & $(76,g^{8})$ & 80.473 & 56 & $(76,g^{9})$ & 76.331 & 66 \\ $(76,g^{10})$ & 74.556 & 74 & $(132,g^{1})$ & 62.13 & 2 & $(132,g^{2})$ & 62.13 & 2 & $(132,g^{3})$ & 62.13 & 3 \\ $(132,g^{4})$ & 66.864 & 15 & $(132,g^{5})$ & 63.314 & 13  & $(132,g^{6})$ & 63.314 & 15 & $(132,g^{7})$ & 64.497 & 21 \\ $(132,g^{8})$ & 62.722 & 12 & $(132,g^{9})$ & 62.13 & 7 & $(132,g^{10})$ & 63.314 & 22 & $(136,g^{5})$ & 63.905 & 3 \\ $(136,g^{8})$ & 62.13 & 2 & $(140,g^{1})$ & 62.13 & 2 & $(140,g^{2})$ & 63.314 & 4 & $(140,g^{3})$ & 63.905 & 7 \\ $(140,g^{4})$ & 68.639 & 21 & $(140,g^{5})$ & 69.231 & 23 &  $(140,g^{6})$ & 71.006 & 23 & $(140,g^{7})$ & 69.822 & 27 \\ $(140,g^{8})$ & 65.68 & 20 & $(140,g^{9})$ & 65.68 & 13 & $(140,g^{10})$ & 69.231 & 32 & $(196,g^{1})$ & 62.13 & 2 \\ $(196,g^{2})$ & 62.13 & 2 & $(196,g^{3})$ & 62.13 & 6 & $(196,g^{4})$ & 68.639 & 22 & $(196,g^{5})$ & 71.006 & 19 \\ $(196,g^{6})$ & 70.414 & 15 & $(196,g^{7})$ & 71.006 & 25 & $(196,g^{8})$ & 71.598 & 31 & $(196,g^{9})$ & 62.722 & 11 \\ $(196,g^{10})$ & 66.864 & 24 & $(200,g^{2})$ & 63.314 & 3 & $(200,g^{3})$ & 63.905 & 6 & $(200,g^{4})$ & 63.905 & 13\\$(200,g^{5})$ & 67.456 & 22 & $(200,g^{6})$ & 68.639 & 30 & $(200,g^{7})$ & 71.006 & 35 & $(200,g^{8})$ & 69.231 & 28 \\ $(200,g^{9})$ & 71.006 & 42 & $(200,g^{10})$ & 76.331 & 70 & $(206,g^{1})$ & 62.13 & 2 & $(206,g^{2})$ & 71.598 & 36 \\ $(206,g^{3})$ & 72.189 & 44 & $(206,g^{4})$ & 77.515 & 64 & $(206,g^{5})$ & 75.74 & 38 & $(206,g^{6})$ & 76.923 & 57\\$(206,g^{7})$ & 72.189 & 65 & $(206,g^{8})$ & 78.107 & 78 & $(206,g^{9})$ & 79.29 & 82 & $(206,g^{10})$ & 81.065 & 72 \\ $(207,g^{1})$ & 62.13 & 2 & $(207,g^{2})$ & 66.864 & 21 & $(207,g^{3})$ & 70.414 & 25 & $(207,g^{4})$ & 72.189 & 37 \\ $(207,g^{5})$ & 70.414 & 27 & $(207,g^{6})$ & 72.189 & 38 & $(207,g^{7})$ & 73.373 & 39 & $(207,g^{8})$ & 76.923 & 53\\$(207,g^{9})$ & 74.556 & 48 & $(207,g^{10})$ & 75.74 & 49 & $(236,g^{1})$ & 62.13 & 3 & $(236,g^{2})$ & 68.639 & 28 \\ $(236,g^{3})$ & 71.006 & 31 & $(236,g^{4})$ & 71.598 & 27 & $(236,g^{5})$ & 73.373 & 44 & $(236,g^{6})$ & 73.373 & 44 \\ $(236,g^{7})$ & 71.598 & 41 & $(236,g^{8})$ & 76.331 & 47 & $(236,g^{9})$ & 72.781 & 48 & $(236,g^{10})$ & 81.065 & 87\\$(237,g^{1})$ & 62.722 & 2 & $(237,g^{2})$ & 63.905 & 11 & $(237,g^{3})$ & 72.189 & 17 & $(237,g^{4})$ & 67.456 & 17 \\ $(237,g^{5})$ & 66.272 & 13 & $(237,g^{6})$ & 66.864 & 11 & $(237,g^{7})$ & 69.231 & 11 & $(237,g^{8})$ & 68.639 & 13  \\ $(237,g^{9})$ & 69.822 & 15 & $(237,g^{10})$ & 70.414 & 20 & $(238,g^{1})$ & 62.13 & 3 & $(238,g^{2})$ & 69.822 & 29\\$(238,g^{3})$ & 68.047 & 21 & $(238,g^{4})$ & 68.047 & 18 & $(238,g^{5})$ & 65.68 & 9 & $(238,g^{6})$ & 66.272 & 12 \\ $(238,g^{7})$ & 65.089 & 14 & $(238,g^{8})$ & 65.089 & 14 & $(238,g^{9})$ & 65.68 & 14 & ($(238,g^{10})$ & 65.68 & 8 \\ $(252,g^{1})$ & 63.905 & 10 & $(252,g^{2})$ & 63.905 & 14 & $(252,g^{3})$ & 62.13 & 3 & $(252,g^{4})$ & 62.13 & 2\\$(252,g^{5})$ & 62.13 & 3 & $(252,g^{6})$ & 62.13 & 2 & $(252,g^{7})$ & 62.13 & 2 & $(252,g^{8})$ & 62.13 & 2 \\ $(252,g^{9})$ & 62.13 & 2 & $(252,g^{10})$ & 62.13 & 2 & $(254,g^{1})$ & 62.722 & 10 & $(254,g^{2})$ & 62.722 & 10 \\ $(254,g^{3})$ & 62.13 & 3 & $(254,g^{4})$ & 62.13 & 2 & $(254,g^{5})$ & 62.13 & 3 & $(254,g^{6})$ & 62.13 & 2\\$(254,g^{7})$ & 62.13 & 2 & $(254,g^{8})$ & 62.13 & 2 & $(254,g^{9})$ & 62.13 & 2 & $(254,g^{10})$ & 62.13 & 2\\\hline
		\end{tabular}
	\end{adjustbox}

\end{table}

\begin{table}[!ht]

	\scriptsize
	\begin{center}
		\caption{Performance of the proposed classifiers using various datasets.}
		\label{tableavg}
		\begin{adjustbox}{width=0.8\columnwidth,center}
			\begin{tabular}{|ccccc|} \hline 
			Datasets & \makecell{ LCA\\ Size} & \makecell{Training \\Accuracy} & \makecell{Testing\\ Accuracy} & \makecell{Proposed\\  LCAs}\\ \hline
			Monk-1 & 11 & 90.983 & 76.798 & ($236,g^{10}$)\\
			Monk-2 & 11 & 82.248 & 72.222 & ($76,g^7$)\\
			Monk-3 & 11 & 95.081 & 95.833 & ($200,g^9$)\\
			Haber man & 9 & 73.469 & 73.717 & ($206,g^5$)\\
			Heart-statlog & 16 & 92.592 & 82.222 & ($206,g^9$)\\
			Tic-Tac-Toe & 18 & 100 & 100 & ($236,g^{17}$)\\
			Hepatitis-1 & 19 & 98.717 & 97.402 & ($206,g^8$)\\
			Hepatitis-2 & 22 & 100 & 98.701 & ($207,g^5$)\\ \hline
			\end{tabular}
		\end{adjustbox}
	\end{center}

\end{table}

\section{LCA based on Maximization}

The underlying concept of this model is to maximize the density of 1s within a block. This is achieved through a maximization function that takes into account the number of 1s in the left and right neighboring blocks. If the density of 1s is lower in the current block compared to the neighboring blocks, additional 1s are added to maximize the density. Conversely, if the current block already has a higher density of 1s compared to the neighboring blocks, no changes are made to the configuration. This approach aims to optimize the distribution of 1s within the blocks to maximize the overall density of 1s in the pattern. 

\begin{figure*}[hbt!]
	\begin{center}
		\scalebox{0.8}{
			\begin{tabular}{ccccc}				
				ECA 8 & ($8,g^{25}$) & ($8,g^{65}$) & ($8,g^{150}$) & ($8,g^{250}$) \\[6pt]
				\includegraphics[width=31mm]{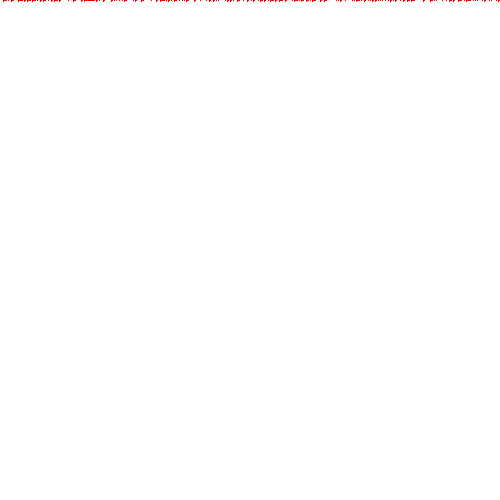} & \includegraphics[width=31mm]{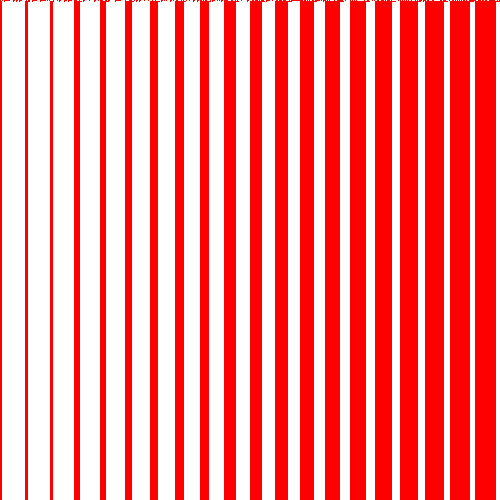} &   \includegraphics[width=31mm]{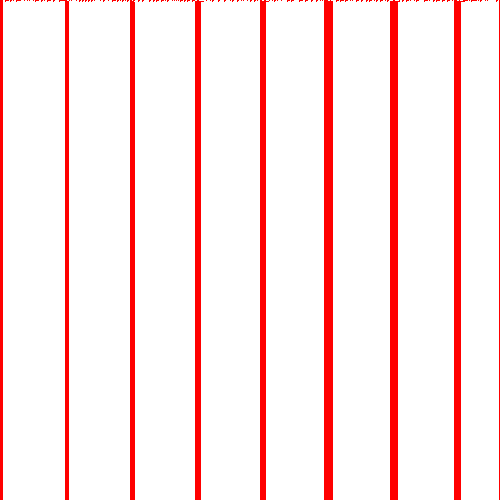} &   \includegraphics[width=31mm]{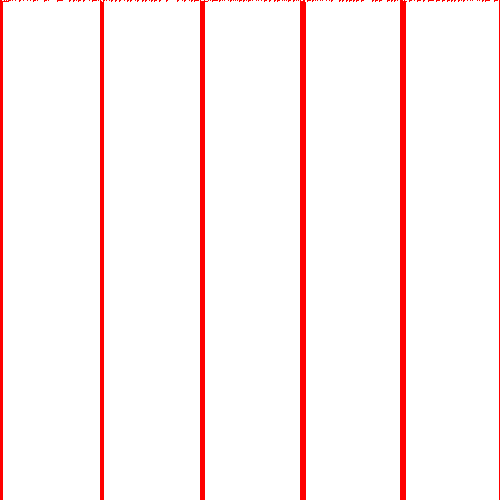} &   \includegraphics[width=31mm]{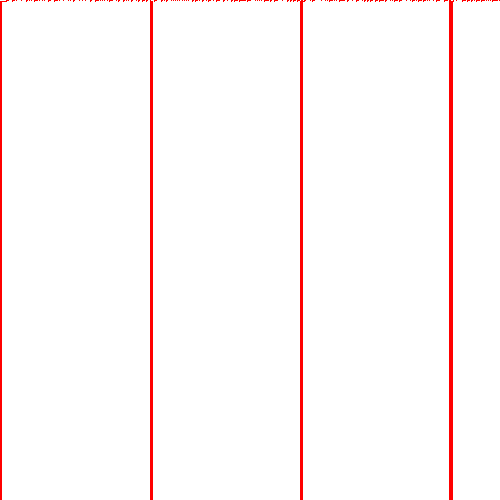} \\
		\end{tabular}}
		\caption{Convergent LCA($8,g^{b}$) dynamics}
		\label{con-max}
	\end{center}
\end{figure*}

Let us consider the LCA($8,g^b$) shown in Figure.~\ref{con-max}. In this case, we select Rule 8 as the rule $f$, which is a Class A rule. When the dynamics of rule $f$ are influenced by rule $g$, the behavior of the LCA transitions to a state of convergence, resulting in a fixed-point configuration.
Figure.~\ref{con-max} provides visual representations of the fixed-point configurations obtained from the LCA($8,g^b$) for different values of $b$, such as $b=25,65,150,250$. These figures illustrate the stable patterns that emerge as the LCA converges to a fixed point, demonstrating the consistent convergence behavior of the LCA across different parameter values.

Convergent LCAs, such as the one described above, play a crucial role in the design of pattern classifiers. Their capacity to consistently converge to fixed-point configurations makes them well-suited for capturing and recognizing specific patterns. 

In our experiment, we did not find a comprehensive list of LCAs that would be convergent for all possible values of $n$ and $b$. Hence, we conducted an exhaustive search for convergent LCAs by considering all possible combinations of $256 \times (n-1)$ LCAs. Here, 256 represents all possible Elementary Cellular Automata (ECA) rules, and $n-1$ represents the number of possible block sizes for a cell size of $n$, excluding a block size of 1. We excluded LCAs with a block size of 1 since we have previously discussed that in such cases, the LCA tends to converge to an all-1s configuration due to the maximization of 1s, which acts as a single attractor.

For the case of $n=11$, where the block size can range from 2 to 11, we have a total of 2560 LCAs. Out of these, 1269 LCAs exhibit convergence behavior. To select LCAs for the proposed pattern classifier, we exclude 865 LCAs that have a single fixed point attractor. This leaves us with a set of 404 candidate LCAs. Among these candidates, 148 LCAs demonstrate maximum efficiency and are considered as entries in the table (refer to Table~\ref{monkset1}). These selected LCAs show promise for serving as effective components of the pattern classifier.

\subsection{Design of pattern classifier}


	Let us assume for the real-time dataset \emph{Monk-1}, where the size of CA is 11. We have considered a $11$-cell convergent LCA($13,g^{3}$) which has $2$ fixed-point attractors. The attractors are $11011111111$,  and $10111011111$. The fixed point $11011111111$ represents Class $1$ because more patterns from the pattern set $P_1$ get converged to this fixed point attractor and fixed point $10111011111$ represents Class $2$. A pattern, say $10101000101$ is given, the LCA runs with $10101000101$ as seed. After some time, the CA reaches to a fixed point attractor - $10111011111$. Since $10111011111$ represents Class $2$, class of $10101000101$ is declared as $2$. Hence, this multiple attractor LCA can act as two class pattern classifier.

\subsection{Training phase}
Candidate LCAs are trained using patterns from datasets $P_1$ and $P_2$, and their accuracy is evaluated to select the best classifier. 
As an example, let us consider the Monk-$1$ dataset ($11$-bit data) for classification. Let us take the  LCA $(9,g^3)$ as a two-class pattern classifier (similar to Fig~\ref{multiple-fixed-point}), with two pattern set $P_1$ and $P_2$ loaded to the  LCA as Class $1$ and Class $2$, respectively. $P_1$ and $P_2$ contain a total of $122$ patterns, out of which $30$ patterns of $P_2$ and $1$ patterns of $P_1$ are wrongly identified as in Class $1$ and Class $2$, respectively. Hence, $91$ patterns are properly classified, which gives training accuracy as $74.59\%$. 

To get the best candidate  LCA, we train all the multiple attractor LCAs on the Monk-$1$ dataset. The result of some of the training accuracy is noted in Table~\ref{monkset1}. We find that the LCA($140,g^{11}$) with training accuracy $86.066\%$, has the highest training accuracy. This LCA is considered our desired classifier. In Table~\ref{monkset1}, we find the best performing $b$ value and corresponding accuracy for each  LCA.

\begin{table}[!htpb]
	\scriptsize
	\centering
	\caption{Accuracies in training utilizing candidate LCAs on the Monk-1 dataset.}
	\label{monkset1}
	\begin{adjustbox}{width=1\columnwidth,center}
		\begin{tabular}{|cccccccccccc|} \hline 
			LCAs & \makecell{ Accuracy\\ (in \%)} & \makecell{ Number of\\ Attractor} & LCAs & \makecell{ Accuracy\\ (in \%)} & \makecell{ Number of\\ Attractor}  & LCAs & \makecell{ Accuracy\\ (in \%)} & \makecell{ Number of\\ Attractor}& LCAs & \makecell{ Accuracy\\ (in \%)} & \makecell{ Number of\\ Attractor}\\\hline
		$(140,g^{11})$ & 86.066 & 67 & $(140,g^{10})$ & 85.246 & 53 & $(172,g^{11})$ & 81.967 & 47 & $(141,g^{11})$ & 80.328 & 45 \\ $(196,g^{10})$ & 78.689 & 40 & $(13,g^{3})$ & 77.869 & 2 & $(141,g^{10})$ & 76.23 & 29 & $(172,g^{10})$ & 76.23 & 34 \\ $(12,g^{10})$ & 75.41 & 18 & $(143,g^{10})$ & 75.41 & 14 & $(197,g^{10})$ & 75.41 & 29 & $(204,g^{10})$ & 75.41 & 50 \\ $(207,g^{10})$ & 75.41 & 25 & $(9,g^{3})$ & 74.59 & 2 & $(136,g^{10})$ & 74.59 & 20 & $(142,g^{11})$ & 74.59 & 20 \\ $(196,g^{11})$ & 74.59 & 36 & $(136,g^{11})$ & 73.77 & 22 & $(197,g^{11})$ & 73.77 & 24 & $(143,g^{11})$ & 72.951 & 21 \\ $(205,g^{10})$ & 72.951 & 49 & $(206,g^{10})$ & 72.951 & 25 & $(208,g^{2})$ & 72.951 & 2 & $(12,g^{11})$ & 72.131 & 18 \\ $(142,g^{10})$ & 71.311 & 14 & $(143,g^{2})$ & 71.311 & 2 & $(143,g^{5})$ & 71.311 & 2 & $(139,g^{2})$ & 69.672 & 2 \\ $(142,g^{9})$ & 69.672 & 7 & $(168,g^{11})$ & 69.672 & 35 & $(212,g^{2})$ & 69.672 & 2 & $(12,g^{8})$ & 68.852 & 15 \\ $(209,g^{5})$ & 68.852 & 2 & $(220,g^{10})$ & 68.852 & 19 & $(12,g^{5})$ & 68.033 & 16 & $(12,g^{9})$ & 68.033 & 15 \\ $(13,g^{6})$ & 68.033 & 11 & $(25,g^{3})$ & 68.033 & 2 & $(132,g^{7})$ & 68.033 & 18 & $(142,g^{2})$ & 68.033 & 2 \\ $(144,g^{2})$ & 68.033 & 2 & $(209,g^{2})$ & 68.033 & 2 & $(214,g^{10})$ & 68.033 & 5 & $(12,g^{6})$ & 67.213 & 14 \\ $(13,g^{10})$ & 67.213 & 12 & $(138,g^{11})$ & 67.213 & 9 & $(12,g^{7})$ & 66.393 & 10 & $(138,g^{2})$ & 66.393 & 2 \\ $(141,g^{8})$ & 66.393 & 12 & $(142,g^{5})$ & 66.393 & 2 & $(159,g^{11})$ & 66.393 & 16 & $(208,g^{5})$ & 66.393 & 2 \\ $(212,g^{10})$ & 66.393 & 5 & $(213,g^{2})$ & 66.393 & 2 & $(8,g^{10})$ & 65.574 & 3 & $(8,g^{11})$ & 65.574 & 3 \\ $(13,g^{11})$ & 65.574 & 12 & $(29,g^{3})$ & 65.574 & 2 & $(132,g^{8})$ & 65.574 & 17 & $(139,g^{10})$ & 65.574 & 9 \\ $(212,g^{5})$ & 65.574 & 2 & $(200,g^{10})$ & 64.754 & 31 & $(210,g^{2})$ & 64.754 & 2 & $(221,g^{10})$ & 64.754 & 21 \\ $(228,g^{10})$ & 64.754 & 14 & $(72,g^{10})$ & 63.934 & 17 & $(132,g^{5})$ & 63.934 & 15 & $(132,g^{9})$ & 63.934 & 12 \\ $(141,g^{2})$ & 63.934 & 2 & $(141,g^{5})$ & 63.934 & 2 & $(148,g^{5})$ & 63.934 & 4 & $(158,g^{10})$ & 63.934 & 15 \\ $(158,g^{11})$ & 63.934 & 21 & $(168,g^{10})$ & 63.934 & 26 & $(211,g^{5})$ & 63.934 & 2 & $(228,g^{11})$ & 63.934 & 14 \\ $(136,g^{8})$ & 63.115 & 11 & $(136,g^{9})$ & 63.115 & 12 & $(138,g^{10})$ & 63.115 & 9 & $(139,g^{11})$ & 63.115 & 8 \\ $(158,g^{9})$ & 63.115 & 7 & $(207,g^{2})$ & 63.115 & 2 & $(207,g^{5})$ & 63.115 & 2 & $(216,g^{10})$ & 63.115 & 12 \\ $(8,g^{8})$ & 62.295 & 4 & $(13,g^{8})$ & 62.295 & 10 & $(24,g^{3})$ & 62.295 & 2 & $(40,g^{9})$ & 62.295 & 2 \\ $(137,g^{2})$ & 62.295 & 2 & $(138,g^{9})$ & 62.295 & 5 & $(140,g^{8})$ & 62.295 & 11 & $(153,g^{9})$ & 62.295 & 5 \\ $(213,g^{5})$ & 62.295 & 2 & $(214,g^{2})$ & 62.295 & 2 & $(4,g^{7})$ & 61.475 & 9 & $(8,g^{3})$ & 61.475 & 2 \\ $(8,g^{5})$ & 61.475 & 4 & $(8,g^{6})$ & 61.475 & 5 & $(8,g^{7})$ & 61.475 & 3 & $(8,g^{9})$ & 61.475 & 4 \\ $(13,g^{7})$ & 61.475 & 12 & $(64,g^{9})$ & 61.475 & 2 & $(72,g^{7})$ & 61.475 & 7 & $(96,g^{10})$ & 61.475 & 2 \\ $(145,g^{2})$ & 61.475 & 2 & $(148,g^{2})$ & 61.475 & 2 & $(173,g^{11})$ & 61.475 & 11 & $(192,g^{9})$ & 61.475 & 8 \\ $(193,g^{6})$ & 61.475 & 2 & $(196,g^{5})$ & 61.475 & 2 & $(196,g^{7})$ & 61.475 & 9 & $(200,g^{11})$ & 61.475 & 20 \\ $(202,g^{10})$ & 61.475 & 16 & $(203,g^{2})$ & 61.475 & 2 & $(203,g^{5})$ & 61.475 & 2 & $(203,g^{10})$ & 61.475 & 19 \\ $(216,g^{2})$ & 61.475 & 2 & $(224,g^{11})$ & 61.475 & 10 & $(72,g^{11})$ & 60.656 & 18 & $(128,g^{5})$ & 60.656 & 6 \\ $(137,g^{5})$ & 60.656 & 2 & $(140,g^{2})$ & 60.656 & 2 & $(140,g^{5})$ & 60.656 & 2 & $(140,g^{9})$ & 60.656 & 10 \\ $(141,g^{9})$ & 60.656 & 12 & $(143,g^{9})$ & 60.656 & 7 & $(145,g^{5})$ & 60.656 & 3 & $(159,g^{2})$ & 60.656 & 2 \\ $(159,g^{5})$ & 60.656 & 2 & $(192,g^{2})$ & 60.656 & 2 & $(192,g^{10})$ & 60.656 & 11 & $(197,g^{8})$ & 60.656 & 9 \\ $(206,g^{5})$ & 60.656 & 2 & $(211,g^{2})$ & 60.656 & 2 & $(217,g^{10})$ & 60.656 & 16 & $(4,g^{8})$ & 59.836 & 5 \\ $(12,g^{3})$ & 59.836 & 2 & $(135,g^{2})$ & 59.836 & 2 & $(138,g^{8})$ & 59.836 & 2 & $(141,g^{7})$ & 59.836 & 6 \\ $(149,g^{2})$ & 59.836 & 2 & $(149,g^{6})$ & 59.836 & 2 & $(158,g^{5})$ & 59.836 & 2 & $(164,g^{5})$ & 59.836 & 5 \\ $(192,g^{5})$ & 59.836 & 2 & $(206,g^{2})$ & 59.836 & 2 & $(213,g^{10})$ & 59.836 & 5 & $(214,g^{5})$ & 59.836 & 2 \\\hline
		\end{tabular}
	\end{adjustbox}
	\vspace{-1 em}
\end{table}

\subsection{Testing phase}
In the testing phase, new patterns are used to evaluate the LCA's convergence behavior and classification accuracy.

Table~\ref{tableMAX} presents a comprehensive overview of the performance of the LCAs on various datasets. It provides detailed information regarding the training accuracy and testing accuracy of each LCA, specifically considering the aspect of maximization. The table serves as a valuable reference for analyzing the effectiveness of the LCAs in accurately classifying patterns based on their performance on different datasets.
\begin{table}[!htpb]
	\scriptsize
	\begin{center}
		\caption{Performance of the proposed classifiers using various datasets.}
		\label{tableMAX}
		\begin{adjustbox}{width=0.8\columnwidth,center}
			\begin{tabular}{|ccccc|} \hline 
			Datasets & \makecell{ LCA\\ Size} & \makecell{Training \\Accuracy} & \makecell{Testing\\ Accuracy} & \makecell{Proposed\\  LCAs}\\ \hline
			Monk-1 & 11 & 86.065 & 70.765 & ($140,g^{11}$)\\
			Monk-2 & 11 & 82.248 & 68.981 & ($140,g^9$)\\
			Monk-3 & 11 & 87.704 & 72.222 & ($140,g^{11}$)\\
			Haber man & 9 & 73.469 & 75.641 & ($72,g^5$)\\
			Heart-statlog & 16 & 92.592 & 85.925 & ($200,g^{15}$)\\
			Tic-Tac-Toe & 18 & 100 & 98.121 & ($12,g^{17}$)\\
			Hepatitis-1 & 19 & 96.153 & 93.506 & ($12,g^{17}$)\\
			Hepatitis-2 & 22 & 98.717 & 100 & ($12,g^{21}$)\\ \hline
			\end{tabular}
		\end{adjustbox}
	\end{center}
\end{table}

\begin{table}[!htpb]
	\scriptsize
	\begin{center}
		
		\label{convergentLCAmax}
		\begin{adjustbox}{width=0.95\columnwidth,center}
			\begin{tabular}{|c|c|c|} \hline 
				$\#b$&Rule $f$&Attr.\\\hline
				2&\makecell{1, 2, 3, 4, 5, 6, 7, 16, 17, 18, 19, 20, 21, 22, 23, 64, 65, 66, 67, 68, \\69, 70, 71, 72, 73, 74, 75, 76, 77, 78, 79, 80, 81, 82, 83, 84, 85, 86, 87, 88, 89, \\90, 91, 92, 93, 94, 95, 160, 161, 162, 163, 164, 165, 166, 167, 168, 169, 170, 171, 172, 173, 174, \\175, 176, 177, 178, 179, 180, 181, 182, 183, 184, 185, 186, 187, 188, 189, 190, 191, 224, 225, 226, 227, \\228, 229, 230, 231, 232, 233, 234, 235, 236, 237, 238, 239, 240, 241, 242, 243, 244, 245, 246, 247, 248, \\249, 250, 251, 252, 253, 254, 255}& Sing\\\hline
				3&\makecell{1, 4, 5, 128, 129, 130, 131, 132, 133, 134, 135, 136, 137, 138, 139, 140, 141, 142, 143, 144, 145, \\146, 147, 148, 149, 150, 151, 152, 153, 154, 155, 156, 157, 158, 159, 160, 161, 162, 163, 164, 165, 166, \\167, 168, 169, 170, 171, 172, 173, 174, 175, 176, 177, 178, 179, 180, 181, 182, 183, 184, 185, 186, 187, \\188, 189, 190, 191, 192, 193, 194, 195, 196, 197, 198, 199, 200, 201, 202, 203, 204, 205, 206, 207, 208, \\209, 210, 211, 212, 213, 214, 215, 216, 217, 218, 219, 220, 221, 222, 223, 224, 225, 226, 227, 228, 229, \\230, 231, 232, 233, 234, 235, 236, 237, 238, 239, 240, 241, 242, 243, 244, 245, 246, 247, 248, 249, 250, \\251, 252, 253, 254, 255}&Sing \\\hline
				4&\makecell{14, 15, 64, 65, 129, 130, 131, 133, 134, 135, 136, 137, 138, 139, 140, 141, 142, 143, 144, 145, 146, \\147, 148, 149, 150, 151, 152, 153, 154, 155, 156, 157, 158, 159, 160, 161, 162, 163, 164, 165, 166, 167, \\168, 169, 170, 171, 172, 173, 174, 175, 176, 177, 178, 179, 180, 181, 182, 183, 184, 185, 186, 187, 188, \\189, 190, 191, 192, 193, 194, 195, 196, 197, 198, 199, 200, 201, 202, 203, 204, 205, 206, 207, 208, 209, \\210, 211, 212, 213, 214, 215, 216, 217, 218, 219, 220, 221, 222, 223, 224, 225, 226, 227, 228, 229, 230, \\231, 232, 233, 234, 235, 236, 237, 238, 239, 240, 241, 242, 243, 244, 245, 246, 247, 248, 249, 250, 251, \\252, 253, 254, 255 }&Sing \\\hline
				5&\makecell{
					64, 162, 163, 166, 167, 168, 169, 170, 171, 172, 173, 174, 175, 176, 177, 178, 179, 180, 181, 182, 183, \\184, 185, 186, 187, 188, 189, 190, 191, 224, 225, 226, 227, 228, 229, 230, 231, 232, 233, 234, 235, 236, \\237, 238, 239, 240, 241, 242, 243, 244, 245, 246, 247, 248, 249, 250, 251, 252, 253, 254, 255
				}& Sing\\\hline
				
				6&\makecell{  137, 138, 139, 142, 143, 146, 147, 152, 153, 154, 155, 156, 157, 158, 159, 162, 163, 166, 167, 169, 170, \\171, 172, 173, 174, 175, 176, 177, 178, 179, 180, 181, 182, 183, 184, 185, 186, 187, 188, 189, 190, 191, \\194, 195, 202, 203, 204, 205, 206, 207, 208, 209, 210, 211, 216, 217, 218, 219, 220, 221, 222, 223, 225, \\226, 227, 229, 230, 231, 233, 234, 235, 236, 237, 238, 239, 240, 241, 242, 243, 244, 245, 246, 247, 248, \\249, 250, 251, 252, 253, 254, 255 
				}&Sing\\\hline
				7&\makecell{ 20, 138, 139, 145, 146, 147, 152, 153, 154, 155, 166, 167, 169, 170, 171, 174, 175, 180, 181, 182, 183, \\184, 185, 186, 187, 188, 189, 190, 191, 194, 195, 201, 202, 203, 204, 205, 206, 207, 208, 209, 210, 211, \\216, 217, 218, 219, 220, 221, 222, 223, 225, 226, 227, 230, 231, 233, 234, 235, 236, 237, 238, 239, 240, \\241, 242, 243, 244, 245, 246, 247, 248, 249, 250, 251, 252, 253, 254, 255
				}& Sing\\\hline
				8&\makecell{20, 32, 145, 146, 147, 165, 166, 167, 169, 170, 171, 172, 173, 174, 175, 180, 181, 182, 183, 184, 185, \\186, 187, 188, 189, 190, 191, 194, 195, 201, 202, 203, 204, 205, 206, 207, 208, 209, 210, 211, 216, 217, \\218, 219, 220, 221, 222, 223, 225, 226, 227, 228, 229, 230, 231, 233, 234, 235, 236, 237, 238, 239, 240, \\241, 242, 243, 244, 245, 246, 247, 248, 249, 250, 251, 252, 253, 254, 255
				}& Sing\\\hline
				9&\makecell{129, 131, 145, 146, 147, 161, 162, 163, 164, 165, 166, 167, 170, 171, 174, 175, 176, 177, 178, 179, 180, \\181, 182, 183, 184, 185, 186, 187, 188, 189, 190, 191, 194, 195, 196, 197, 198, 199, 202, 203, 204, 205, \\206, 207, 208, 209, 210, 211, 212, 213, 214, 215, 216, 217, 218, 219, 220, 221, 222, 223, 224, 225, 226, \\227, 228, 229, 230, 231, 232, 233, 234, 235, 236, 237, 238, 239, 240, 241, 242, 243, 244, 245, 246, 247, \\248, 249, 250, 251, 252, 253, 254, 255
				}&Sing\\\hline
				10&\makecell{32, 84, 85, 147, 150, 151, 170, 171, 174, 175, 184, 185, 186, 187, 188, 189, 190, 191, 226, 227, 230, \\231, 234, 235, 236, 237, 238, 239, 240, 241, 242, 243, 244, 245, 246, 247, 248, 249, 250, 251, 252, 253, \\254, 255 
				}&Sing\\\hline
				11&\makecell{170, 171, 174, 175, 184, 185, 186, 187, 188, 189, 190, 191, 204, 205, 206, 207, 220, 221, 222, 223, 226, \\227, 230, 231, 234, 235, 236, 237, 238, 239, 240, 241, 242, 243, 244, 245, 246, 247, 248, 249, 250, 251, \\252, 253, 254, 255}& Sing\\\hline&&\\\hline
				2&\makecell{ 128, 129, 130, 131, 132, 133, 134, 135, 136, 137, 138, 139, 140, 141, 142, 143, 144, 145, 146, 147, 148, \\149, 150, 151, 152, 153, 154, 155, 156, 157, 158, 159, 192, 193, 194, 195, 196, 197, 198, 199, 200, 201, \\202, 203, 204, 205, 206, 207, 208, 209, 210, 211, 212, 213, 214, 215, 216, 217, 218, 219, 220, 221, 222, \\223}& Mult\\\hline
				3&\makecell{8, 9, 10, 11, 12, 13, 14, 15, 24, 25, 28, 29}&Mult \\\hline
				4&\makecell{4, 5, 8, 9, 12, 13, 68, 69, 128, 132}&Mult\\\hline
				5&\makecell{4, 8, 12, 36, 40, 72, 128, 130, 131, 132, 134, 135, 136, 137, 138, 139, 140, 141, 142, 143, 144, \\145, 146, 147, 148, 149, 150, 151, 152, 153, 154, 155, 156, 157, 158, 159, 160, 161, 164, 165, 192, 193, \\194, 195, 196, 197, 198, 199, 200, 201, 202, 203, 204, 205, 206, 207, 208, 209, 210, 211, 212, 213, 214, \\215, 216, 217, 218, 219, 220, 221, 222, 223}&Mult\\\hline					
				6&\makecell{8, 12, 13, 64, 72, 96, 104, 128, 134, 135, 136, 140, 141, 148, 149, 150, 151, 168, 192, 193, 196, \\197, 198, 199, 200, 201, 212, 213, 214, 215, 224, 228, 232
				}&Mult\\\hline
				7&\makecell{4, 8, 12, 13, 36, 64, 72, 128, 132, 133, 136, 140, 141, 142, 143, 150, 151, 156, 157, 158, 159, \\168, 172, 173, 192, 196, 197, 198, 199, 200, 212, 213, 214, 215, 224, 228, 229, 232
				}& Mult\\\hline
				8&\makecell{4, 8, 12, 13, 64, 96, 128, 132, 135, 136, 138, 139, 140, 141, 142, 143, 150, 151, 152, 153, 154, \\155, 158, 159, 168, 192, 193, 196, 197, 200, 212, 213, 214, 215, 224, 232
				}& Mult\\\hline
				9&\makecell{4, 8, 12, 13, 40, 64, 128, 132, 134, 135, 136, 138, 139, 140, 141, 142, 143, 148, 149, 150, 151, \\152, 153, 154, 155, 158, 159, 168, 169, 172, 173, 192, 193, 200
				}&Mult\\\hline
				10&\makecell{8, 12, 13, 40, 64, 72, 96, 128, 136, 138, 139, 140, 141, 142, 143, 154, 155, 158, 159, 168, 169, \\172, 173, 192, 194, 195, 196, 197, 200, 202, 203, 204, 205, 206, 207, 208, 209, 210, 211, 212, 213, 214, \\215, 216, 217, 218, 219, 220, 221, 222, 223, 224, 225, 228, 229, 232, 233
				}&Mult\\\hline
				11&\makecell{8, 12, 13, 40, 64, 72, 96, 128, 136, 138, 139, 140, 141, 142, 143, 152, 153, 154, 155, 158, 159, \\168, 169, 172, 173, 192, 194, 195, 196, 197, 200, 202, 203, 208, 209, 210, 211, 212, 213, 214, 215, 216, \\217, 218, 219, 224, 225, 228, 229, 232, 233 
				}& Mult\\\hline
			\end{tabular}
		\end{adjustbox}
	\end{center}
	\caption{Convergent LCAs and type of attractor for different block size $b$ where $n=11$ }
\end{table}
\newpage
\section{LCA based on Minimization}

The model operates on the principle of minimizing the density of 1s within each block. It achieves this through a minimization function that considers the number of 1s in the neighboring blocks. If the current block has a higher density of 1s than its neighbors, some 1s are removed to reduce the overall density. On the other hand, if the current block already has a lower density of 1s, no changes are made. This approach ensures an optimized distribution of 1s within the blocks, resulting in a minimized density of 1s in the pattern as a whole. 

\begin{figure*}[hbt!]
	\begin{center}
		\scalebox{0.8}{
			\begin{tabular}{ccccc}				
				ECA 253 & ($253,g^{25}$) & ($253,g^{65}$) & ($253,g^{100}$) & ($253,g^{150}$) \\[6pt]
				\includegraphics[width=31mm]{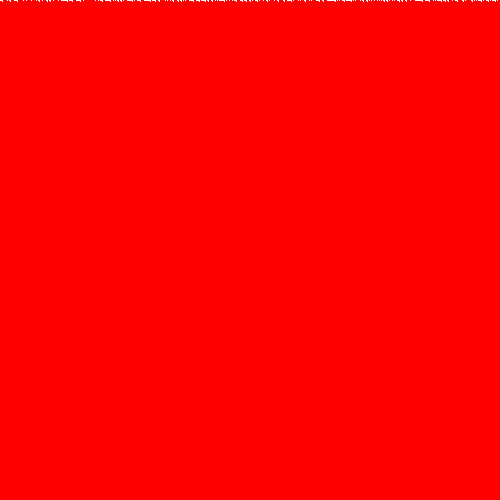} & \includegraphics[width=31mm]{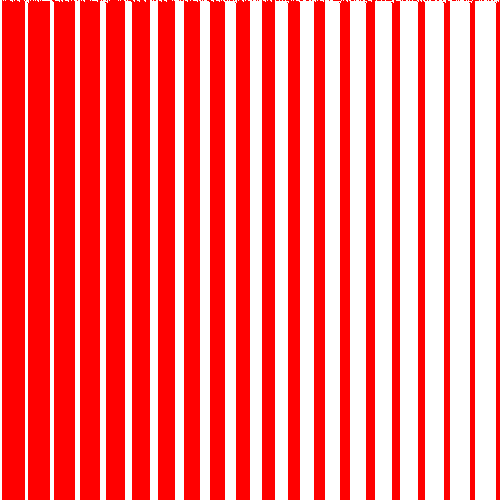} &   \includegraphics[width=31mm]{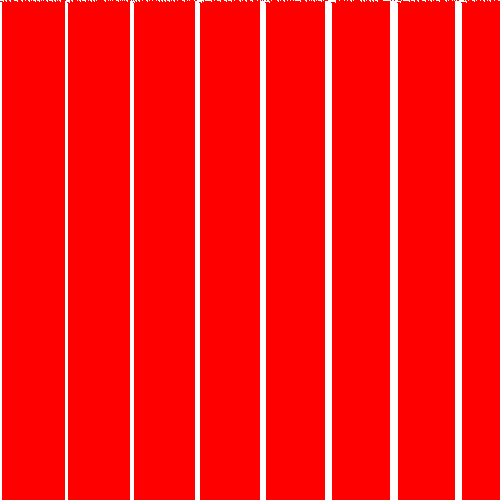} &   \includegraphics[width=31mm]{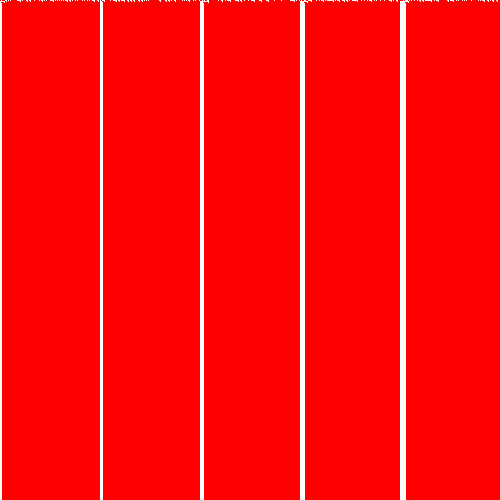} &   \includegraphics[width=31mm]{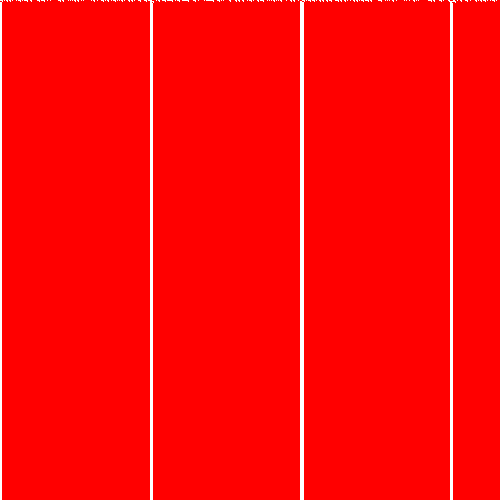} \\
		\end{tabular}}
		\caption{Convergent LCA($253,g^{b}$) dynamics}
		\label{con-min}
	\end{center}
\end{figure*}

Let us discuss an example, consider the LCA($253,g^b$) shown in Figure.~\ref{con-min}. In this case, we select Rule 253 as the rule $f$, which is a Class A rule, where it converges to all-1s configuration. When the dynamics of rule $f$ are influenced by rule $g$, the behavior of the LCA transitions to a state of convergence, resulting in a fixed-point configuration.
Figure.~\ref{con-min} provides visual representations of the fixed-point configurations obtained from the LCA($253,g^b$) for different values of $b$, such as $b=25,65,100,150$. These figures illustrate the stable patterns that emerge as the LCA converges to a fixed point, demonstrating the consistent convergence behavior of the LCA across different parameter values.


In our experiment, similar to maximization, we did not find a comprehensive list of LCAs that would be convergent for all possible values of $n$ and $b$. Hence, we conducted an exhaustive search for convergent LCAs by exploring all combinations of $256 \times (n-1)$ LCAs. 
Similar to maximization, here also we have excluded block size $b=1$. LCAs with a block size of 1 were excluded from the analysis, as we have previously discussed that they tend to converge to an all-0s configuration due to the minimization of 1s, acting as a single attractor.

For the case when $n$ is equal to 11, allowing for block sizes ranging from 2 to 11, there is a total of 2560 LCAs. Among them, 1269 LCAs display convergence behavior. To identify suitable LCAs for the proposed pattern classifier, we eliminate 865 LCAs which are single fixed point attractors. This resulted in a set of 404 candidate LCAs which were eligible candidates. Out of these candidates, 148 LCAs exhibit maximum efficiency and are included as entries in the table (see Table~\ref{monkset2}). These selected LCAs hold promise as potential components of the pattern classifier, showcasing their potential for effective pattern classification tasks.

\subsection{Design of pattern classifier}


	Let us assume for the real-time dataset \emph{Monk-2}, where the size of CA is 11. We have considered a $11$-cell convergent LCA($150,g^{5}$) which has $4$ fixed-point attractors. The attractors are $00000000000$, $00000000001$, $00001000001$ and $00101000101$. The fixed point $00000000000$ and $00000000001$ represents Class $1$ because more patterns from the pattern set $P_1$ get converged to this fixed point attractor and fixed point $00001000001$ and $00101000101$ represent Class $2$. A pattern, say $00001010001$ is given, the LCA runs with $00001010001$ as seed. After some time, the CA reaches to a fixed point attractor - $00001000001$. Since $00001000001$ represents Class $2$, class of $00001010001$ is declared as $2$. Hence, this multiple attractor LCA can act as two class pattern classifier.

\subsection{Training phase}
As an illustrative example, let us consider the Monk-$1$ dataset ($11$-bit data) for classification. Let us take the  LCA $(237,g^6)$ as a two-class pattern classifier, with two pattern set $P_1$ and $P_2$ loaded to the  LCA as Class $1$ and Class $2$, respectively. $P_1$ and $P_2$ contain a total of $169$ patterns, out of which $39$ patterns of $P_2$ and $6$ patterns of $P_1$ are wrongly identified as in Class $1$ and Class $2$, respectively. Hence, $124$ patterns are properly classified, which gives training accuracy as $73.373\%$. 

To get the best candidate  LCA, we train all the multiple attractor LCAs on the Monk-$2$ dataset. The result of some of the training accuracy is noted in Table~\ref{monkset2}. We find that the LCA($206,g^{11}$) with training accuracy $76.923\%$, has the highest training accuracy. This LCA is considered our desired classifier. In Table~\ref{monkset2}, we find the best performing $b$ value and corresponding accuracy for each  LCA.

\begin{table}[!htpb]
	\scriptsize
	\centering
	\caption{Accuracies in training utilizing candidate LCAs on the Monk-2 dataset.}
	\label{monkset2}
	\begin{adjustbox}{width=1\columnwidth,center}
		\begin{tabular}{|cccccccccccc|} \hline 
			LCAs & \makecell{ Accuracy\\ (in \%)} & \makecell{ Number of\\ Attractor} & LCAs & \makecell{ Accuracy\\ (in \%)} & \makecell{ Number of\\ Attractor}  & LCAs & \makecell{ Accuracy\\ (in \%)} & \makecell{ Number of\\ Attractor}& LCAs & \makecell{ Accuracy\\ (in \%)} & \makecell{ Number of\\ Attractor}\\\hline
		$(206,g^{11})$ & 76.923 & 65 & $(220,g^{10})$ & 73.373 & 43 & $(237,g^{6})$ & 73.373 & 10 & $(78,g^{11})$ & 72.781 & 53 \\ $(207,g^{6})$ & 71.598 & 16 & $(220,g^{11})$ & 71.598 & 39 & $(92,g^{11})$ & 71.006 & 25 & $(222,g^{9})$ & 71.006 & 25 \\ $(92,g^{10})$ & 70.414 & 25 & $(202,g^{11})$ & 70.414 & 48 & $(207,g^{11})$ & 70.414 & 25 & $(236,g^{10})$ & 70.414 & 44 \\ $(236,g^{11})$ & 70.414 & 38 & $(207,g^{7})$ & 69.231 & 25 & $(221,g^{11})$ & 69.231 & 15 & $(222,g^{5})$ & 69.231 & 15 \\ $(234,g^{11})$ & 69.231 & 33 & $(237,g^{11})$ & 69.231 & 19 & $(79,g^{7})$ & 68.639 & 16 & $(14,g^{11})$ & 68.047 & 14 \\ $(207,g^{8})$ & 68.047 & 24 & $(237,g^{10})$ & 68.047 & 18 & $(202,g^{10})$ & 67.456 & 26 & $(206,g^{7})$ & 67.456 & 17 \\ $(216,g^{11})$ & 67.456 & 21 & $(237,g^{7})$ & 67.456 & 9 & $(76,g^{10})$ & 66.864 & 33 & $(150,g^{5})$ & 66.864 & 4 \\ $(214,g^{9})$ & 66.864 & 3 & $(222,g^{7})$ & 66.864 & 11 & $(234,g^{10})$ & 66.864 & 25 & $(239,g^{6})$ & 66.864 & 6 \\ $(46,g^{11})$ & 66.272 & 11 & $(134,g^{11})$ & 66.272 & 24 & $(142,g^{11})$ & 66.272 & 17 & $(206,g^{8})$ & 66.272 & 22 \\ $(207,g^{10})$ & 66.272 & 22 & $(216,g^{10})$ & 66.272 & 20 & $(222,g^{8})$ & 66.272 & 16 & $(238,g^{11})$ & 66.272 & 18 \\ $(252,g^{10})$ & 66.272 & 11 & $(46,g^{10})$ & 65.68 & 8 & $(79,g^{6})$ & 65.68 & 10 & $(134,g^{10})$ & 65.68 & 16 \\ $(140,g^{10})$ & 65.68 & 10 & $(206,g^{10})$ & 65.68 & 33 & $(220,g^{6})$ & 65.68 & 9 & $(220,g^{7})$ & 65.68 & 6 \\ $(237,g^{8})$ & 65.68 & 10 & $(238,g^{7})$ & 65.68 & 10 & $(12,g^{10})$ & 65.089 & 7 & $(79,g^{8})$ & 65.089 & 10 \\ $(111,g^{3})$ & 65.089 & 2 & $(142,g^{10})$ & 65.089 & 13 & $(156,g^{7})$ & 65.089 & 3 & $(164,g^{10})$ & 65.089 & 12 \\ $(207,g^{5})$ & 65.089 & 13 & $(207,g^{9})$ & 65.089 & 16 & $(218,g^{5})$ & 65.089 & 6 & $(223,g^{5})$ & 65.089 & 6 \\ $(237,g^{5})$ & 65.089 & 5 & $(238,g^{8})$ & 65.089 & 13 & $(238,g^{10})$ & 65.089 & 15 & $(239,g^{7})$ & 65.089 & 5 \\ $(239,g^{8})$ & 65.089 & 5 & $(239,g^{10})$ & 65.089 & 3 & $(239,g^{11})$ & 65.089 & 3 & $(252,g^{11})$ & 65.089 & 11 \\ $(22,g^{5})$ & 64.497 & 3 & $(28,g^{7})$ & 64.497 & 2 & $(78,g^{8})$ & 64.497 & 10 & $(79,g^{3})$ & 64.497 & 2 \\ $(79,g^{10})$ & 64.497 & 13 & $(79,g^{11})$ & 64.497 & 13 & $(92,g^{6})$ & 64.497 & 7 & $(134,g^{8})$ & 64.497 & 2 \\ $(142,g^{9})$ & 64.497 & 7 & $(148,g^{10})$ & 64.497 & 4 & $(166,g^{11})$ & 64.497 & 10 & $(172,g^{10})$ & 64.497 & 8 \\ $(174,g^{10})$ & 64.497 & 9 & $(196,g^{10})$ & 64.497 & 15 & $(202,g^{9})$ & 64.497 & 7 & $(206,g^{9})$ & 64.497 & 11 \\ $(212,g^{10})$ & 64.497 & 4 & $(223,g^{4})$ & 64.497 & 2 & $(232,g^{10})$ & 64.497 & 12 & $(232,g^{11})$ & 64.497 & 12 \\ $(6,g^{11})$ & 63.905 & 15 & $(14,g^{8})$ & 63.905 & 2 & $(20,g^{10})$ & 63.905 & 4 & $(78,g^{7})$ & 63.905 & 12 \\ $(92,g^{7})$ & 63.905 & 6 & $(103,g^{3})$ & 63.905 & 2 & $(148,g^{11})$ & 63.905 & 4 & $(174,g^{9})$ & 63.905 & 5 \\ $(174,g^{11})$ & 63.905 & 11 & $(219,g^{7})$ & 63.905 & 4 & $(222,g^{4})$ & 63.905 & 2 & $(223,g^{7})$ & 63.905 & 6 \\ $(228,g^{10})$ & 63.905 & 13 & $(238,g^{9})$ & 63.905 & 14 & $(6,g^{8})$ & 63.314 & 2 & $(6,g^{9})$ & 63.314 & 6 \\ $(38,g^{11})$ & 63.314 & 8 & $(44,g^{10})$ & 63.314 & 12 & $(74,g^{10})$ & 63.314 & 9 & $(74,g^{11})$ & 63.314 & 13 \\ $(78,g^{10})$ & 63.314 & 27 & $(86,g^{9})$ & 63.314 & 4 & $(100,g^{10})$ & 63.314 & 16 & $(102,g^{11})$ & 63.314 & 10 \\ $(132,g^{10})$ & 63.314 & 8 & $(148,g^{8})$ & 63.314 & 2 & $(158,g^{9})$ & 63.314 & 3 & $(166,g^{10})$ & 63.314 & 8 \\ $(172,g^{11})$ & 63.314 & 3 & $(219,g^{5})$ & 63.314 & 3 & $(220,g^{8})$ & 63.314 & 8 & $(223,g^{8})$ & 63.314 & 6 \\ $(228,g^{11})$ & 63.314 & 5 & $(236,g^{9})$ & 63.314 & 10 & $(239,g^{9})$ & 63.314 & 3 & $(248,g^{6})$ & 63.314 & 4 \\ $(248,g^{8})$ & 63.314 & 6 & $(248,g^{10})$ & 63.314 & 10 & $(6,g^{10})$ & 62.722 & 13 & $(14,g^{9})$ & 62.722 & 6 \\ $(20,g^{2})$ & 62.722 & 2 & $(20,g^{5})$ & 62.722 & 2 & $(22,g^{2})$ & 62.722 & 2 & $(30,g^{5})$ & 62.722 & 4 \\ $(30,g^{9})$ & 62.722 & 3 & $(36,g^{10})$ & 62.722 & 12 & $(38,g^{9})$ & 62.722 & 5 & $(46,g^{9})$ & 62.722 & 5 \\ $(62,g^{5})$ & 62.722 & 3 & $(68,g^{10})$ & 62.722 & 10 & $(70,g^{7})$ & 62.722 & 4 & $(71,g^{3})$ & 62.722 & 2 \\ $(74,g^{7})$ & 62.722 & 2 & $(78,g^{9})$ & 62.722 & 10 & $(79,g^{4})$ & 62.722 & 4 & $(79,g^{9})$ & 62.722 & 7 \\ $(88,g^{7})$ & 62.722 & 2 & $(92,g^{8})$ & 62.722 & 6 & $(94,g^{7})$ & 62.722 & 5 & $(100,g^{11})$ & 62.722 & 4  \\\hline
		\end{tabular}
	\end{adjustbox}
	
\end{table}

\subsection{Testing phase}

Table~\ref{tableMIN} provides detailed information on the LCAs employed in the pattern classification task. It showcases the LCAs selected based on the minimization criterion, along with their corresponding training and testing accuracy for different datasets. This comprehensive table offers insights into the performance and effectiveness of the LCAs in accurately classifying patterns.

\begin{table}[!htpb]
	\scriptsize
	\begin{center}
		\caption{Performance of the proposed classifiers using various datasets.}
		\label{tableMIN}
		\begin{adjustbox}{width=0.8\columnwidth,center}
			\begin{tabular}{|ccccc|} \hline 
			Datasets & \makecell{ LCA\\ Size} & \makecell{Training \\Accuracy} & \makecell{Testing\\ Accuracy} & \makecell{Proposed\\  LCAs}\\ \hline
			Monk-1 & 11 & 85.245 & 77.958 & ($220,g^{11}$)\\
			Monk-2 & 11 & 76.923 & 65.185 & ($206,g^{11}$)\\
			Monk-3 & 11 & 89.344 & 83.564 & ($202,g^{11}$)\\
			Haber man & 9 & 73.469 & 75.641 & ($77,g^4$)\\
			Heart-statlog & 16 & 93.333 & 80.74 & ($206,g^{16}$)\\
			Tic-Tac-Toe & 18 & 98.329 & 86.221 & ($220,g^{18}$)\\
			Hepatitis-1 & 19 & 98.717 & 92.207 & ($207,g^{12}$)\\
			Hepatitis-2 & 22 & 100 & 97.402 & ($207,g^{18}$)\\ \hline
			\end{tabular}
		\end{adjustbox}
	\end{center}
\end{table}

\begin{table}[!htpb]
	\scriptsize
	\begin{center}
		
		\label{convergentLCAmin}
		\begin{adjustbox}{width=0.95\columnwidth,center}
			\begin{tabular}{|c|c|c|} \hline 
				$\#b$&Rule $f$&Attr.\\\hline
				2&\makecell{1, 2, 3, 4, 5, 6, 7, 16, 17, 18, 19, 20, 21, 22, 23, 64, 65, 66, 67, 68, 69, \\70, 71, 72, 73, 74, 75, 76, 77, 78, 79, 80, 81, 82, 83, 84, 85, 86, 87, 88, 89, 90, \\91, 92, 93, 94, 95, 160, 161, 162, 163, 164, 165, 166, 167, 168, 169, 170, 171, 172, 173, 174, 175, \\176, 177, 178, 179, 180, 181, 182, 183, 184, 185, 186, 187, 188, 189, 190, 191, 224, 225, 226, 227, 228, \\229, 230, 231, 232, 233, 234, 235, 236, 237, 238, 239, 240, 241, 242, 243, 244, 245, 246, 247, 248, 249, \\250, 251, 252, 253, 254, 255}& Sing\\\hline
				3&\makecell{2, 4, 6, 8, 10, 12, 14, 16, 18, 20, 22, 24, 26, 28, 30, 32, 34, 36, 38, 40, 42, \\44, 46, 48, 50, 52, 54, 56, 58, 60, 62, 64, 66, 68, 70, 72, 74, 76, 78, 80, 82, 84, \\86, 88, 90, 92, 94, 95, 96, 98, 100, 102, 104, 106, 108, 110, 112, 114, 116, 118, 120, 122, 124, \\126, 127, 128, 130, 132, 134, 136, 138, 140, 142, 144, 146, 148, 150, 152, 154, 156, 158, 160, 162, 164, \\166, 168, 170, 172, 174, 176, 178, 180, 182, 184, 186, 188, 190, 192, 194, 196, 198, 200, 202, 204, 206, \\208, 210, 212, 214, 216, 218, 220, 222, 223, 224, 226, 228, 230, 232, 234, 236, 238, 240, 242, 244, 246, \\248, 250, 252, 254, 255}&Sing \\\hline
				4&\makecell{2, 4, 6, 8, 10, 12, 14, 15, 16, 18, 20, 22, 24, 26, 28, 30, 32, 34, 36, 38, 40, \\42, 44, 46, 48, 50, 52, 54, 56, 58, 60, 62, 64, 66, 68, 70, 72, 74, 76, 78, 80, 82, \\84, 86, 88, 90, 92, 94, 96, 98, 100, 102, 104, 106, 108, 110, 112, 114, 116, 118, 120, 122, 124, \\125, 126, 128, 130, 132, 134, 136, 138, 140, 142, 143, 144, 146, 148, 150, 152, 154, 156, 158, 160, 162, \\164, 166, 168, 170, 172, 174, 176, 178, 180, 182, 184, 186, 188, 190, 192, 194, 196, 198, 200, 202, 204, \\206, 208, 210, 212, 214, 216, 218, 220, 224, 226, 228, 230, 232, 234, 236, 238, 240, 242, 244, 246, 248, \\250, 252, 253, 255 }&Sing \\\hline
				5&\makecell{
					2, 8, 10, 16, 18, 24, 26, 32, 34, 40, 42, 48, 50, 56, 58, 64, 66, 72, 74, 80, 82, \\88, 96, 98, 104, 106, 112, 114, 120, 128, 130, 136, 138, 144, 146, 152, 154, 160, 162, 168, 170, 176, \\178, 184, 186, 192, 194, 200, 202, 208, 210, 216, 224, 226, 232, 234, 240, 242, 248, 253, 255
				}& Sing\\\hline
				
				6&\makecell{ 2, 4, 6, 8, 10, 12, 14, 16, 18, 24, 26, 32, 34, 36, 38, 40, 42, 44, 46, 48, 50, \\52, 54, 56, 58, 60, 64, 66, 68, 70, 72, 74, 76, 80, 82, 88, 96, 98, 100, 102, 104, 106, \\110, 112, 114, 116, 120, 128, 130, 132, 134, 136, 138, 140, 142, 144, 146, 152, 154, 160, 162, 164, 166, \\168, 170, 172, 174, 176, 178, 180, 182, 184, 186, 188, 192, 194, 196, 198, 200, 202, 204, 208, 210, 224, \\226, 228, 230, 240, 242, 244, 255 
				}&Sing\\\hline
				7&\makecell{ 2, 4, 8, 10, 12, 16, 18, 24, 26, 32, 34, 36, 38, 40, 42, 44, 46, 48, 52, 54, 56, \\60, 64, 66, 68, 72, 76, 80, 82, 96, 98, 100, 102, 104, 106, 108, 112, 116, 118, 120, 128, 130, \\132, 136, 138, 140, 144, 146, 152, 154, 160, 162, 164, 166, 168, 170, 172, 174, 176, 180, 182, 184, 188, \\192, 194, 196, 200, 204, 208, 210, 215, 224, 226, 228, 230, 240, 244, 255
				}& Sing\\\hline
				8&\makecell{2, 4, 8, 10, 12, 16, 18, 24, 26, 32, 34, 36, 40, 42, 44, 48, 52, 54, 56, 60, 64, \\66, 68, 72, 74, 76, 80, 82, 88, 90, 96, 98, 100, 104, 106, 108, 112, 116, 118, 120, 128, 130, \\132, 136, 138, 140, 144, 146, 152, 154, 160, 162, 164, 168, 170, 172, 176, 180, 182, 184, 188, 192, 194, \\196, 200, 202, 204, 208, 210, 215, 216, 224, 226, 228, 240, 244, 251, 255
				}& Sing\\\hline
				9&\makecell{2, 4, 8, 10, 12, 16, 18, 20, 24, 26, 28, 32, 34, 36, 40, 42, 44, 48, 50, 52, 54, \\56, 58, 60, 62, 64, 66, 68, 72, 76, 80, 82, 84, 88, 90, 92, 96, 98, 100, 104, 112, 114, \\116, 118, 120, 122, 126, 128, 130, 132, 136, 138, 140, 144, 146, 148, 152, 154, 156, 160, 162, 164, 168, \\170, 172, 176, 178, 180, 182, 184, 186, 188, 192, 194, 196, 200, 204, 208, 210, 212, 216, 218, 220, 224, \\226, 228, 232, 240, 242, 244, 248, 255
				}&Sing\\\hline
				10&\makecell{2, 8, 10, 16, 22, 24, 32, 34, 40, 42, 48, 54, 56, 64, 66, 72, 80, 85, 96, 98, 112, \\128, 130, 136, 138, 144, 150, 152, 160, 162, 168, 170, 176, 184, 192, 194, 200, 208, 213, 224, 226, 240, \\251, 255
				}&Sing\\\hline
				11&\makecell{2, 4, 8, 10, 12, 16, 24, 32, 34, 40, 42, 48, 56, 64, 66, 68, 72, 76, 80, 96, 98, \\112, 128, 130, 132, 136, 138, 140, 144, 152, 160, 162, 168, 170, 176, 184, 192, 194, 196, 200, 204, 208, \\224, 226, 240, 255}& Sing\\\hline&&\\\hline
				2&\makecell{ 128, 129, 130, 131, 132, 133, 134, 135, 136, 137, 138, 139, 140, 141, 142, 143, 144, 145, 146, 147, 148, \\149, 150, 151, 152, 153, 154, 155, 156, 157, 158, 159, 192, 193, 194, 195, 196, 197, 198, 199, 200, 201, \\202, 203, 204, 205, 206, 207, 208, 209, 210, 211, 212, 213, 214, 215, 216, 217, 218, 219, 220, 221, 222, \\223}& Mult\\\hline
				3&\makecell{15, 47, 71, 79, 103, 111, 143, 175, 199, 207, 231, 239}&Mult \\\hline
				4&\makecell{79, 93, 95, 111, 207, 221, 222, 223, 239, 254}&Mult\\\hline
				5&\makecell{4, 6, 12, 14, 20, 22, 28, 30, 36, 38, 44, 46, 52, 54, 60, 62, 68, 70, 76, 78, 84, \\86, 90, 92, 100, 102, 108, 110, 116, 118, 122, 124, 132, 134, 140, 142, 148, 150, 156, 158, 164, 166, \\172, 174, 180, 182, 188, 190, 196, 198, 204, 206, 207, 212, 214, 218, 219, 220, 222, 223, 228, 230, 235, \\236, 237, 238, 239, 244, 246, 250, 252, 254}&Mult\\\hline					
				6&\makecell{20, 22, 28, 30, 78, 79, 84, 86, 92, 108, 124, 148, 150, 156, 158, 206, 207, 212, 214, 216, 220, \\232, 233, 234, 236, 237, 238, 239, 248, 249, 252, 253, 254
				}&Mult\\\hline
				7&\makecell{6, 14, 20, 22, 28, 70, 74, 78, 79, 84, 88, 92, 94, 134, 142, 148, 150, 156, 198, 202, 206, \\207, 212, 216, 219, 220, 222, 223, 232, 234, 236, 237, 238, 239, 248, 252, 253, 254
				}& Mult\\\hline
				8&\makecell{6, 14, 20, 22, 30, 38, 46, 78, 79, 84, 92, 102, 124, 134, 142, 148, 150, 166, 174, 206, 207, \\212, 220, 222, 223, 230, 232, 234, 236, 238, 239, 248, 249, 252, 253, 254
				}& Mult\\\hline
				9&\makecell{6, 14, 22, 30, 38, 46, 74, 78, 79, 86, 102, 106, 124, 134, 142, 150, 158, 166, 174, 202, 206, \\207, 214, 222, 223, 230, 234, 235, 236, 238, 239, 252, 253, 254
				}&Mult\\\hline
				10&\makecell{4, 6, 12, 14, 20, 36, 38, 44, 46, 52, 60, 68, 74, 76, 78, 79, 84, 88, 92, 100, 104, \\106, 116, 120, 132, 134, 140, 142, 148, 164, 166, 172, 174, 180, 188, 196, 202, 204, 206, 207, 212, 216, \\220, 228, 232, 234, 235, 236, 237, 238, 239, 244, 248, 249, 252, 253, 254
				}&Mult\\\hline
				11&\makecell{6, 14, 20, 36, 38, 44, 46, 52, 60, 74, 78, 79, 84, 88, 92, 100, 102, 104, 106, 116, 120, \\134, 142, 148, 164, 166, 172, 174, 180, 188, 202, 206, 207, 212, 216, 220, 228, 230, 232, 234, 235, 236, \\237, 238, 239, 244, 248, 249, 252, 253, 254
				}& Mult\\\hline
			\end{tabular}
		\end{adjustbox}
	\end{center}
	\caption{Convergent LCAs and type of attractor for different block size $b$ where $n=11$ }
\end{table}

\section{LCA based on ECA with modified neighborhood}

In this model, each cell applies Elementary Cellular Automaton (ECA) rules based on its adjacent neighbors in layer 0. Additionally, the cell considers its extended neighbors in layer 1 to determine the next generation. This is achieved through a blocking mechanism, where the cell's left neighbor and right neighbor cells are positioned in the left block and right block, respectively. By incorporating information from both layer 0 and layer 1, the model enhances the rule application process and influences the cell's behavior in the next generation.
\vspace{-2em}
\begin{figure}[hbt!]
	\subfloat[]{
		\begin{minipage}[c][1\width]{
				0.40\textwidth}
			\label{von}
			\centering
			\includegraphics[width=1\textwidth]{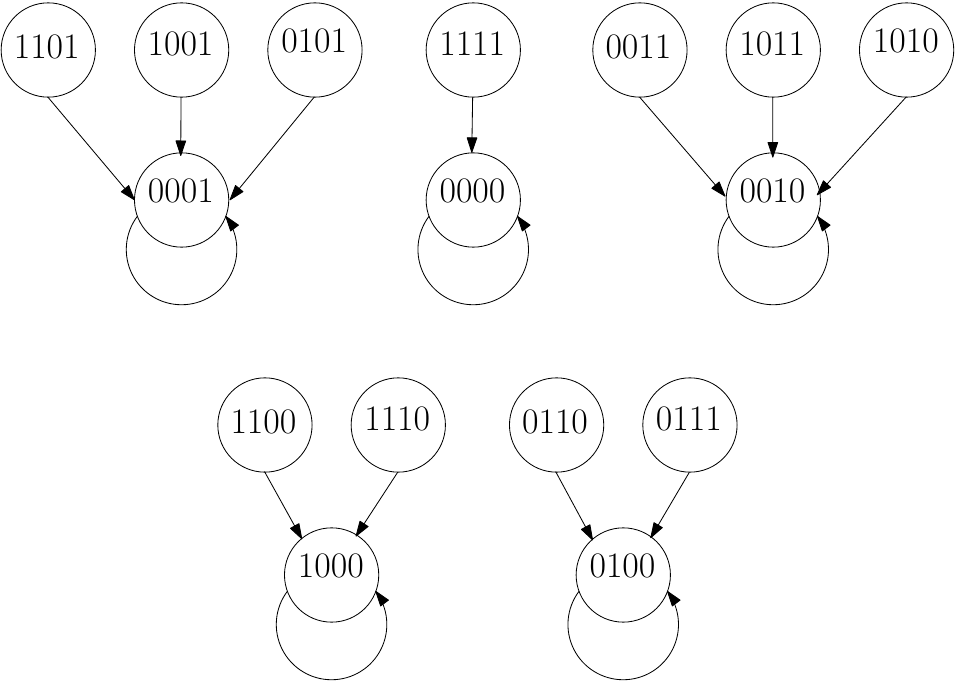}
	\end{minipage}}
	\hfill 	
	\subfloat[]{
		\begin{minipage}[c][1\width]{
				0.48\textwidth}
			\label{moor}
			\centering
			\includegraphics[width=1\textwidth]{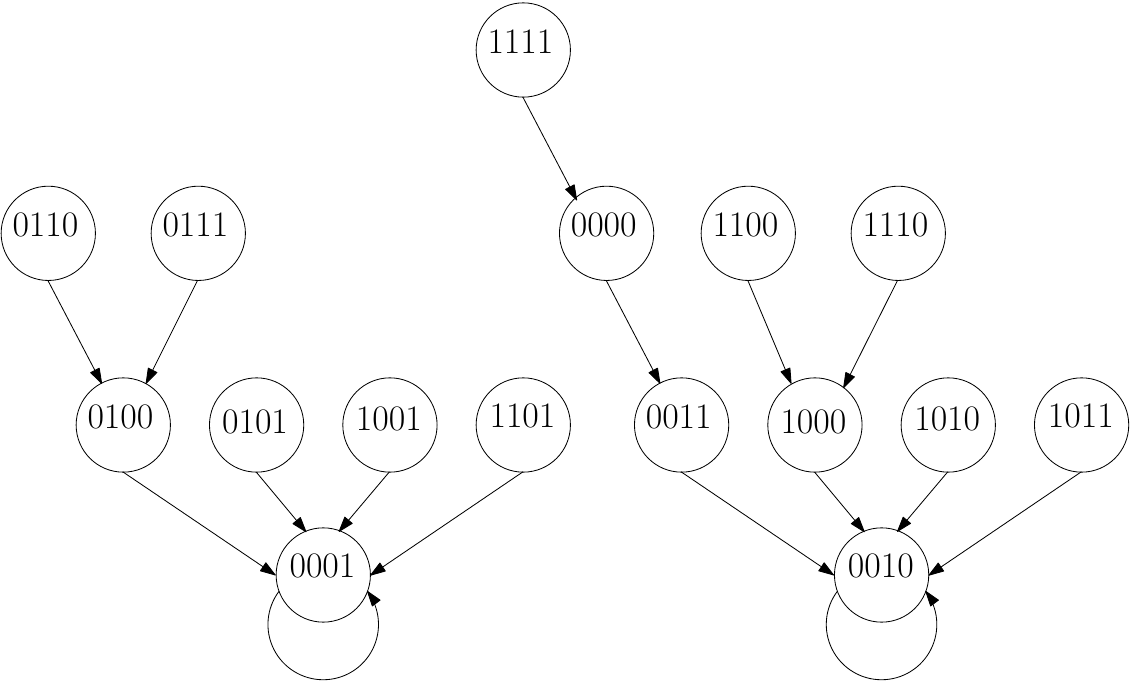}
	\end{minipage}}
	\caption{Transition diagram of LCAs. (a) LCA(12,$76^2$); (b) LCA(13,$76^2$)}
\label{attractor}
\end{figure}

\begin{figure*}[hbt!]
	\begin{center}
		\scalebox{0.8}{
			\begin{tabular}{ccccc}				
				ECA 23 & ECA 45 & ($23,45^{20}$) & ($23,45^{50}$) & ($23,45^{100}$) \\[6pt]
				\includegraphics[width=31mm]{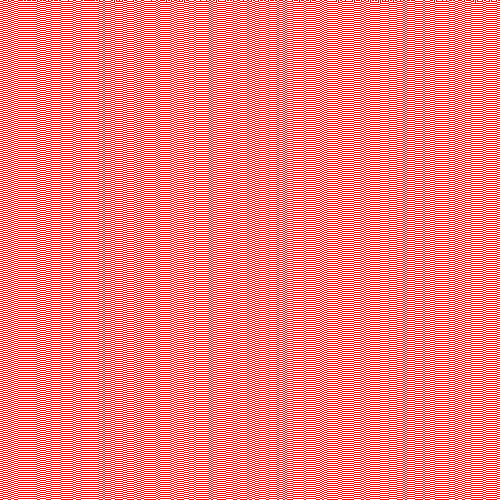} & \includegraphics[width=31mm]{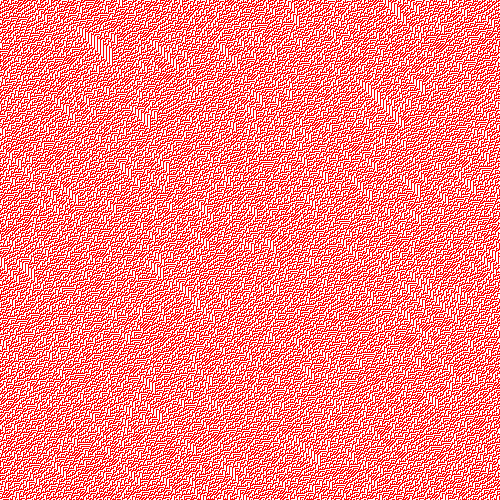} &   \includegraphics[width=31mm]{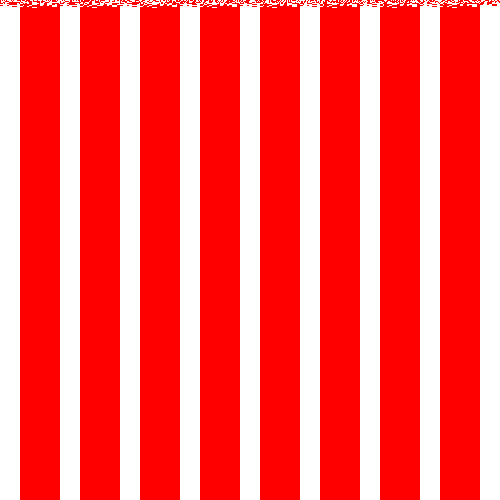} &   \includegraphics[width=31mm]{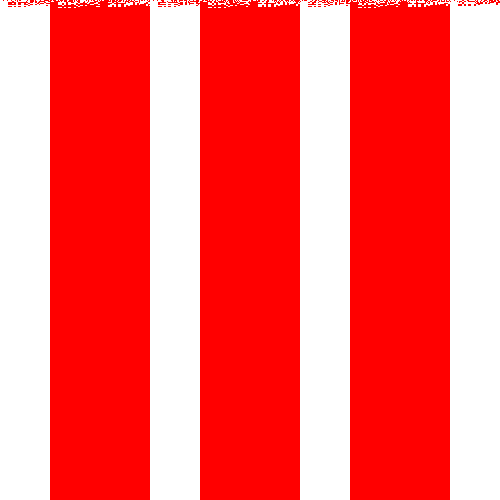} &   \includegraphics[width=31mm]{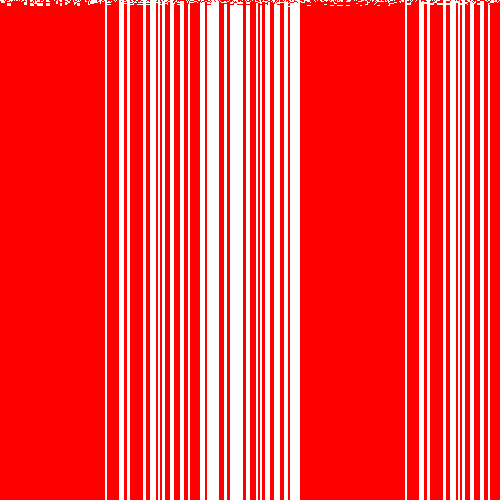} \\
		\end{tabular}}
		\caption{Convergent LCA($23,45^{b}$) dynamics}
		\label{con-eca}
	\end{center}
\end{figure*}

As an example, consider the LCA($23,45^b$) shown in Figure.~\ref{con-eca}. In this case, we select Rule 23 as the rule $f$, which is a periodic rule, and Rule 45 as the rule $g$, which is a chaotic rule. When the dynamics of rule $f$ are influenced by rule $g$, the behavior of the LCA transitions to a state of convergence, resulting in a fixed-point configuration.
Figure.~\ref{con-eca} provides visual representations of the fixed-point configurations obtained from the LCA($23,45^b$) for different values of $b$, such as $b=20, 50, 100$.


Let us consider $n=11$ for which the possible block size are $1$ and $11$. Hence, out of $131072$ LCAs, $60898$ LCAs are convergent LCAs. We have excluded $46211$ LCAs that are block sensitive such as LCAs which show \emph{phase} transition or \emph{class} transition dynamics and the remaining $14687$ LCAs are convergent LCA for $n=11$ which are block insensitive.

We found 6020 LCAs which are convergent irrespective of cell size ($n$) and block size ($b$). After we exclude the LCAs which are associated with single fixed point attractor and a large number of attractors from these LCAs, we have a few of $840$ LCAs that are the candidate of the proposed pattern classifier (see Table~\ref{monk3}, having 148 out of 840 LCAs as entries that are showing maximum efficiencies).

\subsection{Design of pattern classifier}


	Let us assume for the real-time dataset \emph{Monk-3}, where the size of CA is 11. We have considered a $11$-cell convergent LCA($128,246^{1}$) which has $8$ fixed-point attractors. The attractors are $00000000000$, $00000001110$, $00000111110$, $00001111110$, $00111111110$, $01111111110$, $11111111110$ and $11111111111$. The fixed points $00000000000$, $00001111110$, $01111111110$ and $11111111110$ represent Class $1$ because more patterns from the pattern set $P_1$ gets converged to these fixed point attractor and the rest of fixed points $00000001110$, $00000111110$, $00111111110$ and $11111111111$ represents Class $2$. A pattern, say $11101100111$ is given, the LCA runs with $11101100111$ as seed. After some time, the CA reaches to a homogeneous state $11111111111$. Since $11111111111$ represents Class $2$, class of $11101100111$ is declared as $2$. Hence, this multiple attractor LCA can act as two class pattern classifier.

\subsection{Training phase}
let us consider the Monk-$3$ dataset ($11$-bit data) for classification. Let us take the  LCA $(236,232^1)$ as a two-class pattern classifier, with two pattern set $P_1$ and $P_2$ loaded to the  LCA as Class $1$ and Class $2$, respectively. $P_1$ and $P_2$ contain a total of $122$ patterns, out of which $4$ patterns of $P_2$ and $7$ patterns of $P_1$ are wrongly identified as in Class $1$ and Class $2$, respectively. Hence, $111$ patterns are properly classified, which gives training accuracy as $90.984\%$. 

To get the best candidate  LCA, we train all the multiple attractor LCAs on the Monk-$3$ dataset. The result of some of the training accuracy is noted in Table~\ref{monk3}. We find that the LCA($192,202^{1}$) with training accuracy $98.361\%$, has the highest training accuracy. This LCA is considered our desired classifier. In Table~\ref{monk3}, we find the best performing $b$ value and corresponding accuracy for each  LCA. Note that possible block sizes for the dataset are 1 and 11 only.

\begin{table}[!htpb]
	\scriptsize
	\centering
	\caption{Accuracies in training utilizing candidate LCAs on the Monk-3 dataset.}
	\label{monk3}
	\begin{adjustbox}{width=1\columnwidth,center}
		\begin{tabular}{|cccccccccccc|} \hline 
			LCAs & \makecell{ Accuracy\\ (in \%)} & \makecell{ Number of\\ Attractor} & LCAs & \makecell{ Accuracy\\ (in \%)} & \makecell{ Number of\\ Attractor}  & LCAs & \makecell{ Accuracy\\ (in \%)} & \makecell{ Number of\\ Attractor}& LCAs & \makecell{ Accuracy\\ (in \%)} & \makecell{ Number of\\ Attractor}\\\hline
		$(192,202^{1})$ & 98.361 & 73 &	$(200,200^{1})$ & 98.361 & 79 &  $(192,234^{1})$ & 98.361 & 73 & $(196,234^{1})$ & 98.361 & 89 \\ $(192,238^{1})$ & 98.361 & 73 & $(196,238^{1})$ & 98.361 & 94 & $(200,128^{11})$ & 98.361 & 79 & $(200,132^{11})$ & 98.361 & 79 \\ $(200,136^{11})$ & 98.361 & 79 & $(200,140^{11})$ & 98.361 & 79 & $(200,144^{11})$ & 98.361 & 79 & $(200,160^{11})$ & 98.361 & 79 \\ $(200,164^{11})$ & 98.361 & 79 & $(200,168^{11})$ & 98.361 & 79 & $(200,172^{11})$ & 98.361 & 79 & $(200,176^{11})$ & 98.361 & 79 \\ $(200,184^{11})$ & 98.361 & 79 & $(200,192^{11})$ & 98.361 & 79 & $(200,196^{11})$ & 98.361 & 79 & $(200,200^{11})$ & 98.361 & 79 \\ $(200,202^{11})$ & 98.361 & 79 & $(200,206^{11})$ & 98.361 & 79 & $(200,208^{11})$ & 98.361 & 79 & $(200,210^{11})$ & 98.361 & 79 \\ $(200,216^{11})$ & 98.361 & 79 & $(200,218^{11})$ & 98.361 & 79 & $(200,220^{11})$ & 98.361 & 79 & $(200,222^{11})$ & 98.361 & 79 \\ $(200,224^{11})$ & 98.361 & 79 & $(200,226^{11})$ & 98.361 & 79 & $(200,228^{11})$ & 98.361 & 79 & $(200,232^{11})$ & 98.361 & 79 \\ $(200,234^{11})$ & 98.361 & 79 & $(200,236^{11})$ & 98.361 & 79 & $(200,238^{11})$ & 98.361 & 79 & $(200,240^{11})$ & 98.361 & 79 \\ $(200,242^{11})$ & 98.361 & 79 & $(200,244^{11})$ & 98.361 & 79 & $(200,246^{11})$ & 98.361 & 79 & $(200,248^{11})$ & 98.361 & 79 \\ $(200,250^{11})$ & 98.361 & 79 & $(200,252^{11})$ & 98.361 & 79 & $(200,254^{11})$ & 98.361 & 79 & $(196,226^{1})$ & 97.541 & 85 \\ $(196,230^{1})$ & 97.541 & 93 & $(200,232^{1})$ & 96.721 & 66 & $(200,236^{1})$ & 96.721 & 66 & $(76,128^{11})$ & 96.721 & 113 \\ $(76,132^{11})$ & 96.721 & 113 & $(76,136^{11})$ & 96.721 & 113 & $(76,140^{11})$ & 96.721 & 113 & $(76,160^{11})$ & 96.721 & 113 \\ $(76,164^{11})$ & 96.721 & 113 & $(76,168^{11})$ & 96.721 & 113 & $(76,172^{11})$ & 96.721 & 113 & $(76,180^{11})$ & 96.721 & 113 \\ $(76,192^{11})$ & 96.721 & 113 & $(76,196^{11})$ & 96.721 & 113 & $(76,200^{11})$ & 96.721 & 113 & $(76,224^{11})$ & 96.721 & 113 \\ $(76,228^{11})$ & 96.721 & 113 & $(76,232^{11})$ & 96.721 & 113 & $(76,236^{11})$ & 96.721 & 113 & $(76,244^{11})$ & 96.721 & 113 \\ $(76,246^{11})$ & 96.721 & 113 & $(76,248^{11})$ & 96.721 & 113 & $(76,252^{11})$ & 96.721 & 113 & $(76,254^{11})$ & 96.721 & 113 \\ $(192,194^{1})$ & 95.902 & 69 & $(192,226^{1})$ & 95.902 & 69 & $(205,128^{11})$ & 95.902 & 103 & $(205,144^{11})$ & 95.902 & 103 \\ $(205,192^{11})$ & 95.902 & 103 & $(205,200^{11})$ & 95.902 & 103 & $(205,202^{11})$ & 95.902 & 103 & $(205,206^{11})$ & 95.902 & 103 \\ $(205,208^{11})$ & 95.902 & 103 & $(205,210^{11})$ & 95.902 & 103 & $(205,216^{11})$ & 95.902 & 103 & $(205,218^{11})$ & 95.902 & 103 \\ $(205,220^{11})$ & 95.902 & 103 & $(205,222^{11})$ & 95.902 & 103 & $(205,224^{11})$ & 95.902 & 103 & $(205,232^{11})$ & 95.902 & 103 \\ $(205,234^{11})$ & 95.902 & 103 & $(205,236^{11})$ & 95.902 & 103 & $(205,238^{11})$ & 95.902 & 103 & $(205,248^{11})$ & 95.902 & 103 \\ $(205,250^{11})$ & 95.902 & 103 & $(205,252^{11})$ & 95.902 & 103 & $(205,254^{11})$ & 95.902 & 103 & $(220,136^{1})$ & 92.623 & 89 \\ $(220,168^{1})$ & 92.623 & 89 & $(252,136^{1})$ & 91.803 & 72 & $(220,152^{1})$ & 91.803 & 94 & $(252,168^{1})$ & 91.803 & 72 \\ $(252,172^{1})$ & 91.803 & 72 & $(76,196^{1})$ & 91.803 & 85 & $(205,206^{1})$ & 91.803 & 83 & $(236,236^{1})$ & 91.803 & 80 \\ $(236,128^{11})$ & 91.803 & 80 & $(236,132^{11})$ & 91.803 & 80 & $(236,136^{11})$ & 91.803 & 80 & $(236,140^{11})$ & 91.803 & 80 \\ $(236,144^{11})$ & 91.803 & 80 & $(236,160^{11})$ & 91.803 & 80 & $(236,164^{11})$ & 91.803 & 80 & $(236,168^{11})$ & 91.803 & 80 \\ $(236,172^{11})$ & 91.803 & 80 & $(236,176^{11})$ & 91.803 & 80 & $(236,180^{11})$ & 91.803 & 80 & $(236,184^{11})$ & 91.803 & 80 \\ $(236,192^{11})$ & 91.803 & 80 & $(236,196^{11})$ & 91.803 & 80 & $(236,200^{11})$ & 91.803 & 80 & $(236,202^{11})$ & 91.803 & 80 \\ $(236,206^{11})$ & 91.803 & 80 & $(236,208^{11})$ & 91.803 & 80 & $(236,216^{11})$ & 91.803 & 80 & $(236,218^{11})$ & 91.803 & 80 \\ $(236,220^{11})$ & 91.803 & 80 & $(236,222^{11})$ & 91.803 & 80 & $(236,224^{11})$ & 91.803 & 80 & $(236,226^{11})$ & 91.803 & 80 \\ $(236,228^{11})$ & 91.803 & 80 & $(236,232^{11})$ & 91.803 & 80 & $(236,234^{11})$ & 91.803 & 80 & $(236,236^{11})$ & 91.803 & 80 \\ $(236,238^{11})$ & 91.803 & 80 & $(236,240^{11})$ & 91.803 & 80 & $(236,242^{11})$ & 91.803 & 80 & $(236,244^{11})$ & 91.803 & 80 \\ $(236,246^{11})$ & 91.803 & 80 & $(236,248^{11})$ & 91.803 & 80 & $(236,250^{11})$ & 91.803 & 80 & $(236,252^{11})$ & 91.803 & 80 \\ $(236,254^{11})$ & 91.803 & 80 & $(76,132^{1})$ & 90.984 & 80 & $(76,140^{1})$ & 90.984 & 86 & $(220,184^{1})$ & 90.984 & 96 \\ $(236,200^{1})$ & 90.984 & 64 & $(236,232^{1})$ & 90.984 & 64 & $(78,132^{1})$ & 90.164 & 69 & $(205,200^{1})$ & 89.344 & 84 \\ $(206,220^{1})$ & 89.344 & 62 & $(206,222^{1})$ & 89.344 & 62 & $(76,236^{1})$ & 89.344 & 87 & $(197,234^{1})$ & 88.525 & 64  \\\hline
		\end{tabular}
	\end{adjustbox}
	\vspace{-1 em}
\end{table}

\subsection{Testing phase}

Table~\ref{tableECA} provides an overview of the LCAs based using a modified neighborhood approach, along with their corresponding training and testing accuracy for different datasets. The table presents valuable information on the performance of these LCAs in terms of their classification accuracy when trained and tested on various datasets.
\begin{table}[!htpb]
	\scriptsize
	\begin{center}
		\caption{Performance of the proposed classifiers using various datasets.}
		\label{tableECA}
		\begin{adjustbox}{width=0.8\columnwidth,center}
			\begin{tabular}{|ccccc|} \hline 
			Datasets & \makecell{ LCA\\ Size} & \makecell{Training \\Accuracy} & \makecell{Testing\\ Accuracy} & \makecell{Proposed\\  LCAs}\\ \hline
			Monk-1 & 11 & 97.54 & 84.686 & ($196,238^{1}$)\\
			Monk-2 & 11 & 95.266 & 81.481 & ($196,238^1$)\\
			Monk-3 & 11 & 98.361 & 97.685 & ($192,202^{1}$)\\
			Haber man & 9 & 75.182 & 75.641 & ($76,168^3$)\\
			Heart-statlog & 16 & 95.556 & 88.888 & ($200,132^{8}$)\\
			Tic-Tac-Toe & 18 & 100 & 100 & ($12,76^{3}$)\\
			Hepatitis-1 & 19 & 98.717 & 93.506 & ($236,12^1$)\\ \hline
			\end{tabular}
		\end{adjustbox}
	\end{center}
\end{table}

\begin{table}[htbp]
	\begin{center}
		
		\begin{adjustbox}{width=\columnwidth,center}
			\begin{tabular}{|ccc|cccc|} \hline 
                \multicolumn{3}{|c|}{ } & \multicolumn{4}{c|}{Efficiency of proposed LCA models} \\
				Datasets & \makecell{Algorithm} & \makecell{Efficiency in \%} &\makecell {Averaging} &\makecell {Maximization} &\makecell {Minimization} &\makecell {Modified ECA \\neighborhood} \\
				\hline
				Monk-1 & Bayesian & 99.9 & 76.798 & 70.765 & 77.958 & 84.686\\
				&C4.5 & 100 & LCA($236,g^{10}$) & LCA($140,g^{11}$) & LCA($220,g^{11}$) & LCA($196,238^{1}$)\\
				&TCC & 100 &&&&\\
				&MTSC & 98.65 &&&&\\
				&MLP & 100 &&&&\\
				&Traditional CA & 61.111 &&&&\\
				&Asynchronous CA & 81.519 &&&&\\\hline				
				Monk-2 & Bayesian & 69.4 & 72.222 & 68.981 & 65.185 & 81.481\\
				&C4.5 & 66.2 & LCA($76,g^{7}$) & LCA($140,g^{9}$) & LCA($206,g^{11}$) & LCA($196,238^{1}$)\\
				&TCC & 78.16 &&&&\\
				&MTSC & 77.32 &&&&\\
				&MLP & 75.16 &&&&\\
				&Traditional CA & 67.129 &&&&\\
				&Asynchronous CA & 73.410 &&&&\\\hline				
				Monk-3 & Bayesian & 92.12 & 95.833 & 72.222 & 83.564 & 97.685 \\
				&C4.5 & 96.3 & LCA($200,g^{9}$) & LCA($140,g^{11}$) & LCA($202,g^{11}$) & LCA($192,202^{1}$)\\
				&TCC & 76.58 &&&&\\
				&MTSC & 97.17 &&&&\\
				&MLP & 98.10 &&&&\\
				&Traditional CA & 80.645 &&&&\\
				&Asynchronous CA & 83.749 &&&&\\\hline
				Haber-man & Traditional CA  & 73.499 & 73.717 & 75.641 & 75.641 & 75.641\\
				&Asynchronous CA & 77.493 & LCA($200,g^{9}$) & LCA($72,g^{5}$) & LCA($77,g^{4}$) & LCA($76,168^{3}$)\\\hline
				Tic-Tac-Toe & Sparce grid  & 98.33 & 100 & 98.121 & 86.221 & 100 \\
				&ASVM & 70.00 & LCA($236,g^{17}$) & LCA($12,g^{17}$) & LCA($220,g^{18}$) & LCA($12,76^{3}$)\\
				&LSVM & 93.330 &&&&\\
				&Traditional CA & 93.330 &&&&\\
				&Asynchronous CA & 99.721 &&&&\\\hline
				Heart-statlog&Bayesian&82.56 & 82.222 & 85.925 & 80.74 & 88.888\\
				&C4.5&80.59 & LCA($206,g^{9}$) & LCA($200,g^{15}$) & LCA($206,g^{16}$) & LCA($200,132^{8}$)\\
				&Logit-boost DS &82.22&&&&\\ \hline
				Hepatitis-1&Bayesian&84.18 & 97.402 & 93.506 & 92.207 & 93.506\\
				&C4.5&82.38&LCA($206,g^{8}$) & LCA($12,g^{17}$) & LCA($207,g^{12}$) & LCA($236,12^{1}$)\\
				&Logit-boost DS &81.58&&&&\\ \hline
				Hepatitis-2&-&- & 98.701 & 100 & 97.402 & - \\
				&&&LCA($207,g^{5}$) & LCA($12,g^{21}$) & LCA($207,g^{18}$) & \\
				
				\hline
			\end{tabular}
		\end{adjustbox}
		\caption{Classification accuracy compared to other well-known classifiers}
		\label{table6}
	\end{center}
	
\end{table}

\section{Comparison}
In order to evaluate the performance and efficiency of the proposed two-class pattern classifier, a comprehensive analysis was conducted using a diverse set of eight datasets. These datasets included Monk-1, Monk-2, Monk-3, Haber-man, Heart-statlog, Tic-Tac-Toe, Hepatitis-1, and Hepatitis-2. Prior to conducting the analysis, the datasets were carefully preprocessed to ensure that the input features were appropriately transformed and aligned with the requirements of the classifier. This preprocessing step was crucial to guarantee accurate and reliable results in the subsequent evaluation of the classifier's performance on each dataset. 	

To assess the effectiveness of the proposed classifier, a comparative analysis was conducted, evaluating its classification accuracy against various established standard algorithms. These algorithms included Bayesian, C4.5 (a decision tree algorithm) \cite{Salzberg1994}, MLP (Multilayer Perceptron), TCC, MTSC, ASVM, LSVM, Sparse grid, Traditional CA \cite{DasMNS09} and Asynchronous CA \cite{Sethi2016}. The performance of the proposed classifier was carefully measured and compared with these existing algorithms, providing valuable insights into its capabilities and potential advantages over the established approaches. This comparative evaluation aimed to determine the competitiveness and effectiveness of the proposed classifier in the context of pattern classification tasks.


In Table~\ref{table6}, the performance of our proposed LCA-based classifier is compared to other widely recognized classifiers. The results demonstrate that our proposed LCA-based two-class pattern classifier outperforms traditional CA-based classifiers. Furthermore, the LCA classifier showed remarkable competitiveness and consistently performed better than other well-known classifier algorithms considered in the evaluation. These findings highlight the effectiveness and reliability of the LCA-based classifier as a powerful approach for pattern classification tasks.

\section{Summary}

In this chapter, we developed a two class pattern classifier using LCA (Layered Cellular Automata) based on different models such as averaging, maximization, minimization, modified ECA neighborhood. These LCA models leverage the convergence behavior of cellular automata to capture and classify patterns effectively. Convergent LCAs, which exhibit stable convergence to fixed-point configurations, are particularly relevant in pattern classification tasks. These LCAs possess the ability to reliably capture specific patterns, regardless of the initial configuration and certain parameters.

The training phase involved using patterns from separate datasets to identify the most efficient attractor sets and LCAs. The convergence behavior and accuracy of the LCAs are evaluated during training to determine the best classifier.

In the testing phase, the selected LCAs are evaluated using new pattern sets. The patterns are inputted into the LCAs, and their convergence behavior is observed as they approach specific fixed point attractors. This evaluation phase assesses the classifier's performance in accurately classifying patterns.

The performance of LCA-based classifiers is compared to other established classifier algorithms. The results often demonstrate the superiority of LCA-based classifiers in terms of classification accuracy and competitiveness. The unique capabilities of LCAs, such as capturing complex patterns through local interactions, contribute to their effectiveness in pattern classification.
\chapter{Conclusion}
\label{chapconclusion}
The primary objective of this chapter is to provide a concise overview of the significant contributions made in this thesis regarding the subject of layered cellular automata. Furthermore, it aims to explore potential avenues for future research in this field, which has the capacity to attract scientists from diverse academic disciplines worldwide.

\section{Main Contribution}
The primary aim of our research is to develop an innovative model of computation with the idea to represent the hierarchy in society by introducing an additional layer of computation in traditional cellular automata. We have named this model as layered cellular automaton (LCA). In LCA, the system is divided into two layers, with each layer following its own set of rules. This layered approach allows for more complex and dynamic simulations, enabling the study of intricate systems and phenomena. The first layer, referred to as Layer 0, represents the working of a predefined model such as Elementary Cellular Automaton (ECA) or the Game of Life. The second layer, known as Layer 1, corresponds to the proposed model and introduces an additional level of computation. Various models of LCA have been explored, including those based on counting, where different counting models are used to design the LCA system. These models involve concepts such as averaging, maximization, and minimization to balance or manipulate the density of certain elements in a block based on neighboring blocks. Another approach is the LCA based on ECA with modified neighborhood, where ECA rules are applied to both Layer 0 and Layer 1. Layer 0 follows the traditional ECA rules using adjacent neighboring cells, while Layer 1 considers adjacent neighboring blocks that are a specific distance away from the current cell. The implementation of LCA in the well-known Game of Life cellular automaton has also been explored. In this case, Layer 0 follows the rules of the Game of Life, while Layer 1 incorporates an averaging concept based on von Neumann's neighborhood for updating the current block.

Chapter~\ref{chap2} provides a concise survey of several important topics that are relevant to this research work. It covers cellular automata, elementary cellular automata, non-uniformity in cellular automata, temporally stochastic cellular automata, and artificial life. The purpose of this survey is to establish a solid foundation and understanding of these concepts as they relate to the research conducted.

In Chapter~\ref{chap3}, the focus shifts to the introduction of Layered Cellular Automata (LCA) as a novel model of computation. The chapter explores various models of LCA, including LCAs based on averaging, maximization, minimization, and modified ECA neighborhood. Each of these models is discussed in detail, highlighting their unique characteristics and applications. Additionally, the chapter presents an LCA model based on the famous Game of Life cellular automaton, showcasing how LCA can be implemented in different existing models.

Overall, Chapters~\ref{chap2} and~\ref{chap3} serve as crucial sections in the thesis, providing a comprehensive survey of relevant concepts and introducing the fundamental concepts and models of Layered Cellular Automata. These chapters lay the groundwork for the subsequent chapters, where further analysis and exploration of LCAs and their applications will be conducted.

Chapter~\ref{chap4} delves into the examination of various classes and dynamics exhibited by different Layered Cellular Automata (LCA) models. The chapter begins by identifying that certain LCAs are unaffected by the application of rule $g$, while others undergo dynamic changes in response to rule $g$. Moreover, it is observed that the sensitivity of LCAs to the block size ($b$) also plays a significant role in their behavior. The study reveals that some LCAs remain robust in the face of alterations to $b$, showcasing consistent dynamics. Conversely, LCAs that are sensitive to changes in $b$ exhibit fascinating phenomena such as phase transition and class transition, highlighting the complex and varied nature of these systems. It emphasizes the importance of considering both the influence of rule $g$ and the block size ($b$) in understanding the behavior and evolution of LCAs.

In Chapter~\ref{chap5}, the focus shifts to the identification of convergent Layered Cellular Automata (LCA) models that can be employed in the design of two-class pattern classifiers. Through extensive analysis and experimentation, specific LCAs that converge to fixed points from various initial configurations are identified.
The convergent LCAs demonstrate promising potential for pattern classification tasks. In fact, when compared to existing common algorithms, the proposed design of an LCA-based two-class pattern classifier exhibits competitive performance. This suggests that LCAs have the ability to effectively capture and represent patterns in a manner that rivals traditional approaches. These findings highlight the value and utility of LCAs in tackling pattern recognition challenges and encourage further exploration and refinement of LCA-based classification approaches.

\section{Future Directions}
Based on the findings and contributions of our current work, several intriguing scopes for future research emerge. These potential research initiatives include:
\begin{enumerate}
	\item Design and Analysis of New Interlayer Rules: Investigate the development of new interlayer rules for LCAs. We can use different rules other than ECA and game of life to explore how different interaction mechanisms between Layer 0 and Layer 1 can affect the dynamics and emergent behavior of the system. Analyze the impact of various interlayer rules on pattern formation, stability, and complexity.
	\item Multilayered LCAs: Extend the concept of LCAs beyond two layers and explore the dynamics and properties of multilayered LCAs. Investigate how the addition of more layers affects the behavior and complexity of the system. Analyze the interplay between different layers and the emergence of higher-order patterns. For example, Multilayered CA can simulate weather patterns and climate dynamics. Each layer can represent different atmospheric variables (e.g., temperature, humidity, wind speed), and the interlayer rules can capture the interactions between them, such as energy transfer, cloud formation, and air circulation. LCA can help in understanding the behavior of weather systems, predicting catastrophic events, and studying the impact of climate change.
	\item Theoretical Analysis of LCA Properties: Conducting in-depth theoretical analysis of Layered Cellular Automata can provide valuable insights into their properties and behavior. Exploring mathematical properties, stability analysis, and computational complexity of LCAs can contribute to a deeper understanding of their capabilities and limitations. 
	\item Exploration of LCA Dynamics in Higher Dimensions: While our current work primarily focuses on one and two-dimensional LCA models, future research can extend the analysis to higher-dimensional LCA systems. Investigating the dynamics and behavior of LCAs in three or more dimensions can reveal unique patterns, emergent properties, and intricate dynamics not observed in lower-dimensional models.
 \item Development of Hybrid LCA Models: Hybrid models that combine Layered Cellular Automata with other computational models or machine learning techniques in order to further enhance pattern recognition and classification tasks. Investigating the integration of LCAs with deep learning, neural networks, or genetic algorithms can lead to the development of hybrid models with improved performance.
	\item LCA and Real-World Applications: Apply LCAs to real-world problems in various domains, such as biology, physics, social sciences, and engineering. Investigate how LCAs can be used to model and analyze complex systems in these domains. Explore their potential for solving specific problems and generating insights.
\end{enumerate}




\cleardoublepage
\phantomsection
\addcontentsline{toc}{chapter}{Author's Publications}
\chapter*{Author's Statements}
\label{chap8}

\hspace{0.1in}\textbf{Published Paper}\\
\begin{itemize}
\item Abhishek Dalai and Subrata Paul, ``Layered Cellular Automata and Pattern Classification'' \textit{Second Asian Symposium on Cellular Automata Technology}, 2023\cite{dalai2023layered}.
\end{itemize}

\vspace{0.4in}
\hspace{0.0in}\textbf{Submitted Paper}\\
\begin{itemize}
\item Abhishek Dalai, Subrata Paul and Sukanta Das, ``Layered Cellular Automata and Pattern Classification'' \textit{AUTOMATA 2023}, 2023.
\end{itemize}

\pagestyle{plain}
\bibliographystyle{IEEEtran}

\cleardoublepage
\phantomsection
\addcontentsline{toc}{chapter}{Bibliography}
\bibliography{Thesis}
\end{document}